\documentclass[final,a4paper,usenglish]{lipics-v2019}
\usepackage{times}

\title{Solutions of Word Equations over Partially Commutative Structures}

\titlerunning{Word Equations over Partially Commutative Structures}

\author{Volker Diekert}{University of Stuttgart, Germany}{diekert@fmi.uni-stuttgart.de}{https://orcid.org/0000-0002-5994-3762}{}

\author{Artur Je\.z}{University of Wroc{\l}aw, Poland}{aje@cs.uni.wroc.pl}{https://orcid.org/0000-0003-4321-3105}{}

\author{Manfred Kuf\-leitner}{University of Stuttgart, Germany}{kufleitner@fmi.uni-stuttgart.de}{https://orcid.org/0000-0003-3869-416X}{}

\author{Alexander Thumm}{University of Siegen, Germany}{alexander.thumm@uni-siegen.de}{https://orcid.org/0009-0005-4240-2045}{}

\acknowledgements{The authors thank the anonymous referee for valuable comments and suggestions that helped improve the paper.
They are also grateful to the GAGTA community for its sustained interest in this work over the years.
Finally, they thank the Leibniz Association for its hospitality and support. 
This project was initiated in 2014 during a stay at Schloss Dagstuhl and was brought to completion in 2026 at the Mathematical Research Institute Oberwolfach.}

\authorrunning{V. Diekert, A. Je\.z, M. Kuf\-leitner, A. Thumm}

\Copyright{V. Diekert, A. Je\.z, M. Kuf\-leitner, A. Thumm}



\nolinenumbers
\hideLIPIcs

\usepackage[utf8]{inputenc}
\usepackage{verbatim,wrapfig}

\usepackage{bold-extra}

\usepackage{amsmath,amssymb,bm}
\usepackage{mathtools}

\usepackage{enumerate}
\usepackage{multirow}
\usepackage{xspace}
\usepackage{tikz}
\usetikzlibrary{decorations.pathreplacing}
\usetikzlibrary{automata, positioning, arrows}

\usepackage{xcolor}

\usepackage{orcidlink}

\usepackage{index}
\newcommand{\emphind}[1]{{\emph{\index*{#1}}}}
\makeindex

\usepackage[ruled]{algorithm}
\usepackage[noend]{algpseudocode}

\newtheorem{convention}[theorem]{Convention}
\newtheorem{invariant}[theorem]{Invariant}
\newtheorem{ruli}[theorem]{Rule}

\theoremstyle{plain}
\newtheorem*{raag_theorem}{Theorem}

\newcommand{\geo}[1]{\mathop{\mathrm{Geo}_{\pi}(#1)}}

\newcommand{\li}{\mathop{\mathrm{li}}}

\renewcommand{\implies}{\Rightarrow}
\renewcommand{\ast}{*}
\renewcommand{\colon}{:}
\newcommand{\cent}{\text{\large\textcent}}
\newcommand{\centime}{\text{\textcent}}

\newcommand{\Csize}{2^{\Oh({\abs \Rs}^2)}n}

\renewcommand{\iff}{\mathrel{\Leftrightarrow}}

\newcommand\SNF{step-normal-form\xspace}
\DeclareMathOperator{\snf}{snf}

\DeclareMathOperator{\nf}{\mathrm{nf}}
\DeclareMathOperator{\nfslex}{\mathrm{nf}_{\small{\mathrm{slex}}}}
\newcommand{\Nop}{N_{\mathrm{op}}}

\newcommand\fg{f.g.\xspace}

\newcommand{\wrt}{w.r.t.\xspace}
\newcommand{\inner}{inner\xspace}

\newcommand{\fresgro}{free resource group\xspace}

\newcommand{\Tflat}{$T$-flat\xspace}

\newcommand{\Tarc}{$T$-arc\xspace}

\newcommand{\SSparc}{$S$-$S'$-arc\xspace}
\newcommand{\SSparcs}{$S$-$S'$-arcs\xspace}
\newcommand{\STarc}{$S$-$T$-arc\xspace}
\newcommand{\TSarc}{$T$-$S$-arc\xspace}

\newcommand{\TTarc}{$T$-$T$-arc\xspace}
\newcommand{\STarcs}{$S$-$T$-arcs\xspace}
\newcommand{\TTarcs}{$T$-$T$-arcs\xspace}

\newcommand{\Sarc}{$S$-arc\xspace}

\newcommand{\edtol}{EDT0L\xspace}
\newcommand{\EDTOL}{\mathop{\mathrm{EDT0L}}}

\newcommand{\lds}{, \ldots ,}

\newcommand{\Rs}{\mathfrak{R}}

\newcommand{\pio}{\wh\pi}
\newcommand{\selfie}{selfie\xspace}
\newcommand{\selfies}{\selfie{s}\xspace}

\newcommand{\sus}[1]{\sig^*(#1)}

\newcommand{\IFF}{if and only if\xspace}
\renewcommand{\hom}{homomor\-phism\xspace}

\newcommand{\Endo}{endo\-mor\-phism\xspace}
\newcommand{\Endos}{endo\-mor\-phisms\xspace}

\newcommand{\morph}{mor\-phism\xspace}
\newcommand{\iso}{isomor\-phism\xspace}

\newcommand{\epi}{epimor\-phism\xspace}

\newcommand{\morphs}{mor\-phisms\xspace}

\newcommand{\stepnf}{step normal form\xspace}
\newcommand{\symnf}{symmetric normal form\xspace}

\DeclareMathOperator{\End}{\mathrm{End}}

\DeclareMathOperator{\pos}{\mathrm{pos}}
\newcommand{\pr}[1]{\mathop{\mathrm{pr_{#1}}}}
\newcommand{\Bold}{B_{\mathrm{old}}} 
\newcommand{\Bpos}{B_{\mathrm{pos}}}
\newcommand{\Bneg}{B_{\mathrm{neg}}}

\newcommand{\cXTfl}{\cX_{\mathrm{Tflat}}}
\newcommand{\cXpTfl}{\cX'_{\mathrm{Tflat}}}

\newcommand{\tra}{transition\xspace} 
\newcommand{\tras}{transitions\xspace}
\newcommand{\epstra}{$\eps$-transition\xspace} 
\newcommand{\epstras}{$\eps$-transitions\xspace} 

\newcommand{\ie}{i.e.,\xspace}
\newcommand{\eg}{e.g.\xspace}

\newcommand{\Ip}{In parti\-cu\-lar,\xspace}
\newcommand{\ip}{in parti\-cu\-lar,\xspace}
\newcommand{\invol}{involution\xspace}

\newcommand{\solu}{solu\-tion\xspace}
\newcommand{\solus}{solu\-tions\xspace}
\newcommand{\esolu}{entire solu\-tion\xspace}

\newcommand{\entipa}{entire path\xspace}

\newcommand{\subst}{sub\-sti\-tu\-tion\xspace}
\newcommand{\comp}{com\-pres\-sion\xspace}

\usepackage{prettyref}
\newcommand{\prref}[1]{\prettyref{#1}}
\newrefformat{thm}{Theorem~\ref{#1}}
\newrefformat{lem}{Lemma~\ref{#1}}
\newrefformat{def}{Definition~\ref{#1}}
\newrefformat{claim}{Claim~\ref{#1}}
\newrefformat{conv}{Convention~\ref{#1}}
\newrefformat{cor}{Corollary~\ref{#1}}
\newrefformat{inva}{Invariant~\ref{#1}}
\newrefformat{prop}{Proposition~\ref{#1}}
\newrefformat{sec}{Section~\ref{#1}}
\newrefformat{kap}{Chapter~\ref{#1}}
\newrefformat{ex}{Example~\ref{#1}}
\newrefformat{eq}{Equation~(\ref{#1})}
\newrefformat{rem}{Remark~\ref{#1}}
\newrefformat{fig}{Figure~\ref{#1}}
\newrefformat{par}{Paragraph~\ref{#1}}
\newrefformat{obs}{Observation~\ref{#1}}
\newrefformat{pro}{Property~\ref{#1}}
\newrefformat{ruli}{Rule~\ref{#1}}
\newrefformat{ruls}{Rules~\ref{#1}}

\newcommand{\wt}[1]{\widetilde{#1}}

\newcommand{\wh}[1]{\smash{\widehat{#1}}}
\newcommand{\wideh}[1]{\widehat{#1}}

\newcommand{\arc}[1]{\overset{#1\hspace*{2pt}}\ra}

\newcommand{\darc}{\leftrightarrow}

\newcommand{\set}[2]
{\{#1\mid #2\}}

\newcommand{\os}[1]{\{#1\}}
\newcommand{\sm}{\setminus}
\newcommand{\es}{\emptyset}
\newcommand{\sse}{\subseteq}
\newcommand{\ssne}{\subsetneq}
\newcommand{\nsse}{\not\sse} 
\newcommand{\ssneq}{\subsetneq}
\newcommand{\ssupe}{\supseteq}

\newcommand{\ssupneq}{\varsupsetneq}

\newcommand{\abs}[1]{|{#1}|}
\newcommand{\Abs}[1]{\Vert{#1}\Vert}

\newcommand{\gen}[1]{\left< {#1} \right>}

\newcommand{\B}{\ensuremath{\mathbb{B}}}

\newcommand{\M}{\ensuremath{\mathbb{M}}}

\newcommand{\N}{\ensuremath{\mathbb{N}}}
\newcommand{\Z}{\ensuremath{\mathbb{Z}}}

\newcommand{\F}{\ensuremath{\mathbb{F}}}
\newcommand{\G}{\ensuremath{\mathbb{G}}}

\newcommand{\id}{{\mathrm{id}}}

\newcommand{\PSPACE}{\ensuremath{\mathsf{PSPACE}}}

\newcommand{\NP}{\ensuremath{\mathsf{NP}}}

\newcommand{\NSPACE}{\ensuremath{\mathsf{NSPACE}}}

\renewcommand{\phi}{\varphi}
\newcommand{\eps}{\varepsilon}

\newcommand{\oo}{\omega}
\newcommand{\alp}{\alpha}
\newcommand{\bet}{\beta}

\newcommand{\del}{\delta}
\newcommand{\lam}{\lambda}
\newcommand{\kap}{\kappa}
\newcommand{\sig}{\sigma}

\newcommand{\Gam}{\Gamma}

\newcommand{\Del}{\Delta}

\newcommand{\Oh}{\mathcal{O}}

\newcommand{\cA}{\mathcal{A}}
\newcommand{\cB}{\mathcal{B}}

\newcommand{\cF}{\mathcal{F}}

\newcommand{\cL}{\mathcal{L}}

\newcommand{\cS}{\mathcal{S}}

\newcommand{\cU}{\mathcal{U}}

\newcommand{\cX}{\mathcal{X}}

\newcommand{\fin}{\mathrm{fin}}
\newcommand{\init}{\mathrm{init}}

\newcommand{\Ninit}{N}
\newcommand{\muinit}{\mu}

\newcommand{\cXinit}{\cX_{\mathrm{init}}}

\newcommand{\Uinit}{U_{\mathrm{init}}}

\newcommand{\Winit}{W_{\mathrm{init}}}
\newcommand{\cSinit}{\cS_{\mathrm{init}}}
\newcommand{\siginit}{\mathop{\sig_{\mathrm{init}}}}

\newcommand{\Wfin}{W_{\mathrm{fin}}}

\newcommand{\ra}{\rightarrow}

\newcommand{\ovv}{\overline{\phantom a}}
\newcommand{\ov}[1]{\overline{#1}}
\newcommand{\oi}[1]{{#1}^{-1}}

\newcommand{\Prob}[1]{\mathrm{{Pr}}\left[\, #1 \,\right]}
\newcommand{\E}[1]{\mathrm{\mathbf{E}}{\mathbf[}#1 {\mathbf]}}

\newcommand{\Rat}{\mathrm{Rat}}
\newcommand{\Rec}{\mathrm{Rec}}
\newcommand{\Reg}{\mathrm{Reg}}
 
\newcommand{\NReg}{\mathrm{NReg}}

\newcommand{\lex}{\mathrm{lex}}
\newcommand{\slex}{\mathop{\leq_\mathrm{slex}}}
\newcommand{\lexi}{\leq_\mathrm{lex}}

\newcommand{\nflex}{\mathop{\mathrm{nf_{\mathrm{lex}}}}}

\newcommand{\DG}{dependence graph\xspace}
\newcommand{\HD}{Hasse diagram\xspace}
\newcommand{\HA}{Hasse arc\xspace}
\newcommand{\HAs}{Hasse arcs\xspace}

\begin{document}
\maketitle
\begin{abstract}
  Let $M(A,I)$ be a free partially commutative monoid with involution and $G(A,I)$ its quotient group (for example, a right-angled Artin or Coxeter group).
  We show that for any system of word equations over $M(A,I)$ with recognizable constraints, the solution set~--~in $M(A,I)$ or in $G(A,I)$~--~is an EDT0L language. 
  It is given by an NFA~$\mathcal{A}$ recognizing endomorphisms over some extended monoid. 
  Furthermore, if the input size is $n$, then the automaton $\mathcal{A}$ can be constructed effectively by an $\NSPACE(n\log n)$-transducer.

As a consequence, both \textit{Satisfiability} (whether the system admits a solution) and \textit{Finiteness} (whether the solution set is infinite) are decidable in $\NSPACE(n \log n)$.
For a natural subclass of constraints, we conjecture that these problems are $\NP$-complete.

\end{abstract}


\tableofcontents

\section{Introduction}\label{sec:intro}

Free partially commutative monoids (a.k.a.~\emph{trace monoids}) and groups (a.k.a.~\emph{graph groups} or~\emph{RAAGs}, for \emph{right-angled Artin groups}) are well-studied objects. 
In computer science trace theory serves as an algebraic foundation to study concurrency since the pioneering work of Keller and Mazurkiewicz \cite{Kel73,maz77}. 
Cartier and Foata studied free partially commutative monoids in \cite{cf69} from a combinatorial viewpoint thereby 
generalizing the Master Theorem of MacMahon. 
Graph groups were introduced by 
Droms in \cite{dro85} and since then many deep results have been discovered, see for example \cite{Charney2007,CasalsK2011,Wise2012}.
For years, decidability of the \emph{satisfiability problem} (\ie the problem whether a given equation is solvable) over these structures was open. 
A positive solution for trace monoids was obtained by Matiyasevich~\cite{mat97lfcs}, see also~\cite{dmm97tr}. For graph groups (resp.~RAAGs) Diekert and Muscholl showed that their existential theory is decidable in \PSPACE, see~\cite{dm06}. 
Most of these papers did not address the \emph{finiteness problem}
(\ie the problem whether a given equation has only finitely many solutions). Casals and  Kazachkov showed in \cite{CasalsK2011} that the
\emph{finiteness problem} is decidable for RAAGs using a sophisticated generalization of Razborov-Makanin diagrams and geometric methods, but without complexity estimates.

Our main contribution is the \edtol description of 
the full solution set for systems of equations over free partially commutative monoids and groups;
in the monoid case such a description is given by 
an NFA which is labeled by \Endos over a large enough `extended' alphabet.
Trimming the NFA tells us whether or not the full solution set is finite: it is finite if and only if the trimmed NFA is without directed cycles. 
The group case is done by a reduction 
to the monoid case with the help of reduced normal forms which are a recognizable trace language. 
More generally, our results cope with 
normalized regular constraints, which are closely related 
to the concept of quasiconvex subsets in the Cayley graph of a RAAG~\cite{KMW2017}.

Our complexity results concern the uniform problem variant, where
description of the free partially commutative monoid resp.~group is part of the input and size of the description is included in the input size.
In the non-uniform variant (as in \cite{DiekertJK2016short}) the monoid (resp.~group) is fixed,
and the input is a system of equations alone, we obtain an upper bound of $\NSPACE(n \log n)$
for both satisfiability and finiteness~--~each problem for trace monoids as well as for RAAGs. These problems are $\NP$-hard, but under some natural assumptions about the constraints, we conjecture that 
these problems are $\NP$-complete. 
Even for satisfiability our $\NSPACE(n \log n)$ complexity result improves previously known upper bounds.

To obtain our results we use the \emph{recompression technique} \cite{Jez16jacm_stacs},
which Je\.z used as a~simple method to solve word equations in 
free monoids without \invol. His technique uses two simple compression operations: compress blocks $a^n$ or pairs $ab$ into a letter $c$; and modify the equation by popping out letters from variables so that such operations are sound.
Hence, there is an alternation which makes equations shorter (compression) 
and longer (popping letters). If the length of an equation exceeds a certain threshold, compressions dominates and the length of the equation falls back below the threshold. 

An algebraic setting enables a shift of perspective:
the inverse operation, replacing $c$ by $ab$ or $c$ by $a^n$, is an endomorphism.
Thus, the set of all solutions of an equation (solvable or not) can be represented as a graph,
whose nodes are labeled with equations and edges by endomorphisms of free monoids.
This graph can also be seen as a nondeterministic finite automaton (NFA)
that accepts a rational set of endomorphisms over a free monoid.
(Recall that a subset in a monoid $M$ is \emph{rational} 
if it is accepted by some NFA whose transitions have labels from $M$.)
It is known that applying a rational set of \Endos to a letter yields an \edtol language~\cite{Asveld1977}.
The construction in our setting of partial commutation and monoids with \invol is more involved. Still it guarantees that
the obtained \edtol language describes exactly the set of all solution of the given equation.
Moreover, as usual in automata theory, the structure of the NFA reflects (after trimming) whether the solution set is finite.
Last but not least, our method is conceptually simpler than all previously known approaches to solving equations over free partially commutative groups or monoids with involution. 

\subsection{Related work}\label{Related work}
Studying word equations has been a part of combinatorics on words for more than half a century \cite{BerstelPerrin2007Origins}. From the very beginning, motivation came partly from group theory:  the goal was to understand and parametrize solutions for equations in free groups. 
For example, Lyndon and Sch\"utzenberger needed sophisticated combinatorial arguments to give a parametrized solution to the equation $a^{m}=b^{n}c^{p}$ in a free group~\cite{LyndonSchutzenberger1962}.
On the other hand, it is known that a parametric description of the solution set is
not always possible~\cite{hme76}.
The satisfiability of word equations in free monoids and free groups 
became a main open problem due to its connection with Hilbert's tenth problem.
The problem was solved affirmative 
by Makanin in his seminal papers \cite{mak77,mak83a}.
His algorithms became famous also due to the difficulty of the termination proof and the extremely high complexity.  
A~breakthrough to lower the complexity was initiated
by Plandowski and Rytter~\cite{pr98icalp},
who were the first to apply compression techniques on word equations.
More precisely, they showed that every solution is highly compressible.
Since this result, compression is a key tool for solving word equations.
Indeed compression was essential in showing that the satisfiability of word equations is in $\PSPACE$ \cite{pla04jacm}.
This approach was further developed in~\cite{Jez16jacm_stacs} using the `recompression technique',
which simplified all existing proofs for solving word equations;
in particular, it provided an effective description of all solutions;
a similar representation was given earlier by Plandowski~\cite{Plandowski06stoc}.
In free groups, an algorithmic description of all solutions was known much earlier due to Razborov~\cite{raz87}. 
His description became known as a \emph{Makanin-Razborov diagram}, a major tool in the positive solution of Tarski's conjectures about the elementary theory in free groups \cite{KMIV06,sela13}.
None of these results provided a structural result in terms of formal language theory on the set of all solutions.
Interest in such results was explicitly expressed in~\cite{JainMS2012Lics}.
The first step in that direction was done for quadratic equations in~\cite{FerteMarinSenizerguesTocs14}.
The general case was established in~\cite{CiobanuDiekertElder2016ijac}.

Since the publication of the two papers \cite{DiekertJK2016short,CiobanuDiekertElder2016ijac}
there is a growing awareness in combinatorial and geometric 
group theory that \edtol languages allow to put a rational control on endomorphisms which is a powerful tool. 
We list a few examples without pretending completeness. 
In \cite{CiobanuElder2021} it is shown that all hyperbolic groups have EDT0L solutions to systems of equations with quasi-isometrically embedded rational constraints. 
In \cite{DuncanEvettsHoltRees2023} it is shown that in some cases the solution to equations in the solvable Baumslag-Solitar groups are represented by EDT0L languages.
The article \cite{EvettsLevine2022} proves that the set of solutions of any system of equations over a virtually abelian group is accepted by a multivariable finite-state automaton; hence, it is an EDT0L language. 
In \cite{Levine2023} it is shown that the class of groups where EDT0L languages can be used to describe solution sets to systems of equations satisfies various natural closure properties, including closure under passing to finite index subgroups.
The paper~\cite{CiobanuLevine2024GAGTA3} surveys work on the characterization of solutions to equations in groups in terms of formal (EDT0L) languages between 2014 and 2023. 
More is to come.\footnote{Indeed, in Michael Ende's (1929-1995) visionary land of Fantastica EDT0L is a truly 
neverending story.}

\subsection{Reading the paper}\label{sec:htrtp}
This is a long and technical, but elementary paper. 
Let us give some suggestions on how to read it.
Authors should read every sentence from the first page to the last, pretending that they don't know the next one, but readers are not the authors.\footnote{Authors should not care how to read a mathematical paper, they should care how to write mathematics, \cite{steenrod1973write}.}
As a reader, skip this subsection if you wish to understand all details or if you are eager to find all remaining mistakes. 
If you are not that hungry, then what to read pretty much depends on your knowledge in the field. 
For example, you (as a reader) know what is meant by a free partially commutative monoid or by a free partially commutative group and you know that free partially commutative groups have normal forms in free partially commutative monoids with \invol. 
If you also know that the best way to represent elements in a free partially commutative monoid or group is to use their \DG{s} or even better their \HD{s}, then a good starting point might be \prref{sec:equations} on page~\pageref{sec:equations}. 
If not, then you better read everything up to \prref{sec:equations}.
The next step is to understand what \prref{thm:uro} actually says.
In a first reading, forget about all complexity issues; decidability or \PSPACE is good enough.
Once, the assertion of \prref{thm:uro} is understood, jump to \prref{sec:PPthmuro} on page~\pageref{sec:PPthmuro}.
Carefully read the soundness-part between page~\pageref{sec:PPthmuro} and \prref{sec:vispos}.
Stop to read for at least one day.
Resume with \prref{sec:stproc} which is easy to understand.
All the energy is back.
The heart of the paper is how to lift so-called \STarc{s} where the resource set $S$ 
is less $T$.
This covers \prref{sec:stproc} and  \prref{sec:prospli}.
After that you happily arrived \prref{sec:eliTT} on page~\pageref{sec:eliTT}.
\prref{sec:eliTT} is unpleasant because explicit constants are calculated.
It is better and justified to take \prref{prop:doors} and \prref{cor:doors} for granted.
In a slightly different context similar calculations have been done and verified by other trustworthy people.
Thus, \prref{prop:doors} and \prref{cor:doors} are signed as \emph{proof by authority}.
\footnote{A famous Nobel laureate (2016) warns: `Don't follow leaders. Watch the parkin' meters.'}
Just fly high across the cruel domain to constants and details.
Relax with the now short proof of \prref{thm:uro} in \prref{sec:compT}.

\subsection{An application for RAAGs and RACGs}\label{sec:aplraacgss}
For readers primarily interested in right-angled Artin and Coxeter groups (RAAGs and RACGs), we state the main consequence of \prref{thm:uro} right away. 
Here $A$ is a finite set with involution $a\mapsto \bar a$, and $I\sse A\times A$ is an irreflexive symmetric relation compatible with the involution in the sense that $(a,b)\in I\iff  (\bar a,b)\in I$. 
The \emph{free partially commutative monoid with involution} $M(A, I)$ and its associated quotient group, the \emph{free partially commutative group} $G(A, I)$, are then defined as follows.
\[
  M(A, I) = A^*/\set{ab=ba}{(a,b)\in I}
  \quad\text{and}\quad
  G(A, I) = M(A,I)/\set{a\bar a=1}{a\in A}
\]

\begin{raag_theorem}[\prref{cor:mostint}]
Suppose that~$M(A,I)$ is a free partially commutative monoid with involution and $G(A,I)$ its associated group, \eg a right-angled Artin or Coxeter group.
    Given a system of word equations over~$M(A,I)$ with recognizable constraints, the set of all solutions in~$M(A,I)$ or $G(A,I)$ is an EDT0L language.
    Moreover, in case $(A, I)$ is fixed, satisfiability and finiteness of systems of word equations can be decided in $\NSPACE(n \log n)$ where $n$ is the input size.
  \end{raag_theorem}

 We obtain the above as a corollary of \prref{thm:uro} tracing a long and winding road.\footnote{It is not tracing the B842, a quiet road on the Kintyre Peninsula in Scotland.} 
 For example, \prref{thm:uro} is not stated in the notation $M(A,I)$: it uses the notion of \emph{resource monoids} and the notation $M(A,\rho)$ explained briefly in \prref{sec:resvsI} and also in \prref{ex:fpcmres}.

The statements of results and their proofs have a strong combinatorial flavor, even though the primary interest in RAAGs and RACGs stems from geometric group theory.
Moreover, our main focus will be on the monoid $M(A,I)$.
The reason for its central role is that we obtain the group $G(A,I)$ as a quotient by a strongly confluent and length-reducing trace-rewriting system, which is \emph{special} as all right-hand sides are~$1$: the set rules is $\set{a \bar a\to 1}{a\in \Gam}$.
We can then identify $G(A,I)$ with the set of irreducible traces in 
$M(A,I)$ which is the recognizable set of traces without any factor $a \bar a$ with $a \in A$. 
By covering systems of equations \emph{with recognizable constraints} in $M(A, I)$, we thus also cover systems of equations in $G(A, I)$.

\subsection{Resource formalism for partial commutation}\label{sec:resvsI}
As in \prref{sec:aplraacgss}, standard definitions introduce a partial commutation relation
using a finite undirected graph $(A,I)$ where 
$I \sse A \times A$ is an irreflexive and symmetric relation, called the \emph{independence} relation, with the convention that independent elements of $A$ commute.
Sometimes this convention is stated in terms of the complementary graph $(A,D)$ instead, where $D=A \times A\sm I$ 
is the reflexive and symmetric \emph{dependence} relation.

Here we favor a resource model where we fix a finite set $\Rs$ of \emph{resources}
and a mapping $\rho:A \to 2^\Rs\sm\es$ such that every letter $a\in A$ 
comes with a nonempty subset $\rho(a)\sse \Rs$. The idea is that
a letter $a$ represents an \emph{event} (in the sense of an \emph{event structure}
as in \cite{npw81}). 
Two letters $a,b \in A$ are independent in this setting, that is $(a,b)\in I$, \IFF
$\rho(a)\cap \rho(b)=\es$. 
Conversely, given $(A, I)$, a possible choice for $\rho$ is to 
define $\rho(a)=\set{b}{(a,b)\in D}$. 
This makes clear that the resource model is mathematically equivalent to the standard approach.\footnote{The paper~\cite{dm06} uses a \emph{clan decomposition}, as defined in \cite[pages 75-90]{EhrenfeuchtRH2Structrures1999}, to cope with a dynamic change of the independence relation. Still, the resource model is more suitable for our purpose.} 

However, the resource model has an intrinsic power of compression, which is a main issue in our paper. 
Imagine, for instance, that we want to specify a direct product 
$A_1^* \times A_2^*$ where both $A_i$ are large. 
Then both of the corresponding graphs $(A,I)$ and $(A,D)$ are very large, but we can use $\Rs=\os{1,2}$ to specify the direct product in the resource model.
This intrinsic ability of compression is particularly important to us, since our constructions take place in a highly dynamic environment: even though we frequently add or remove letters from our alphabet, we are never forced to change $\Rs$; it is chosen once and then fixed throughout.\footnote{Charles de Gaulle,  ville de Montr{\'e}al le 24 juillet 1967: \emph{Vive le mono{\"i}de libre de ressources~!}}
More information and details 
are in \prref{sec:foatamon}.

\section{Preliminaries}\label{sec:prel}

\subsection{Sets and \invol{s}}\label{sec:seti}
We follow standard notation. 
If $A$ and $B$ are sets, then $A\sse B$ means set inclusion, while
$A\ssneq B$ means $A\sse B\wedge A\neq B$. 
By $B^A$ we mean the set of mappings from $A$ to $B$ and, in case $B = \os{0,1}$, we identify~$\os{0,1}^A$ with the power set $2^A=\set{C}{C \sse A}$. 
Moreover, we view $A$ as the subset of singletons in $2^A$, \ie $A\sse 2^A$ whenever convenient.

A mapping $\phi\colon S\to T$ is called a \emph{$\Del$-mapping}
for some $\Del\sse S \cap T$ if $\phi(x) = x$ for all $x \in \Del$.
The notion is used for \morphs in other categories, too.
 Frequently, if $\phi\colon S\to T$ is a mapping and $S'\sse S$ is a subset,
then the restriction of $\phi'\colon S'\to T$ is still denoted by $\phi$. 
This is also used for extensions: if $\phi$  is  extended to a larger domain $S''$, then the extended mapping $\phi''$ might be still denoted by $\phi$. 

An \emphind{involution}  is a mapping $x \mapsto \bar x$ such that 
$\overline{\overline{x}} = x$ for all $x$ in the set. The identity map is an \invol for every set.
A \emphind{morphism} $\phi\colon S\to T$ between sets with involution is a mapping respecting the involutions, \ie $\phi(\bar x) = \ov{\phi(x)}$ holds for all $x \in S$. 
Let $s\in S$ and $t\in T$ where $S\sm\os{s,\bar s}\sse T$. 
We say that a mapping
\emph{$\phi\colon S\to T$ is defined by}\index{phi is defined by@$\phi$ is defined by} $s\mapsto t$, if $\phi(s)= t$, $\phi(\bar s) = \bar t$, and $\phi(r)=r$ for all
$r\in S\sm \os{s,\bar s}$. 
 Thus, if $\phi$ is the unique \morph from~$S$ to~$T$ which maps~$s$ to~$t$ and which leaves $S\sm \os{s,\bar s}$ invariant.

\subsection{Monoids with \invol}\label{sec:mo}
Recall that a \emphind{semigroup} is a set $M$ with an associative operation $(x,y) \mapsto x\cdot y$. If a semigroup $M$
contains an element $1$ such that $1\cdot x = x \cdot 1 = x$ for all $x\in M$, then we say that $M$ 
is a \emphind{monoid} and $1$ is the \emphind{neutral} element.  

A \emphind{zero} in a semigroup $M$ is an element $0$ such that $0\cdot x = x \cdot 0 = 0$ for all $x\in M$. If a monoid $M$ contains a zero, then $0\neq 1$ unless $\abs M= 1$.

For $u,v\in M$ we say that $u$ is a \emphind{factor} of $v$, sometimes written as $u\leq v$, if we can write
$v= xuy$ for some $x,y \in M$. 
If we can write $v= uy$ (resp.~$v= xu$), then we say that $u$ is a \emphind{prefix} (resp.~\emphind{suffix}). 
These concepts are trivial for groups, as any two elements of a group are prefixes, suffixes, and factors of one another.

An \emphind{ideal} of $M$ is a subset $I\sse M$ such that $MIM=I$.
For example $I=\es$ and $I=M$ are ideals. 
More generally, $I \sse M$ is a \emph{left} (resp.\ \emph{right}) \emph{ideal} if $MI \sse I$ (resp.\ $IM \sse I$).

A \emphind{semigroup with \invol} is a semigroup where the underlying set is equipped with an \invol such that $\overline{xy}=\overline{y}\,\overline{x}$. For monoids this implies $\bar 1= 1$ (and $\bar 0= 0$ in case there is a zero).
If $G$ is a group, then it is a monoid with \invol by taking $\bar g = \oi g$ for all $g \in G$. By default, we choose $\bar g$ to be $\oi g$ in groups. 

If $L$ is a subset in a semigroup $S$, then $L^+$ denotes the subsemigroup which is generated by $L$.  If $S$ is a monoid, then we let $L^*=\os{1} \cup L^+$, which is the generated submonoid of $L$. Note that we can define 
$L^+$ in terms of $L^*$ by $L^+=L^*L$.  

A \emphind{morphism} between semigroups with \invol is a \hom $\phi\colon M' \to M$ such that 
$\phi(\bar x) = \ov{\phi(x)}$. Note that every group \hom is a \morph of monoids with involution. The identity is a \morph and 
composition of \morphs is a \morph. So, semigroups (resp.~monoids) with \invol form a category. 
If $M$ is a commutative semigroup, then the identity is an \invol.
For example, the power set $(2^S,{\cup},\id)$ is a monoid with \invol: 
the empty set $\es$ is neutral and 
$S$ behaves as a zero. For a commutative group there are two natural \invol{s}:
$\bar x=\oi x$ or $\bar x=x$. They are equal \IFF the group is a vector space over $\F_2=\Z/2\Z$. The monoids $2^S$ and $\F_2^S$ have the same underlying sets, but they are algebraically different unless $S=\es$.  

\subsection{Free monoids with \invol and \selfies}\label{sec:fmi}
Let $\Gam$ by a set, then $\Gam^*$ denotes the free monoid over $\Gam$. 
The elements of $\Gam^*$ are called \emph{\index*{word}s} and the \emphind{empty word} is denoted by $1$ as in other monoids and sometimes it is also denoted by $\eps$ as done frequently in formal language theory. 
The elements of $\Gam$ are called letters. 
If~$w$ is a word, then $|w|_a$ counts the number of occurrences of a  
letter~$a$ in~$w$, and $|w|=\sum_{a \in \Gam}|w|_a$ denotes the \emphind{length} of~$w$. 

Without restriction $\Gam$ is endowed with \invol. If no other \invol
is specified, then we use the identity $\id_\Gam$ as an \invol of $\Gam$. 
We say that a letter~$a$ is \emphind{self-involuting} or a \emphind{selfie} for short, if~$\bar a=a$. Thus, 
every letter is a \selfie \IFF $\id_\Gam$ is the \invol of $\Gam$.
The \invol extends to the free monoid $\Gam^*$ by the law $\ov{uv}= \bar v\, \bar u$. \Ip $\ov{a_1\cdots a_\ell}= \bar a_\ell\cdots\bar a_1$ if the $a_i$'s are letters. 
If the involution on $\Gam$ is the identity, then $\Gam^*$ is the usual free monoid over the set $\Gam$, and reading $\bar w$ means to read the word~$w$ from right-to-left. 

\subsection{Formal language theory} 
Although it is not mandatory for understanding our results, 
we expect that the reader has some basic knowledge in formal language theory and seen the notion of
\emphind{regular subset} and \emphind{regular expression} before. 
In the case of a free monoid $M=\Gam^*$ for some finite set~$\Gam$ the regular languages are denoted
by $\Reg(M)$ or simply as $\Reg(\Gam)$ as mostly done in formal language theory.
However, the situation is not that simple for other monoids than finitely generated monoids.
\Ip in general it is crucial to distinguish between recognizable and rational subsets. 
In the following $M$ denotes any monoid (with or without \invol).

\subsubsection{Recognizable sets in monoids} 
A subset $L\sse M$ is \emphind{recognizable}
if there is a \hom $\mu\colon M\to N$ to a finite monoid~$N$ such that 
$L=\oi \mu (\mu(L))$. The family of recognizable subsets in $M$ is 
denoted by $\Rec(M)$. 

If $h\colon M\to M'$ is a \hom, then 
$\oi h(\Rec(M'))\sse \Rec(M)$; but $L\in \Rec(M)$ does not imply 
$h(L)\in\Rec(M')$, in general.
For example, $(ab)^*\in \Rec(\os{a, b}^*)$, but
neither its image in $\N \times \N \cong \os{a, b}^*/\os{ab=ba}$ nor in 
$\Z\cong \os{a, b}^*/\os{ab = ba = 1}$ are recognizable.
 
\subsubsection{Recognizable sets in monoids with \invol}\label{sec:recinv}
Let $N$ be a monoid. By $N^{\mathrm{op}}$ we denote its \emphind{dual monoid},
which uses the same set (and neutral element) as 
$N$,
but it is equipped with a new multiplication: $x \circ y= yx$. 
If $N$ is a group then $g\mapsto \oi g$ defines an 
\iso between $N$ and $\Nop$. However, in contrast to groups, $N$ and $N^{\mathrm{op}}$ are not isomorphic, in general.

A straightforward calculation shows that the
direct product of monoids $N\times N^{\mathrm{op}}$ becomes a monoid with involution 
by $\ov{(x,y)} = (y,x).$
Let $\eta\colon N\times N^{\mathrm{op}} \to N$ denote the projection onto the first component, then we have the following observation: if $M$ is any monoid with involution and $\mu\colon M \to N$ is a \hom of monoids, then 
we can lift $\mu$ uniquely to a \morph $\wh \mu\colon M \to N\times N^{\mathrm{op}}$ of monoids with \invol such that 
we have $\mu = \eta \wh \mu$. Indeed, it is sufficient and necessary to define 
$\wh \mu(x) = (\mu(x),\mu(\bar x))$. Having this, we can conclude the following.
\begin{proposition}[\cite{CiobanuDiekertElder2016ijac,DiekertElder2020ijac,DiekertJP16,dgh05IC}]\label{prop:rhotomu}
Let $L$ be a subset in a monoid $M$ with involution. Then $L\sse M$ is recognizable by a \hom $\nu$ to a monoid $N$ \IFF there is a \morph of monoids with \invol 
$\mu\colon M \to N\times N^{\mathrm{op}}$ such that $L=\oi{\mu}(\mu(L))$. \Ip we have 
$L\in \Rec(M)$ \IFF there is a \morph $\mu\colon M\to N$ to a finite monoid 
$N$ with \invol such that $L=\oi \mu (\mu(L))$.
\end{proposition}
\subsubsection{Rational sets in monoids and NFAs}\label{sec:rats}
The family of \emphind{rational sets} $\Rat(M)$ is least family  satisfying: 
\begin{itemize}
\item All finite subsets of $M$ belong to $\Rat(M)$.
\item If $L,K\in \Rat(M)$, then $L\cup K, LK = \set{xy\in M}{x\in L \wedge y \in K}, L^* \in \Rat(M)$.
\end{itemize}
There is an alternative definition using nondeterministic automata.  A \emphind{nondeterministic automaton} 
over $M$ is a tuple 
$\cA= (Q,\del,I,F)$ where $Q$ is a set of \emph{states}, 
$I \sse Q$ is a subset of \emph{initial states}, $F \sse Q$ is a subset of \emph{final states}, and $\del\sse Q\times M \times Q$ is a transition relation. A transition of the form $(p,1,q)\in \del$ 
is traditionally called an \epstra. 

A nondeterministic automaton $\cA$ 
over $M$ can be represented as an arc-labeled directed graph where a \tra
$(p,m,q)\in \del$ is identified with an arc  $p \arc{m}q$.
If $p_0 \arc{m_1} \cdots \arc{m_s} p_s$ is a path with $(p_{i-1},m_i,p_i)\in \del$, then $m= m_1 \cdots m_s$ is called its \emphind{label}. 
The \emphind{accepted language} of $\cA$ is defined as 
\[
  L(\cA)=\set{m\in M}{m \text{ labels a path from an initial to a final state}}
\]

The automaton $\cA$ is called \emphind{trim} if every state is on some accepting path.
Essentially, we are interested in the accepted languages of automata, only.  
If $Q$ and $\del$ are finite, then $\cA$ is a  \emphind{nondeterministic finite automaton} (or \emph{$M$-NFA}\index{M-NFA@$M$-NFA} for short). The following proposition is easy to show and a standard fact in 
formal language theory, see \eg~\cite{eil74}. 
\begin{proposition}
\label{prop:ratnfa}
Let $M$ be a monoid. Then a subset $L \sse M$ belongs to 
$\Rat(M)$ \IFF $L$ is accepted by some (trim) $M$-NFA.
\end{proposition}

If $\Gam$ is a finite set, then Kleene's theorem \cite{kle56} implies 
$\Rec(\Gam^*)= \Reg(\Gam^*)= \Rat(\Gam^*)$. This in turn implies
$\Rec(M) \sse \Rat(M)$ in all finitely generated monoids; and there is an algebraic characterization: $\Rec(M) \sse \Rat(M)$ \IFF $M$ is finitely generated.\footnote{This characterization is attributed to   McKnight. It is shown in his paper~\cite{mck64}.} 

\subsubsection{EDT0L languages} \label{sec:EDT0L_lang}
\emphind{Lindenmayer system}s (or \index{L-system@$L$-system}$L$-systems) are formal grammars.
They were introduced in the late 1960's
by Aristid Lindenmayer to model the growth processes of plant development.
There are various types of $L$-systems,
see \eg~the book by Rozenberg and Salomaa~\cite{RozS86}.
An important subclass is defined by the acronym \emphind{EDT0L} or \emphind{HDT0L}.
It stands for \emph{\textbf{E}xtended} (or \emph{\textbf{H}omomorphic}), \emph{\textbf{D}eterministic, \textbf{T}able, 
\textbf{0} interaction, and \textbf{L}in\-den\-mayer system}.
We have EDT0L=HDT0L and from the perspective of this paper the most suitable definition
is by a characterization due to Asveld~\cite{Asveld1977}. It shows that the concept of EDT0L has a purely algebraic interpretation of an NFA-controlled way to iterate \morphs to a single symbol in order to define an EDT0L language. It allows to define EDT0L languages in every 
monoid. Here we use EDT0L languages in free partially commutative monoids. 

\begin{definition}[EDT0L language]\label{def:etol}
Let $L$ be a subset in a monoid $M$ (with \invol) which is generated 
by a finite set $A$ (with \invol) and let $\pi\colon A^*\to M$ 
the canonical \morph (\ie its presentation) which maps each letter $a\in A$ to itself. 
Then $L\sse M$ is an \emph{\edtol language\index{EDT0L language@\edtol language}} if
there is some finite set $\wh C$ (with \invol) containing~$A$, a rational set $\cL$ of 
$A$-\Endos of the free monoid $\wh C^*$ (with \invol), and a symbol $\cent\in \wh C$ such that 
\begin{equation*}
  \set{h(\cent)}{h \in \cL}\sse A^*  
  \quad\text{and}\quad 
  L=\set{\pi h(\cent)}{h \in \cL}
\end{equation*}
The family of \edtol languages in $M$ is denoted by $\EDTOL(M)$.
\end{definition}

The family $\Rat(\End(\wh C^*))$ of rational subsets is
defined as in every other monoid. Here we mainly use the
fact that we have  $\cL \in \Rat(\End(\wh C^*))$
\IFF there is a NFA where the \tras are labeled 
by endomorphisms and which accepts $\cL$. 
Since we deal with $A$-\Endos, we require that each label leaves $A$ invariant (and respects the \invol).

In \prref{def:etol} the set 
 $\wh C$ is the \emphind{extended alphabet}, an $A$-\Endo of $\wh C^*$ is specified by a `table' 
representing the set of pairs $(c,h(c))$ where $c\in \wh C\sm A$ and 
$h(c)\neq c$. 
It is a `deterministic' table because for each $c \in \wh C$ there is at most one entry $(c,h(c))$. Finally, we have `$0$ interaction' because an application of an
\Endo does not depend on any context. 
However, it depends on a `rational control' which is the rational set $\cL$ of \Endos, typically given by some NFA. 
In general, the monoid $\End(\wh C^*)$ has torsion and is not finitely generated; however, only finitely many \Endos appear as labels of the NFA so that $\cL$ is contained in a finitely generated submonoid. 

Note that $\Rat(M)\sse \EDTOL(M)$. 
\Ip the membership problem for \edtol languages in undecidable if the membership problem for rational sets is undecidable.
For example, the membership problem for rational sets is undecidable
in the direct product of two free (non-abelian) groups by \cite{Mihailova58}. 
Direct products of free groups are free partially 
commutative groups in the sense of the following section. 
The underlying set of these groups can  naturally be identified with a recognizable set in some free partially 
commutative monoid.
Recognizability is important here because the intersection-emptiness problem for rational subsets in free partially commutative monoids is undecidable \cite{ah89}.\footnote{More precisely,  
\cite{ah89} shows that the intersection-emptiness problem for rational subsets is decidable if and only if the associated partial commutation graph $(\Gam,I)$ is a \emph{transitive forest}.}

\section{Partially commutative monoids and groups}\label{sec:foatamon}

\subsection{Classical definition}\label{sec:classdef}
The classical way (see, e.g., \cite{cf69}) to construct a free partially commutative monoid over a generating set~$\Gam$ uses an irreflexive and symmetric relation $I\sse \Gam\times \Gam$,
which is called an \emph{independence} relation\index{independence relation}. The complement $D=\Gam\times \Gam\sm I$
is a reflexive and symmetric and is called an \emphind{dependence} relation.
This defines the \emphind{free partially commutative monoid} $M(\Gam,I)$ 
by $M(\Gam,I)= \Gam^*/\set{ab=ba}{(a,b)\in I}$. If $\Gam$ is finite, then according to \cite{maz77} we also say that $M(\Gam,I)$ is a \emphind{trace monoid} and an element of $M(\Gam,I)$ is called a \emphind{trace}.
If the set $\Gam$ is equipped with an \invol, we are more restrictive: we require that~$I$ is irreflexive, symmetric, and that $(a,b)\in I$ implies $(\bar a, b)\in I$.
If all letters are selfies, then the additional requirement is vacuous. If $\phi\colon \Gam \to M$ is a mapping to a monoid~$M$ with \invol such that $\phi(\bar a) =\ov{\phi(a)}$ for all $a\in \Gam$, 
then $\phi$ can be extended to a unique \morph
$\wh \phi\colon M(\Gam,I)\to M$ as soon as $(a,b)\in I$ implies
$\phi(a)\phi(b)= \phi(b)\phi(a)$ for all $a,b\in \Gam$. 
A \emphind{partially commutative monoid} (not necessarily free) 
is a monoid with \invol together with a surjective 
\morph (its \emphind{presentation}) $\pi\colon M(\Gam,I)\to M$.

\begin{example}[Free partially commutative group]\label{ex:RAAG}
Let $I\sse \Gam\times \Gam$ be an independence relation over a set $\Gam$ with \invol. Then the \emphind{free partially commutative group}
$G(\Gam,I)$ is defined by the quotient monoid 
\begin{equation}\label{eq:fpcg}
G(\Gam,I)=  M(\Gam,I)/\set{a\bar a=1}{a\in \Gam}
\end{equation}

If $\Gam$ is without selfies, then $G(\Gam,I)$ is also known as a \emphind{graph group} \cite{dro85} or as a \emphind{right-angled Artin group} (\emphind{RAAG}), see for example \cite{Charney2007}. 
If, on the other hand, $\bar a= a$ for all $a\in \Gam$, then the generators of $G(\Gam,I)$ have order two and $G(\Gam,I)$ is a \emphind{right-angled Coxeter group} \cite{AS}. \hspace*{\fill}$\diamond$\end{example}
 
It is not difficult to see that every group $G(\Gam,I)$ can be embedded as a subgroup (typically of infinite index) into a right-angled Coxeter group. Actually, 
Davis and Januszkiewicz \cite{DavisJ2000JPAA} showed that for each right-angled Artin group there is some right-angled
Coxeter group containing it as a subgroup of finite index.
However, we will not use their result in our paper. 
In our view, $G(\Gam,I)$ is a graph product over RAAGs and RACGs.

\subsection{Resource monoids and $\Rs$-monoids}\label{sec:rmoni}

Let us say more about our choice to work with $\Rs$-monoids, where, as earlier, 
$\Rs$ is a finite set of \emphind{resources}. We view $2^\Rs= (2^{\Rs},{\cup},\id)$ as a monoid with the identity as the \invol.  A \emphind{resource alphabet} is a pair $(\Gam, \rho)$ where
$\Gam$ is a set with \invol and $\rho\colon\Gam \to 2^\Rs$ a \morph of sets with \invol. 
That is, $\rho$ is a mapping with $\rho(a)=\rho(\bar a)$ for all $a\in \Gam$.

Let $M$ be a monoid with \invol 
and $\rho\colon M\to 2^\Rs$ be a \morph such that $\rho(x)=\rho(\bar x)$ and 
$\rho(x)\cap \rho(y)=\es$ implies $xy=yx$ in $M$. Then  
we say that $\rho$ is a 
\emphind{resource mapping} and $M$ together with $\rho$ is called
a \emphind{resource monoid} (or more precisely, an \emph{$\Rs$-monoid}\index{R-monoid@$\Rs$-monoid}).
The definition of $\rho$ forces a partial commutation 
between elements in~$M$. We also fix a linear order $\leq$ on $2^\Rs$ such that $|S|< |T|$ implies $S< T$ in the linear order. By $T_{\min}$ we denote
the least nonempty subset of $\Rs$ in that order.

Of course, every monoid is an $\Rs$-monoid for some $\Rs$, just as every \fg~monoid is a monoid of the quotient of some $M(\Gam,I)$. Typically we are interested here in monoids with \invol $M$ which came with an \epi $\pi:\Gam^* \to M$ where $\Gam=\ov{\Gam}$ is a finite subset in $M\sm \os 1$ and $\pi$ is induced by the inclusion $\Gam\sse M$. 
Then, we wish that $\rho:\Gam^* \to 2^\Rs$ is surjective. This yields an upper bound on $|\Rs|$. More importantly we might wish that $\rho$ reflects the partial commutation as good as 
possible. Thus, we wish for a given $\Rs$ that $\pi$ induces a \morph 
of monoids with \invol $\psi:M(\Gam,\rho)\to M$. Thus, for a given $\Rs$   
we wish to maximize the subsets $\rho(a)\sse \Rs$ for all $a\in \Gam$ such that 
for all $u,v\in \Gam^*$ the following implication still holds $\rho(u)\cap \rho(v)=\es\implies
\pi(uv)=\pi(vu)$. The main advantage of using $\rho$ is that we can keep $\Rs$ as an invariant under the operation of
introducing a fresh letters $c$ and $\bar c$ into $\Gam$ by a \morph $h$ such that
$h(c)=u$ for some $u\in \Gam^*$. Then we define $\rho'(c)=\rho(u)$
Then $h$ induces a \morph $h:M(\Gam',\rho')\to M(\Gam,\rho)$. Moreover, 
if we have an equation $X=YZ$ and $\rho(X)$ is fixed, then we can guess  $\rho(Y)$ and $\rho(Z)$
such that $\rho(X)=\rho(Y)\cup\rho(Z)$. 
 
A \morph between $\Rs$-monoids $\phi\colon M\to M'$ with resource 
mappings $\rho\colon M\to 2^\Rs$ and $\rho'\colon M'\to 2^\Rs$ is a \morph
of monoids with \invol such that 
$\rho'\phi(x)\sse \rho(x)$.
Thus, the image $\phi(x)$ may use 
fewer
resources than $x$.

It is clear that the class of $\Rs$-monoids forms a category of monoids with \invol. The category has free objects as follows. 
Let $(\Gam,\rho)$ be a resource alphabet. Then the \emphind{free
resource monoid over $\Rs$} (or the \emphind{free
$\Rs$-monoid}) is denoted  by $M(\Gam,\rho)$. It is a $\Rs$-monoid with \invol
$\ovv$ and defined as the  
quotient monoid 
\begin{equation}\label{eq:freerho}
M(\Gam,\rho)= \Gam^*/\set{ab=ba}{a,b\in \Gam \wedge \rho(a)\cap \rho(b)=\es}
\end{equation}
It is a free object because for every $\Rs$-monoid
$M'$ with a resource mapping $\rho'\colon M'\to 2^\Rs$ and every
mapping $\phi\colon \Gam \to M'$ satisfying  $\phi(\bar a)=\ov{\phi(a)}$ and $\rho'(\phi(a))\sse \rho(a)$
there is unique \morph of $\Rs$-monoids 
$\wh \phi\colon  M(\Gam,\rho)\to M'$ with $\wh \phi(a)= \phi(a)$
for all $a\in \Gam$.
\begin{example}[Free partially commutative monoids]\label{ex:fpcmres}
If $M(\Gam,I)$ is a free partially commutative monoid, then it can be defined as an $\Rs$-monoid for the resource mapping $\rho\colon \Gam\to 2^\Rs\sm \es$ where
$\Rs = \Gam^2 \setminus I$ and
$\rho(a)=\set{(a_1,a_2)\in \Gam^2 \setminus I}{a \in \{a_1,a_2\}}$. Thus, given an independence   $I\sse \Gam\times \Gam$ we can 
choose $|\Rs|\leq |\Gam|^2$.
It might be however possible to choose $\Rs$ much smaller than $|\Gam|^2$. 
For the other direction: 
every free resource monoid can be written as a monoid
$M(\Gam,I)$ for $I=\set{(a,b)\in \Gam\times \Gam}{\rho(a)\cap \rho(b)=\es}$. Thus, the class of resource monoids captures the class of free partially commutative monoids.\footnote{The restriction `$\forall a\in \Gam:\rho(a)\neq \es$' is easy to achieve and  technically convenient.}
\hspace*{\fill}$\diamond$\end{example}
The following notion is inspired by the \emphind{Foata normal form} of a trace \cite{cf69}. For convenience we define `steps' in the setting of resource monoids.
\begin{definition}\label{def:step}
Let $(\Gam,\rho)$ be a resource alphabet. 
A \emphind{step}
is a subset $s\sse \Gam$ such that 
$s=\os{a_1\lds a_r}$ satisfies  $\rho(a_i)\cap \rho(a_j)=\es$ for all $1\leq i < j\leq r$. 
\end{definition}
Thus, the elements in a step are different and pairwise commuting letters which are viewed as single symbol. Each step~$s\sse \Gam$ defines a unique trace $[s]\in M(\Gam,\rho)$ by 
$[s]=\prod_{i=1}^ra_i$, and we extend $\rho$ to steps by 
$\rho(s)=\rho([s])$. We let $\bar s= \os{\ov{a_1}\lds \ov{a_r}}$, which implies   $[\bar s]= \ov{[s]}$.
The number of letters in step $s\in M(\Gam,\rho)$ is bounded by $|\Rs|$ and this number does not depend on $\Gam$. This is one of the many reasons that it is more convenient (for us) to work with resource mappings rather than with (in-)dependence alphabets.

\section{The covering alphabet}\label{sec:cover}

Since we are interested in equations, we consider
$\Rs$-monoids in which
there 
is a clear partition into disjoint sets of variables $\cX$ and constants $B$.
Both are resource alphabets of the form $(\cX,\rho)$, resp.~$(B,\rho)$. These sets change dynamically, but at the starting point 
with an initial system of equations there is a set
$\cXinit$ of initial variables and an initial resource alphabet $(A,\rho)$ where $A$ is the alphabet of constants. If necessary, we add a new resource to $\Rs$
such that all letters $a\in A$ satisfy $\rho(a)\ssneq \Rs$.
Since later we add more and more resources to letters, the original letters in $A$ will eventually vanish in equations and their \solu{s}.

All resource alphabets of constants $(B,\rho)$ encountered during the process of computing all solutions are included 
in some rather huge but 
finite resource alphabet $(\smash{\wh C},\rho)$, \ie we maintain the invariant
$(A,\rho)\sse (B,\rho)\ssneq (\smash{\wh C},\rho)$. 

The basic idea of compression is to replace factors $u$ which appear in a given \solu using an alphabet $B\ssneq \smash{\wh C}$ by a fresh letter~$c$ such that 
$\rho(c) = \rho(u)$, $\mu(c) = \mu(u)$, and $\bar c = c \iff \bar u = u$. 
As we add more and more letters we also need a procedure to remove constants from the alphabet $B$ in order to reuse names 
for fresh letters in $\smash{\wh C}$.
However, the size of $\smash{\wh C}$ is limited due to our space bound. 
Typically we intend to remove a letter~$c$ 
if it is used in the \solu $\sig$ but not `visible' in the equation $W$. For that reason we will maintain, for a current alphabet of constants $B$, a \morph $\bet\colon B \to M(A,\rho)$
such that $\bet\sig$ solves the initial system of equations. As we will see, we can remove $c$ from the \solu by replacing $c$ by $\bet(c)$.
This does not change the equation and does not change $\bet\sig$.
However, there is a technical problem because $\bet(c)$ may contain letters which use strictly less resources than $c$;
and our strategy would not work if we were allowed to do so. 
Therefore we  embed $(A,\rho)$ into a larger 
resource alphabet $(\widehat A,\rho)$ inside $(B,\rho)$. 
More precisely, if a resource alphabet $(A,\rho)$ is smaller
than $(B,\rho)$, then we assume 
$(\widehat A,\rho)\sse (B,\rho)$. 
We call $(\widehat A,\rho)$ a \emphind{covering alphabet} because 
each $a\in A$ is `covered' by several letters in $\widehat A$ (with the exception of $a\in \os{\cent,\#}$, which are treated in a special way).
The  letter outside $A$ are denoted by 
$(a,T)$ where $\rho(a)\sse T \sse \Rs$ and $\rho(a,T)=T$. This larger resource alphabet comes canonical 
$A$-\epi $\wh \pi:M(\widehat A,\rho)\to M(A,\rho)$  where $\wh \pi(a,T)=a$. This enables us to lift $\bet$ to a \morph $\alp\colon  B \to M(\widehat A,\rho)$ with $\bet = \wh \pi \alp$. 
In addition, we need for each $T$ a normal form mapping $\nf_T: M(A,\rho)\hookrightarrow M(\widehat A,\rho)$ 
respecting the involution which maps a trace $u$ with 
$\rho(u)\sse T$ to some trace every letter of which is of the 
form $(a,T)\in \widehat A$. 

\subsection{The definition of $(\wh A,\rho)$ and the \morph 
$\pio\colon M(\wh A,\rho) \to M(A,\rho)$}\label{sec:whApi}
The definition of the alphabet $\smash{\wh A}$ uses the concept of a step (\prref{def:step}). 
Firstly, we let
\[A_s = \set{s\sse A}{s \text{ is a nonempty step}}\] and extend $\mu_\init$ to the elements of $A_s$ by 
$\muinit(s)=\mu_\init([s])$ for $s\in A_s$. 

We then embed $(A\cup A_s,\rho)$ into $(\smash{\wh A},\rho)$ by a defining 
\begin{align}\label{eq:defwhA}
\wh A= A
\cup \set{(s,T)\in A_s\times 2^\Rs}{\rho(s)\sse T} &\quad\text{ This is a disjoint union!}
\\
\label{eq:mumuini}
\mu(a)=\mu_\init(a) &\quad\text{ for all $a\in A$}
\\
\label{eq:defwhAT}
     \ov{(s,T)}=(\bar s,T),\;\rho(s,T) =T, &\text{ and }  \mu(s,T) =\mu(s)
\end{align}
Since we have $(A,\rho)\sse (\wh A,\rho)$ such that $\mu_\init$ is the restriction of $\mu:M(\wh A,\rho) \to N$, we write 
$\mu$ for $\mu_\init$, too.
For every letter $a\in A$ 
there is a letter $(a,T)\in \wh A$ and the homomorphism 
$\mu\colon M(\smash{\wh A},\rho)\to \Ninit$ is defined
such that $\mu(\smash{\wh A})\sse \mu(A^*)$. We 
obtain a chain of embeddings of free $\Rs$-monoids:
\begin{align}\label{eq:hooks}
M(A,\rho) \hookrightarrow M(A\cup \cX,\rho)\hookrightarrow M(\wh A \cup\cX,\rho).
\end{align}
The alphabet $\smash{\wh A}$ is implicitly 
specified by $A$ and $\rho$, so we do not need any extra memory for the representation of $\smash{\wh A}$.
The bit-encoding of letters in $\smash{\wh A}$ is described in \prref{sec:bit}.

The resource monoid $M(A,\rho)$ is a \emphind{retract} of $M(\smash{\wh A},\rho)$.
That is, firstly we have an embedding $\iota\colon M(A,\rho) \hookrightarrow  M(\smash{\wh A},\rho)$ which maps~$a$ to 
$(a,\rho(a))$ and secondly there is an \epi $\wh\pi$ of $M(\smash{\wh A},\rho)$ onto $M(A,\rho)$ such that $\wh \pi \,\iota$ is the identity on $M(A,\rho)$ and which maps 
a pair $(s,T)$ to $\wh \pi(s,T)= [s]$.

\subsection{Step normal forms with respect to  $T$}\label{sec:snf}
Consider the natural projection
$\pi:A^*\to  M(A,\rho)$. For solving equations in $M(A,\rho)$
we rely on a \emphind{normal form mapping}
$\nf: M(A,\rho)\to A^*$ such that $\nf(\bar x)=\ov{\nf(x)}$ for all
$x\in  M(A,\rho)$. Such a normal form was constructed in \cite{dm06} and did not involve recompression technique.
However, using $\wh A$ simplifies the construction.

For all $T\sse\Rs$ we define a subset 
$\smash{\wh A}_T \sse\smash{\wh A}$ by 
$\smash{\wh A_T}=\set{a\in \smash{\wh A}}{\rho(a)= T}$.
Note that the free monoid $\smash{\wh A}_T^*$ is a submonoid of $M(\smash{\wh A},\rho)$. Moreover, $\Reg(A_T)\sse   \Rec(M(\wh A,\rho))$.
We call a mapping \(\nf:  M(A,\rho) \to  M(\wh A,\rho)
\)
a \emph{\symnf\index{symmetric normal form@\symnf}} \wrt~$T$ given that
\[
  \nf(\set{u\in M(A,\rho)}{\rho(u)\sse T}) \sse \wh A_T^*,
  \quad 
  \nf(\bar u)= \ov{\nf(u)},
  \quad
  \wh \pi (\nf (u))=u.
\]

Our aim is to show that \symnf{s} exist,
to this end we define a specific one.
Recall that by putting a linear order $\leq$ on $A$ we can represent every trace 
in  $u\in M(A,\rho)$ by its lexicographical normal form
$\nflex(u)\in A^*$. 
\begin{definition}\label{def:stepnf}
Let $u\in M(A,\rho)$. If $u=1$ or $\rho(u)\sm T\neq \es$, then 
we define  $\snf_T(u)=u$. 
In the other case, where $u\neq 1$ and 
$\rho(u)\sse T$, we also assume that 
$\nflex(u)\lexi \nflex(\bar u)$. 

If $a_1\cdots a_\ell= \nflex(u)\neq \nflex(\bar u)$ with $a_i\in A$ for $1\leq i \leq \ell$, then we have $u\neq \bar u$ and 
we define the \emph{\stepnf{s}\index{step normal form@\stepnf{s}}} $\snf_T(u)$ and $\snf_T(\bar u)$ simultaneously 
as follows: 
\begin{align}\label{eq:Tsnfnos}
\snf_T(u) &= (\os{a_1},T) \cdots (\os{a_\ell},T), \\ 
\snf_T(\bar u) &= (\os{\ov{a_\ell}},T) \cdots (\os{\ov{a_1}},T).
\label{eq:Tsnfnov}
\end{align}
If $\nflex(u)= \nflex(\bar u)$, then $u=\bar u$; and $u$ can be written as
$u=a_1\cdots a_m\,[s] \,\ov{a_m}\cdots \ov{a_1}$ with $a_i\in A$ for $1\leq i \leq m$ such that
$s$ is a possibly empty step of selfies. 
In this case we define the \emph{\stepnf} $\snf_T(u)$ as follows  (with $(1,T)=1$):
\begin{align}\label{eq:Tsnf}
\snf_T(u) &= (\os{a_1},T) \cdots (\os{a_m},T) (s,T)  (\os{\ov{a_m}},T) \cdots (\os{\ov{a_1}},T).
\end{align}
\end{definition}
In general there are many words $(\os{a_1},T) \cdots (\os{a_m},T)$ that represent a given trace $a_1 \cdots a_m$, and we choose one of arbitrarily.
\begin{lemma}\label{lem:snf}
The \stepnf is a \symnf \wrt~$T$.
\end{lemma}
\begin{proof}
Let $u\in M(A,\rho)$  with $a_1\cdots a_\ell= \nflex(u)\leq \nflex(\bar u)$ with $a_i\in A$ for $1\leq i \leq \ell$. If $\nflex(u)\neq \nflex(\bar u)$, then we obtain
$\snf_T(u) = u$ and $\snf_T(\bar u) = \bar u$. Since $\wh \pi(u)=u$ for all 
$u\in M(A,\rho)$, we may assume that $\nflex(u) = \nflex(\bar u)$.
In this case $\snf_T(u)$ is given by \prref{eq:Tsnf}; and 
we have $\wh \pi\snf_T(u) =u$ and $\wh \pi\snf_T(\bar u) =\bar u$ because $\ov{(s,T)}=(\bar s,T)$ for all steps~$s$. 
\end{proof}
\begin{remark}\label{rem:snf}
The set $\snf_T(M(A,\rho))$ is not a regular subset of ${\wh A_T}^*$, in general. 
Indeed, suppose that there are selfies $c,d,f\in A$ with $\rho(c)\cap \rho(d)=\es$, $c\leq_\lex d\leq_\lex f$,  and 
$ \rho(c)\cup \rho(d)\sse \rho(f)= T$.
Let $a=(c,T)$, $b=(d,T)$, and $e=(f,T)$.
Consider the word
$u=(eabe)^m\, e(eabe)^n$. 
For $m\neq n$ the word is a step normal form, but for 
$m=n$ it is not because $u\neq \bar u$ although $v=\bar v$ for $v=(fcdf)^m\, f(fcdf)^m$. Thus $\snf_T(M(A,\rho))$ is not a regular in
${\wh A_T}^*$ and therefore not recognizable in $M(A,\rho)$.  
Similar words can be constructed if there are non-selfies 
in $\os{c,d,f}$.   
\hspace*{\fill}$\diamond$\end{remark}

\subsection{Resource monoids with type}\label{sec:tfrm}
During the process of solving equations in free resource monoids (inside the so-called `$T$-block-compression') we encounter more general monoids than free partially commutative monoids by giving some letters and/or variables a `type'. 
In such a situation we work with a finite alphabet which is a disjoint union of a set $B$ 
of constants and a set $\cX$ of variables. 
Both sets are closed under \invol and, as usual, $\cX$ is without \selfies.
For every $\es\neq T\ssneq \Rs$ we define $B_T=\set{a\in B}{\rho(a)=T}$ and $\cX_T=\set{X\in \cX}{\rho(X)=T}$.
With this setting we let $\Gam=B\cup \cX$ and $\Gam_T=B_T\cup 
\cX_T$.
\begin{definition}\label{def:type}
Let $a \in B_T \sm \smash{\wh A}$ and $b \in \os{1} \cup (B_T \sm \smash{\wh A})$ with $a \neq b = \bar b$.
  An \emph{$(a,b,\Del)$-\emphind{type}} is a set $\theta\sse\Gam_T^+ \times \Gam_T^+$ of additional defining relations subject to the following conditions.
  There is a mapping $\delta\colon \Del \to \os{a,\bar a}$ of sets with involution where $\Del \sse \Gam_T \sm (\os{a,\bar a, b} \cup \smash{\wh A})$ such that for all constants $\wt a \in \os{a,\bar a}$ and $c \in B \cap \Del$, and all variables $X \in \cX \cap \Del$
  \begin{align}
    (\wt a b c, c b \wt a) \in \theta &\iff \wt a = \delta(c), \label{eq:type_ac}\\
      (X b \wt a, \wt a b X) \in \theta &\iff \delta(X) = \wt a, \label{eq:type_xa}\\
      (X b c, c b X) \in \theta &\iff \delta(c) = \delta(X). \label{eq:type_xc} 
  \end{align}
  Moreover, each element of $\theta$ is of (exactly) one of the above forms.
\end{definition}
Note that the empty set is an $(a,b,\es)$-type for all $a, b$ as above.
It is not difficult to see that if $\theta$ is not empty, then $\theta$ determines $(a,b,\Del)$ as well as $\delta \colon \Del \to \os{a, \bar a}$.
As such, we may refer to $\theta$ simply as a \emph{type}.
Moreover, $\theta$ can be specified
using $\Oh(|\Del|)$ symbols. 

The conditions appearing in the above definition are shaped according to our application. 
We use types only in \prref{sec:blocked} where we give
concrete examples for them.

Given a type $\theta$ we denote by $M(B,\cX,\rho,\theta)$ (or by
$M(B\cup \cX,\rho,\theta)$)
the quotient 
monoid 
\begin{equation}\label{eq:montype}
M(B,\cX,\rho,\theta)= M(\Gam,\rho)/ \set{u=v}{(u,v) \in \theta}.
\end{equation}
We say that $M(B,\cX,\rho,\theta)$ is a \emphind{resource monoid with type}.
For the empty relation $\theta = \es$, we recover the resource monoids $M(B,\cX,\rho)$ and $M(\Gam,\rho)$ as $M(B,\cX,\rho,\es)$ and $M(\Gam,\rho,\es)$, respectively. 
Thus, every statement which holds for all resource monoid with type is also true for all resource monoids.
Since types are not used outside \prref{sec:blocked}
we neither use nor define a notion of resource group with type.

Let us explain some consequences. 
$M(\Gam,\rho,\theta)$ is a monoid with a \invol because $(u,v)\in \theta$ implies $(\bar v,\bar u)\in \theta$. The defining relations are length preserving and letter-homogeneous. Hence, $|u|_a$, the length $|u|$, and 
 $\rho(w)$ are well-defined for all $a\in \Gam$ and $w\in M(\Gam,\rho,\theta)$ by choosing any representing word in $\Gam^*$ for~$w$. \Ip $M(\Gam,\rho,\theta)$ is a 
resource monoid. Moreover, the 
$\Gam$ is the least set of generators 
as $\Gam \cap  \Gam^+\Gam = \es$.
\begin{remark}\label{rem:embed}
If $\Gam'\sse \Gam \sm (\os{a, \bar a} \cup \Del)$, then $M(\Gam',\rho)$ embeds into $M(\Gam,\rho,\theta)$.
\hspace*{\fill}$\diamond$\end{remark}
Let $(\Gam,\rho)$ be a resource alphabet. Then $\rho$ induces 
a \morph  $\N^\Gam\to 2^\Rs$, and $2^\Rs$ itself is a resource monoid with
$\rho(S)=S$ for $S\sse \Rs$. As a consequence there is a natural sequence of canonical \morph{s} of resource monoids with a factorization of $\rho$ as follows.
\begin{align}\label{eq:canseq}
\rho\colon\Gam^* \to M(\Gam,\rho) = M(\Gam,\rho,\es)\to M(\Gam,\rho,\theta)\to M(2^\Rs,\rho)\to 2^\Rs.
\end{align}
\Ip we can always represent an element~$w$ of $M(\Gam,\rho,\theta)$ either by a word in $\Gam^*$ or by a trace in $M(\Gam,\rho)$.

\subsection{Bit encoding a resource monoid with type}\label{sec:bit}

There are several ways to encode a resource monoid with type $M(\Gam,\rho,\theta)$ (or $M(\Gam,\rho)=M(\Gam,\rho,\es)$)
as a bit string;  
and practical encodings are mostly polynomially equivalent.
Throughout, we count sizes assuming that describing a symbol $x$ in~$\Gam$
uses~$\Oh(\log\abs \Gam)$ space. 

The relation~$\theta$ can be encoded as a sequence of letters in $\Gam$ of length $\Oh(\abs \Gam)$. 
Generators are encoded by listing them in some linear order.
The involution is specified by a list of corresponding pairs.
Specifying the commutation is done through resources:
for each~$a \in \Gam$ we keep a bit vector of length~$|\Rs|$.
In addition, a letter $x\in \Gam$ may have a constraint $\mu(x)\in N$ for some finite monoid~$N$. 
Hence, we reserve $\Oh(\log\abs N)$ bits for each $x$ to denote its value $\mu(x)$. 
Such an encoding allows an efficient \solu
of the uniform factor problem.
\begin{remark}\label{rem:facalg}
The \emphind{uniform factor problem} for a resource monoid with type
$M(\Gam,\rho,\theta)$ asks on input
$M(\Gam,\rho,\theta)$ (encoded as described above)
and $u,v \in \Gam^*$ whether there are $p,q\in \Gam^*$
such that $puq= v$ is an identity in the quotient $M(\Gam,\rho,\theta)$.
This problem can be solved in nondeterministic linear space by guessing, step by step, how to transform $puq$ into~$v$.
\hspace*{\fill}$\diamond$\end{remark}

\subsection{Hasse diagrams and their positions}\label{sec:DGHA}
Our proofs heavily rely on the representation of a trace by its Hasse diagram.\footnote{More drastically, our proofs are close to incomprehensible when working without that representation.} 
More generally, 
let $M(\Gam,\rho,\theta)$ be a resource monoid with type where 
$\rho(a) \neq \es$ for all $a \in \Gam$.
An element $w = a_1\cdots a_m \in M(\Gam,\rho,\theta)$ with $a_i \in \Gam$ has many word representations, but we can always choose a representation as a vertex-labeled directed acyclic graph which, as an abstract graph, is unique if $\theta=\es$, see \cite{maz77} or \eg~\cite{dr95}. 

Given~$w$ as above we view $V(w)=\os{1, \ldots, m}$
as the set of vertices for a concrete graph representation.
Each vertex $i$ is labeled by the letter $\lam(i)=a_i$. 
We define the set of \emph{\index*{position}s} of~$w$ by
\begin{equation}\label{eq:defpos}
\pos(w)= \set{(a,i)\in \Gam\times \N}{0\leq i \leq |w|_a}.
\end{equation}
The set $\pos(w)$ is well-defined even if $\theta\neq \es$ because there are no commutation rules to interchange the ordering of the same letter.
Given a position $p=(a,i)$ we define its resources by
$\rho(p)=\rho(a)$, which is never the empty set, and its label
by $\lam(p)\in \Gam$.

In general there is no unique dependence graph representation for the elements of $M(\Gam,\rho,\theta)$ if $\theta\neq \es$.
However we can choose for each $w\in M(\Gam,\rho,\theta)$
some trace $\wh w =w$ in the free resource monoid $M(\Gam,\rho)$ which represents~$w$, and we use the unique dependence graph of $\wh w$ as a graphical representation for~$w$. 

It remains to define dependence graphs and their Hasse diagrams for traces in free resource monoids. 
Given a trace $w = a_1\cdots a_m\in M(\Gam,\rho)$ we choose any concrete labeled vertex set $V = V(w)=\os{1, \ldots, m}$ and its labeling $\lam\colon V\to \Gam$ as above. 
The set of arcs in the \DG $D(w)$ is given by 
\begin{equation}\label{eq:DGarc}
E= \set{(i,j)\in V\times V}{i< j \wedge \rho(a_i)\cap \rho(a_i) \neq \es}
\end{equation}
We view $D(w)= [V,E,\lam]$ as an abstract graph: hence, $D(W)$ is defined up to \iso and, as such, it is no restriction to realize 
$V$ as the set $\pos(w)$. 
The dependence graph $D(w)$ induces a \emphind{vertex-labeled partial order} $P(w)= [V,\prec,\lam]$ (where $\prec$ is irreflexive) and we denote by $H(w)$ to be \emph{\HD\index{Hasse Diagram@\HD}} of the induced partial order.
That is: $p\to q$ is an arc in $H(w)$ \IFF both, $p\to q$ is an arc in $D(w)$ and there is no position $r$ with 
is $p\prec r \prec q$. By a standard result in trace theory~\cite{dr95} due to Mazurkiewicz~\cite[Chapt.~1]{maz77}
we have $w = w'$ in $M(\Gam,\rho)$ \IFF $D(w) = D(w')$ \IFF $H(w) = H(w')$. 

As usual in graph theory, we say that a trace $w \in M(\Gam,\rho)$ is \emphind{connected} if its dependence 
graph $D(w)$, viewed as an undirected graph, is connected.
\begin{example}\label{ex:dephas}
Let $\Rs=\os{r,s}$ and $a,b,c$ be letters with $\rho(a) = \rho(\bar a) = \os{r}$,
$\rho(b) = \rho(\bar b) = \os{s}$, and
$\rho(c) = \rho(\bar c) = \os{r,s}$. Then the word $ab\bar aca \bar b abca\bar b$ represents a connected trace~$w$.
The \DG $D(w)$ and its \HD $H(w)$ with thick 
\HA{s} are depicted in \prref{fig:hade}. 
\begin{figure}[h]
	\begin{center} 
		\begin{tikzpicture}[node distance=14mm, xscale=20, yscale=8]
		\node[circle] (a4) {$c$};
		\node[circle] (a3) [below left of=a4] {$ \bar a$};
		\node[circle] (a2) [above left of=a4] {$ b$};
		\node[circle] (a5) [below right of=a4] {$ a$};
		\node[circle] (a6) [above right of=a4] {$ \bar b$};
		
		\node[circle] (a1) [left of=a3] {$ a$};
		
		\node[circle] (a31) [right of=a5] {$a$};
		\node[circle] (a21) [right of=a6] {$\bar b$};
		
		\node[circle] (a41) [below right of=a21] {$ c$};
		\node[circle] (a51) [below right of=a41] {$ a$};
		\node[circle] (a61) [above right of=a41] {$ \bar b$};

		\draw (a1) edge[->,dotted, bend left] (a4); 
		
		\draw (a1) edge[->,dotted, bend right=90] (a41); 
		\draw (a3) edge[->,dotted, bend right=90] (a41); 
		\draw (a5) edge[->,dotted] (a41); 
\draw (a4) edge[->,dotted] (a41);
\draw (a4) edge[->,dotted] (a21);
\draw (a4) edge[->,dotted] (a31);
\draw (a4) edge[->,dotted] (a51);
\draw (a4) edge[->,dotted] (a61);

		\draw (a1) edge[->,dotted, bend right] (a5);
		\draw (a1) edge[->,dotted, bend right] (a31);
		\draw (a1) edge[->,dotted, bend right] (a51);

		\draw (a3) edge[->, dotted] (a5);		
		\draw (a3) edge[->,dotted, bend right] (a31);
		\draw (a3) edge[->,dotted, bend right] (a51);

		\draw (a5) edge[->, dotted, bend right] (a51);

		\draw (a31) edge[->, dotted] (a51);

		\draw (a2) edge[->,dotted, bend left=90] (a41);
		\draw (a6) edge[->,dotted] (a41);

		\draw (a2) edge[->,dotted] (a6);
		\draw (a2) edge[->,dotted, bend left] (a21);
		\draw (a2) edge[->,dotted, bend left] (a61);
		
		\draw (a6) edge[->,dotted, bend left] (a61);

		\draw (a21) edge[->,dotted] (a61);
		
		\draw (a1) edge[->,very thick] (a3);
		\draw (a2) edge[->,very thick] (a4);

		\draw (a3) edge[->,very thick] (a4);
		\draw (a4) edge[->,very thick] (a5);
		\draw (a4) edge[->,very thick] (a6);
		
		\draw (a6) edge[->,very thick] (a21);
		\draw (a5) edge[->,very thick] (a31);
		\draw (a21) edge[->,very thick] (a41);
		\draw (a31) edge[->,very thick] (a41);
		\draw (a41) edge[->,very thick] (a61);
		\draw (a41) edge[->,very thick] (a51);
\end{tikzpicture}\end{center}
	\vspace{-1cm}
	\caption{Dependence graph of $w=ab\bar aca \bar b abca\bar b$. The dotted arcs disappear in $H(w)$.}\label{fig:hade}
\end{figure}
\hspace*{\fill}$\diamond$\end{example}
\begin{definition}\label{def:Tletarc}
Let $(\Gam,\rho)$ be a resource alphabet and $T\sse \Rs$.
\begin{itemize}
\item A letter~$a\in \Gam$ is  called a \emph{$T$-letter}\index{T-letter@$T$-letter}, if~$\rho(a)=T$.
\item Let $w\in M(\Gam,\rho)$ be trace and $p\to q$ a \HA of~$w$. 
It is called a \emph{\Tarc\index{T-arc@\Tarc}} if one of its endpoints is labeled by a $T$-letter, \ie if $T \in \os{\rho(p),\rho(q)}$;  and it is called an \emph{$S$-$T$ arc\index{S-T arc}} if $\os{\rho(p),\rho(q)} = \os{S,T}$.\\
(Hence, every \STarc is a \TSarc, an \Sarc, and a \Tarc.)
\item A \HA $p\to q$ is called \emph{unbalanced} if $\rho(p)\neq \rho(q)$.
\end{itemize}
\end{definition}
\begin{remark}\label{rem:cancan}
If $M = M(B,\cX,\rho,\theta)$ is a resource monoid, where $\theta$ in an $(a,b,\Del)$-type according to \prref{def:type}, then $M$
is cancellative (\ie $uw= vw$ or $wu= wv$ implies $u=v$).
For $b=1$ this follows \eg from the representation by Hasse diagrams 
of elements in $M$ because then
$M$ is a free partially commutative monoid, see \cite{dr95}.
In case $b\neq 1$ it is enough to consider the 
case where $M(\Gam,\rho)$ is a free monoid.
We put a linear order on the set $\Gam_\Del=\set{a,\wt a, b, x}{x\in \Del}$ such that $a \leq \wt a < b < c <X$ if $c\in B\cap \Del$ and $X\in \cX\cap \Del$.
Having this $S=\set{(u,v)\in \theta \cup \oi \theta}{v <_{\lex} u}$ forms a confluent and terminating (a.k.a.~\emph{convergent})
semi-Thue system for the monoid $\Gam_\Del^*$.
The claim now follows from standard arguments: 
we represent two different elements in different irreducible normal forms, multiply a letter on the left (resp.~on the right) and show that their irreducible normal forms remain different.
\hspace*{\fill}$\diamond$\end{remark}
\begin{remark}\label{rem:HAs}
To compute $H(\bar w)$ from $H(w)$ we simply reverse the direction of all arcs and we replace every position label~$a$ by $\bar a$. 
Frequently, for better reading, we use a letter~$a$ to point to a specific position which has label~$a$. This convention was used, for example, in \prref{fig:hade} where we 
drew 
$a_p \to a_q$ to denote the actual \HA $p\to q$
where~$p$ is the position (in $H(w)$) of $a_p$ and~$p$ is the position of $a_q$.
\hspace*{\fill}$\diamond$\end{remark}

\subsubsection{Subtraces in free resource monoids and their positions }\label{sec:stha}

It is crucial to understand the correspondence between 
factors of a trace~$w$ and subsets in the induced labeled order $[V,\preceq, \lam]$
of $H(w)$, resp.~$D(w)$. 
Let $w \in M(\Gam,\rho)$ be given by some word $w=a_1\cdots a_m \in\Gam^*$ with $a_i \in \Gam$ and $V=\os{1, \ldots, m}$. Each subset $U\sse V$ defines a labeled linear order and hence a trace which we denote by $w[U]$. However, in general $w[U]$ is not a factor of~$w$. 
We say that a subset $U$ of positions is a \emphind{subtrace} if
 for all $i,j,k\in V$ with $i \preceq k \preceq j$ and $i,j\in U$ we have $k\in U$, too. 
 If $U$ is a subtrace, then $u=w[U]$ is a factor of~$w$.
 Furthermore, $U$ defines three disjoint subtraces: 
 $P=\set{p\in V \sm U}{\exists i \in U: p\prec i}$, $Q=\set{q\in V \sm U}{\exists  i \in U: i\prec q}$, and 
 $R=V\sm (P\cup U \cup Q)$. Thus, $P$ is the set of positions less than $U$, 
 $Q$ is the set of positions greater than $U$, and $R$ is the set of positions which are independent of $U$: we have $\rho(a)\cap \rho(b)$ for all labels $a \in U$ and 
 $b\in R$. Thus, we have four traces $p= w[P]$, $q= w[Q]$, $r = w[R]$, and $u=w[U]$ and each factorization $r=r'r''$ defines factorizations 
 $$ w = pr' u r'' q = pruq= purq.$$ 
 Conversely, let $w=p'uq'$ by any factorization where $u$ is a factor of~$w$. 
 Then we find disjoint subtraces $P'$, $U$, and $Q'$ such that $p'= w[P']$, $q'= w[Q']$, and $u= w[U]$. Thus, every occurrence of a factor $u$ defines a unique subtrace $U$. Moreover, we can factorize $p'= pr'$ and $q'= r''q$ such that 
 we find $p= w[P]$, $q= w[Q]$, $r'r' = w[R]$, and $u=w[U]$ 
as above. 
Note that defining a subtrace $U$ by its positions is more precise than saying that the label of $U$ by the factor $u= \lam(U)$, as the factor $u$ may have many different occurrences in the trace~$w$. 
 
As we work mainly with \HD{s}, we identify a subtrace $U$ of~$w$ with the induced 
subgraph in $H(w)$. It is clear that every arc in $H(w[U])$ appears as an arc in $H(w)$. However, it is not true that every induced subgraph in $H(w)$ defines a subtrace, see \prref{fig:STHD}. 
\begin{figure}[h]
	\begin{center} 
		\begin{tikzpicture}[node distance=8mm]
		\node[circle] (w) at (-0.7,0) {$w=\quad$}; 
		\node[circle] (a) at (0,0) {$a$}; 
		\node[circle] (b1) at (3,0.5) {$b$};
		\node[circle] (br) at (3,-0.5) {$c$};
		\node[circle] (c) at (6,0) {$d$};
		\draw (a) edge[->, >=latex,thick] (b1);
		\draw [thick,->, >=latex] (a) to (br);
		\draw (b1) edge[->, >=latex,thick] (c);
		\draw (br) edge[->, >=latex,thick] (c);
		\end{tikzpicture}\end{center}
	\vspace{-0.5cm}
	\caption{The \HD $a\to e \to d$ for $e\in \os{b,c}$ is not a subtrace of~$w$ because every convex subset of $w$ containing the endpoints $a$ and $d$ must contain both of the positions of $b$ and $c$.}\label{fig:STHD}
\end{figure}
\HA{s} play a special role. As we mentioned above, sometimes a \HA $(p,q)$ are drawn as $a\to b$ where $a=\lam(p)$ and $b=\lam(q)$.  
This implies $\rho(a)\cap \rho(b)\neq \es$. Every \HA defines a subtrace, and there is a one-to-one correspondence between occurrences of factors $ab$ of~$w$ (\ie subtraces with label $ab$) and \HA{s} $a\to b$ in $H(w)$. 
\prref{fig:STHD} shows that a sequence $a\to b\to d$ of \HA{s} does not induce a subtrace, in general. 
 
A letter~$a \in \Gam$ is \emph{minimal\index{minimal letter}} in a trace~$w$ if it is minimal in its Hasse diagram,
which means that $w= av$ for some trace $v$.
We denote the set of minimal positions (or their labels) of~$w$ by $\min(w)$. Maximal positions are left-right dual; they are denoted by $\max(w)$. 
Note that $\abs {\min(w)}, \abs {\max(w)}\leq \abs{\rho(w)}$; moreover, $\ov{\max(w)} = \min(\bar w)$. Letters in $\min(w)$ (resp.~$\max(w)$) are pairwise independent: they use pairwise disjoint sets of resources. 
If $m\sse \Gam$ is any subset of pairwise independent letters, then we simply write 
 $m = \prod_{a\in m}a$ where the product can be taken in any order. 
 Hence,
we can identify $\min(w)$ and $\max(w)$ with traces;
and we can write $w = \min(w) v= u \max(w)$ for some traces $u,v$. 
\begin{remark}\label{rem:extTarc}
We extend the notion of a subtrace to elements in free resource monoids with type as follows. 
Given $w\in M(\Gam,\rho,\theta)$, we say that a subset $u\sse \pos(w)$ is a subtrace 
if~$w$ has a representing trace $\wh w\in M(\Gam,\rho)$ where 
$u\sse \pos(\wh w)=\pos(w)$ is a subtrace. Hence, $u$ is a factor of $\wh w$.
\Ip we can speak that a position~$w$ is a $T$-letter or that
a factor $ab$ of~$w$ with $a,b\in \Gam$ is a \Tarc because this does not depend on its representation $\wh w$. 

Let us explain this important aspect in more detail.
Recall that a type $\theta$ is relation where, for some $R\sse \Rs$, every $(u,v)\in \theta$ 
satisfies $|u|=|v|$ and $|uv|_a\geq 1\implies \rho(a)=R$ for all $a\in \Gam$. 
For example, we might have $(cab,bac)\in \theta$ for some $a,b,c\in \Gam$ with $\rho(a) = \rho(b) = \rho(c) = R$.
Now, consider the \HD $H = H(\wh w)$ of a representing trace $\wh w\in M(\Gam,\rho)$ and, for the sake of explanation, let $r$ be a fresh selfie with $\rho(r) = R$. 
If we replace in $H$ every position of an $R$-letter by the fresh letter $r$, then we obtain a \HD $H'$ in $M(\Gam\cup \os r,\rho)$.
The \HD{s} $H$ and $H'$ have exactly the same set of positions and \HAs. 
The only information lost in $H'$ are the labels of $R$-positions. 
The crucial point is that $H'$ is well-defined for every $w \in M(\Gam, \rho, \theta)$. 
For a slightly more abstract explanation we refer to 
factorization in (\ref{eq:canseq}). The canonical \morph {}from
$M(\Gam,\rho,\theta)$ to $M(2^\Rs,\rho)$ maps an element
in $M(\Gam,\rho,\theta)$ to a trace in $M(2^\Rs,\rho)$ without changing the \HAs or positions. For a position $p$ it just replaces its label $\lam(p)$ by a new label $\rho(\lam(p))$.  
\hspace*{\fill}$\diamond$\end{remark}

\subsection{Mapping the positions along a \morph forth an back}\label{sec:mapos}

Let $\phi:M(\Gam,\rho)\to M(\Gam',\rho')$ be a \morph between free $\Rs$-monoids.
Then $\phi$ maps a labeled partial order $w= [V,\leq,\lam]$, representing a trace in $M(\Gam,\rho)$ where $V=\pos(w)$, to a labeled partial order $\phi(w)=[V',\leq,\lam']$, representing a trace in $M(\Gam',\rho')$ where $V'=\pos(w')$.
But there is also a canonical 
mapping $\phi^*$ in the other direction which maps the set of positions~$V'$ to~$V$.
Informally, $\phi^*$\index{pre-imae of a position}\index{phi@$\phi*$} assigns a pre-image of a position;
formally the definition of $\phi^*$ is given by induction of $\abs{V}$. For $\abs{V}=0$ the notation $\phi^*$ refers to the identity of the empty set,
\ie no positions from $V'$ are mapped.
Let $V=\os{1\lds \ell}$ with $\ell\geq 1$ and $\lam(i)=a_i$.
Hence, we can write $w=a_1\cdots a_\ell\in M(\Gam,\rho)$. 
Since $\phi$ is a \morph of
free resource monoids, $\rho(a)\cap \rho(b)=\es$ implies that
$\rho'(\phi(a))\cap \rho'(\phi(b))=\es$.
Thus, $h(w)$ has a (unique) factorization $\phi(a_1) u$.
(Note that $\phi(a_1)=1$ is possible and allowed.) 
By induction, $\phi^*$ maps the set of positions $U$ of $u$ to $\os{2\lds \ell}$. 
Hence, it maps $U$ to $\os{1\lds \ell}$ by the inclusion of $\os{2\lds \ell}$ into $\os{1\lds \ell}$. It remains the define $\phi^*$ on the positions of $\phi(a_1)$:
we map all positions of this set to the position $1\in \os{1\lds \ell}$. 
Actually, the induction shows that $\phi^*$ is a mapping between partially ordered sets, $\phi^*:(V',\leq)\to (V,\leq)$. That is, 
$i\leq j$ implies $\phi^*(i)\leq  \phi^*(j)$. 

Note that $\phi^*$ is injective \IFF $|\phi(a_i)|\leq 1$ for all $i\in V$. It is surjective \IFF $|\phi(a_i)|\geq 1$ for all $i\in V$. Moreover, for $\phi = \psi \sig$ we have $\phi^* = \sig^*\psi^*$ which is the reason to use an upper-star notation rather than the lower star.
\begin{figure}[h]
	\begin{center} 
		\begin{tikzpicture}
		\node[circle] (w) at (-0.7,0) {$w=\quad$}; 
		\node[circle] (a) at (0,0.3) {$a$}; 
		\node[circle] (b) at (0,-0.3) {$b$};
		\node[circle] (c) at (1,0) {$c$};
		\node[circle] (ar) at (2,0.3) {$a$};
		\node[circle] (d) at (2,-0.3) {$d$};
		\draw (a) edge[->, >=latex,thick] (c);
		\draw [thick,->, >=latex] (b) to (c);
		\draw (c) edge[->, >=latex,thick] (ar);
		\draw (c) edge[->, >=latex,thick] (d);
		\node[circle] (phiw) at (4,0) {$\mapsto \quad bcbc= \phi(w)$};
		\end{tikzpicture}\end{center}
	\vspace{-1cm}
	\caption{Let $w=abcad$ depicted on the left by its \HD. Define $\phi(a)=1$, 
	$\phi(b)=b$, $\phi(c)=c$, and $\phi(d)=bc$. Then we can write
	 $\phi(w)=bcbc$ as a sequence of positions $pqrs$. In $\pos(w)$ 
	 we have $b=\phi^*(p)$, $c=\phi^*(q)$, and $d=\phi^*(r)= \phi^*(s)$.
	 Thus, $\phi^*(\phi(w))$ is the subtrace $bcd$ in $w$.}\label{fig:upstar}
\end{figure}

\section{Equations in $\Rs$-monoids}\label{sec:equations}

\subsection{Systems with recognizable constraints}\label{sec:syrecco}
We begin with 
an initial system $\cS$ of word equations over a finite resource alphabet $(A,\rho)$ and set of variables 
$\cXinit=\os{X_1,\ov{X_1}\lds X_k,\ov{X_k}}$ with \invol without fixed points for variables. Since the family of \edtol languages is closed under finite unions,
we fix without restriction a resource $\rho(X)\sse \Rs$ from the very beginning for all $X\in \cXinit$. Hence $(A\cup \cXinit,\rho)$ becomes a resource alphabet. 
The resource mapping $\rho:A\cup \cXinit\to 2^\Rs$ 
has to satisfy $\es\neq \rho(a)\neq \Rs$ for all $a\in A$
and $\rho(X)\neq \Rs$ for all $X\in \cXinit$. We also suppose that $2\leq |\Rs| \leq 2 +|A|^2$ 
which is no restriction by \prref{ex:fpcmres}.

More generally, let $\Gam$ be a disjoint union of two resource alphabets with involution $(B,\rho)$ and $(\cX,\rho)$. 
We call $B$ the set of constants and $\cX$ the set of variables.
A \emphind{system of word equations} in $M(\Gam,\rho)$ is  given as a finite set $\cS=\set{U_i{=}V_i}{i=1\lds s}$ where 
$U_i,V_i$ are elements in $M(\Gam,\rho)$ which are conveniently represented by their \HD{s}, which in turn can be written as representing words in
$\Gam^*$.

Since the equations are in $\M= M(\Gam,\rho)$ we have a natural interpretation in quotient monoids. For that let 
$S\sse M(\Gam,\rho)\times M(\Gam,\rho)$ be a relation and 
$\psi:M(\Gam,\rho)\to \M/\set{u=v}{(u,v)\in S}$ be the canonical 
projection onto the quotient monoid. 
Then we we have a natural interpretation 
of $\psi(\cS)$ in the quotient.

In particular, this concept applies to resource monoids with type $M(\Gam,\rho,\theta)$, which we need to solve equations in free resource monoids $M(\Gam,\rho)$.
Working with $M(\Gam,\rho)$ is also the main tool  
to describe (all) \solu{s} of equations in a free partially commutative group $G(\Gam,\rho)$ because we are able to reduce the problem of solving equations in $G(\Gam,\rho)$ to 
solving equations in $M(\Gam,\rho)$ with recognizable constraints.
These constraints allow us to keep solutions in reduced normal form, see \prref{sec:tgr}. 
There are various ways to specify $\Rec(M(\Gam,\rho))$. 
Here, a \emphind{recognizable constraint} is specified by a \morph $\mu_\init:M(A\cup \cX,\rho)\to N$ to a finite monoid  with \invol.  
It is convenient to assume that the monoid $(N,\cdot)$ always has a zero~$0$. 
This is true if $|N|=1$ (with $1=0$) but for $|N|>1$ we have $\bar 1= 1\neq 0=\bar 0$. 
{We extend $\mu_\init$ to a \morph $\mu:M(\smash{\wh A}\cup \cX,\rho)\to N$ by $\mu(a,T)=\mu_\init(a)$ for all $(a,T)\in \smash{\wh A}\sm A$. Thus, we can write $\mu_\init=\mu\wh \pi$ which factorizes as follows. 
\begin{align}\label{eq:muinit}
\mu_{\init}:M(A\cup \cXinit,\rho)\hookrightarrow 
M(\wh A\cup \cXinit,\rho)\arc {\mu}
\Ninit
\end{align}
}
For complexity issues we assume that $\Ninit$ is `admissible' in the following sense. 
\begin{definition}\label{def:negligibleN}
A finite monoid $N$ is called \emph{admissible\index{admissible monoid}} if every element of $N$ can be encoded by 
$\Oh(\log|N|))$ bits in such a way that monoid computations (equality testing, involution and multiplication)
can be performed by a transducer in $\Oh(\log|N|))$ space. 
\end{definition}
\begin{example}
Let $n=|A|+|\cXinit|$. Suppose 
we define the initial recognizable constraints $\mu$
by using an $M(A\cup \cXinit,\rho)$-NFA with $m$ states and where the \tras are triples $(p,a,q)$ with $p,q\in \os{1\lds m}$ and $a\in A$. 
Then we can realize $\mu$ (which has to respect the \invol)
by a \morph $\mu:M(A\cup \cXinit,\rho)\to N$ with $N= \B^{2m \times 2m}$ such that
the \invol on the set of Boolean matrices $\B^{2m \times 2m}$ is the transposition, see \eg~\cite{dgh05IC}.
The specification of $\mu$ as a list 
which assigns to each $x\in A\cup \cXinit$ a Boolean matrix is clearly admissible. 
Moreover, we have $\log|N|\in n^{\Oh(1)}\iff m\in n^{\Oh(1)}$ since $\log|N|=4m^2$.
\hspace*{\fill}$\diamond$\end{example}
\begin{definition}\label{def:soluCS}
Let $\psi:M(B,\rho)\to M$ be a \morph of monoids with \invol
and $\cS=\set{U_i{=}V_i}{1\leq i \leq s}$ be a system of word equations 
in $M(B\cup \cXinit,\rho)$ with recognizable constraints given by \morph $\mu:M(B\cup \cX,\rho)\to N$. 

A \emph{solution\index{solution of system of word equtions}} of $\cS=\set{U_i{=}V_i}{1\leq i \leq s}$ with a recognizable constraint $\mu$ is 
a $B$-\morph $\sig: M(B\cup \cX,\rho)\to M(B,\rho)$ such that firstly, 
$\mu(X)= \mu\sig(X)$ for all $X\in \cX$
(\ie the constraints are met) and secondly, $\psi\sig(U_i) = \psi\sig(V_i)$ holds in $M$ for all $1\leq i\leq s$.

The \emph{size\index{size of system of word equtions}} $\Abs{\cS}$ is defined 
by 
\begin{align}\label{eq:sizecS}
\Abs{\cS} = \abs \Rs + \abs B+ \abs \cXinit + \sum_{i=1}^s(1+\abs{U_iV_i})
\end{align}

The pairs $(U_i=V_i)\in \cS$
are called the \emphind{inner equations} of $\cS$.
\end{definition}
We speak about word equations since 
we can represent traces in $M(B\cup \cXinit,\rho)$
by words and then
$\rho$ and the set if \inner equations $\cS$ define \epi{s}.
\[(B\cup \cXinit)^*\to M(B\cup \cXinit,\rho)\to M_\cS= M(B\cup \cXinit,\rho)/\set{(U_i=V_i)\in \cS}{1\leq i \leq s}\]
such that a \solu $\sig$ corresponds to a $B$-\morph
$\sig:(B\cup \cXinit)^*\to M_\cS$ respecting~$\mu$.

\section{The main results of the paper}
\label{sec:mainMG} 
\subsection{Free resource monoids}\label{sec:trm} 
For stating complexity results, we suppose that $N$ is admissible 
(as in \prref{def:negligibleN})
and  
we use the following notion of \emphind{input size} for $\cS$ (as defined in \prref{eq:sizecS}).
\begin{theorem}\label{thm:uro} There is an effective construction of an $\NSPACE(2^{\Oh({\abs \Rs}^2)}n \log (n\cdot |N|))$-transducer which performs the following task.

{\noindent \textbf{Input:}} A set $\Rs$ of resources, a system of (\inner) equations $\cS$ over $(A\cup \cXinit,\rho)$  of size $n=\Abs{\cS}$ with constraints given by a \morph $\muinit:M(A\cup \cX
,\rho)\to \Ninit$ where $\Ninit$ is admissible according to \prref{def:negligibleN}.

{\noindent \textbf{Output:}}
An extended resource alphabet $(\wh C,\rho)$ with $|\wh C\sm \wh A| \in \Csize$ with an inclusion $(A,\rho)\sse (\wh C,\rho)$, distinguished letters $d_1\lds d_k \in \wh C$ with $d_i\neq \ov{d_i}$ for all $1\leq i \leq k$, and an NFA $\cA$ accepting a rational set $\cL$ of $A$-\Endo{s} over the free resource monoid $M(\wh C,\rho)$ such that
we obtain the following equality of \edtol languages. 
\begin{align}\label{eq:EQh}
\begin{split}
&\set{(h(d_1),\ldots ,h(d_k))}{h \in \cL} \\
&=\set{\sig(X_1),\ldots ,\sig(X_k))}{\sig \text{ solves } \cS\text{ in $M(A,\rho)$ with $\muinit$-constraints}}.
\end{split}
\end{align}
Furthermore, $\cS$ has a solution \IFF the NFA $\cA$ accepts a nonempty set; and $\cS$ has infinitely many solutions \IFF 
 $\cA$ has a directed cycle. \\
 Moreover, these conditions can be tested in $\NSPACE(2^{\Oh({\abs \Rs}^2)}n \log (n\cdot |N|))$, too. 
 \end{theorem} 

\begin{corollary}\label{cor:uro1}
If, in \prref{thm:uro}, the set of resources $\Rs$ or the alphabet $A$ is of constant size, then we obtain an $\NSPACE(n \log(n\cdot |N|))$ transducer, and the decision problems `emptiness' and `finiteness' can be decided in $\NSPACE(n \log(n\cdot |N|))$. 
\end{corollary}

\begin{proof}
  If the alphabet $A$ is of constant size, then we can chose 
$\Rs$ such that $|\Rs|\leq |A|^2$. Hence $\Rs$ is of constant size, too.  
  In both cases $2^{\Oh({\abs \Rs}^2)}$ is thus a constant, and the corollary is a special case of the theorem.
\end{proof}

\subsection{Free partially commutative groups}\label{sec:tgr}
`La raison d'{\^e}tre' for \prref{thm:uro} is to a great extent that 
we aim to apply the result to free partially commutative groups as stated in \prref{cor:mostint}.

We let $(A,\rho)$ be resource alphabet.  
$\M=M(A,\rho)$ denotes the 
free $\Rs$-monoid and 
$\G=M(A,\rho)/\set{a\bar a=1}{a\in A}$ is the associated free partially 
commutative group.\footnote{If $B=\bar B\sse A$, then everything would work 
for $\G_B=M(A,\rho)/\set{a\bar a=1}{a\in B}$ as well.} Note that special cases of $\G$ include the following.
\begin{itemize}
\item Right-angled Artin groups (graph groups or RAAGs), where $A$ is without selfies.
\item Right-angled Coxeter groups (RACGs), where every $a\in A$ is a selfie.
\end{itemize}

It is well-known that we can embed $\G$ as a set into $\M$ by representing every group element in $\G$ by a trace without a subtrace in 
the finite set of (forbidden) factors\index{forbidden factors} $\cF=\set{a\bar a}{a\in A}$; for a proof based on trace-rewriting systems see \cite{die90}. 
Thus, if $\psi:\M\to\G$ is the canonical projection, 
then there is a bijection
\begin{align}\label{eq:nfG}
\nf: \G\to \M \sm A^*\cF A^*
\end{align}
such that $\psi\nf=\id_\G$. 
It is also well-known that
$\nf(\G)$ is a recognizable subset in $\M$. In order to see this
we recognize $\nf(\G)$ by a \morph to a finite monoid $N_\G$ with \invol. 
The elements in $N_\G$ are the
triples $(\min(x),\rho(x),\max(x))$ with $x\in \M$ together with an extra 
zero~$0\in N_G$. 
This yields a mapping $\mu_A$ as follows.
\begin{align}\label{eq:muGG}
\mu_A:\M\to N_\G \quad x\mapsto (\min(x),\rho(x),\max(x))
\end{align}
Having the same image under $\mu_A$ defines an equivalence relation 
of finite index which is easily seen to be a congruence. 
The multiplication in $N_\G$ between non-zero elements is given by
letting $x,y\in \M$ and 
\begin{align}\label{eq:multNG}
\mu_A(x) \cdot 
\mu_A(y)=\begin{cases}
\mu_A(xy) &\text{if $\max(x)\cdot\min(y)\notin A^\ast\cF A^\ast$}
\\0 &\text{otherwise}
\end{cases}
\end{align}
Note that the multiplication in $N_\G$ is well-defined
and $N_\G$ is another example of an admissible monoid with \invol given by
$\ov{(s,r,t)}= (\bar t,r,\bar s)$ and $\ov{0}=0$.
The neutral element is the triple $(\es,\es,\es)$. 
The size of $N_\G$ is bounded above by  
\begin{align}\label{eq:sizeNG}
|N_\G|\leq |\wh A \times 2^\Rs \times \wh A|= 2^{|\Rs|\cdot (2\log |A| +1)} 
\end{align}

Using the morph $\mu_A$ allows to express an inequality in $\M$ or in $\G$
 by equalities with constraints in $N_\G$. This follows from 
 \prref{lem:inequ}.
\begin{lemma}\label{lem:inequ}
Let $x,y\in \M$ be traces, then we have
$x\neq y$ if there are some $p,u,v\in \M$
with $x=pu$, $y=pv$, and $\mu_A(u)\neq \mu_A(v)$.
We have $\psi(x)\neq \psi(y)$ in $\G$ \IFF 
there are some $p,u,v\in \M$
with $\nf(x)=pu$, $\nf(y)=pv$, and $\mu_A(u)\neq \mu_A(v)$.
\end{lemma}
\begin{proof}
We have $x\neq y$ if there are some $p,u,v\in \M$
such that $x=pu$, $y=pv$, and $\min(u)\neq \min(v)$. 
Now, $\min(u)\neq \min(v)$ implies $\mu_A(u)\neq \mu_A(v)$ and that implies $u\neq v$. This implies $x\neq y$ because $\M$ is cancellative.
\end{proof}

\begin{convention}\label{conv:NG}
When considering word equations in $\G$ we assume that we also use for the recognizable constraints an admissible 
finite monoid $N$ which can be written as a direct product 
$N=N_\G \times N'$. Note, if $|N'|\neq 1$, then $(0,1)$ is not a zero in $N$.
\end{convention}
Be aware of the following obstacle.
If $\psi(x)=\psi(yz)\in \G$, then this does not imply
$\mu_A(x)=\mu_A(y)\mu_A(z)$, in general.
To deal with this issue, we triangulate the system of word equations as described in the following section.
\subsubsection{Triangulating a system of word equations}
Let $(\cX,\rho)$ be a resource alphabet of variables and 
$\cS$ be a system of (\inner) trace equations over 
$M(A\cup \cX,\rho)$ with recognizable constraints 
$\mu:M(A\cup \cX,\rho) \to N$ according to \prref{conv:NG}
such that $\mu(X)\notin \os{0}\times N'$ for all $X\in \cX$. (Thus, $\mu(X)\neq 0$ in the Rees quotient $N/(\os{0}\times N')$.)

A \solu $\cS$ is defined (as usual) by an $A$-\morph $\sig:M(A\cup \cX) \to \M$ such that firstly $\psi\sig(U)=\psi\sig(V)$ for all 
\inner equations $U=V$ 
and secondly $\mu(X)=\mu\sig(X)$ for all  $X\in \cX$.
Note that $\mu(x)$ is defined for $x\in M(A\cup \cX,\rho)$, but not 
for $x\in \G$.

The idea is replace $\psi\sig(U)=\psi\sig(V)$ by $\sig(U)= \nf\psi\sig(U) = \nf\psi\sig(V) =\sig(V)$.
But before we can do so we need to triangulate the system $\cS$. 
The procedure is based on how to compute the normal form of a product of normal forms, \ie on the following fact (which is easy to see by induction): let $y,z\in \nf(\G)$,  
then there are $p,r,q\in \nf(\G)$ such that 
$y=pr$, $z=\bar r q$, and $\nf(yz) = pq$.

The triangulation is now defined by the following steps which introduce various fresh variables. We keep the invariant that for each variable 
$X$ there is another variable $\bar X\neq X$. 
After each step we rename the new system again as 
$\cS$. 
\begin{enumerate}
\item Replace each $U=V$ in $\cS$ 
by two equations $X=U$ and $X=V$ where $X$ is fresh. 
\item Replace each $X=xU$ in $\cS$ with $x\in A\cup \cX$ and $|U|\geq 2$ 
by two equations $X=xY$ and $Y=U$ where $Y$ is fresh.
\item Replace each $X=yz$ in $\cS$ with $y,z\in A\cup \cX$ 
by three equations $X=PQ$, $y=PR$, and $z=\bar R Q$, where
$P,Q,R$ are fresh variables. 
\end{enumerate}

Let $\sum_{(U=V)\in \cS}|UV|\leq n$ and 
$\cS'$ denote the final system,
then we can compute $\cS'$ by an $\NSPACE(n \log n)$-transducer.
It is not deterministic since for each fresh variable $X$
we have to guess $\mu(X)\in N$ with $\mu(X)\notin {0}\times N'$. 
\begin{proposition}\label{prop:trias}
The nondeterministic procedure to replace $\cS$ by the triangulated 
system $\cS'$ is sound: 
if $\cS'$ has a \solu defined by $\sig':\cX'\to \M$, then the restriction of $\sig'$ defines a \solu $\sig$ of $\cS$.
It is complete: if $\cS$ has a \solu, then 
there is a possible output $\cS'$ such that for all \inner equations $U'=V'$ we have $\mu(U')=\mu(V')$ and there is a \solu $\sig'$ satisfying and $\mu\sig'(U')=\mu\sig'(V')$.
\end{proposition}
\begin{proof}
This follows from the construction and that $y,z\in \nf(\G)$ we can write (as stated above)
$y=pr$, $z=\bar r q$, and $\nf(yz) =pq$.
\end{proof}

\subsubsection{Normalized regular languages}\label{sec:nrl}
The notion of \emphind{normalized regular} languages was introduced 
in \cite{dm06} to solve word equations in graph groups with 
constraints in that class. It can be defined in every 
finitely generated monoid $M$ with a regular set of normal forms. 
In this case it yields a Boolean algebra which sits between $\Rec(M)$ and $\Rat(M)$ containing all finite and co-finite subsets of $M$. \Ip if $M$ is an infinite group, then this class is strictly larger than $\Rec(M)$. 
\begin{definition}\label{def:NREG}
Let $A$ be finite and $\psi:A^*\to M$ be a surjective \hom onto~$M$.
A word $u\in A^*$ is called a \emph{geodesic\index{geodesic word}} if 
$|u|\leq |v|$ for all $v\in \oi{\psi}(\psi(u))$.

Suppose that there is a mapping $\nf:M\to A^*$ such that 
$\psi\nf=\id_M$. We say that $\nf$ is \emph{geodetic\index{geodetic normal form}} if every word in 
$\nf(M)$ is geodesic.

If $\nf(M)$ is a regular language, 
then $\NReg(M)$ denotes the class of \emph{normalized regular languages\index{normalized regular language}} and is defined by
\(\NReg(M) = \set{L\sse M}{\nf(L) \in \Reg(A)}\).
\end{definition}
\begin{proposition}\label{prop:NREG}
\(\set{F\sse M}{\abs F < \infty}\cup \Rec(M) \sse \NReg(M) \sse \Rat(M)
\sse \EDTOL(M)\) and $\NReg(M)$ is a Boolean algebra.
\end{proposition}
\begin{proof}
The definition 
of \edtol languages for finitely generated monoids in \prref{sec:EDT0L_lang} implies $\Rat(M)\sse \EDTOL(M)$. Since $A$ is finite, we see that 
$\Reg(A)=\Rat(A^*)=\Rec(A^*)$ is closed under homomorphic images. This yields the line of inclusions as stated in the proposition. 
Since $\Rec(A^*)$ is a Boolean algebra, the same property holds for 
$\NReg(M)$.
\end{proof}
Choosing any linear order on $A$ we can define for every 
surjective \hom $\psi:A^*\to M$ 
the \emphind{short-lex normal form} $\nfslex(x)$ for elements $x\in M$ by choosing the lexicographic first word (\wrt to the chosen linear order of $A$) among all geodesic words~$w$ which satisfy $\psi(w)= x$. If $\oi \psi(x)$ is finite for all $x\in M$, then another normal form is $\nflex$ which chooses for $x\in M$
just the lexicographic first word.
A result of Ochma{\'n}ski \cite{och85} says\footnote{Proofs also can be found, for example, in \cite[Thm.~2.3.1]{die90}) or \cite[Thm.~6.3.16]{dr95}} that the language $\nflex(\M)$ is 
regular for 
the canonical \hom $\pi:A^*\to \M$. It is geodetic since all words in  
$\oi{\pi}(x)$ have the same length. Now consider the canonical \hom $\phi:\M\to \G=\M/\set{a\bar a =1}{a\in A}$, 
then for each element $g\in\G$ the trace $\nf(g)$ (defined in \ref{eq:nfG}) is the uniquely defined
shortest trace $u_g\in \M$ such that $\phi(u_g)=g$. 
As stated above, it is known that $\nf(\G)$ is a recognizable subset
of traces. This implies that $\slex$ defines a 
geodetic normal form for $\psi:A^* \arc{\pi}\M \arc{\phi}\G$ such that $\slex(\G)\sse A^*$ is regular; and it is defined by the set of words $u\in\nflex(\M)=\nfslex(\M)$ such that the trace $\pi(u)$ is without a factor
$a\bar a$ with $a\in A$.

When referring to normalized regular languages of $\M$ and $\G$, we subsequently assume these to be with respect to the above normal forms.
As such, we write
\begin{align}\label{eq:NREGRAAG}
  \NReg(\M)&=\set{L\sse \M}{\nflex(L) \in \Reg(A)} \\
  \NReg(\G)&=\set{L\sse \G}{\nfslex(L) \in \Reg(A)} 
\end{align}
Let us give a few explicit examples for $\NReg(M)$ where $M=\M$ or $M=\G$.
\begin{itemize}
\item For the free resource monoid
$\M=M(A,\rho)$ we have $\NReg(\M) =\Rec(\M)$.  This is 
Ochma{\'n}ski's Theorem \cite{och85} together with \prref{prop:rhotomu}.
\item If $M=\N \times \N$, then $\Rec(M) =\NReg(M)$ because 
$M$ is a free resource monoid. 
 We have $\NReg(M) \neq \Rat(M)$ because 
$\set{(n,n)}{n\in \N}\in \Rat(\N \times \N) \sm \Rec(\N \times \N)$.
\item If $F=A^*/\set{a\bar a =1}{a\in A}$ is a free group $F$ where
$A = \bar A \neq\es$ is without selfies, then we have $\Rec(F) \varsubsetneq \NReg(F) = \Rat(F)$. The first inequality is valid in every infinite group, because finite sets are normalized regular but never recognizable in an infinite group. 
The equality $\NReg(F) = \Rat(F)$ is due to Benois \cite{ben69}. 
\item Finally, for $M=\Z \times \Z$ we have 
\begin{equation}\label{eq:sepNREG}
\Rec(Z \times \Z) \varsubsetneq \NReg(Z \times \Z) \varsubsetneq \Rat(Z \times \Z).
\end{equation}
\end{itemize}
Finite subsets and the diagonal $\set{(n,n)}{n\in \N} \sse \Z \times \Z$ show the two strict inclusions in (\ref{eq:sepNREG}). 

The reason to define normalized regular languages in the context of a graph group $\G$ is guided by the search for a natural class of constraints which still allows
the decidability of the existential theory with constraints of $\G$ by a reduction to the existential theory with constraint of $\M$. In general 
we cannot include $\Rat(\G)$ because a result of Muscholl in  
\cite[Prop. 2.9.2 and 2.9.3]{mus99habil} says that 
that the existential theory with constraints of a trace monoid 
$\M$ is decidable \IFF the monoid $\M$ is a free product of free commutative monoids.
\begin{theorem}\label{thm:guro}
Let $\M= M(A,\rho)$ and $\G= G(A,\rho)$ be the associated free resource group. Let 
$\muinit:M(A\cup \cX
,\rho)\to N_\G \times \Ninit$ where $\Ninit$ is admissible according to \prref{def:negligibleN}. (Hence, $N_\G \times \Ninit$ is admissible, too.)
Let $\cS$ be an initial system of word equations over $\M$
which is viewed as a set of defining equations over the group $\G$.
Let $n=\Abs{\cS}$.
There is effectively an $\NSPACE(2^{\Oh({\abs \Rs}^2)}n \log( n \cdot |N|))$
transducer which, on the input above, produces the following output:

An extended resource alphabet $(\wh C,\rho)$ with $|\wh C\sm \wh A| \in \Csize$ with an inclusion $(\wh A,\rho)\sse (\wh C,\rho)$, distinguished letters $d_1\lds d_k \in \wh C$, and an NFA $\cA$ accepting a rational set $\cL$ of $A$-\Endo{s} over the free resource monoid $M(\wh C,\rho)$ such that
we obtain the following equality of \edtol languages. 
\begin{align}\label{eq:gEQh}
\begin{split}
&\set{(h(d_1),\ldots ,h(d_k))}{h \in \cL} \\
&=\set{\sig(X_1),\ldots ,\sig(X_k))}{\sig \text{ solves } \cS \text{ in the group $\G$ with $\muinit$-constraints}}.
\end{split}
\end{align}
Furthermore, $\cS$ has a solution \IFF the NFA $\cA$ accepts a nonempty set; and $\cS$ has infinitely many solutions \IFF 
 $\cA$ has a directed cycle. \\
 These conditions can be tested in $\NSPACE(2^{\Oh({\abs \Rs}^2)}n \log (n\cdot |N|))$, too. 
\end{theorem} 
\begin{proof}
The proof is a consequence of \prref{thm:uro}, since we can identify
$\G$ as the recognizable set $\nf(\G)$ of traces in $\M$. 
In order to cope with constraints we triangulate the system. 
This yields a linear blow-up in the size of the initial system, only. 
For a triangulated system we use \prref{prop:trias}. 
\end{proof}
Just as for $\M$ we can derive the very same non-uniform complexity.
\begin{corollary}\label{cor:guro1}
If, in \prref{thm:guro}, the set of resources $\Rs$ or the alphabet $A$ is of constant size, then we obtain an $\NSPACE(n \log(n\cdot |N|))$ transducer, and the decision problems `emptiness' and `finiteness' can be decided in $\NSPACE(n \log(n \cdot |N|))$. 
\end{corollary}

Combining \prref{thm:uro} and \prref{thm:guro} (as well as \prref{cor:uro1} and \prref{cor:guro1}), we obtain the following result, which was stated in \prref{sec:aplraacgss}.

\begin{corollary}\label{cor:mostint}
Let~$M(A,I)$ be a free partially commutative monoid with involution, and let~$G(A,I)$ be its associated quotient group, \eg a right-angled Artin or Coxeter group.
Then, given a system of word equations over~$M(A,I)$ with recognizable constraints, the set of all solutions in~$M(A,I)$ or $G(A,I)$ is an EDT0L language.
Moreover, in case $(A, I)$ is fixed, satisfiability and finiteness of systems of word equations can be decided in $\NSPACE(n \log n)$ where $n$ is the input size.
\end{corollary}

\subsubsection{Quasiconvex and normalized regular subsets and subgroups}\label{sec:quasi}
Quasiconvexity is a natural relaxation of being convex in a metric space.
For RAAGs, it was defined in \cite{KMW2017}. 
Let us show that being normalized rational is closely related to this notion of being
quasiconvex. Let $M$ be a finitely generated monoid with \invol and 
$\pi: A^*\to M$ be an \epi where $A$ is a finite alphabet with \invol.
For a subset $L\sse M$ let 
$\geo{L} \sse A^*$ denote the set of geodesic words
in $\oi{\pi}(L)$. That is 
\begin{align}\label{eq:defgeo}
  \geo{L}=\set{u\in \oi{\pi}(L)}{\forall v\in A^*: \pi(u) =\pi(v) \implies \abs u \leq \abs v}
\end{align}
\begin{definition}\label{def:gromqc} 
A subset $L$ in $M$ is called \emph{quasiconvex\index{quasiconvex subset}} \wrt~$\pi$ if there is some $d\in \N$ such that  
for all $u,v\in A^*$ with $uv\in  \geo{L}$ there is some $p\in A^*$ with $|p|\leq d$ satisfying 
$\pi(up) \in L$.  
If the parameter $d$ is given, then $L$ is also called \emph{$d$-quasiconvex\index{d-quasiconvex subset@$d$-quasiconvex subset}}.
\end{definition}

For $d=0$ the definition says that $\geo{L}$ is prefix closed.
Moreover, it is well-known and easy to see that a quasiconvex subgroup $H$ of a finitely generated group $G$ is finitely generated.
This statement is shown in the proof of \prref{prop:dm06}. 
The hypothesis of that proposition requires 
a certain normal form $\nf$ such that $\nf(G)$ is regular, but the reader can check that the finiteness of the NFA accepting $\nf(G)$
is not used to show that $H$ is finitely generated. 

Note also that the assertion fails for submonoids in groups.
For example, $(0,1) + \N\times \N$ is a $1$-quasiconvex submonoid in $\Z\times \Z$ which is not finitely generated. 
\begin{lemma}\label{lem:nreginqc}
Let $\pi: A^*\to M$ as above with a normal form mapping 
$\nf:M\to A^*$ such that $\nf(M)$ is a regular language of geodesic words in $A^*$.
If $L\sse M$ is normalized regular \wrt~$\pi$, then 
$L$ is  quasiconvex \wrt~$\pi$.
\end{lemma}
\begin{proof}
Since $\nf(L)\sse A^*$ is regular, $\nf(L)$ is accepted by some NFA with $d$ states. Now, for every 
$uv\in \nf(L)$, after reading the prefix $u$ we are in some state and there is a word $p\in A^*$ of length at most $d$ such that $\pi(up)\in L$. As such, $L$ is $d$-quasiconvex. 
\end{proof}

Consider $L=\set{(a,b)\in \N \times \N}{a = 0 \lor (a=1 \text{ and } b \text{ is a power of } 2)}$ with the natural presentation and $\nf(\N \times \N)= a^*b^*$. Then $L$ is quasiconvex but not normalized regular.
Thus, the converse of \prref{lem:nreginqc} does not hold, in general.
This changes if $\pi: A^*\to G$ is an \epi  onto a group with $\pi(\bar a) = \oi {\pi(a)}$, and $H$ is a subgroup of $G$. 
In this case we will show that $H$ is normalized regular \IFF $H$ is quasiconvex.
The property of a subgroup to be quasiconvex depends on $\pi$, in general.\footnote{This is contrast to hyperbolic groups, see \cite{gro87}.}
Indeed, let $G=\Z\times \Z$ be generated (as a group) by 
$a=(1,0)$ and $b=(0,1)$. The cyclic subgroup $H= \set{a^nb^n}{n\in \Z}$ is not quasiconvex since the prefix 
$a^n$ of $a^nb^n$ is at distance $|n|$ from the nearest element in $H$. 
However, if we replace $b$ by a new generator $c=(1,1)$,
then $H= \set{c^n}{n\in \Z}$ becomes quasiconvex \wrt~$\pi$ which is induced by $\os{a,c}\sse G$. 

The following proposition shows that 
a subgroup~$H$ of \fresgro $\G= G(A,\rho)$ is quasiconvex \IFF the set $H$ is normalized regular \wrt~the canonical \epi $\pi:A^*\to \G$ and 
the short-lex ordering~$\nflex$. All we need for $\G$ is Ochmanski's Theorem \cite{och85} which is stated above and says that $\nflex(\G)$ is regular. \prref{prop:dm06} was shown \cite{dm06} for \fresgro{s}, but its proof is valid in a more general setting.
\begin{proposition}\label{prop:dm06}
Let $\pi: A^*\to G$ be an \epi as above with a normal form mapping 
$\nf:G\to A^*$ such that $\nf(G)$ is a regular language of geodesic words in $A^*$.
Then, a subgroup $H$ of $G$  is quasiconvex \wrt~$\pi$ \IFF the set $H$ is normalized regular \wrt~$\pi$. Moreover, if 
$H$ is $d$-quasiconvex, then $H$ is  generated 
by at most $|A|^{2d +1}$  generators.
\end{proposition}
\begin{proof}
First, let $H$ be $d$-quasiconvex for some $d\in \N$. As an intermediate step we construct an NFA $\cA$ which accepts a super set of $\geo{H}$ such that 
$\pi(L(\cA))= H$. The state set of $\cA$ is the set of geodesic words  $p$ where 
$p\in A^*$ and $\abs p \leq d$. The initial state is equal to the final state which is the empty word. The invariant of the construction is that after reading the prefix $u$ of a geodesic word 
$uv \in \geo{H}$ we are in some  state $p\in Q$ such that $\pi(up)\in H$. Moreover, we make sure that $\pi(u)\in H\iff p=1$.
Let us define the \tra relation  $\del\sse Q\times A \times Q$ by 
\[(p,a,q)\in \del \iff \pi(\bar p a q) \in H\]

Now, consider $a_1\cdots a_m\in \geo{H}$ where $a_i\in A$.
For every $0\leq i \leq m$ there is some $p_i\in Q$ such that $\pi(a_1\cdots a_i p_i) \in H$. We choose $p_0=p_m=1$ which is the initial and final state. Since $\pi(\bar a)=\oi{\pi(a)}$ we have
\[\pi(a_1\cdots a_i p_i)= \pi(a_1\cdots a_{i-1}p_{i-1})\cdot \pi(\ov{p_{i-1}}a_ip_i)\in H\]
This implies, $\pi(\ov{p_{i-1}}a_ip_i)\in H$ for all $1\leq i \leq m$.
Hence, we can choose the sequence $p_0 \lds p_m$ in such a way that
$(\ov{p_{i-1}},a_i,p_i)\in \del$ for all $1\leq i \leq m$.
\Ip the geodesic word $a_1\cdots a_m\in \geo{H}$ is accepted; and 
$H$ is generated $\set{paq\in A^{2d+1}}{\pi(paq)\in H}$ which is a set 
of size at most $|A|^{2d +1}$. 
We conclude $\geo{H}\sse L(\cA) \sse \oi \pi(H)$.
Since $\nf(G)$ is regular by hypothesis, the set $\nf(G)\cap L(\cA)$ is regular, too. Hence, $H$ is normalized regular according to \prref{def:NREG}. 
The converse is \prref{lem:nreginqc}.
\end{proof}

\section{Preparing the proof of \prref{thm:uro}}\label{sec:PPthmuro}
The proof of \prref{thm:uro} covers the rest of the paper.  

\subsection{Representing a systems of word equations by a single word}\label{sec:sys2word}
For convenience we transform a system of word equations with constraints $\cS$ in a free resource monoid $M(\Gam,\rho)$ into a single equation~$W_\cS$ with a very specific 
form which is shaped by the initial (input) system. 
As above $\Gam$ is a disjoint union of a set $\cX$ of variables and a set $B$ of constants 
which contains the covering alphabet $\wh A$. 
Furthermore, $B$ contains two special self-involuting symbols (selfies) $\cent$ ,$\#$. The idea is that the selfie $\#$ separates \inner equations and that $\cent$ is the central position of $\sig(W)$ if $\sig$ is a \solu.
Symbols in $\wh A\sm A$ do not appear~$\Winit$. 
We define $\mu(\cent)=\mu(\#)=0\in N$ where~$0$ is the zero in~$N$.
In addition we let $2k =|\cXinit|$ and we choose $2k$ fresh \emph{distinguished}\index{distinguished letters} letters 
in $d_1,\ov{d_1}\lds d_{k},\ov{d_{k}}$ 
with $d_i\neq \ov{d_i}$,  $\rho(d_i)=\Rs$ and $\mu(d_i)=1$.
We assume $k\geq 1$ which is the only case of interest.
The recognizable constraints are given by
a \morph $\mu:\Gam \to N$ as defined in \prref{sec:whApi}. 
\Ip $\mu(a)=\mu_{\init}\wh \pi (a)$ for all $a\in \wh A$.
The representation of such a system $\cS$ is given by a word
$W_\cS$ which has for 
$u_1\lds u_k\in\Gam^*$
the following form. 
\begin{align}
W_\cS &=U \,\cent \,{\bar{V}}
\label{eq:WcS}
\\
\label{eq:UCCS}
U
&= \# d_1u_1 \# \cdots d_ku_k \# U_1 \# \cdots U_s \# V_1 \# \cdots V_s \#\\
\label{eq:BVCS}
V &=  \# d_1u_1\# \cdots d_ku_k \# V_1 \# \cdots V_s \# U_1 \# \cdots U_s\#
\end{align}
For simplicity we call the element $W_\cS\in M(\Gam,\rho,\theta)$ an \emph{equation} (or a \emph{word equation}). 
\begin{definition}\label{def:soli}
A \solu of $W_\cS$ is 
a $B$-\morph $\sig: M(B\cup \cX,\rho,\theta)\to M(B,\rho,\theta)$
such that $\sig(\bar W)= \ov{\sig(W)}$ and $\mu\sig(X)=\mu(X)$ for all variables $X\in \cX$. 
\end{definition}
Let 
$\sig:M(B\cup\cX,\rho,\theta)\to M(B,\rho,\theta)$ be any \solu of the system $\cS$ as above according to \prref{def:soluCS}. 
Then have the following equivalence.
\begin{align}\label{eq:thesame}
\sig(W) = \sig(\bar W) \iff \sig(\bar W)= \ov{\sig(W)} \iff \forall 1\leq i\leq s: \sig(U_i) = \sig(V_i).
\end{align}
We can recover the system $\cS$ from the word $W$ since $\#$ is not used in any equation $U_i{=}V_i$.  Moreover, $\sig(W)= \sig(\bar W)$ implies $\sig(\bar X_i)=\ov{\sig(X_i)}$ for all $1\leq i\leq s$.
Consequently, given $\cS$ or $W_\cS$ leads to equivalent views.
We continue to use \inner equations $U_i=V_i$ of $\cS$ in examples and in some proofs.\footnote{The reader may wonder why we repeat the $U_i$'s and also the $V_i$'s in both Equations (\ref{eq:UCCS}) and (\ref{eq:BVCS}). We do so because  this implies that for every \HA 
$p\to q$ in $\Winit$ there is a dual \HA $\bar q \to \bar p$ with labels $\lam(\bar p) = \ov{\lam(p)}$ and  $\lam(\bar q) = \ov{\lam(q)}$. Otherwise this would not be true, in general. For example, suppose there is single \inner equation $X=ab$ where~$a$ and $b$ are constants
sharing a resource, then the \HAs $a\to b$ and $\bar b\to \bar a$ both appear in $\Winit$, but 
there is no \HA $a\to b$ in the word
$\# X \# X \#\cent\# \ov{ab}\# \bar X\#$.}

\subsection{The initial state $\cSinit$ and the equation $\Winit$}\label{sec:Winit}

We begin with an input system over $M(A\cup \cXinit,\rho)$ of input size $n$. 
During our process to an \edtol description for the full \solu set, the initial variables in $\cXinit$ disappear because, eventually, they are replaced by words over constants using all resources in $\Rs$. Fresh variables may also appear in \inner equations. Hence, the set of variables changes. We also have to enlarge the alphabet $A$ and to fix a (large enough) extended resource alphabet $(\wh C,\rho)$.
 
The resource alphabet $\wh C$ is a strict superset of $\wh A$.
It contains the distinguished letters $d_1\lds d_k$ and the 
symbols $\cent,\#$. Recall that each of these symbols $x$ satisfies $\rho(x)=\Rs$ and $\mu(x) \in \os{0,1}$. 
The condition that $\rho(X)\neq \Rs$ for all variables $X$ prevents that
any such a special symbol appears in $\sig(X)$ even if $N$ is trivial. 
As we did in \prref{sec:sys2word} we encode the defining equations in input system 
by a single element $\Winit$. 
\begin{align}
\label{eq:Winit}
\Winit &=U_\init \,\cent\,{\ov{V_\init}}
\\
\label{eq:Uinit}
U_\init
&= \# d_1 X_1\# \cdots d_k X_k\# U_1 \# \cdots U_s \# V_1 \# \cdots V_s \#  \quad\text{with $k\geq 1$} \\
\label{eq:Vinit}
V_\init &=  \# d_1 X_1\# \cdots d_k X_k \#  V_1 \# \cdots V_s \# U_1 \# \cdots U_s\#
\end{align}
Having this we define 
$\cSinit$ as the following tuple, which we call the \emphind{initial state}.
\begin{align}\label{eq:defCsinit}
\cSinit=(\Winit,A_\init,\cXinit,\mu,\rho,\es)
\end{align}
Here, $A_\init$ is the smallest alphabet with \invol, 
containing $\wh A$, the symbols $\cent,\#$, and the distinguished letter $d_1\lds d_k$ with $\rho$ and $\mu$ as defined above. 
The last component $\es$ will be changed to a nonempty subset of $\Rs$ 
when moving to another state later.
According to \prref{eq:WcS} all initial variables $X_i$ and $\ov{X_i}$ appear in $\Winit$ and in every $W_S$ the factors $\#d_1$,\ldots,$\#d_k$, $\ov{d_k}\#$\lds$\ov{d_k}\#$
appear as well.\footnote{Note that \prref{thm:uro} refers to the set of variables in $\set{X_i\in \cX}{\#d_i\leq \Uinit}$.
} 
Although $\Winit$ does not use any symbol in $\wh A\sm A$, we view $\Winit$ as an element in $M(\wh C\cup \cXinit,\rho)$ where $\mu(a)= \mu_{\init}\wh \pi(a)$ for all $a\in A$.
The idea is to transform an initial equation $\Winit$ into 
some equation $W_{\text{fin}}$ where $\# d_1\# \cdots d_k \#$ is a prefix of 
$W_{\text{fin}}$. The transformation encodes a $\wh A$-\Endo 
$h\in \End(\wh C,\rho)$ which leads to the \edtol result we are aiming for.

We also need an ambient set of variables $\wh \cX$ with  $\wh \cX\cap \wh C=\es$. 
It is required that $X\neq \bar X$ for all $X\in \wh\cX$, and $\cXinit\sse \wh \cX$. Moreover we need that
the size $|\wh C\cup \wh \cX|$ depends on the input size of the system $\cS$, but not on $N$. 

We do not define $\mu(x)$ for $\wh C\cup \cX$, and 
we do not define $\rho(X)$ for $X\in \wh \cX$. 
This allows us to change $\mu(c)$, $\rho(X)$, and $\mu(X)$ dynamically during the process without changing the names $c$ or $X$.
An important restriction is that changing $\rho(X)$ to 
$\rho'(X)$ is allowed only if $\rho'(X)\sse \rho(X)$. 
For $a\in \wh A \cup \os{\cent,\#}$ the element $\mu(a)\in N$ is always fixed with $\wh \pi \mu(a)=\mu_\init$ for all $a\in \wh A$ and $\mu(\cent)= \mu(\#) = 0$. 
The $\mu$-values of  distinguished letters~$d_i$ 
are not fixed, since later we compress factors $d_ia$ into 
the symbol $d_i$, which forces us to redefine 
$\mu(d_i)$ as $\mu(d_ia)$.

The convention that $\rho$ and $\mu$ are  not fixed for 
variables 
allows to substitute a variable $X$ by $Xx$ with $x\neq 1$ 
and define $\rho'(X)$ and $\mu'(X)$ such that $\rho(X)=\rho'(X)\cup \rho(x)$ and $\mu(X)=\mu'(X)\mu(x)$.
For constants we have a dual effect, 
If we wish to compress a factor $uv\in \wh C^+$ into  letter $c$, then we put $\rho(c)=\rho(uv)$ and we change $\mu(c)$ to $\mu'(c)=\mu(uv)$.
\begin{lemma}\label{lem:Ahut}
It is enough to prove \prref{thm:uro} in the version where the equality of sets in (\ref{eq:EQh}) is replaced by the following equality of sets.
\begin{align}
\label{eq:EQnohut}
\begin{split}
&\set{(h(d_1),\ldots ,h(d_k))}{h \in \cL}\\
{}={} 
&\set{(\wh \pi \sig(X_1),\ldots ,\wh \pi \sig(X_k))}{\sig \text{ solves } \cSinit \text{ in $M(\wh A,\rho)$
with $\muinit$  constraints}}
\end{split}
\end{align}
\end{lemma}
\begin{proof}
If $\sig: M(\wh A\cup \cXinit,\rho) \to M(\wh A,\rho)$ is a $\wh A$-\morph with $\sig(\Winit)=\sig(\ov\Winit)$,
then $\wh\pi\sig\iota$ is an $A$-\morph of $M(A\cup \cXinit,\rho)$ to $M(A,\rho)$ solving $\cS$ over $M(A,\rho)$. Thus, 
the equality in (\ref{eq:EQnohut}) implies the 
equality in (\ref{eq:EQh}). 

Vice versa, if $\sig': M(A\cup \cXinit,\rho) \to M(A,\rho)$ solves $\cS$ over $M(A,\rho)$, then 
$\sig'$ lifts uniquely to a $\wh A$-\morph $\sig: M(\wh A\cup \cXinit, \rho)
\to M(\wh A,\rho)$ with $\sig(\Winit)=\sig(\ov\Winit)$. We obtain
\begin{align}\label{eq:EQCOV}
&\set{(\sig(X_1),\ldots ,\sig(X_k))}{\sig\text{ solves } \cSinit \text{ in } M(A,\rho) \text{ with $\muinit$ constraints}}\\
\label{eq:EQhut}
  {}={}&
\set{(\wh \pi \sig(X_1),\ldots ,\wh \pi \sig(X_k))}{\sig \text{ solves } \cSinit \text{ in } M(\wh A,\rho) \text{ with $\muinit$ constraints}}
\end{align}
Let $L$ the left-hand side in the equation  (\ref{eq:EQh}).
If $L$ is \edtol over $M(A,\rho)$ we are done. 

Assume that
the set $\wh L$ in (\ref{eq:EQhut}) is an \edtol language 
over the alphabet $\wh A$. 
Then $L=\wh \pi(\wh L)$ is \edtol over the alphabet $A \cup \os {\cent,\#}$.

Moreover, since $\abs{\oi{\wh \pi}(w)} < \infty$ for all $w\in M(A,\rho)$, 
there are infinitely many $A$-\morph{s} $\sig': M(A\cup \cXinit,\rho) \to M(A,\rho)$ solving $\cS$ in $M(A,\rho)$ \IFF 
there are infinitely many $\wh A$-\morph{s} $\sig: M(\wh A\cup \cXinit,\rho) \to M(\wh A,\rho)$ solving $\cS$ in $M(\wh A,\rho)$.
\end{proof}

\subsection{Fresh letters, clones, and $T$-clones}\label{sec:CloneT}
The proof of \prref{thm:uro}
requires some more fine tuning which allows us introduce new letters and therefore to work with a resource alphabet
$(B,\rho)$  and which satisfies 
\begin{align}\label{eq:B}
\os {\cent,\#}\cup  \os{d_1,\ov{d_1}\lds d_{k},\ov{d_{k}}} \cup (\wh A,\rho) \sse (B,\rho) \sse (\wh C,\rho)
\end{align}
Here the $d_i$'s are the distinguished letters,
which were introduced in \prref{sec:Winit} and appear in 
the the trace $\Winit$, which has been defined in \prref{eq:Winit}.
We keep (\ref{eq:B}) as an invariant whenever we refer to 
a resource alphabet denoted by $(B,\rho)$.
Moreover, for each $(B,\rho)$ there is a $\wh A$-\morph 
$\alp:M(B,\rho)\to M(\wh A,\rho)$ which satisfies
$\muinit(b)=\muinit \alp(b)$ for all $b\in B$. 
During the process to describe all \solu{s} as an \edtol language 
we also change the set of variables. 
Thus, we actually work with a systems $\cS$ and resource alphabets
$(B\cup \cX,\rho)$ where $\cX$ refers to the current set of variables. 
If a mapping $\sig:\cX\to M(B,\rho)$ defines a \solu of 
a trace $W_\cS\in M(B\cup \cX,\rho)$ (and hence the system $\cS$), then we have
$\muinit(x)=\muinit \alp \sig(x)$ for all $x\in M(B\cup \cX,\rho)$.

We also assume without restriction that $\wh C\cap (B\cup \cX)^* = B$ where $(B,\rho) \ssneq (\wh C,\rho)$ denotes a resource alphabet of constants.
During the process we need to introduce fresh constants~$c\in \wh C\sm B$.
\begin{definition}\label{def:Tclone}
Let $(B,\rho)$ be a resource alphabet with 
$(\wh A,\rho)\sse(B,\rho)\ssneq (\wh C,\rho)$.
A \emphind{fresh letter} is a letter~$c\in \wh C$ such that $\os{c,\bar c}\sse \wh C\sm B$.

Let~$u\in B^+$ be a nonempty word. A \emph{clone\index{clone of $u$}} of~$u$ is fresh letter~$c$ such that 
$\rho(c)= \rho(u)$, $\mu(c)= \mu(u)$, and ${\bar c}= c\iff \bar {u}= u$. 

If the relation~$\theta$ is empty and~$T\in 2^\Rs$ satisfies $\rho(u)\sse T$, then we define the \emph{$T$-clone\index{T-clone of u@$T$-clone of $u$}} of~$u$ 
by a fresh letter~$c$ such that 
$\rho(c)= T$, $\mu(c)= \mu(u)$, and ${\bar c}= c\iff \bar {u}= u$. 
\end{definition}
Introducing a fresh symbol enlarges~$B\cup \cX$.
This implies that we have to choose 
$\wh C \cup \wh\cX$ large enough such that there is room enough to 
find fresh symbols. Moreover, we need ways to shrink $\wh C \cup \wh\cX$
such that we can recycle names for symbols.

\section{The automaton~$\cU$ accepting $\wh A$-\Endos}\label{sec:cGstate}

We define a nondeterministic (and infinite) automaton~$\cU$ which accepts a set of $\wh A$-\Endos in $\End(M(\wh C,\rho))$, and only at the very end 
we define some finite 
subautomaton $\cA$ such that $L(\cA)=L(\cU)$.
Since $\cA$ is finite, it is an NFA accepting a rational set of the monoid $\End(M(\wh C,\rho))$. 
The definition of $\cA$ is postponed because 
most of the technical work uses some \emph{universal\index{universal automaton}} automaton~$\cU$ (and its \emph{unfolding\index{unfolding of an automaton}} $\wh \cU$). The definition of $\cU$ and the proof of its soundness is in this section. Every state in $\cU$ has a loop. In contrast, the unfolding $\wh \cU$ is a directed acyclic graph. It is defined in \prref{sec:unfold}. 
The automaton $\wh \cU$ comes with a natural graph \morph to~$\cU$. 

\subsection{States of~$\cU$, extended monoids, and (entire) \solu{s}}\label{sec:exeqs}

We begin with the definition of the state set in~$\cU$.
\begin{definition}[State]\label{def:status}
The \emphind{initial state} is given by $\cSinit$ as defined in (\ref{eq:defCsinit}) and the \emphind{final state} is the central symbol~$\cent$. Every other state in the universal automaton~$\cU$ is a \emph{$T$-state}\index{T-state@$T$-state}, which is a tuple 
\(
E=(W,B,\cX,\rho,\mu,\theta,T)
\)
satisfying the following conditions: 
\begin{itemize}
\item We have $\es\neq T\sse \Rs$.
\item The pair $(B\cup \cX,\rho)$ is a resource alphabet with $\rho(X) \ssneq \Rs$ for all $X\in \cX$,
where $\wh A\sse B\sse \wh C$ is the set of constants, and $\cX\sse \wh \cX$ is the set of variables such that $(B,\rho)$ satisfies the condition~(\ref{eq:B}). 
\item The element $W\in M(B\cup \cX,\rho,\theta)$ is called the \emph{equation} at $E$. It is an element belonging to the $\Rs$-monoid 
$M(B\cup \cX,\rho,\theta)$ with type. The equation $W$ is represented as a trace in the free $\Rs$-monoid $M(B\cup \cX,\rho)$ or by some word 
in $(B\cup \cX)^+$. 
\item Every \HA in the trace $W$ has one endpoint 
$p$ such that $\rho(p)\geq T$.
\item We have $|W|_\cent=1$, $|W|_\# = |\Winit|_\#$ and $|W|_x = |W|_{\bar x}$ for all $x\in B\cup \cX$. 
\item If $\cX=\es$, \ie there are no variables, then $\bar W =W$ in the monoid $M(B,\rho,\theta)$.
\item By $\mu$ we denote a \morph $\mu: M(B\cup \cX,\rho,\theta)\to N$ 
to the finite monoid $N$ which defines the recognizable constraints for the 
initial state $\cSinit$ and extends the definition
of $\mu:\wh A\to N$ in~(\ref{eq:muinit}).
\item By $\theta$ we denote an $(a,b,\Del)$-type according to \prref{def:type}. 
\end{itemize}

A \emph{standard state}~$E$ is a state $E=(W,B,\cX,\rho,\mu,\es,T)$.
That is, there is no type. Highlighting this situation we often write $(W,B,\cX,\rho,\mu,T)$ instead of 
$(W,B,\cX,\rho,\mu,\es,T)$. 

A standard state $E=(W,B,\es,\rho,\mu,\Rs)$ with an empty set of variables is called \emph{semi-final\index{semi-final state}}
if firstly $W=\bar W$ and secondly $W$ begins with a prefix 
$\#d_1\cdots \#d_k\#$.
\end{definition}
\begin{definition}[Extended monoid]\label{def:extem}
Let $E=(W,B,\cX,\rho,\mu,\theta,T)$ be a state in $\cU$.
\begin{enumerate}
\item By $M(B,\cX,\rho,\mu,\theta)$, resp.~$M(B,\rho,\mu,\theta)$, we denote 
the monoid $M(B\cup\cX,\rho,\theta)$, resp.~$M(B,\rho,\theta)$, with type 
together with the \morph $\mu: M(B,\cX,\rho,\theta)\to N$.
We say that $M(B,\cX,\rho,\mu,\theta)$ is an \emphind{extended monoid}. 
\item A \morph $\phi: M(B,\cX,\rho,\mu,\theta)\to 
M(B',\cX',\rho',\mu',\theta')$ of $\Rs$-monoids with type is a 
\emph{\morph between extended monoids} if $\mu(x)=\mu'\phi(x)$ for all $x\in M(B\cup \cX,\rho,\theta)$.
\end{enumerate}
\end{definition}
Another 
crucial concept is the notion of an `\esolu' being defined next.
\begin{definition}[Solution,  \esolu, and weak \solu]\label{def:statesolu}
Let $E=(W,B,\cX,\rho,\mu,\theta,T)$ be a state and
$M(B,\cX,\rho,\mu,\theta)$ be its associated extended monoid
according to \prref{def:extem}.
\begin{enumerate}
\item A \emph{\solu} at $E$ is a $B$-\morph $\sig:M(B,\cX,\rho,\mu,\theta) \to M(B,\rho,\mu,\theta)$ of extended monoids such that $\sig(W) = \sig(\bar W)$. 
\item 
An \emph{\esolu} at $E$  is a pair $(\alp,\sig)$ such that $\alp:M(B,\rho,\mu,\theta)\to M(\smash{\wh A},\rho,\mu,\es)$ is a $\wh A$-\morph of extended monoids and $\sig$ is a \solu at $E$ such that 
and $\alp(a)\in \wh A_S^+$ for every $a\in B$ and $S=\rho(a)$.\footnote{That is, $\alp(a)$ is a word in the 
free semigroup over $\set{b\in \smash{\wh A}}{\rho(b)=\rho(a)}$.}
\item A \emph{weak \solu} is an \esolu at the state $E=(W,B,\cX,\rho,\mu,\theta,T_{\min})$ where $T_{\min}$ is the least nonempty subset of $\Rs$ 
in the linear order $\leq$ defined on $2^\Rs$, see \prref{sec:rmoni}.
\end{enumerate}
\end{definition}
\begin{remark}\label{rem:weaksolu}
  Let $E=(W,B,\cX,\rho,\mu,\theta,T)\in \cU$ be a state.
  For every $\es \neq S \sse \Rs$ with $S \leq T$, the automaton $\cU$ also contains a state $E_S = (W,B,\cX,\rho,\mu,\theta,S)$.
  Moreover, every \esolu at~$E$ is an \esolu at $E_S$ and the weak solutions at~$E$ are precisely the weak solutions at $E_S$.
  Crucially, weak \solu{s} are \esolu{s} at the initial state $\cSinit$.
\hspace*{\fill}$\diamond$\end{remark}
\begin{lemma}\label{lem:esolu}
Let $\alp:M(B,\rho,\mu)\to M(\wh A,\rho,\mu)$ be a $\wh A$-\morph 
with $1\notin\alp(B)$ and $\sig$ a \solu at a standard state $E=(W,B,\cX,\rho,\mu,T)$. 
Define for each $a\in B$ and $S=\rho(a)$ the word
$\alp'(a)= \snf_{\rho(a)}\wh A_S^+$. 
Then 
$(\alp',\sig)$ is an \esolu at $E$.
\end{lemma}
\begin{proof}
Recall \prref{def:stepnf} and \prref{lem:snf}. Since 
$\snf_{\rho(a)}=a=\alp(a)$ for all $a\in \wh A$, the mapping 
$\alp'$ leaves the letters in $\wh A$ invariant. 
We claim that 
$\alp': M(B,\rho,\mu)\to M(\wh A,\rho,\mu,\es)$ is a $\wh A$-\morph. To see the claim let $a,b\in B$ with 
$\rho(a)\cap \rho(b)=\es$. This implies $\rho(\alp(a))\cap \rho(\alp(b))=\es$
and hence, $\rho(\snf_{\rho(a)}\alp(a))\cap \rho(\snf\alp(b))=\es$. Therefore $\alp'$ is a $\wh A$-\morph with 
$\snf_{\rho(a)}\alp(a)\in \wh A_S^+$. Hence, $(\alp',\sig)$ is an \esolu at $E$.
\end{proof}
\begin{lemma}\label{lem:esthe}
Let $E=(W,B,\cX,\rho,\mu,\theta,T)$ be a state with an \esolu $(\alp,\sig)$ where $\sig(W)$ is represented (as always) 
by a trace in $M(B,\rho)$. Then $\sig(W)$ contains an $S$-letter \IFF $\alp\sig(W)$ does. Moreover,
there is a natural bijection between the sets of unbalanced \HAs in $\sig(W)$ and in $\alp\sig(W)$.
\Ip if $\sig(W)$ is without any \SSparc with $S,S'<T$, then 
$\alp\sig(W)$ is without any \SSparc with $S,S'<T$.
\end{lemma}
\begin{proof}
The property whether an $S$-position $p$ appears in a trace depends only on $\rho(p)$. The same for 
the set of unbalanced \HAs in a trace: it depends only on the labels
of the endpoints.
Thus, we can think that we replace every label of a position $p$ 
in $W$ and $\sig(W)$ just by $\rho(p)$. Then we obtain a trace in $M(2^\Rs,\rho)$.
The images of $W$ and $\sig(W)$ in $M(2^\Rs,\rho)$ do not depend on the trace 
representation of $W$ since $(u,v)\in \theta$ implies $u,v\in B_T^+$. 
The assertion is now clear since $\alp(a)\in \rho(a)^+$ for all $a\in B$. 
\end{proof}
The strategy to solve an equation is based on nondeterministic choices of outgoing \tras in $\cU$:
either substitute a variable $X$ by some $x \in M(B,\cX,\rho,\theta)$ (which typically makes $W$ longer) or compress a suitable set of subtraces in $\sig(W)$ into single positions (which typically makes $W$ shorter). 
Clearly, if we substitute $X\mapsto w$, then, simultaneously, we
substitute $\bar X\mapsto \bar w$; and if we compress a subtrace $u$ into a single position labeled by letter~$c$, then, simultaneously, we compress its `dual' subtrace $\bar u$  into a position labeled by  $\bar c$.
These \subst{s} will be explained in detail later.

\subsection{Weights}\label{sec:weight}

We introduce a partial order on states and 
\esolu{s}. 
We do so by defining a weight functions. Each weight is a tuple of natural numbers. For each $\ell\in \N$ we equip $\N^\ell$ with the canonical lexicographic linear order.
This means, the leftmost component is dominant, then the second etc. 
For example, $(0,42,0) < (1,0,0)<(1,0,1)$. 
The lexicographic order on $\mathbb N^\ell$ is well-founded,
 \ie there are no infinite descending chains. 
\begin{definition}\label{def:weight}
Let $E=(W,B,\cX,\rho,\mu,\theta,T)$ be a state with 
an \esolu $(\alp,\sig)$. The \emphind{weight of a variable} $X\in \cX$ is defined as ${\Abs X}= 4^{\abs{\rho(X)}}$.
The \emph{weight\index{weight of $E$}} $\Abs E$ of~$E$ 
is a tuple $\abs E=(\oo_1\lds \oo_5)\in \N^5$ 
where the $\oo_i$ are defined as follows. The comments on the right side indicate when $\oo_i$ decreases.
\begin{align}\label{eq:weightoo1}
\oo_1 &= |\set{T'\sse \Rs}{T<T'}| &\text{ `Lift $T$ to $T'$.'}
\\
\label{eq:weightoo2}
\oo_2 &= \sum_{X \in \cX} \Abs{X}\cdot |W|_X &\text{ `Split $X$ into something lighter.'}
\\
\label{eq:weightoo3}
\oo_3 &=\sum_{a\in \wideh A} |W|_{a} &\text{ `Make $\wh A$ invisible in the equation.'}
\\
\label{eq:weightoo4}
\oo_4 &=|W|+|B| - 2|\set{a\in B}{|W|_a\geq 1}|  &\text{ `Increase diversity of visible letters.'}
\\
\label{eq:weightoo5}
\oo_5 &=\sum_{a\in B} (|\Rs| - |\rho(a)|) &\text{ `Add resources to letters.'}
\end{align}
The \emph{weight\index{weight of $(\alp, \sig)$ at $E$}} $\Abs{E,\alp, \sig}$ of $(\alp, \sig)$ at~$E$ is defined as $\Abs{\alp,\sig,E}= (\oo_0,\oo_1\lds \oo_5)\in \N^6$ where:
\begin{align}\label{eq:weih}
\oo_0&=
\sum_{X \in \cX\vphantom{\widehat A}} \sum_{a \in \widehat A}(|\Rs| - |\rho(a)|)\cdot|\alp\sig(X)|_{a} &\text{ `Add resources to invisible letters.'}
\end{align}
If the state~$E$ is known by the context, then we may write
$\Abs{\alp, \sig}$ instead of $\Abs{E,\alp, \sig}$.
\end{definition}

\begin{remark}\label{rem:weista}
The weight of state decreases if the $T$-component 
increases \wrt~the linear order on $2^\Rs$. Not changing $T$, but 
making the set of variables smaller (or lighter by reducing the their resources) makes the weight of a state~$E$ smaller, since 
the weight of every variable is at least one.
If $\oo_1$ and $\oo_2$ are not changed, then replacing a letter $a\in \wh A$ with $|W|_a \geq 1$ by a clone
$a'\in \wideh C\sm B$ decreases the weight of the state as well.
If $\oo_1$ and $\oo_2$ are not changed and if
$|W|_{\wh a}=0$ for all $\wh a \in \wh A$, then 
$\oo_4$ decreases the weight of a state if the diversity of visible letters in $W$ increases. That is, if 
we have $|W|_a \geq 2$ for some $a\in B\sm \wh A$, then it decrease 
the weight of~$E$ if we choose a fresh clone $c\in \wideh C\sm B$ 
and replace at least one occurrence  of~$a$  in $W$ by $c$ in such a way that both letters~$a$ and $c$ are visible in the new equation. 
Clearly, such a move makes it harder or impossible to find a \solu at the new state.  Finally, the least important value $\oo_5$
decreases (resp.~increases) if the size of $B$ decreases (resp.~increases). If the size of $B$ is not changed by replacing a 
letter $b\in B$ with $\rho(b)<T$ by some $T$-clone, 
then  $\oo_5$ decreases. 

The dominant value in $\Abs{(\alp,\sig)}$ is $\oo_0$. It 
decreases \IFF we substitute a variable
$X\in \cX$ by some $x\in M(B,\cX,\rho,\mu,\theta,T)$
such that either $|x|_b\geq 1$ for some $b\in B$ 
or the \subst leads to a state with a smaller weight. 
For example, if we substitute $X$ by $YZ$,
where $Y$ and $Z$ are variables such that $\rho(YZ) = \rho(X)$ and $\rho(Y) \neq \rho(X)\neq \rho(Z)$, then the state of the weight decreases~$\oo_2$ without changing~$\oo_1$.
\hspace*{\fill}$\diamond$\end{remark}

\subsection{The \tras of the automaton~$\cU$}\label{sec:Atras}

Recall that a state refers always to a state of~$\cU$.
We prove soundness for~$\cU$, the completeness of~$\cU$ is essentially trivial,\footnote{In all interesting case~$\cU$ in infinite.} but we need to prove it for some effectively computable and finite sub automaton~$\cA$ and for that we use the following notation.  
\begin{definition}\label{def:entist}
A triple $(E,\alp,\sig)$ where $(\alp,\sig)$ is an \esolu at 
a state in $\cU$ is called an
\emphind{entire state}. It is called an entire state $T$-state, if~$E$ is a $T$-state.
\end{definition}
We will begin with an entire state~$(E,\id_{A},\siginit)$ which is connected to 
the initial state in~$\cU$ by some \tra $\cSinit\arc \eps E$. 
We will show that if  $\cA$ is chosen large enough, then it contains a path of \tras from~$E$ to an entire state $(E_t,\alp,\id_{B})$
such that $E_t$ has an empty set of variables and 
$\siginit=\wh \pi \alp$. 
\begin{definition}\label{def:trasif}
The initial state $\cSinit$ has no incoming \tras.\\
For a standard state $E=(\Winit,\wh A,\cXinit,\rho,\mu,\os{r})$ as defined in \prref{sec:exeqs} we define a \tra $\cSinit\arc{\wh \pi} E$ labeled by the $A\cup \cXinit$-\morph $\wh \pi: M(\wh A,\cXinit,\rho,\mu) \arc{\smash{\wh \pi}} M(A,\cXinit,\rho,\mu)$ which maps  $(a,t)\in \wh A$ to $a\in A$.
\\
Every semi-final state $E=(W,B,\es,\rho,\mu,\Rs)$ has exactly one outgoing \tra 
$E\arc {h_{\fin}} \cent$
to the final state~$\cent$, and all \tras incident to~$\cent$ are of this form. (\Ip $\cent$ has no outgoing \tra.)
The label $h_{\fin}$ is the \Endo $h_{\fin}\in \End(M(\wh C,\rho))$ such that $h_{\fin}(\cent) = W=\bar W$ and $h_{\fin}(c)=c$ for $c\neq \cent$. 
\end{definition}
Recall that we have $k\geq 1$ where $2k=|\cXinit|$.
Moreover, we have $\rho(X)\ssneq \Rs$ for all 
$X\in \wh \cX$ and  $\rho(\cent)=\rho(\#)= \rho(d_i) =\Rs$ for $1\leq i \leq k$.
Hence, $|h_{\fin}(\cent)|>2$.
\begin{lemma}\label{lem:semfinfin}
Let $E=(W,B,\es,\rho,\mu,\Rs)$ be a semi-final state in~$\cU$ and
$\alp:M(B,\rho,\mu) \to M(\wh A,\rho,\mu)$ be a $\wh A$-\morph. 
Then $(\alp,\id_{B})$ is an \esolu at $E$. 
\end{lemma}
\begin{proof}
The fact that $(\alp,\id_{B})$ is an \esolu at~$E$ follows from $W=\bar W$, $\cX=\es$, $T=\Rs$, and \prref{def:statesolu}.
\end{proof}
The \tras which are not captured by \prref{def:trasif} are between the states of the form $E= (W,B,\cX,\rho,\mu,\theta,T)$.
They are either \epstra{s}, defined by a \subst of variables, or compressions. We define \subst{s} and \epstras first.
\begin{definition}[Substitution, \epstra]\label{def:trascA}
Let $E= (W,B,\cX,\rho,\mu,\theta,T)$ be a state. 
\begin{enumerate}
\item For~$E$ there is an \epstra which is a loop $E\arc \eps E$.
Here $\eps$ is the identity.
\item If~$E$ is a standard state, $B'\sse B$, and 
$T \leq T'$ such that $(B,T)\neq (B',T')$ and if~$W$ is without any \SSparc where $S,S'<T'$, then there is 
an \emph{\epstra}
\[(W,B,\cX,\rho,\mu,T)\arc \eps (W,B',\cX,\rho,\mu,T')\]
Here, $\eps$ refers to the inclusion of $M(B',\rho)$ into $M(B,\rho)$. 
\item If~$E$ is a standard state, then a \emphind{substitution} is a $B$-\morph of free $\Rs$-monoids
\[\tau: M(B,\cX,\rho,\mu)\to M(B,\cX',\rho',\mu')\] which is defined by a mapping $X\mapsto x$ 
with $|W|_X\geq 1$ such that firstly
$x\in M(B,\cX',\rho',\mu')\sm \cX'$
and secondly $|\tau(W)|_{X'}\geq 1$ for all $X'\in \cX'$. 
\item If~$E$ is a state with an $(a,b,\Del)$-type (according to \prref{def:type}) with $\rho(a)=T$, then a \emph{substitution}  is a $B$-\morph of extended $\Rs$-monoids with type
\[\tau\colon M(B,\cX,\rho,\mu,\theta)\to M(B,\cX',\rho',\mu',\theta')\] which is defined by a mapping $X\mapsto x$ where 
$|W|_X\geq 1$ and
$x\in M(B,\cX',\rho',\mu',\theta')$
satisfies the following additional conditions.
\begin{itemize}
\item We have  $\rho' \sse \rho$ and $\theta' \sse \theta$.
\item Either $x\in \os{a,aba}$ 
and~$\cX'=\cX\sm \os{X,\bar X}$
or $x=Xba$, $\rho'=\rho$, and~$\cX'=\cX$.
\end{itemize}
\item If $\tau$ is a \subst such that 
$E'= (\tau(W),B',\cX',\rho',\mu',\theta',T')$ is a state with $T'=T$ and 
\begin{align}\label{eq:tauW}
\Abs{E'}<\Abs{E}\quad\text{or}\quad\sum_{\set{x\in B'\cup \cX'}{\rho'(x)\neq \es}}|\tau(W)|_x>|W|
\end{align}
then there is an \emph{\epstra} 
\[(W,B,\cX,\rho,\mu,\theta,T)\arc \eps (\tau(W),B',\cX',\rho',\mu',\theta',T).\]
\end{enumerate}
\end{definition}
\begin{lemma}\label{lem:varsub}
Let ${E}= (W,B,\cX,\rho,\mu,\theta,T) \arc{\eps} (\tau(W),B',\cX',\rho',\mu',\theta',T')= {E}'$ be an \epstra 
and $(\alp,\sig')$ a weak \solu at $E'$. 
If $T\neq T'$, then $(\alp,\sig')$ is a weak \solu at $E$, too.
If $\tau(W)=W$ and either $B'\ssneq B$ or $T\ssneq T'$,
then $\Abs{E'}<\Abs{E}$. 
If  $E\neq E'$, $T= T'$, and the \epstra is defined by a \subst $\tau$, 
then $\sig= \sig'\tau$ is a \solu
at~$E$ and $(\alp, \sig)$ is a weak \solu at $E$. 

Moreover, suppose that $\tau$ is defined by $X\mapsto x$ with 
$x=Xba$ or $x\in \os{a,aba}$ and where $X\,bc = cb \,X$ belongs to $\theta$. 
Then firstly, the relation $x\,bc = cb \,x$ still holds in $M(B,\cX',\rho',\mu',\theta')$
and secondly, if $X \not\in \cX'$ or $\sig'(X)\in a(ba)^*$, then $\sig(X)\in a(ba)^*$.
\end{lemma}
\begin{proof}
If $\tau(W)=W$ and either $B'\ssneq B$ or $T\ssneq T'$, then 
we let $\sig=\eps\sig'$ where $\eps$ denotes the inclusion 
of $M(B',\rho)$ into $M(B,\rho)$. 
The assertions are therefore trivial in this case.
Hence, without restriction we have $B'=B$, $T=T'$ and $\eps$ is the identity on $M(B,\rho)$.
Since $\sig'$, and $\tau$ are $B$-\morph{s}, so is their
 composition 
$\sig= \eps\sig'\tau= \sig'\tau$. Since $\sig'$ is a \solu of $\tau(W)$, we have
\[\sig(W)=\sig'(\tau(W))=\sig'(\ov{\tau(W)})= \sig'(\tau(\bar {W}))=\sig(\bar W).\] 
See also~\prref{fig:varsub}. As a consequence, $(\alp,\sig)$ is a weak \solu. 
If $\tau$ is defined by $X\mapsto Xba$ for a typed variable, then
$(B,\cX',\rho',\mu',\theta')=(B,\cX,\rho,\mu',\theta)$ and the result is clear since $cbx=cbXba=Xbcba= Xbabc=x bc$. Finally, 
if $x\in \os{a,aba}$ with $abc=cba$ in $\theta'$, then we have $abc=cba$ in $\theta$. Hence, either $x=a$ and $xbc=cbx$ or $x bc= ababc=cbaba =cbx$. 
It remains to show that if $X \not\in \cX'$ or $\sig'(X)\in a(ba)^*$, then $\sig(X)\in a(ba)^*$.
  In case $X \not\in \cX'$, then $\sig(X) \in \os{a, aba} \sse a(ba)^\ast$.
  On the other hand, if $X \in \cX'$, then $\sig(X) = \sig'(\tau(X)) = \sig'(X)ba$ and thus $\sig'(X) \in a(ba)^\ast$ implies $\sig(X) \in a(ba)^+ \sse a(ba)^\ast$.
\end{proof}
\begin{figure}[H]
\begin{center}
\begin{tikzpicture}[
	xscale=6, yscale=2
]

\path (0,1) node (1) {$M(B',\cX',\rho',\mu',\theta')$};
\path (-1,1) node (2) {$M(B,\cX,\rho,\mu,\theta)$};
\path (0,0) node (3) {$M(B',\rho,\mu,\theta)$};
\path (-1,0) node (4) {$M(B,\rho,\mu,\theta)$};

\draw [->, >=latex] (2) -- (1) node[midway, above] {$\tau$};
\draw [->, >=latex] (1) -- (3) node[midway, right] {$\sigma'$};
\draw [->, >=latex] (2) -- (4) node[midway, right] {$\sigma$};
\draw [->, >=latex] (3) -- (4) node[midway, above] {$\eps= \id_{M(B',\rho)}$};
\end{tikzpicture}
\end{center}
\vspace{-0.5cm}
	\caption{The definition of $\sig$ in \prref{lem:varsub} 
	where $\eps$ might be an inclusion.}\label{fig:varsub}
\end{figure}
A compression \tra is defined by a $\wh A \cup \cX$-\morph 
$h:M(B',\cX,\rho',\mu',\theta') \to M(B,\cX,\rho,\mu,\theta)$.
We encounter both situations: $B\sse B'$ or $B'\sse B$;
the former intuitively corresponds to the case when we introduce new
constants that represents compressed factors or letters where we increase its set of resources
while the latter allows the removal of unused letters from $B$.
\begin{definition}[
Compression \morph]\label{def:compmo}
An $(\wh A\cup \cX)$-\morph $h$ of free $\Rs$-monoids with type {}from $M(B',\cX,\rho',\mu',\theta')$ to $M(B,\cX,\rho,\mu,\theta)$ is called a \emphind{
compression \morph} 
if $h$ is defined by $c\mapsto h(c)\in M(B,\rho)$ for some $c\in B'\sm \wideh A$
such that the following conditions hold. 
\begin{itemize}
  \item We have $h(c)\neq c$ and $\rho(h(c)) \sse \rho (c)$. 
  \item If $\theta'\neq \es$, then $B=B'$ and $\theta'$ is an $(a,b,\Del)$-type with $ab\in B_T^+$ according to \prref{def:type}. 
    Moreover, we then require that $\del(c) = a$ and either $h(c)=aba$ or $h(c)=abc$ (or both if $c=a$).
\end{itemize}
\end{definition}

Next, we define \comp \tras which are labeled by 
\comp \morphs.
\begin{definition}[Compression transition]\label{def:comptra}
Let  $E'=(W',B',\cX,\rho',\mu',\theta',T)$ be a state 
 and $h:M(B',\cX,\rho',\mu',\theta')\to 
M(B,\cX,\rho,\mu,\theta)$ be a compression \morph (according to \prref{def:compmo}) which is defined by $c \mapsto h(c)$ for $c\in B'$ and $T\leq \rho(c)$. 
If, under these assumptions, $E=(h(W'),B,\cX,\rho,\mu,T)$ is a state
with $\Abs{E'}<\Abs{E}$, then $h$ defines the following
\emph{\comp \tra\index{compression transition}} between $T$-states. 
\[E=(h(W'),B,\cX,\rho,\mu,\theta,T)\arc h (W',B',\cX,\rho',\mu',\theta',T) = E'.\]
\end{definition}
If the context is clear, we simply refer to the \comp \morph $h$ as a \emph{\comp} when we actually mean the corresponding \comp \tra. There should be no risk of confusion.
The rule of  thumb is that \epstras tend to make the equation longer and \comp \tras make them shorter. 
The condition  $\Abs{E'}<\Abs{E}$ implies that after at most 
$\Abs{E}$ steps we must use an \epstra which makes $\alp\sig(W)$ shorter than before. 
\begin{lemma}\label{lem:stacomp}
  Let $E = (h(W'),B,\cX,\rho,\mu,\theta, T) \arc{h} (W',B',\cX,\rho',\mu', \theta' ,T) = E'$ be a \comp \tra and $(\alp',\sig')$ be a weak \solu at a state $E'$.\footnote{Remember that a weak \solu refers to an \esolu.}
Then there is a unique $B$-morphism \[\sig : M(B, \cX, \rho, \mu, \theta) \to M(B, \rho, \mu, \theta)\] with $\sig h = h\sig'$ (depicted in the commutative diagram in \prref{fig:comcon}). It defines a weak \solu $(\alp,\sig)$ at $E$ such that 
$\wh \pi \alp(a) = \wh \pi \alp'(a)$ for all $a\in B$.
In other words, we obtain an \esolu $(\alp,\sig)$ by letting $\sig(X)=h\sig'(X)$ 
for $X\in \cX$ and $\sig(a)=a$ for $a\in B$.
\end{lemma}
\begin{proof}We split the proof into the several cases. For the notation of $\theta$, $\Del$, and $\del$ see \prref{def:type}. The assertion $\wh \pi \alp(a) = \wh \pi \alp'(a)$ for all $a\in B$ is guaranteed by \prref{lem:esolu}. So we concentrate on~$\sig$.
\begin{enumerate}
\item 
$\theta'=\theta= \es$. Then $M(B',\cX,\rho',\mu',\theta')$ and $M(B,\cX,\rho,\mu,\theta)$ are free $\Rs$-monoids.
 Letting 
$\sig(X)=h\sig'(X)$ defines
a $B$-\morph $\sig:M(B,\cX,\rho,\mu)\to M(B,\rho,\mu)$. Since $h$ leaves~$\cX$ invariant and~$\sig$ leaves~$B$ invariant, we have $\sig(h(x))=h(\sig'(x))$
for all $x\in B'\cup X$. Therefore $\sig h=h \sig'$. Hence, the 
diagram in \prref{fig:comcon} commutes.
\item 
If $\theta'= \es$ and $\theta$ is a nonempty $(a,b,\Del)$-type, then $\sig':M(B',\cX,\rho',\mu',\es)\to M(B',\rho',\mu',\es)$
is a \morph between free $\Rs$-monoids, and  the $B$-\morph $h\sig':M(B',\cX,\rho',\mu',\es)\to M(B,\rho,\mu,\theta)$ is well defined.
In $M(B,\cX,\rho,\mu,\theta)$ there are no typed variables. It follows that $h\sig'$ factorizes through 
$M(B,\cX,\rho,\mu,\theta)$ and therefore we can define $\sig:M(B,\cX,\rho,\mu,\theta)\to 
M(B,\rho,\mu,\theta)$ such that diagram in \prref{fig:comcon} commutes.
\item 
$\theta'\neq \es$ and $h$ is defined by
$h(c)=\wt a bc =\del(c)bc$, then we have  $(B',\cX,\rho',\mu',\theta')= (B,\cX,\rho,\mu,\theta)$ and the 
diagram in \prref{fig:comcon} commutes. 
\item 
If $(B'\sm \Del,\cX,\rho',\mu',\theta')=(B,\cX,\rho,\mu,\theta)$ and $h(c)=\wt ab\wt a$ for all $c\in \Del$ where $\wt a=\del(c)$, then the 
diagram in \prref{fig:comcon} commutes.
\item 
$(B'\sm (\Del\cup \os{c_1,\bar{c_1}}),\cX,\rho',\mu',\theta')=(B,\cX,\rho,\mu,\theta)$
with $c_1\notin \Del$, $h(c_1)= a$,  and $h(c)=\del(c)b\del(c)$ for all $c\in B\cap \Del$. It follows that the 
diagram in \prref{fig:comcon} commutes.
\end{enumerate}\vspace{-0.5cm}
\end{proof}
\begin{figure}[h]
\begin{center}
\begin{tikzpicture}[
	xscale=6, yscale=1.5
]

\path (0,1) node (1) {$M(B',\cX,\rho',\mu',\theta')$};
\path (-1,1) node (2) {$M(B,\cX,\rho,\mu,\theta)$};
\path (0,0) node (3) {$M(B',\rho',\mu',\theta')$};
\path (-1,0) node (4) {$M(B,\rho,\mu,\theta)$};

\draw [->, >=latex] (1) -- (2) node[midway, above] {$h$};
\draw [->, >=latex] (1) -- (3) node[midway, right] {$\sigma'$};
\draw [->, >=latex] (2) -- (4) node[midway, right] {$\sigma$};
\draw [->, >=latex] (3) -- (4) node[midway, above] {$h$};

\draw [dashed,->, >=latex] (1) -- (4) node[midway, above] {$h\sig'$};
\end{tikzpicture}
\end{center}
\vspace{-0.5cm}
	\caption{The diagram used for the statement in \prref{lem:stacomp} and in \prref{thm:backstage}.}\label{fig:comcon}
\end{figure}
\begin{theorem}\label{thm:backstage}
Let  
$\cSinit\arc{\wh \pi}{E}_0\arc{h_{1}} {E}_1\cdots \arc{h_{t}} {E}_{t}\arc{h_\fin}\cent$
be a path in~$\cU$ {}from the initial state $\cSinit$ to the final state $\cent$ for some $t\geq 1$. Then  $E_t=(W_t,B_t,\es,\rho,\mu,T)$ is a standard state, the equation 
$W_t$ satisfies 
$W_t=\ov{W_t}$, and $\sig_t=\id_{B_t}$ is its unique \solu. 
The standard state ${E}_0$ has a weak \solu $(\id_{\wh A},\sig)$ with
 $\sig(W_0)= h_1 \cdots h_t(W_t)$
and, more generally, any ${E}_i$ (for $0 \leq i \leq t$)
 has a weak \solu $(\alp_i,\sig_i)$ with
 $\alp_i\sig_i(W_i)= \wh \pi h_1 \cdots h_t(\Wfin)$ where
 $W_i$ is the equation at state $E_i$. 
\end{theorem}
\begin{proof}
For $0 \leq i \leq t$ let ${E}_i = (W_i,B_i,\cX_i,\rho_i,\theta_i,\mu_i,T_i)$ and $\alp_i = h_1 \cdots h_i$ Hence, each $\alp_i:M(W_i,\cX_i,\rho_i,\theta_i,\mu_i)\to M(A,\rho)$ is a $\wh A$-\morph{s} because all $h_j$ leave $\wh A$ invariant. Thus, whenever $\sig_s$ is a solution
 at $E_s$, then $(\alp_s,\sig_s)$ is a weak \solu at $E_s$. We have $\ov{W_t} = W_t$ by the definition of \tras to the final state $\cent$. Since no variables occur in $W_t$, we conclude that $\sig_t = \id_{B}$ is the (unique) solution at $E_t$. 
For all $1\leq i \leq t$ there are two cases. If $h_i=\eps$, then there is 
an \epstra $E_{i-1} \arc \eps E_i$ defined by a \subst $\tau_{i-1}$ such that $W_i=\tau_{i-1}(W_{i-1})$. In the other case 
$E_{i-1} \arc {h_i} E_i$ is a compression. 

 Now, let $0\leq i \leq s \leq t$ and consider the infix 
${E}_{i}\arc{h_{i+1}} \cdots \arc{h_{s}} {E}_s$.
We strengthen the assertion of \prref{thm:backstage} as follows.
There is a \solu
$\sig_i$ at $E_i$ such that $\sig_i(W_i) = h_{i+1} \cdots h_s\sig_s(W_s)$ and if there is a weak \solu 
$(\alp_i,\sig_i)$ such that 
\begin{align}\label{eq:willy}
\alp_i \sig_i (W_i) = \id_{\wh A} h_1 \cdots h_i h_{i+1} \cdots h_s\sig_s(W_s)
\end{align}
 at $E_s$, then there exists a weak \solu $(\id_{\wh A},\sig)$ at 
 $E_0$ such that we have
\begin{align}\label{eq:wonderc}
\sig(W_0) = h_1 \cdots h_s\sig_s(W_s). 
\end{align} 
Equation \eqref{eq:wonderc} holds trivially for $s=0$.
For $s>0$ the claim follows by induction using \prref{lem:varsub} (if $W_s= \tau_{s_1}(W_{s-1})$) or by \prref{lem:stacomp} (if $W_{s-1}= h_s(W_{s})$) where $E_{s-1}$ and $E_{s}$ are standard states. 
If $\cX_s$ has a typed variable and $W_s\neq \tau_{s_1}(W_{s-1}$, then $h_s(W_s)=W_{s-1}$
we the existence of the \tras implies that 
$(B_{s-1},\cX_{s-1},\rho_{s-1},\mu_{s-1},\theta_{s-1},T_{s-1})= (B_{s},\cX_{s},\rho_{s},\mu_{s},\theta_{s},T_{s})$ with $\theta_{s}\neq \es$.
Moreover, the \tra is a compression and $h_s$ is defined 
by $h(c)= a b a$ or $h(c)= abc$ or $h(a)=aba$.
In this case we can apply \prref{lem:stacomp} in case
that we have $\sig(X)\in \wt a (b\wt a)^*$ for 
all typed variable. To see this, we observe that 
every typed variable $X \in\cX_{s}$ vanishes in some 
$\cX_{s'}$ for $s<s'\leq t$. When $X$ vanishes, then we cannot use 
a \subst defined by $X\mapsto Xb\wt a$ because using it implies $\rho_{s}(X)=\rho(a)=T$. Hence, $X$ vanishes with an \epstra
defined by $X\mapsto \sig_{s'}(X)= x$ with $x\in \os{a, aba}\sse a(ba)^*$. Therefore we can use induction to keep the invariant 
$\sig_{s'}(X)\in \wt a (b\wt a)^*$ for all $k\leq s'<t$ backwards to reach 
the state $E_{s-1}$. 
This implies that ${E}_{\init}$ has a weak \solu $(\wh\pi,\sig)$ with
 $\sig(\Winit)= h_1 \cdots h_t(W_t)$.
\end{proof}

Recall that the distinguished letters $d_i$ are given by the definition of the final states.

\begin{corollary}[Soundness of~$\cU$]\label{cor:sound}
The following assertions hold.
\begin{itemize}
\item If $h\in L(\cU)$, then $h$ is a $\wh A$ \Endo of $M(C,\rho)$ such that 
$h(d_i) \in M(\wh A,\rho)$ and hence $\wh \pi h(d_i) \in M(A,\rho)$ for all distinguished letters $d_i$,
where $\wh\pi$ is as in \prref{sec:cover}
\item We have inclusions: 
\begin{align}\label{eq:EQwhA}
\set{(h(\cent)
}{h \in L(\cU)} &\sse \{(\sig(X_1),\ldots ,\sig(X_k))\mid \sig \text{ solves } \cSinit \text{ in } M(\wh A,{\rho})\},\\
\label{eq:EQA}
\{(\wh \pi h(\cent)
\mid h \in L(\cU)\}&\sse \{(\sig(X_1),\ldots ,\sig(X_k))\mid \sig \text{ solves } \cSinit \text{ in } M(A,{\rho})\}.
\end{align}
\item If $L(\cU)\neq \es$, then $\cSinit$ has an \esolu. 
\end{itemize}
\end{corollary}
\begin{proof}
Every weak \solu at $\cSinit$ is an \esolu, see \prref{rem:weaksolu}.
The proof is therefore straightforward from \prref{thm:backstage} and the definition of final states,
where the distinguished letters are introduced and used in the equation of final states. 
\end{proof}
\prref{cor:sound} shows the inclusions in (\ref{eq:EQwhA})
and (\ref{eq:EQA}), which are needed for soundness.
For completeness we also need the converse of these inclusions. This is trivial 
if the right-hand side of (\ref{eq:EQwhA}) is empty. 
\prref{cor:sound} implies that it is not empty
if $L(\cU)\neq \es$. 

\subsection*{Intermezzo}\label{sec:inter}
Where we are? Since soundness has been established 
the remaining task is the completeness proof. For that we start with 
a given \solu $\sig$ at the initial state $\cSinit$.
Whatever $\sig(\Winit)$ is, it has finite length. As we will see, we can make 
a finite sub automaton $\cA$ of~$\cU$ large enough such that we can track the \solu by using a path in the sub automaton~$\cA$. 
Of course, we need a concrete size bound for $\cA$ which fits to our main theorem and \ip
which does not depend on $\sig$, but only of $\Abs{\cSinit}$. 
We postpone the calculation of an upper bound on the size 
in order to show that 
$\cA$ is large enough for all \solu{s}. That is why we work in the `universal' and infinite 
automaton~$\cU$. 

Actually, there an effective way by a naive dove tailing algorithm, to calculate the least NFA $\cA$ such that
every \solu $\sig$ at the initial state $\cSinit$ defines
via the identity on $A$ an \esolu $(\id_A,\sig)$.
We simply enumerate  all $\sig$ and all possible 
$\cA$ by the length of their bit encodings. 
If a current pair $(k,\ell)\in \N^2$ bounds the bit encoding of $\cA$ by  $k$ and that of $\sig$
by $\ell$, then we check whether $\cA$ witnesses that 
$(\id_A,\sig)$ is an \esolu. 
If not, then we replace $k$ by $k+1$. If $k$ is good enough for all \solu{s} $\sig$ with a bit encoding of length at least 
$\ell$, then we replace $\ell$ by $\ell+1$. This is a trivial algorithm, but the complicated part is to prove termination. 
The benefit of this naive algorithm is that it will construct a minimal size NFA~$\cA$ inside 
$\cU$ 
for the \edtol-description of all \solus.\footnote{Actually, our approach is not too far away. Some comments on Makanin's original algorithm say that the algorithm is complicated. 
This is highly debatable: the algorithm is very clever, but not complicated and easy to implement, 
see \cite{abd92}. Using some standard tools from linear integer programming, it is actually hard to 
find examples where it does not give the answer quickly (although these examples exist if SAT is hard). What is complicated, and took Makanin in \cite{mak77} more than $80$~pages, is 
to prove termination on every input.}

\section{Visibility and the equivalence of subtraces}\label{sec:vispos}

The intuitive idea to define equivalent subtraces is as follows: if we modify a \solu at a position which belongs to some subtrace, then we must modify all equivalent subtraces at the corresponding position in order to obtain a modified \solu. Consider a position~$p$ with label~$a$. Assume that we choose a fresh clone $c\in \wh C\sm B $ of~$a$.
Then it is possible to define a new \solu $\sig'$ which uses the letter $c$ instead of~$a$ at the position~$p$. In order to define $\sig'$, various positions labeled by~$a$ (resp.~$\bar a$) need to be relabeled by $c$ (resp.~$\bar {c}$).
There is a least equivalence relation of positions where we have to 
 use the label $c$ or $\bar {c}$ to guarantee that $\sig'$ is still a \solu. The strategy to replace the letter~$a$ by $c$ using this least equivalence relation gives more freedom and is more versatile. 
The formal definition of the equivalence relation $\equiv$ between subtraces is guided by this idea and it uses the mapping $\sig^*$ (defined in \prref{sec:mapos}) to make the idea rigorous. 
Recall that $\sig^*$ maps the positions of $\sig(W)$ to positions of $W$, and 
that $\sig(\sig^*(p))$ is a set of positions: it is the set of positions 
$\set{q\in \pos(\sig(W))}{\sig^*(p)=\sig^*(q)}$. It is clear that
$\sig(\sig^*(p))$ is a subtrace in $\sig(W)$ and therefore every subtrace 
of $\sig(\sig^*(p))$ is also a subtrace in $\sig(W)$.

Consider subtraces $u$ and $u'$ of $\sig(W)$ (which are defined by their positions) such that $\sus u$ and $\sus {u'}$ are single occurrences of the same variable in $\cX$. 
For positions $p\in \sig(\sus{u})$ and $p'\in \sig(\sus {u'})$ we define $p \sim p'$, if it holds for all $a\in B$ that~$p$ is the $i$-th position labeled by~$a$ in $\sig(\sus{u})$ \IFF $p'$ is the $i$-th position labeled by~$a$ in $\sig(\sus{u'})$. 
If the relation $p \sim p'$ induces a bijection between the set of positions $\pos(u)$ and $\pos(u')$, then we let $u \sim u'$.

Note that this is well-defined for traces:
all positions labeled by~$a$ are comparable and so the partial order restricted to positions labeled by~$a$ is a linear order.
Note that $u \sim u'\neq u$ implies that $u$ and $u'$ are disjoint.
Clearly, $\sim$ is an equivalence relation.
For each position $p\in \sig(W)$ there is a canonical \emphind{dual position} 
$\bar p$ in $\sig(W)$. 
For the definition of $\bar p$ we use $\sig(W)= U\cent \bar U= \ov{\sig(W)}$. Let $a\in B$ and suppose that~$p$ is the $i$-th position from the left in $\sig(W)$ which is labeled by~$a$. Then we define $\bar p$ as the $i$-th position from the right in $\sig(W)$ which is labeled by $\bar a$. 
Note that $p=\bar p$ is true for the central position labeled by $\cent$, only.
We extend the duality of positions to a duality between certain subtraces of $\sig(W)$ which do not use the central position as follows. 
Let $u$ be a subtrace in $\sig(W)$ with 
$|u|_\cent=0$, then we define $\bar u= \set{\bar p}{p\in u}$ to be the \emphind{dual subtrace} in $\sig(W)$, and we write $u \darc \bar u$ in that case. 
If we write $\sig(W)=w \cent\bar w$, then $u \darc \bar u$ and $|u|_\centime=0$
imply that either $u$ or $\bar u$ is a subtrace in~$w$, but not both are subtraces of~$w$.  The definition fits to the idea to visualize a \inner equation 
$U=\bar V$. Since $\sig$ is a \solu, we can write 
$\sig(U)=\sig(V)$, and the idea is that $\sig(U)$ and $\sig(V)$ share the same 
set of positions. In $\sig(W)$ the equality is reflected by the 
duality of the subtraces of the nonempty subtraces $\ov{\#\sig(U)}= \sig(\bar  V)\#$. 

Note that $u \darc \bar u \darc u'$ implies
$u=u'$.
Moreover, the set of positions of $u$ and $\bar u$ are disjoint unless $u$ is the subtrace defined by the central position, which is self-dual. 

By $\approx$ we denote the equivalence relation between subtraces
of $\sig(W)$ which is generated by $\sim$, and by $\equiv$ we denote the equivalence relation generated by the union of ${\approx}$ and ${\darc}$. 
The relation $\equiv$ is the transitive closure of ${\approx}\cup{\darc}$.

Clearly, if $u \equiv u'$, then their labelings satisfy $\lam(u')=\lam(u)$ or $\lam(u')=\lam(\bar u)$. 
The obstacle to compress a subtrace $u\in \sig(W)$ into a letter is that $u$ `crossing' in the following sense: there exists some $u'\equiv u$ and two positions $p,q\in \pos(u')$ such that $\sus{p}\neq \sus{p}$ 
and $\sus{p}$ is a variable. That means that $\sus {u}$ crosses the border of a variable. 
An equivalent characterization is part of the following definition. 
\begin{definition}\label{def:fiv}
Let~$E$ be a state with an equation $W$ and a \esolu $(\alp,\sig)$.
A nonempty subtrace $u$ of $\sig(W)$ is called \emphind{invisible} (resp.~\emphind{visible}) in $W$ if $\sus {u}$ is a variable (resp.~no position in $\sus {u}$
is a variable). 
A subtrace $u$ is called \emphind{fully invisible} 
(resp.~\emphind{fully visible}) in~$W$ if $u\equiv u'$ implies that 
$\sus {u'}$ is \emph{invisible} (resp.~\emph{visible}).
A subtrace $u$ is called \emphind{crossing} if there exists $u\equiv u'$ 
such that $u'$ is neither visible nor invisible. 
\end{definition}
\begin{lemma}\label{lem:whAinv}
$E=(W,B,\cX,\rho,\mu,\theta,T)$ be a state with an \esolu $(\alp,\sig)$ and
$p\in \pos(\sig(W))$ a position which has a visible equivalent 
position in $\pos(\sig(W))$. If by following an \epstra $E\arc \eps E'$ 
we reach the \esolu $(\alp',\sig')$, then the position $p\in \pos(\sig(W))$ still  has a visible and equivalent 
position in $\pos(\sig'(W'))=\pos(\sig(W))$.
\end{lemma}
\begin{proof}
Consider the shortest sequence $p=p_0,p_1\lds p_m=q$ of relations in $\sim$ and $\darc$ which links $p$ to a visible position $q$. Following an \epstra 
may break this sequence into pieces, as some position cease to be in $\sim$ relation. If $p_i, p_{i+1}$ were in $\sim$ in $\pos(\sig(W))$
but not in $\pos(\sig'(W'))$ then this means that $p_i$ is a visible position in $\pos(\sig'(W'))$.
Choosing the smallest possible index $i$ yields the claim.
For example assume that $p_0\lds p_{j-1}$ remain invisible in $\sig'(W')$ and  
$r=p_j$  becomes visible in $\sig'(W')$, then $p$ and $r$ are equivalent 
in $\pos(\sig'(W'))$ \wrt~$\sig'$. 
\end{proof}
We also use a relaxed notion of semi-visibility: 
\begin{definition}\label{def:semi-visible}
Let $E=(W,B,\cX,\rho,\mu,\theta,T)$ be any state with a \solu $\sig$. A position~$p$ in~$\sig(W)$ is called \emphind{semi-visible on the left} (resp.~\emphind{semi-visible on the right})
if~$p$ is a minimal (resp.~maximal) position in $\set{q \in \pos(\sig(W))}{\sig^*(q) = \sig^*(p)}$.
The position~$p$ is called \emphind{semi-visible} if it is semi-visible on the left or on the right. 
(Every visible positions is semi-visible.)
\end{definition}
\begin{remark}\label{rem:visov}
Remember that an equation $W$ at state~$E$ can also represented
as a system of \inner equations of the form $U=V$. 
If $\sig$ is a \solu at $E$, then the traces 
$\sig(U)$ and $\sig(V)$ are equal.
If $u\sse \pos(\sig(U))$ is a visible subtrace in $\sig(U)$ 
and $u$ contains a position $p$ 
such that $\sus p\sse V$ is a variable, then $u$ is crossing.
This follows because the identity 
$\pos(\sig(U))=\pos(\sig(V))$ is realized in $W$  
trough the relation ${\darc}$ between $u\sse \pos(\sig(U))\sse \sig(W)$ 
and $\bar u \sse \pos(\sig(\bar V))\sse \pos(\sig(\bar W))$. 
\hspace*{\fill}$\diamond$\end{remark}
\begin{example}\label{ex:croissant}
Let~$E$ be a state as above having a single \inner equation $U=V$ with $U=XbY$ and $V=abba$ where $\rho(a)=\rho(b)$. It can be written as
\begin{align*}
X\to b \to Y = (a \to b) \to b \to (a) 
\end{align*}
The equation $U=V$ has a \solu $\sig(X)=ab$ and $\sig(Y)=a$
which is shown by the brackets.
All positions are semi-visible.
The \HA $a \to b$ in $\sig(U$ is invisible, but it not fully invisible 
$V=abba$. The \HA $b \to a$ in $\sig(U$ is crossing, but 
$a \to b$ and $b \to a$ are not equivalent, even if~$a$ and $b$ are selfies. Choosing a fresh clone $c\in \wh C\sm B$ of $ab$ we find 
a compression \tra defined by $h(c)=ab$ to a standard state $E'$
with a \solu $\sig'$ such that $\sig=h\sig'$ by letting 
$\sig'(X)=c$ and $\sig'(Y)=a$.
The \inner equation at $E'$ has the form $XbY=cba$.
\hspace*{\fill}$\diamond$\end{example}
\begin{figure}[t]
	\begin{center} 
		\begin{tikzpicture}[xscale=2, yscale=0.5]
		
		\node[circle] (bb) at (2,0){$b\sig(X)abd$};
		\node[circle] (eq) at (2.8,0) {$=$};

		\node[circle] (d_1) at (3,1) {$d$};
		\node[circle] (c1) at (3,-1) {$b$};
		\node[circle] (a1) at (4,0) {$a$};
		\node[circle] (e1) at (4,1.5) {$a'$};
		\node[circle] (e) at (5,3) {$e$};
		\node[circle] (a2) at (5,0) {$\bar a$};
		\node[circle] (a3) at (6,0) {$a$};
		\node[circle] (d_2) at (7,1) {$d$};
		\node[circle] (c2) at (7,-1) {$b$};

		\draw (d_1) edge[->, >=latex,thick] (a1); 
		\draw (c1) edge[->, >=latex,thick] (a1); 
		\draw (d_1) edge[->, >=latex,dashed] (e1);
		\draw (e1) edge[->, >=latex,dashed] (e);
		\draw (d_1) edge[->, >=latex,thick,bend left=40] (e);
		\draw (e) edge[->, >=latex,thick,bend left=40] (d_2); 
		\draw (c1) edge[->, >=latex,dashed] (e1);
		\draw (a1) edge[->, >=latex,thick] (a2); 
		\draw (e1) edge[->, >=latex,dashed] (a2); 
		\draw (a2) edge[->, >=latex,thick] (a3); 
		\draw (a3) edge[->, >=latex,thick] (d_2); 
		\draw (a3) edge[->, >=latex,thick] (c2); 
		
		\node[circle] (qe) at (7.3,0) {$=$};
		\node[circle] (cc) at (8,0){$d\sig(Y)abd$};
		
		\draw[decorate,decoration={brace,mirror},thick] (3.85,-0.7) --node[below] {$h(c)=a\bar a$} (5.15,-0.7); 
		\end{tikzpicture}\end{center}
	\vspace{-0.5cm}
	\caption{A compressible \TTarc $a\bar a$ becomes $c$ for $\sig(bXabd)=  bda\bar a abd= \sig(dYabd)$.}
\label{fig:deco}
\end{figure}
\begin{example}\label{ex:difffirst}
Let us begin with a system having two \inner equations $bXabed=bda\bar aabed$ and $dYabed= bda\bar aabed$ as depicted in \prref{fig:deco} 
where $\bar a\neq a$ and $b,d,e$ are selfies, but the letter 
$a'$ is not present. 
Suppose further $T=\rho(a)< \rho(d)$. Then 
there is a unique \solu $\sig(X)=da\bar a$ and $\sig(Y)=ba\bar a$. 
At this state, there are several options for how to proceed.

For example, the compression of the \TTarc $a\to \bar a$ into a fresh letter $c$ is possible, since this \HA is not crossing. 
Hence, there is a \comp \tra to a state $E'$ defined by $h(c)=a\bar a$ where $c$ is a selfie. The new \inner equations 
are given by $bXabed=bd\, c\, abed=dYabed$. The \solu at $E'$ is
$\sig(X)=d\, c\, a$ and $\sig(Y)=b\, c\,  a$.
Note that introducing the selfie~$c$ made the weight of the system decrease because we have $|bd\, c\, abed|<|bd a\bar a abed|$.

Instead of compressing the \HA $a\to \bar a$ we can also lift the left-most~$a$ by replacing it 
with a $(T\cup \rho(d))$-clone~$a'$. This is realized 
by \comp \tra to a state $E''$ defined by $h(a')=a$.
The new \inner equations 
are given by $bXabed=bd\, a'\, \bar a abed=dYabed$. The \solu at $E'$ is
$\sig(X)=d\,a'\, \bar a$ and $\sig(Y)=b\,a'\, \bar a$.
Introducing the fresh letter~$a'$
made the weight of the system decrease because we have $\rho(a)\ssneq \rho(a')$.
The \HD of  $bd\, a'\, \bar a abed$ is obtained by performing the following two steps in \prref{fig:deco}.
First remove the position of the leftmost letter~$a$ and the incident \HAs. Second remove the \HA $d\to e$ and introduce~$a'$ and the dashed \HAs. 
Actually, we can lift both $a$-positions using the same fresh letter~$a'$, but we cannot lift $\bar a$ to $\ov{a'}$. 

So there are several options, but giving a preference is subtle. 
If we intend to remove \STarcs at a $T$-state where $S<T$, then we give the priority to lifting: for example if we have $\rho(b)<T$, then
$b\to a$ is an \STarc which is replaced by the \HA $b\to a'$ which is not a \Tarc. 
If there are no \STarcs at a $T$-state where $S<T$, then we lift only fully invisible $T$-positions.
\hspace*{\fill}$\diamond$
\end{example}

\section{The unfolding $\wh \cU$ defined by $\cU$} 
\label{sec:unfold}

The concept of \emphind{unfolding} has its origins in algebraic topology in the framework of covering spaces. In Computer science unfoldings are applied, for example, for Model Checking 
state \tra systems, as in the textbook \cite{EsparzaH08}.
A \emphind{state \tra system} is a directed graph where the vertices are called states, the directed edges are labeled and called \tras. There is also a set of initial states, but (typically) there are no final states. 

Our definition of unfolding is in the spirit of unfoldings of state \tra systems, but not the same. 
In our case the underlying state \tra system is given 
by the set of states $E\in \cU$ with their \esolu{s} which are reachable from the initial system $\cSinit$ and an initial \solu $\siginit$
together with the \tras between reachable states.\footnote{The first question is whether we can decide if $\cSinit$ has a \solu. 
The answer is yes, see \cite{dm06}. 
However we do not need this. 
Thus, we might explore a large portion of $\cU$ before we possibly detect that there was no \solu at $\cSinit$.}

More precisely, the state set $V(\wh{\cU})$ is a subset the set of entire states
$(E,\alp,\sig)$ where~$E$ is a $T$-state in $\cU$ for some $T\sse \Rs$ and $(\alp,\sig)$ is an \esolu at $E$. 
The set of initial states $\wh I$ in our unfolding $\wh{\cU}$ is given by $(\cSinit,\id_A,\siginit)$ 
where $\siginit$ is some initial \solu. 
For every $(E,\alp,\sig)$ in $V(\wh{\cU})$ and every 
outgoing labeled \tra $E\arc h E'$ (including $\eps$-loops) with $E=(W,B,\cX,\rho,\mu,\theta,T)$ and 
$E' = (W',B',\cX',\rho',\mu',\theta',T')$, we stipulate that $(E',\alp',\sig')$ is in $V(\wh{\cU})$ and that there is a labeled \tra $(E,\alp,\sig) \arc h (E',\alp',\sig')$ in $\wh \cU$ 
\IFF the following conditions are satisfied.
\begin{enumerate}
\item The pair $(\alp',\sig')$ is an \esolu at $E'$ with $\Abs{E',\alp',\sig'}< \Abs{E,\alp,\sig}$.
\item If $E= E'$, then $\wh \pi \alp' \sig'= \wh \pi \alp \sig$.
\item If $T\neq T'$, then $(\alp,\sig)=(\alp',\sig')$.
\item If $E\neq E'$, $T= T'$ and $h=\eps$, then 
there is a \subst $\tau:\cX\to M(B'\cup \cX',\rho')$ such that $W'=\tau(W)$ and 
$(\alp,\sig)=(\alp,\sig'\tau)$.
\item If $h\neq \eps$, then there is a \comp \tra $E\arc h E'$ in $\cU$
(and hence $T'=T$, $\cX'=\cX$, and $W=h(W')$) such that $\wh \pi \alp'\sig'(x)= \wh \pi \alp h\sig'(x) = \wh \pi \alp \sig h(x)$ for all $x\in B'\cup \cX$. 
\end{enumerate}
Let $(E,\alp,\sig)\in \wh \cU$, then 
the projection $\pr 1$ onto the first component 
yields a \morph of $\wh \cU$ to $\cU$ 
which respects (initial) states, it maps an $\eps$-loop to an empty path in $\cU$, the other \tras to \tras are respecting their labeling and their incident relation: the  \morph $\pr 1$ is therefore a \morph of state \tra systems. It is not an \epi, in general. 
However, for the completeness proof we are interested 
in $\pr1(\wh \cU)$, only. 
In order to distinguish states and paths in $\wh \cU$, in $\pr1(\wh \cU)$, and in $\cU$ we use the following terminology. 
\begin{definition}\label{def:entipath}
A state or a path in the unfolding $\wh \cU$ is called an \emphind{entire state} resp.~\entipa\index{entire path}.\footnote{Recall that states in $\wh \cU$ are entire states of the form $(E,\alp,\sig)$.}
An entire state $(E,\alp,\sig)$ is a $T$-state, if~$E$ is a $T$-state.

A state or a path in $\cU$ which is in  $\pr1(\wh \cU)$ is called \emph{covered\index{covered state}}. 
\end{definition}
\begin{proposition}\label{prop:entia}
The following properties hold.
\begin{enumerate}
\item A path $\pi$ with a label $h:M(B',\rho')\to M(B',\rho')$ from a state~$E$ to $E'$ is covered 
by an \entipa $\wt \pi$ with the same label~$h$.
\item If a semi-final state $E=(W,B,\es,\rho,\mu,\Rs)$ in $\cU$ is covered by
an entire state $(E,\alp_{\textrm{fin}},\id_\es)$, then $\siginit=\alp_{\textrm{fin}}$ solves the
initial system $\cSinit$.
\item 
There are no infinite (entire) paths in the unfolding $\wh \cU$. 
\end{enumerate}
\end{proposition}
\begin{proof}
The first item follows from the definition of $\wh \cU$. 	
	
The second item follows from \prref{thm:backstage} and that every path in $\cU$ from $\cSinit$ to 
the final state $\cent$ defines a path from $\cSinit$ to 
the some semi-final state and vice versa. 

The third item is trivial because we have $\Abs{E',\alp',\sig'}< \Abs{E,\alp,\sig}$
for all \tras in $\wh \cU$ from $(E,\alp,\sig)$ to  $(E',\alp',\sig')$
and all paths in $\wh\cU$ are entire.
\end{proof}
\begin{remark}\label{rem:entia}
The second item in \prref{prop:entia} shows that the unfolding is an acyclic \tra system as every unfolding should be. 
Later we will show the converse of the first item: that is, if $\siginit$ solves 
$\cSinit$, then there is a path in $\pr1(\cU)$ which can be `unfolded' 
to an \entipa in $\wh \cU$ from $(\cSinit,\id_A,\siginit)$ to some entire state $(E,\siginit,\id_B)$
where~$E$ is semi-final and~$B$ is its set of constants.
\hspace*{\fill}$\diamond$\end{remark}
\begin{convention}\label{conv:enzian}
In the following whenever we consider $E=(W,B,\cX,\rho,\mu,\theta,T)$ with an \esolu $(\alp, \sig)$, then actually refer to the entire state $(E,\alp,\sig)$ in $\wh \cU$. 
Following a path in $\cU$ at a state with an \esolu means to  follow the corresponding \entipa in $\wh \cU$. 
It is this \entipa which controls the flow of \esolu{s}.
\end{convention}

\section{Basic $T$-reductions}\label{sec:stproc}

\subsection*{Notation in this section}
We denote by $E=(W,B,\cX,\rho,\mu,\theta,T)$ a state with an \esolu $(\alp, \sig)$
and throughout we use \prref{conv:enzian} which refers the entire state $(E,\alp, \sig)$ in the unfolding.
In some cases we impose the condition that~$E$ is a standard state.

We describe some outgoing \tras $E\arc h E'$ which we call \emphind{basic $T$-reductions}.
Each of them defines an \esolu $(\alp', \sig')$ at~$E'$ such that $\Abs{(E',\alp', \sig')}<\Abs{(E,\alp, \sig)}$ and such that the equation $W'$ at $E'$ is never longer than $W$.
\Ip following an \epstra defined by $X\mapsto \tau(X)$ with
$|\tau(X)|\geq 2$ is not a basic $T$-reduction. 

The general idea of a \emphind{reduction} is to reduce the 
problem to find a \esolu at a state~$E$ to the problem to find a \esolu 
at a state $E'$ where less \solus exist. 
Thus, finding a \solu at~$E'$ is typically harder than at~$E$.
The reduction is a guess of an outgoing \tra. Due to soundness, we cannot move from an unsolvable state to a solvable one. Hence, making a wrong guess never leads to a false \solu,
\ie a `{\textsc{Yes}}'-answer of our algorithm is always correct.
 
\subsection{Removing a variable $X$ as soon as $|\sig(X)|\leq 1$}\label{sec:Xnores}
A variable $X\in \cX$ is called a \emphind{dummy}, if $\rho(X)=\es$.
This implies $\sig(X)=1$ when $\sig$ is \solu. Being a dummy is a syntactic condition.
More generally, if $|\sig(X)|\leq 1$, then we typically remove $X$ {}from $W$ and $\cX$ by using an \epstra which makes $\cX$ smaller.\footnote{Note that
$|\sig(X)|\leq 1$ refers to a semantic condition, we cannot check in general.} 
This leads to a state 
$E'= (W',B,\cX',\rho,\mu,\theta,T)$ which satisfies $\abs{W'}\leq \abs{W}$ and $\Abs{W'}<\Abs{W}$.
Moreover, the restriction $\sig'$ of $\sig$ yields an \esolu $(\alp,\sig')$ at $E'$.
Thus, whenever convenient we may assume that $|\sig(X)|\geq 2$ for all $X\in \cX$.

\subsection{Removing `useless' letters outside $\wh A$}\label{sec:useless}
Let ~$E$ satisfy $|\sig(W)|_c = 0$ for some $c\in B\sm\wh A$.
Such a letter is called \emphind{useless}, as we can remove it without affecting neither the equation nor the solution. Let 
$B'= B\sm \os{c,\bar c}$. Let $\alp'$ be the restriction of $\alp$ to $B'$ and $\sig'(X) = \sig(X)\in B'^*$.  Then there is an \epstra reducing the weight of the state:  
\(
E=(W,B,\cX,\rho,\mu,\theta,T)\arc {\eps} E'=(W,B',\cX,\rho,\mu,\theta,T).
\)
The \esolu at $E'$ is $(\alp',\sig')$. Thus, for showing completeness we are interested at a standard state 
only in those \solus where no letter in $B\sm A$ is useless. 
Remember that letters in $\wh A$
are never useless. 

\subsection{Making letters in $\wh A$ invisible}\label{sec:whAinvis}
We cannot avoid that~$E$ is a state where a letter $a \in \wh A$ is visible. This is typically true at the initial of it appears in the equation when we follow a \subst \tra popping out of a variable a letter from $\wh A$. As a matter of fact, this will happen only if~$E$ is a standard state. 
If $E=(W,B,\cX,\rho,\mu,T)$ is a standard state where some $a \in \wh A$ is visible, then 
we repair this by a fresh clone $c\notin B$ of~$a$, and by defining a $B\cup \cX$-\morph
$h:M(B',\cX,\rho',\mu') \to M(B,\cX,\rho',\mu')$ by $h(c)=a$ where 
$B'=B\cup \{c,\bar {c}\}$. This, means 
we consider all positions~$p$ in $\sig(W)$ which are not fully invisible and labeled 
by~$a$ (resp.~$\bar a$). For each such position~$p$ relabel~$p$ the by $c$ (resp.~$\bar c$).
This is realized by following a 
\tra 
$E\arc{h} E'= (W',B',\cX,\rho',\mu',T)$ with $W=h(W')$. It is clear that $|W'|=|W|$ and $E'$ has an \esolu $(\alp h,\sig')$ such that
$\sig=h\sig'$. (Note that $\alp \sig= \alp h\sig'= \alp hh\sig'$.) Moreover, the weight of $(\alp h,\sig')$  at $E'$ is smaller than the weight of $(\alp,\sig)$ at $E$.
Thus, at standard states it is enough to deal with states~$E$ and \esolu{s} where all letters from $\wh A$ are invisible.

\subsection{Internal alphabetic $T$-reduction at standard states}\label{sec:inTred}
Let $T\ssneq \Rs$ and $E=(W,B,\cX,\rho,\mu,T)$ be a standard state with an \esolu $(\alp,\sig)$ which is 
without any useless letter and where no letter of $\wh A$ is visible in the equation $W$.
That is, every letter in $B\sm\wh A$ appears somewhere in $\sig(W)$, 
and $|W|_a=0$ for all $a\in \wh A$,
see \prref{sec:useless} and \prref{sec:whAinvis}.  

Suppose that there is some $c \in B \sm \wh A$ such that $|W|_c = 0$ and $\rho(c)=T'\geq T$. 
Then for $B'=B\sm \os{c,\bar c}$ there is an \epstra
\(
E\arc{\eps}E'=(W,B',\cX,\rho,\mu,T)).
\)
Moving to the state $E'$ is sound, but we loose completeness in general.
The idea, to avoid this, is to replace the letter $c\in B'$ by 
its \SNF $\snf_{T'}(\alp(c))\in \wh{A}^+$, and to define another
\esolu $(\alp,\sig')$ at~$E$ such that $|\sig'(W)|_c=0$ and 
$\wh \pi \alp \sig(W)=
\wh \pi \alp \sig'(W)$. We have to apply this idea in a more general form to fully invisible positions rather than 
to letters.
Suppose that $\sig'(W)$ contains a fully invisible 
position $p$ with $\rho(p)=T'\geq T$.\footnote{If a fully invisible position 
satisfies $\rho(p)\leq T$, then we do nothing with~$p$ at this point.}
Let $c=\lam(p)=c\in B \sm \wh A$ (whether or not $|W|_c = 0$). 
Then we replace every equivalent position $q\equiv p$ by the $T'$-sequence $u_c$ 
where $u_c$ is the \SNF $\snf_{T'}(\alp(c))\in \wh{A}^+$ if $\lam(q)=c$ 
and by $\ov{u_c}$ if $\lam(q)=\bar c$. 
So we still have $\alp(a)\in \wh A_S^+$ for all $a\in B$.

This procedure can be implemented without loosing completeness 
by the following procedure.
\begin{enumerate}
\item Choose a fully invisible 
position $p$ with $\rho(p)=T'\geq T$ and label $\lam(p)=c\in B \sm \wh A$
\item Consider all positions $q$
in $\sig(W)$ which which are equivalent to $p$
\\(None of these positions is visible because $p$ is fully invisible.) 
\item Choose a fresh clone $d$ of 
$c$ and replace all invisible positions labeled by $c$ (resp.~$\bar c$)
by $d$ (resp.~$\bar d$). Define $B'' =B\cup \os{d,\bar d}$,
\item Define a $B$-\morph $\bet:B''\to B$ defined 
by $\bet(d)=\snf_{T'}(\alp(c))$.\\
(Recall that $\snf_{T'}(\alp(c))$ is a word in 
$\wh A_{T'}^+$ of $T'$-letters such that $\ov{\snf_{T'}(\alp(c))}= \snf_{T'}(\alp(\bar c))$.
Moreover, the fresh letter $d$ does not appear in $\bet(B'')$.)
\item Replace the \esolu $(\alp,\sig)$ at~$E$ by $(\alp,\bet\sig)$ without changing the state. 
(Note that we have $\mu(c)=\mu(\alp(c))= \mu(\snf_{T'}(\alp(c)))$.)
This is \tra in $\wh\cU$ from $(E,\alp,\sig)$
to $(E,\alp,\bet\sig)$ which covers an $\eps$-loop in $\cU$.
\item If the letter $c$ does not appear in $\bet\sig(W)$, then 
follow another \epstra in $\wh\cU$ to the entire state $(E',\alp,\bet\sig)$ where
$E'=(W,B\sm{c,\bar c},\cX,\rho,\mu,T))$. This implies that $\Abs{E'}<\abs{E}$, and the  \epstra covers an \epstra in $\cU$. 
\end{enumerate}
The procedure leads to the following lemma which is crucial to compute an a priori bound on the ambient alphabet 
$\wh C$ which depends on $\Abs\cSinit$ and not on the length of an  initial entire \solu. 
\begin{lemma}\label{lem:alTred}
Let $E=(W,B,\cX,\rho,\mu,T)$ be a standard state with an entire state $(E,\alp,\sig)$ which is 
without any useless letter and where no letter of $\wh A$ is visible in the equation $W$. Then there is another 
\esolu $(\alp,\sig')$ at~$E$ such that firstly $\wh \pi \alp \sig(W)=\wh \pi\alp \sig'(W)$ and secondly there is no fully invisible position $p$ 
in $\sig'(W')$ where $\rho(p)\geq T$ and $\lam(p)\in B\sm \wh A$.
\end{lemma}
\begin{proof}
By the third step in the procedure above  may assume that there is some $c\in B\sm \wh A$
such that $|W|_c=0$, $|\sig(W)|_c\geq 1$, and $T\leq \rho(c)$. 
Assuming this, the procedure does nothing but replacing the \esolu $(\alp,\sig)$ by the pair $(\alp,\bet\sig)$. We know that $\sig(W)=\sig(\bar W)$. This implies 
$\bet\sig(W)=\bet \sig(\bar W)$, and therefore $\sig'=\bet\sig$ is 
a \solu at $E$. Since $\sig$ is an \esolu, there are no $S$-$S'$-\HAs neither in $\sig(W)$ in nor in
$\alp\sig(W)$ for $S,S'<T$. Replacing a position in $\sig(W)$ by the nonempty word in $T'$-letters with $T\leq T'$ cannot introduce $S$-$S'$-\HAs. Hence, $\bet\sig(W)$ is without $S$-$S'$-\HAs 
for $S,S'<T$, too. This implies that $(\alp,\sig')$ is an \esolu 
at~$E$ and if $X$ is a variable, then $\sig'(X)$ does not use the letter $c$. This reduces the weight of the entire state.

Now it is enough to show $\wh\pi\alp\bet= \wh \pi\alp$ because this 
implies $\wh\pi\alp\sig'= \wh \pi\alp\sig$. 
As $\bet(a) = a$ for $a \notin \{ c, \bar c\}$,
it is enough to consider the value on $c$ (the argument for $\bar c$ is the same).
We have $\bet(c) = \wh \pi(\snf_{T'}(c))= \wh \pi\alp(c)$ and as $\wh\pi\alp= \wh\pi\alp\wh\pi\alp$
we conclude that
\(
\wh\pi\alp= \wh\pi\alp\bet
\) as desired. 
\end{proof}

\subsection{$T$-lifting of steps at standard states}\label{sec:steplift}

The process of $T$-lifting is another main feature and rather powerful, but technical.
We enter \prref{sec:steplift} only at a standard  state $E=(W,B,\cX,\rho,\mu,T)$ with an \esolu $(\alp,\sig)$. 
If there are still \STarc{s} with $S<T$, then we `lift-as-lift-can' steps with are connected to one or two $T$-positions whenever the lifting is possible and leads to some entire state $(E',\alp',\sig')$ with less \STarc{s} and where $\Abs{E',\alp',\sig'} < \Abs{E,\alp,\sig}$.
Behold, a $T$-lifting is not always possible: this depends on an entire state $(E,\alp,\sig)$ with a standard state $E=(W,B,\cX,\rho,\mu,T)$ and the choice (which depends on the \solu) of a nonempty step $s$ which appears as a subtrace in $\sig(W)$ and which we wish to lift. 

The formal definition is as follows.
We consider a step $s=\prod_{1\lds r}s_i$ in $\sig(W)$ such that there is a $T$-position 
$p$ in $\sig(W)$ with \HAs from $p$ to every position in $s$. 
Additionally, for $r=1$ we require that $\rho(s_1)\ssneq \rho(s_1) \cup T$, and for $r\geq 2$ we require that there is also a second $T$-position $p'$ with \HAs from every position in $s$ to $p'$.

Let $T'=\rho(s)\cup T$; and choose a fresh $T'$-clone
of $s$. Define $B'=B\cup \os{c,\bar c}$ and a $B\cup \cX$-\morph $h:M(B',\cX,\rho',\mu',T)\to M(B,\cX,\rho,\mu,T)$ by $h(c)=s$. 
For a possible lift we require that the following conditions are satisfied.
\begin{enumerate}
\item There is some $W'\in M(B'\cup\cX,\rho)$ such that 
$h(W')=W$. 
\item There is a $B$-\morph $\sig':M(B',\cX,\rho,\mu,T)\to M(B,\rho,\mu,T)$ such  that $h\sig'=\sig$. That is, the diagram in \prref{fig:comcon} commutes. 
\item We have $\sig'(W') \neq \sig(W)$ which implies $\Abs{W'}\leq \Abs{W}$ and $\Abs{\sig'(W')}\leq \Abs{\sig(W)}$.
\item We have $\sig'(\bar W')= \sig'(W')$. That is, $(\alp',\sig')$ is an \esolu.
\item No fully invisible position in $\sig'(W')$ is labeled by $c$ (resp.~by $\bar c$); instead, such positions are replaced by the \SNF $\snf_{T'}(c)$ (resp.~by $\snf_{T'}(\bar c)$). 
\item If $\alp\sig(W)$ is without any \STarc where $T \nsse S$, then $T' \neq T$.
  That is, once $\alp\sig(W)$ is without any \STarc where $T \nsse S$, then we do not allow to introduce new $T$-positions. 
\end{enumerate}
If a $T$-lifting of a step is possible, \ie the above conditions are satisfied, then there are two cases. 
Firstly, if there remains in $\sig(W)$ a visible position labeled by 
the $T'$-clone $c$, then we can realize the $T$-lifting by a \comp \tra.
Secondly, if there in $\sig(W)$ there is no visible position labeled by the $T'$-clone $c$, then we switch directly from the entire state $(E,\alp,\sig)$ to $(E,\alp',\sig')$ without changing the state $E$ where in $\alp'\sig'(W)$ the step is replaced by $\snf_{T'}(s)$.

If we can lift the step $s$, then we implicitly 
apply the following lemma. Its proof is straightforward and left to reader.
\begin{lemma}\label{lem:alpbet}
Let $E=(W,B,\cX,\rho,\mu,\theta,T)$ be a state with an \esolu $(\alp,\sig)$ and $s$ be fully invisible subtrace in $\sig(W)$ which is a step
in $\wh A$. 
Let $(s,T')\in \wh A$ with $\rho(s)\cup T\sse T'$. 
Suppose further that for each $X\in \cX$ there 
is some factorization 
$\sig(X)= u_{X,0}s_{X,1}u_{X,1}\cdots  s_{X,k}u_{X,k}$
such that a subtrace $s'$ of $\sig(X)$ is  equivalent 
to $s$ (in $\sig(W)$) \IFF $s'\in \os{s_{X,1}\lds s_{X,k}}$.
Assume that letting 
$\sig'(X)= u_{X,0}s_{X,1}u_{X,1}\cdots  s_{X,k}u_{X,k}$ for each 
$X\in \cX$ defines a \solu at~$E$ such that $\sig'(W)\neq \sig(W)$. 
Then $(\alp,\sig')$ is an \esolu~$E$ such that 
$\wh \pi \alp \sig' =\wh \pi \alp \sig$ with $\Abs{\alp \sig'}<\Abs{\alp \sig}$.\footnote{Here $\wh \pi: \wh A\to A$ is the $A$-\morph with $\wh \pi(a,T)=a$
as defined in \prref{sec:whApi}.}
\end{lemma}

\subsubsection{Some examples for $T$-liftings}\label{sec:Tliftex}
\begin{figure}[t]
\begin{center} \vspace{-1cm}
		\begin{tikzpicture}[xscale=1, yscale=0.5]
    \node[circle] (X) at (-2.3,0) {$\exists \sig: \sig(X) =\mathrlap{ab} \qquad \mathclap{\wedge} \quad \mathrlap{\sig(Xd\,\bar a)= \sig(ad \bar X)={}}$}; 
    \node[circle] (Xc) at (-2.3,-1.8) {$\forall \sig:\sig(X) =\mathrlap{ac} \qquad \mathclap{\implies} \quad \mathrlap{\sig(Xd\,\bar a)\neq \sig(ad \bar X)}$};
    \node[circle] (ra) at (3,0) {$a$}; 
		\node[circle] (rd) at (5,1) {$d$};
		\node[circle] (rb) at (5,-1) {$b$};
		\node[circle] (raa) at (7 ,0) {$\bar a$};
		\draw (ra) edge[->, >=latex,thick] (rd);
		\draw (ra) edge[->, >=latex,thick] (rb);
		\draw (rd) edge[->, >=latex,thick] (raa);
		\draw (rb) edge[->, >=latex,thick] (raa);
		
\end{tikzpicture}\end{center}
\vspace{-1.8cm}
	\caption{Mission impossible: replace $b$ by a letter $c$ with $\rho(a) \sse \rho(c)$.}
		\label{fig:bddiam}
\end{figure}
\begin{example}\label{ex:hcab}
Let $E=(W,B,\cX,\rho,\mu,T)$ be a standard state with a single \inner equation 
$Xab Y= Zab X$ and 
$a,b$ be two constants with $\rho(a)\sse\rho(b)=T$. The \inner equation has infinitely many \solu{s}. 
For example, $\sig(X)=aab$, $\sig(Y)=abaab$, and $\sig(Z)=aabab$
yields a \solu as shown by the following bracketing.
\[
\sig(Xab Y)=(aab) (ab) (ab aab) = (aab ab) (ab) (aab)= \sig(Zab X)
\]
Let $c\in \wh C\sm B$ be fresh a clone of $ab$. 
Then $h:M(B\cup \os{c,\bar c},\rho)\to M(B,\rho)$, defined by 
$h(c)=ab$, is a $B$-\morph. It yields a \comp \tra 
to the standard state $E'=(W',B',\cX,\rho,\mu,T)$ where
$B'=B\cup \os{c,\bar c}$. Letting $\sig'(X)=ac$, $\sig'(Y)=cac$, and $\sig'(Z)=acc$ is a \solu at $E'$ such that 
$\sig=h \sig'$. Thus, following this \tra is sound and complete \wrt~$(\alp,\sig)$ if $\alp:B\to \wh A$ is a $\wh A$-\morph. 
The new \inner equation is $Xc Y= Zc X$ and, since there is no other \inner equation, we have $|W'|_a =0$ although the letter~$a$ is used by the \solu.
However, replacing $a\in B$ by $\snf_T(a)\in \wh A$ enables us to follow an \epstra from $E'$ to 
$E''= (W',B'\sm \os{a,\bar a},\cX,\rho,\mu,T)$ without loosing completeness. This makes it possible to use the letter $a\in \wideh C$ later again. Remember that we can recycle~$a$ with a different image in the finite monoid $N$ but we may not change $\rho(a)$. 
\hspace*{\fill}$\diamond$\end{example}
Splitting a variable $X$ at a state $(W,B,\cX,\rho,\mu,\theta,T)$ by a \subst
$X\mapsto x$ where $|x|\geq 2$ cuts the equation in pieces. 
It makes the equation longer and puts more restrictions on the set of \solus. 
However, sometimes it helps to enable a $T$-lifting which is more important 
since, in any case,  at some point we have to relabel the 
$S$-position in every \STarc where $S<T$ by a letter $c$ such that $T \leq \rho(c)$. 
The following example shows that a $T$-lifting is impossible unless we perform a splitting. 
\begin{example}\label{ex:adca}
Let~$E$ and its \esolu $(\alp,\sig)$ be as above. Suppose that there 
are letters $a,b,d\in B$ and that $W$ is given by two \inner equations $Xd\,\bar a= ad \bar X$ 
and $X=Yb$. 
If the \HD of $\sig(Xd\,\bar a)$ looks like the \HD  
depicted in \prref{fig:bddiam}, then we must have $\sig(X)=ab$ with $\bar b =b$ and $\sig(Y)=a$; 
and a $T$-lifting, where $T = \rho(a)$, of the $b$-position in $\sig(Xd\,\bar a)= \sig(ad \bar X)$ is impossible.
The reason is that 
every relabeling of the $b$-position by a letter $c$ has to satisfy 
$T\sse \rho'(c)$, and there is no \solu~$\sig'$ which solves both equations $Xd\,\bar a= ad \bar X$ and $X=ac$, see \prref{fig:bddiam}.
Hence, we begin with a splitting by using a \subst $X\mapsto Xb$. 
Then we have two new \inner equations $Xbd\,\bar a= ad b\bar X$  and 
$Xb=Yb$. The latter equation is trivial and we forget about it. We content ourselves to lift
the $b$-position in the \inner equation $Xbd\,\bar a= ad b\bar X$ which is still  impossible. But fortunately there is an alternative. We compress the step $bd$ into a single fresh letter 
$c\in \wh C\sm B$ such that $\rho'(c)=\rho(bd)>\rho(b)$, $\mu(c)=\mu(bd)$,
and $c=\bar c\iff d=\bar d$. The automaton ~$\cU$ has a compression \tra labeled by $h(c)=bd$ to a standard state $E'$
with an \esolu $(\alp,\sig')$ such that $\sig=h\sig'$ and the 
 $\sig'(X)= a$. It yields the only \solu $\sig'$ for the \inner equation $Xc\bar a= ac \bar X$, which is fine since $h(c)=bd=db$, but we don't have 
 $T \\sse \rho'(c)$, in general. If $T \ssupneq \rho'(c)$, then it is still an improvement, since $\sig'(W)$ has less \STarcs with $S<T$ than $\sig(W)$. 
In our concrete example, we can go even further: 
instead of letting $\rho'(c)=\rho(bd)$ we redefine 
$\rho'(c)=\rho(abd)$. Then there are no \STarcs in $\sig(Xc\bar a)$ anymore. 
As expected, the mission impossible
was successful.  
\hspace*{\fill}$\diamond$\end{example}

\subsection{Moving to a semi-final state and to the final state}\label{sec:movefin}
Recall that a semi-final state is standard $\Rs$-state with an empty set of variables, $W=\bar W$, and  $W$ has a prefix 
$\#d_1\cdots \#d_k\#$ for 
$\cXinit= \set{X_i,\bar X_i}{1\leq i \leq k}$. The final state is just the symbol~$\cent$. 
Now, let $E=(W,B,\es,\rho,\mu,T)$ be any standard state 
without 
variables such that $W=\bar W$ which is not yet semi-final. Then 
either we can switch directly to $E'=(W,B,\es,\rho,\mu,\Rs)$ and $E'$ is semi-final or there is a distinguished letter
$d_i$ for some $1\leq i \leq k$ and a letter in $c\in B\sm \os{\cent,\#}$ such that $W$ has a \HA  $d_i\to c$.
Apart from its dual \HA  $\bar c\to \ov{d_i}$ there no other equivalent \HA{s}.
Hence there is \comp \tra 
\begin{align*}
E=(W,B,\es,\rho,\mu,\Rs)&\arc {h} E'=(W',B,\es,\rho,\mu,\Rs)
\end{align*}
which is defined by $h(d_i) = d_ic$ 
such that $\ov{W'}=W'$, 
$h(W')=W$.
Crucially, $W'$ is shorter than $W$. 
Repeating this process leads to the following observation.
\begin{proposition}\label{prop:movie}
Let $\pi$ be an \entipa{} from $(\cSinit,\id_A,\siginit)$ 
to an entire state $(E,\alp,\id_{B})$ where~$E$ is standard state, $B$ is the set of constant, and the equation $W$ is without variables. Then 
we can continue the \entipa $\pi$  to the final state $\cent$ without increasing the length $|W|$ on that path. 
\end{proposition}
The following corollary is a direct consequence of \prref{prop:movie}.
It implies that we can reach a semi-final state inside a finite subautomation of $\cU$ \IFF 
we can reach a standard state~$E$ with an 
empty set of variables inside a finite subautomaton of $\cU$. 
\begin{corollary}\label{cor:movie}
Let $f:\N \to \N$ be any function and 
$\pi$ an \entipa{} from $(\cSinit,\id_A,\siginit)$ 
to an entire state $(E,\alp,\id_{B})$ where~$E$ is standard state
without variables.
Suppose that the path~$\pi$ can be realized in a subautomaton 
$\cU_f \sse \cU$ which is defined by the states having equations such that their length is at most $f(\Abs{\cSinit})$, then 
$\cU_f$ contains an \entipa{} from $(\cSinit,\id_A,\siginit)$ 
to a semi-final state which is labeled by an $A$-\Endo~$h$ 
of $\wh C$ such that $h(d_i)= \siginit(X_i)$ for all $1\leq i \leq k$.
\end{corollary}
\begin{remark}\label{rem:movie}
It is easy to see that every subautomaton $\cU'$ of $\cU$, which is defined
by the set of states~$E$ in~$\cU$ with a given length bound on the equation at~$E$, is finite. 
This follows because the sets $\wh C$, $\wh \cX$, $\Rs$, $N$ are finite and the relation 
$\theta$ is a subset of $(\wh C\cup \wh \cX)^3\times (\wh C\cup \wh \cX)^3$.
\hspace*{\fill}$\diamond$\end{remark}

\subsection{Reduced \solus}\label{sec:redsoli}

The previous subsections give rise to the following notation. 
\begin{definition}\label{def:redusol}
Let $\es\neq T\sse \Rs$.
An \esolu $(\alp,\sig)$ (or a \solu $\sig$) at a $T$-state  $E=(W,B,\cX,\rho,\mu,\theta,T)$ is called \emphind{reduced} if none of the basic $T$-reductions is applicable. 
\end{definition}
Applying the standard reductions as long as possible shows the following result. 
\begin{proposition}\label{prop:redsol}
Let $\es\neq T\sse \Rs$ and $E=(W,B,\cX,\rho,\mu,\theta,T)$ be a state with an \esolu $(\alp,\sig)$. Then there is a path 
in~$\cU$  to a state $E'=(W',B',\cX',\rho',\mu',\theta',T)$ which is labeled by a $\wh A$-\Endo 
$h\in \End(M(\wh C,\rho))$. The state $E'$ has an \esolu $(\alp h,\sig')$ such that $(\alp h,\sig')$ is reduced, we have $h\sig'=\sig$, and 
$\wh \pi \alp h\sig'= \wh \pi \alp \sig$.
Moreover, on this path every \tra reduces the weight of the 
corresponding \esolu. 
\end{proposition}

\section{Various types of traces and variables}\label{sec:bptv}

  In general, \HA{s} in a solution $\sig(W)$ need not originate from \HA{s} in the equation $W$, but only from paths in its Hasse diagram $H(W)$; an example of this is shown in \prref{fig:adYb}.
  Moreover, the correspondence of \HA{s} in $\sig(W)$ to paths in $H(W)$ can be rather complicated.
  The matter can be simplified, to some extent, by imposing conditions on the solution $\sig(X)$ of every variable $X$.

  In the example shown in \prref{fig:adYb} we can split the variable $Y$ by a \subst $Y\mapsto Y_1Y_2$ such that $\rho(Y_i)\ssneq \rho(Y)$ for $i=1,2$.
This means that $Y$ is decomposable in the sense of \prref{def:decca}.
\begin{figure}[h]
	\begin{center} 
		\begin{tikzpicture}[xscale=2, yscale=0.5]
		\node[circle] (aa) at (-0.8,0){$\sig(a$};
		\node[circle] (dd) at (0,0){$d$};
		\node[circle] (Y) at (0.8,0) {$Y$};
		\node[circle] (bb) at (1.6,0) {$b)$};
		\draw (aa) edge[->, >=latex,thick] (dd);
		\draw (dd) edge[->, >=latex,thick] (Y);
		\draw (Y) edge[->, >=latex,thick] (bb);
		\node[circle] (eq) at (2.3,0) {$=$};

		\node[circle] (a) at (3,1) {$a$};
		\node[circle] (d) at (3.86,1) {$d$};
		\node[circle] (e) at (3,-1) {$e$};
		\node[circle] (b) at (4.6,-1) {$b$};
		\node[circle] (f) at (4.6,1) {$f$};
		\draw (a) edge[->, >=latex,thick] (d); 
		\draw (a) edge[->, >=latex,thick] (b); 
		\draw (e) edge[->, >=latex,thick] (b); 
		\draw (e) edge[->, >=latex,thick] (f); 
		\draw (d) edge[->, >=latex,thick] (f); 
		
		\end{tikzpicture}\end{center}
	\vspace{-0.5cm}
	\caption{The \HA $a\to b$ in the solution is not visible as a \HA in the equation. }\label{fig:adYb}
\end{figure}

\subsection{Decomposable traces and variables}\label{sec:decvar}
A standard definition says that a trace~$w$ in free partially 
commutative monoid is \emph{connected\index{connected trace}} if its \HD $H(w)$ (or equivalently its \DG $D(w)$)
is connected in the usual graph theoretical sense, i.e., the induced undirected graph defined by the directed graph $H(w)$ (and/or by $D(w)$) is connected. 
Here we use a stronger notion than connectedness.

\begin{definition}[Indecomposability]\label{def:decca}
Let $E=(W,B,\cX,\rho,\mu,\theta,T)$ a state.
\begin{itemize}
\item  
An element~$w$ in the extended monoid $M(B,\cX,\rho,\theta)$ is called \emphind{decomposable} if we can factorize 
$w=uv$ such that $\rho(u)\neq \rho(w)\neq \rho(v)$. 
Otherwise it is called \emphind{indecomposable}.
\item Let $(\alp,\sig)$ be an \esolu at~$E$ and 
$X\in \cX$ be a variable. Then $X$ called \emph{decomposable}
(resp.~\emph{indecomposable}) if $\sig(X)$ has this property. 
\end{itemize}
\end{definition}
\begin{remark}\label{rem:decca}
Every indecomposable trace is connected in the usual sense. 
The converse does not hold as soon as there are letters 
$a$ und $b$ such that $\es \neq \rho(a)\cap \rho(b)\ssneq \rho(a)$ and $\rho(b)\sm\rho(a)\neq \es$. The connected trace $w=ab$ is decomposable. 
\hspace*{\fill}$\diamond$\end{remark}
\begin{proposition}\label{prop:Tim}
Let $w\in M(B\cup \cX,\rho)$ be a trace, $p,q$ positions in~$w$, and~$y$ be a subtrace of~$w$ such that 
$y$ is indecomposable.
If there are paths $p \arc{*}r \in \pos(y)\ni s\arc{*} q$ in~$w$
with $\pos(y)\cap \os{p,q}=\es$. 
Then there exists a position $r'$ in~$y$ and a path $p\arc{+}r' \arc{+}q$.
\end{proposition}
\begin{proof}
We proceed by contradiction. 
Consider all positions in~$y$ which appear on some path from~$y$ to $q$. These positions form a prefix closed subset of positions  $u\sse y$ which is nonempty because $s\in u$. Thus, we can write 
$y=uv$. 
Let $q'\to q''$ be a \HA on the path from $s$ to $q$ with $q'\in \pos(y)$ and $q''\notin \pos(y)$. Such a \HA exists since $q\notin \pos(y)$.
Then we have $q'\in u$ by definition of $u$ and $\rho(u)\cap \rho(q'')\neq \es$. 
On the other hand we must have $\rho(q'')\cap \rho(v)=\es$ because there can be no path from any position $r'\in \pos(v)$ to $q''$ (by definition of $v$ and $q''$), and there
can be no path from $q''$ to any position $r'\in v$ (since~$y$ is subtrace and $q''\notin \pos(y)$). 
We conclude that $\rho(v)\sse \rho(y)\sm \rho(q'')$. 

Let $p \arc{*}r$ be any path in~$w$. Since $p\notin \pos(y)$, the path is nonempty.
If it uses some position in $u$, then we are done.  Hence, $r\in \pos(v)$ and $v\neq 1$. 
By symmetry, there is a \HA $p''\to p'$ on the path $p \arc{+}r$ with 
$p''\notin \pos(y)$ and $p'\in \pos(y)$. If $\rho(p'')\cap \rho(u)\neq\es$, then we find the position $r'$ inside $u$. 
Thus, we may assume
$\rho(p'')\cap \rho(u)=\es$. 
Since $u\neq \es$ we conclude (as we did above for the \HA $q'\to q''$) that $\rho(u)\sse \rho(y)\sm \rho(p'')$. This shows that~$y$ is a decomposable subtrace in~$w$. Contradiction.
\end{proof}

\subsection{$T$-disjoint, bordered, conic traces and variables}\label{sec:bordvar}

\begin{definition}[$T$-borders]\label{def:bord}
  Let $x \in M(B, \rho)$ be a trace and $T \sse \Rs$.
\begin{itemize}
\item  The trace $x$
is called \emphind{$T$-disjoint} if $\rho(x)\cap T=\es$.
\item  The trace $x$
is called \emphind{left-$T$-bordered} (resp.~\emphind{right-$T$-bordered}) if every position in $\min(x)$ (resp.~in $\max(x)$) uses a resource from $T$.
It is called \emph{$T$-bordered}\index{T-bordered@$T$-bordered} if $x$ and $\bar x$ are left-$T$-bordered. 
\item The trace $x \in M(B,\rho)$ is called \emphind{strongly left-$T$-bordered} (resp.~\emphind{strongly right-$T$-bordered}) if it is a left-$T$-bordered (resp.\ right-$T$-bordered) trace and every minimal (resp.\ maximal) position $q$ of $x$ satisfies $\rho(q)\sm T \neq \es$.
  It is called \emphind{strongly $T$-bordered} if $x$ and $\bar x$ are strongly left-$T$-bordered.
\item A trace $x \in M(B,\rho)$ is called \emphind{left-$T$-conic} (resp.\ \emphind{right-$T$-conic}), or simply a \emphind{left $T$-cone} (resp.\ \emphind{right $T$-cone}), if it has a unique 
minimal (resp.\ maximal) position $q$ such that $T\sse \rho(q)$.
It is called \emphind{$T$-conic}, or simply a \emphind{$T$-cone}, if $x$ and $\bar x$ are left-$T$-conic.
\footnote{The notion of \emphind{cone} was introduced in \cite{die90} for traces which have exactly one minimal and one maximal position which have (in our terminology) the same resource set.
Our notion of a $T$-cone allows these resource sets to differ.}
\end{itemize}
  Let $E=(W,B,\cX,\rho,\mu,T)$ be a standard state with a \solu $\sig$. 
  A variable $X\in \cX$ is said to have one of the properties above (\wrt $\sig$) if $\sig(X) \in M(B,\rho)$ 
  has the corresponding property. 
\end{definition}
\begin{figure}[h]
	\begin{center} 
		\begin{tikzpicture}
		\node[circle] (w) at (-1,0) {$H(w)=$};
		\node[circle] (a) at (0,0) {$a$};
		\node[circle] (b) at (2,0.6) {$b$};
		\node[circle] (d) at (2,-0.6) {$d$};
		\node[circle] (c) at (4,0) {$c$};
		\node[circle] (bb) at (6,0.6) {$b$};
		\node[circle] (e) at (6,0) {$e$};
		\node[circle] (ee) at (6,-0.6) {$e'$};
		\node[circle] (E) at (8,0) {$e$};
		\node[circle] (Ee) at (8,-0.6) {$e'$};

		\node[circle] (bbb) at (8,0.6) {$b$};
		\node[circle] (ddd) at (10,-0.6) {$d$};

		\draw (a) edge[->, >=latex,thick] (b); 
		\draw (a) edge[->, >=latex,thick] (d); 
		\draw (b) edge[->, >=latex,thick] (c); 
		\draw (d) edge[->, >=latex,thick] (c); 
		\draw (c) edge[->, >=latex,thick] (bb); 
		\draw (c) edge[->, >=latex,thick] (e); 
		\draw (c) edge[->, >=latex,thick] (ee); 
		\draw (bb) edge[->, >=latex,thick] (bbb); 
		\draw (e) edge[->, >=latex,thick] (E); 
		\draw (ee) edge[->, >=latex,thick] (Ee); 
		\draw (E) edge[->, >=latex,thick] (ddd); 
 \draw (Ee) edge[->, >=latex,thick] (ddd); 
		\end{tikzpicture}\end{center}
	\vspace{-0.5cm}
	\caption{The picture shows the Hasse-diagram of a connected trace~$w$. Let $T=\rho(a)$. Then~$w$ is $T$-bordered, but it is neither $T$-conic nor $\rho(d)$-bordered. The prefix 
	$abdc$ is $T$-conic \IFF $T\sse \rho(c)$. 
	We have $\rho(ee'd)\neq T$ because $\rho(b)\cap \rho(ee'd)=\es$. If $\rho(e)\sm \rho(adc)\neq \es$, then~$w$ is decomposable.	}\label{fig:bord}
\end{figure}
\begin{corollary}[to \prref{prop:Tim}]\label{cor:TimX}
  Let $E = (W, B, \cX, \rho, \mu, T)$ be a standard state with an \esolu $(\alp,\sig)$ where every decomposable variable $X$ is $T$-bordered or satisfies $T \sse \rho(X)$.
Moreover, suppose that $p \to q$ is a Hasse arc in $\sig(W)$ such that $\rho(p) = T$ and $q$ is visible.
Then, for every path $\sig^*(p) \to y_1 \to \cdots \to y_{m} \to \sig^*(q)$ in $H(W)$, we have $\rho(y_1 \cdots y_m) \cap T = \emptyset$.
\end{corollary}
\begin{proof}
  Clearly $\sig(y_1 \cdots y_m)$ cannot contain any position $\tilde p$ with $\rho(p) \cap \rho(q) \sse\rho(\tilde p)$. 
  \Ip none of the positions $y_1, \dotsc, y_m$ is labeled by a variable $X$ with $T \sse \rho(X)$.
  So we can assume that for $1\leq i \leq m$ every decomposable trace $\sig(y_i)$ is $T$-bordered.
  By contradiction, let us assume that $\rho(y_i) \cap T \neq \emptyset$ for some $1 \leq i \leq m$.
  Choose $i$ to be maximal with this property.
  Then, in particular, $\rho(y_j) \cap T = \emptyset$ for all $i < j \leq m$.
  Note that this implies that $y_j$ is indecomposable for all $i < j \leq m$.

  We claim that there exists some path $p \arc{+} t_i \arc{*} \cdots \arc{*} t_m \arc{+} q$ in $H(\sig(W))$ with $t_j \in \pos(\sig(y_j))$ for all $i \leq j \leq m$.
  Note that the existence of such a path contradicts the assumption that $p \to q$ is a Hasse arc.
  First, choose $r_j,s_j \in \pos(\sig(y_j))$ where $i \leq j \leq m$ such that there exist paths 
  \[
    p \arc{+} r_i,\quad s_j \arc{+} r_{j+1} \;(i \leq j < m),\quad s_m \arc{+} q
  \]
  in $H(\sig(W))$.
  The existence of such an $r_i$ follows from $\rho(p) \cap \rho(y_i) = T \cap \rho(y_i) \neq \emptyset$ and the existence of a path $\sig^*(p) \smash{{}\arc{+}{}} y_i$ in $H(W)$.
  The remaining positions $s_i, \dotsc, s_m$ and $r_{i+1}, \dotsc, r_m$ exists because of the corresponding Hasse arcs in $H(W)$.
  Repeatedly applying \prref{prop:Tim}, we obtain positions $t_j \in \pos(\sig(y_j))$ for $i < j \leq m$ admitting a path $s_i \smash{{}\arc{*}{}} t_{i+1} \smash{{}\arc{*}{}} \cdots \smash{{}\arc{*}{}} t_m \smash{{}\arc{+}{}} q$ in $H(\sig(W))$.
  If the trace $\sig(y_i)$ is indecomposable, then we apply \prref{prop:Tim} again to obtain the remaining position $t_i \in \pos(\sig(y_i))$ so that there exists a path $p \smash{{}\arc{+}{}} t_i \smash{{}\arc{*}{}} \cdots \smash{{}\arc{*}{}} t_m \smash{{}\arc{+}{}} q$ in $H(\sig(W))$.
  Otherwise the trace $\sig(y_i)$ is $T$-bordered and we can choose any $t_i \in \min(\sig(y_i))$ with $t_i \smash{{}\arc{*}{}} s_i$ a path in $H(\sig(W))$.
  Then $p \smash{{}\arc{+}{}} t_i$ in $H(\sig(W))$, as $\rho(p) \cap \rho(t_i) \neq \emptyset$; and $\sig^*(p) \smash{{}\arc{+}{}} y_i$ is a path in $H(W)$.
\end{proof}
An important property of strongly $T$-bordered traces, upon which we will heavily rely, is that strongly $T$-bordered subtraces retain this property during $T$-lifting as defined in \prref{sec:steplift}.
\begin{lemma}\label{lem:alexie}
Let $s = \prod_{i=1}^rs_i\in w\in M(B,\rho)$ be a step with $s_i\in B$ for all $1\leq i \leq r$ which is defined by $\pos(s)$.
Let $c$ be a fresh $T'$-clone of~$s$ with $T'=T\cup \rho(s)$.
Let $z$ be a strongly left-$T$-bordered subtrace of~$w$; and consider any trace $w' \in M(B\cup\os{c,\bar c},\rho)$ which is obtained by relabeling $\pos(s)$ by the clone~$c$.
Thus, we can write $w=h(w')$ where the \morph 
is defined by $h(c)=s$. Then $h^\ast(z)$ defines a strongly left-$T$-bordered subtrace $z'$ of~$w'$ with $\min(z') \sse h^\ast(\min(z))$.
\end{lemma}
\begin{proof}
  For every path $p \arc + q$ in $H(w)$ there is corresponding path $h^\ast(p) \arc + h^\ast(q)$ in $H(w')$.
  Hence, we have $\min(z') = \min(h^\ast(\pos(z))) \sse h^\ast(\min(z))$.
  Let $p'$ be the unique position of $w'$ labeled~$c$.
  Clearly, every $q' \in \min(z') \sse h^\ast(\min(z))$ other than $p'$ satisfies $\rho(q') \cap T \neq \es$ and $\rho(q') \sm T \neq \es$.
  \Ip if $p' \not \in \min(z')$, then $z'$ is strongly left $T$-bordered.
  On the other hand, if $p' \in \min(z')$, then there is a position $p \in \min(z) \cap \pos(s)$ and, hence, $\rho(s) \sm T \ssupe \rho(p) \sm T \neq \es$.
  It then follows that $\rho(p') \cap T = T \neq \es$ and $\rho(p') \sm T \neq \es$. 
  Therefore $z'$ is strongly left $T$-bordered.
\end{proof}

\subsection{Immunity and protection of traces and variables}\label{sec:sandy}
The splitting procedure is based on the following notation which will be used for subtraces in $\sig(W)$. Thus, we define it relatively to an ambient 
trace~$w$. 
\begin{definition}\label{def:sanim}
Let $E=(W,B,\cX,\rho,\mu,T)$ be a standard state with an \esolu $(\alp,\sig)$ and $w\in M(B,\rho)$ a trace. 
\begin{enumerate}
\item The \emphind{interior} of~$w$ is the subtrace~$y$ of~$w$ such that $\pos(y)\cap (\min(w)\cup \max(w))=\es$.
\item A strongly $T$-bordered trace $z \in M(B, \rho)$ is called a \emph{$T$-sandwich}\index{T-sandwich@$T$-sandwich} (or, if $T$ is known from the context, simply a \emphind{sandwich}) if it is without $T$-positions. (Recall that strongly $T$-bordered means that $\es \neq \rho(q) \cap T \neq \rho(q)$ for all $q \in \min(z) \cup \max(z)$.)
\item A subtrace~$y$ of~$w$ is called \emph{$T$-immune}\index{T-immune@$T$-immune} in either case:
\begin{itemize}
\item The trace~$y$ is $T$-disjoint. That is $\rho(y)\cap T=\es$. 
\item There is some $T$-sandwich~$z$ which appears in~$w$ and~$y$ is inside the interior of~$z$.
\end{itemize}
\item A variable $X\in \cX$ is called \emph{$T$-immune} (\wrt~$\sig(W)$) if $\sig(\sig^*(q))$ is a $T$-immune subtrace in $\sig(W)$ for all positions $q\in \pos(\sig(W))$, where $\sig^*(q)$ is labeled by $X$ or $\bar X$.
\end{enumerate}
\end{definition}
\begin{lemma}\label{lem:timmu}
Let $p\to q$ be a \Tarc in $\sig(W)$. 
Then neither $\sig(\sus{p})$ nor $\sig(\sus{q})$ is $T$-immune.
\end{lemma}
\begin{proof}
By symmetry we may assume $\rho(p)=T$. 
Then $\sig(\sus{p})$ contains a $T$-position and is therefore not $T$-immune. 
By contradiction assume that $\sig(\sus{q})$ is $T$-immune.
Then $\rho(q)\neq T$ and $\sus{p}\neq\sus{q}$.
Since $\rho(q)\cap T\neq \es$, the trace $\sig(\sus{q})$ is not $T$-disjoint. 
Hence, $\sig(\sus{q})$ is the interior of a $T$-sandwich~$z$. 
A sandwich is without $T$-positions and, as such, $p \not \in \pos(z)$.
Since $q$ is an interior position of $z$, there exists $q' \in \min(z)$ and a path $q' \arc{+} q$ in $H(\sig(W))$.
Moreover, there is also a path $p \arc{+} q'$ in $H(\sig(W))$ as $z$ is $T$-bordered.
So $p \arc{+} q' \arc{+} q$ in $H(\sig(W))$, which is a contradiction.
\end{proof}
Note that $T$-sandwiches are not defined for variables. The notion of $T$-immune subtrace is defined \wrt{} some ambient trace. The notion of $T$-immune variable refers to the ambient 
trace $\sig(W)$. Note also that every subtrace of a $T$-immune subtrace is $T$-immune itself. \Ip if $X$ is a $T$-immune variable, then all subtraces of $\sig(X)$ are $T$-immune.

\begin{definition}\label{def:Tprot}
  Let $w \in M(B, \rho)$ be a trace and $x$ be a subtrace of $w$ with $T \sse \rho(x)$.
  Then the subtrace $x$ is called \emphind{left $T$-protected} if the following condition holds.
  \begin{itemize}
    \item If $x = yx'$ such that $\rho(y) \ssneq \rho(x)$ and $x'$ is left $T$-conic, then $x$ is left $T$-conic.
    \item If $x$ is not left $T$-conic, then $x$ is contained in a subtrace $y$ of $w$ such that $\pos(x) \sse \pos(y) \sm \min(y)$ and such that every $q \in \pos(y) \sm \pos(x)$ satisfies $\rho(q) \cap T \neq \es$ and $\rho(q) \cup T \neq \es$.
  \end{itemize}
  The subtrace $x$ is called \emphind{right $T$-protected} if $\bar x$ is left $T$-protected as a subtrace of $\bar w$, and it is called \emphind{$T$-protected} if it is both left and right $T$-protected.

  Moreover, let $E = (W, B, \cX, \rho, \mu, T)$ be a standard state with an \esolu $(\alp, \sig)$.
  Then a variable $X\in \cX$ is called \emph{(left or right) $T$-protected} (\wrt~$\sig(W)$) if $\sig(\sig^*(q))$ is a (resp.\ left or right) $T$-protected subtrace in $\sig(W)$ for all positions $q \in \pos(\sig(W))$, where $\sig^*(q)$ is labeled by $X$.
\end{definition}
Again, the notion of $T$-protected subtrace is defined \wrt{} some ambient trace, and the notion of $T$-protected variable refers to the ambient trace $\sig(W)$.
Note also that every (left or right) $T$-cone is (resp.\ left or right) $T$-protected, regardless of the ambient trace.
\begin{lemma}\label{lem:prot}
  Let $p\to q$ be a \Tarc in $\sig(W)$ with $\sus{p} \neq \sus{q}$. 
  \begin{itemize}
    \item If $\rho(p) = T$ and $T \nsse \rho(q)$, then $\sig(\sus{q})$ is not left $T$-protected.
    \item If $\rho(q) = T$ and $\sig(\sus{q})$ is left $T$-protected, then $q \in \min(\sig(\sus{q}))$ and so $q$ is semi-visible.
  \end{itemize}
\end{lemma}
\begin{proof}
  Suppose that $\rho(p) = T$ and $T \nsse \rho(q)$.
  We claim that $x = \sig(\sus{q})$ is not left $T$-conic.
  Indeed, if $q'$ is the (unique) minimal position of $x$, then $q' \neq q$ as $T \nsse \rho(q)$.
  This implies the existence of a path $\smash{p \arc + q' \arc + q}$ in $H(\sig(W))$, thereby contradicting the assumption that $p \to q$ is a \HA in $\sig(W)$.
  Similarly, if $x$ is contained in a subtrace $y$ of $\sig(W)$ with $\pos(x) \sse \pos(y) \sm \min(y)$ and such that every $q' \in \pos(y) \sm \pos(x)$ satisfies $\rho(q) \cap T \neq \es$ and $\rho(q) \sm T \neq \es$, then we can again find a path $\smash{p \arc + q' \arc + q}$ in $H(\sig(W))$ for some suitable $q' \in \min(y)$.
  This proves the first assertion.

  Suppose now that $\rho(q) = T$ and that $x = \sig(\sus{q})$  is left $T$-protected as a subtrace of $\sig(W)$.
  Consider any resource $r \in \rho(p) \cap \rho(q) \sse T$.
  We can then write $x = yx'$ where $y$ is the maximal prefix of $x$ with $r \not\in \rho(y)$.
  Since $p \not\in \pos(x)$, we obtain $\min(x') = \{q\}$ due to the \HA $p \to q$.
  \Ip $x'$ is left $T$-bordered and, thus, so is the trace $x$.
  Moreover, we then have $\min(x) = \{q\}$ due to the \HA $p \to q$.
  This proves the second assertion.
\end{proof}
\begin{lemma}\label{lem:splinv}
  Let $(E, \alp, \sig)$ be an entire state with $E = (W, B, \cX, \rho, \mu, T)$.
  Further, let $(E', \alp', \sig')$ with $E' = (W', B', \cX', \rho', \mu', T)$ be obtained from $(E, \alp, \sig)$ by basic $T$-reductions.
  \begin{itemize}
    \item If $X \in \cX' \sse \cX$ is $T$-immune \wrt~$\sig(W)$, then $X$ is $T$-immune \wrt~$\sig'(W')$.
    \item If $X \in \cX' \sse \cX$ is left $T$-protected \wrt~$\sig(W)$, then $X$ is left $T$-protected \wrt~$\sig'(W')$ unless there exists a factorization $\sig'(X) = y x'$ with $\rho'(y) \ssne \rho'(X)$ such that $x'$ is left $T$-conic.
  \end{itemize}
\end{lemma}
\begin{proof}
  We can assume that $(E', \alp', \sig')$ is obtained from $(E, \alp, \sig)$ by a single $T$-lifting, as other basic $T$-reductions clearly cannot destroy $T$-immunity and (left or right) $T$-protectedness. 

  It suffices to consider the effect of lifting a single position $q \in \sig(W)$ with an incident \HA $p \to q$ or $q \to p$ where $\rho(p) = T$ and $\rho(q) \cup T \neq \rho(q)$.
  Indeed, we can then think of the transition from $(E, \alp, \sig)$ to $(E', \alp', \sig')$ as being composed of such individual liftings.
  The case of lifting a step $s$ with two or more positions is similar and left to the reader.
  (Note that this case is even more restrictive since we require that there are positions $p$ and $p'$ with $\rho(p) = \rho(p') = T$ such that every position $q \in \pos(s)$ is incident with \HA{s} $p \to q$ and $q \to p'$ in $\sig(W)$.)

  Let $w \in M(B, \rho)$ be a trace with a \HA $p \to q$ where $\rho(p) = T$ and $T' = \rho(q) \cup T \neq \rho(q)$.
  Further, let $w' \in M(B', \rho')$ be any trace obtained from $w$ by replacing the label $a \in B$ of $q$ with a fresh $T'$-clone $c$ where $B' = B \cup \os{c, \bar c}$.
  Thus, $h(w') = w$ where $h$ is defined by $h(c) = a$.

  Consider a subtrace $x$ of $w$, and let $x'$ be the subtrace of $w'$ defined by $\pos(x') = h^\ast(\pos(x))$.
  If the trace $x$ is $T$-disjoint, then so is the trace $x'$ as, clearly, $q \not \in \pos(x)$.
  If the subtrace is $T$-immune in $w$ but $\rho(x) \cap T \neq \es$, then $x$ is contained in the interior of a $T$-sandwich $z$ in $w$.
  Let $z'$ be the corresponding subtrace of $w'$ defined by $\pos(z') = h^\ast(\pos(z))$.
  By \prref{lem:alexie}, the trace $z'$ is strongly $T$-bordered and we have $\min(z') \sse h^\ast(\min(z))$ as well as $\max(z') \sse h^\ast(\max(z))$.
  Hence, the subtrace $x'$ is contained in the interior of the subtrace $z'$.
  It remains to show that $z'$ is without $T$-positions.
  Since $z$ is without such positions, the only possibility is that $q' = h^\ast(q) \in \pos(z')$ satisfies $\rho'(q') = \rho(q) \cup T = T$.
  However, $q' \in \pos(z')$ if and only if $q \in \pos(z)$ which, due to the \HA $p \to q$, is equivalent to $q \in \min(z)$.
  In that case, $\rho(q) \sm T \neq \es$ and thus $\rho(q') \neq T$.
  So $z'$ is indeed a $T$-sandwich containing $x'$ in its interior.
  Therefore the subtrace $x'$ is $T$-immune in $w'$.

  Clearly, if $x$ is left or right $T$-conic, then so is $x'$.
  Suppose now that $x$ is left $T$-protected but not left $T$-conic.
  Then there exists a subtrace $y$ in $w$ with $\pos(x) \sse \pos(y) \sm \min(y)$ such that every $\tilde q \in \pos(y) \sm \pos(x)$ satisfies $\rho(\tilde q) \cap T \neq \es$ and $\rho(\tilde q) \sm T \neq \es$.
  By \prref{lem:alexie}, the corresponding subtrace $y'$ with $\pos(y') = h^\ast(\pos(y))$ satisfies $\min(y') \sse h^\ast(\min(y))$.
  Therefore we have $\pos(x') \sse \pos(y') \sm \min(y')$ and, clearly, every $\tilde q' \in \pos(y') \sm \pos(x') = h^\ast(\pos(y) \sm \pos(x))$ satisfies $\rho(\tilde q') \cap T \neq \es$ and $\rho(\tilde q') \sm T \neq \es$.
  Hence, $x'$ is left $T$-protected unless it admits a factorization claimed by the assertion.
  The case where $x$ is right $T$-protected but not right $T$-conic is similar.
\end{proof}

\section{Elimination of all \STarc{s}}\label{sec:elist}

\subsection{The $T$-splitting procedure at a standard state}\label{sec:prospli}
Suppose that $E = (W, B, \cX, \rho, \mu, T)$ is a standard state with an \esolu $(\alp, \sig)$.
By definition of an \esolu (\prref{def:statesolu}) this means that $\sig(W)$ contains no \SSparcs with $S, S' < T$.
To improve on this condition we now also eliminate all \STarcs with $S < T$ from $\sig(W)$ and, in fact, all \STarcs with $T \nsse S$.
This is achieved by repeatedly splitting variables and lifting positions or steps.
In \prref{sec:steplift} we have already described \emph{how} such liftings can be performed via basic $T$-reductions, but we have not yet described \emph{when} such liftings are possible.
The following definition encompasses a sufficient condition (see \prref{prop:Tper}).
\begin{definition}\label{def:Tper}
  Let $E = (W, B, \cX, \rho, \mu, T)$ be a standard state with an \esolu $(\alp, \sig)$.
  Then $(\alp, \sig)$ is called \emph{$T$-perfect} if, firstly, every decomposable variable $X \in \cX$ is $T$-bordered or satisfies $T \sse \rho(X)$ and, secondly, the following hold for every \STarc $p \to q$ in $\sig(W)$ with $T \nsse S$.
  \begin{itemize}
    \item If $\rho(p) = S$ and $\rho(q) = T$, then $\sus{p} = \sus{q}$ or $p$ is visible and $q$ is semi-visible on the left.
    \item If $\rho(p) = T$ and $\rho(q) = S$, then $\sus{p} = \sus{q}$ or $p$ is semi-visible on the right and $q$ is visible.
  \end{itemize}
\end{definition}

Our strategy for splitting variables, described in \prref{sec:bsp}, is based on the following.

\begin{lemma}\label{lem:good}
  Let $E = (W, B, \cX, \rho, \mu, T)$ be a standard state with an \esolu $(\alp, \sig)$.
  \mbox{If every} $X \in \cX$ is either $T$-protected, or $T$-immune and indecomposable, then $(\alp, \sig)$ is $T$-perfect.
\end{lemma}
\begin{proof}
  To see the indecomposability requirement satisfied recall that, by definition, $T \sse \rho(X)$ holds for every $T$-protected variable $X$.
  The other requirements follow from \prref{lem:timmu} and \prref{lem:prot}.
\end{proof}

\subsubsection{Fully invisible positions have a label in $\wh A$}\label{sec:fiwhA}

We continue with a standard state $E=(W,B,\cX,\rho,\mu,T)$ in~$\cU$ with a reduced \esolu $(\alp,\sig)$ according to \prref{def:redusol}. 
Moreover, we assume throughout that the following invariant holds.
\begin{invariant} \label{inva:invis}
  Every fully invisible position is labeled by a letter in $\wh A$.
\end{invariant}
The invariant holds at the initial state $\cSinit$ for every \esolu since all letters are in $\wh A$.
By \prref{lem:whAinv} it cannot be destroyed by \epstra{s}.
We will keep the invariant by \comp \tras which either compresses a subtrace of $T$-letters into a fresh $T$-letter or by compression of a step into fresh $T'$-letters with $T'\geq T$. 
Recall that \comp \tras introduce a letter $c$ with with $\rho(c) = T' \geq T$. 
If the letter~$c$ appears thereby by a fully invisible position, then we replace it immediately by its $T'$-step-normal form. 
See \prref{sec:inTred}, \prref{sec:steplift}, and \ip \prref{lem:alpbet}.

\subsubsection{Basic splittings}\label{sec:bsp}
Let $E=(W,B,\cX,\rho,\mu,T)$ be a standard state with an \esolu $(\alp, \sig)$.
Our goal is to obtain a reduced solution where every variable $X \in \cX$ is $T$-immune and indecomposable if $\sig(X)$ contains no $T$-position, and is $T$-protected otherwise. 
We begin by making all variables $T$-protected.
\begin{lemma}[Giant splitting]\label{lem:gsp}
  Let $x \in M(B,\rho)$ be a trace without any \SSparc where $S,S'<T$ and where $\rho(x) \cap T \neq \es$.
  Then there is a tuple $g(x) = (y_1,u,v,x_1)$, which we henceforth refer to as a \emph{giant splitting of $x$}, with the following properties.
\begin{enumerate}
\item There is a factorization $x=y_1uvx_1$.
\item We have $\rho(y_1) \ssne \rho(x)$ and $y_1$ is without any $T$-letter.
\item The factors $u$ and $v$ are (possibly empty) steps.
\item If $v=1$, then $x_1$ is left $T$-conic or the empty trace as well.
\item If $v\neq 1$, then $\min(vx_1) = \pos(v)$ and $vx_1$ is strongly left $T$-bordered.
\end{enumerate}
\end{lemma}
\begin{proof}
To begin with we assign to each $r\in R = \rho(x) \cap T$ the unique minimal position $q_r\in \pos(x)$ with $r\in \rho(q_r)$.
Consider the factorization $x = z s z'$ where~$zs$ is the minimal prefix of $x$ containing all these positions $q_r$, i.e.,
\begin{align}\label{eq:lemFSP1b}
    \pos(zs) = \set{p \in \pos(x)}{\exists r\in R: p \arc* q_r}
    \text{ with $s=\max(zs)$}
\end{align}
Then there is some $r\in R$ with $r\notin \rho(z)$,  \ip $\rho(z)\ssneq \rho(zs)\sse \rho(x)$.
Indeed, consider any $q = q_r \in \pos(\max(z))$ and assume that there is another position $p \in \pos(zs)$ with $r \in \rho(p)$. 
  Then $p \in \pos(z)$, as $s=\max(zs)$ is a step.
  This contradicts the minimality of $q = q_r$. Thus, there is some $r\in T$ with $r\in \pos(s)\sm \pos(z)$ and $\rho(z)\ssneq \rho(zs)\sse \rho(x)$. 
  
Next, we consider the factorization of the suffix $sz'= y'sx'$ such that 
$sx'$ is left $T$-bordered with $s= \min(sx')$. Then we cannot have
$r\in \rho(y')$. This implies that $y_1=zy'$ satisfies 
$r\notin \rho(y_1)$. Hence, $\rho(y_1)\ssneq\rho(x)$ and $y_1$ is without any $T$-letter. 

Let $u=\pos\set{q\in \pos(s)}{\rho(u)\ssneq T}$. We obtain 
$sx'= ux''$ and $u$ is a (possibly empty) step. 
If $x''=1$, then we let $v=x_1=1$ and we are done. 
Thus, we may assume that $x''\neq 1$. 
If $x''$ is left $T$-conic, then we let $v=1$ and $x_1=x''$.

If $x''$ is not left $T$-conic, then we write 
$x''=vx_1$. 
where $\pos(v)=\min(vx_1)$. Since for $S,S'<T$ the trace $x$ is without \SSparc{s}, 
every minimal position $q\in \pos(v)$ satisfies 
$\es \neq \rho(q)\cap T \ssneq \rho(q)$. To see this let $q\in \pos(v)$.
Since $1\neq x''$ is not left $T$-conic, we have $\rho(q)\neq T$. 
If $q\in \pos(s)$, then $q\notin \pos(u)$ and therefore 
$\es \neq \rho(q)\cap T\ssneq \rho(q)$. If $q\notin \pos(s)$, then 
$q$ is an endpoint for a \HA with origin in $u$. Thus, $\es \neq \rho(q)\cap T$ and $T\leq \rho(q)$. If $T=\rho(q)$, we are done. In the other case we have
$\rho(q)\sm T\neq \es$, and therefore $\rho(q)\cap T\ssneq \rho(q)$.
\end{proof}
The input for the following procedure is a standard state $E=(W,B,\cX,\rho,\mu,T)$ with an \esolu $(\alp, \sig)$.
The procedure \textsc{GiantSplit} is executed only once for each $T \sse \Rs$. 
It turns every variable with a $T$-position into a $T$-protected variable.
\\

\noindent\textsc{\textbf{BeginProcedure}} \textsc{GiantSplit}\\
\noindent\textsc{\textbf{ForAll}} $X\in \cX$ where $\sig(X)$ has a $T$-position, but it is not left $T$-protected  
\textsc{\textbf{Do}}
\begin{enumerate}
\item 
Let $g(X) = (y,u,v,x')$ be a giant splitting of $\sig(X)$ according to the notation in \prref{lem:gsp}.
Follow an \epstra defined by $X \mapsto Y uv X$ for a fresh variable $Y$.
This leads to a new entire state $(E', \alp', \sig')$ such that $\sig'(Y) = y$ and $\sig'(X) = x'$.
Note that $\sig'(Y)$ has no $T$-position, and that $X$ is left $T$-protected \wrt~$\sig'(W')$ where $W'$ is the equation at $E'$.
\item Rename $(E', \alp',\sig')$ as $(E, \alp, \sig)$.
\end{enumerate}
\noindent\textsc{\textbf{EndForAll}}\\
\noindent\textsc{\textbf{EndProcedure}}\\

Notably we do not perform any basic $T$-reductions at this point so, in particular, the resulting entire state $(E, \alp, \sig)$ may contain dummies (although it would have been safe to remove these).
Our next step is to turn every variable $X$ such that $\sig(X)$ is without $T$-positions into a $T$-immune variable.
\begin{lemma}[Baby splitting]\label{lem:bsp}
  Let $x \in M(B, \rho)$ be a trace without any \SSparcs where $S, S' < T$ and where $x$ has no $T$-position.
  Then there is a tuple $b(x) = (y_1,u,v,x_1)$, which we henceforth refer to as a \emph{baby splitting of $x$}, with the following properties.
\begin{enumerate}
  \item There is a factorization $x=y_1uvx_1$.
  \item We have $\rho(y_1) \cap T = \es$ and, in particular, $\rho(y) \ssne \rho(x)$.
  \item The factors $u$ and $v$ are (possibly empty) steps.
  \item If $v=1$, then $x_1$ is $T$-disjoint (and possibly the empty trace as well).
  \item If $v\neq 1$, then $\min(vx_1) = \pos(v)$ and $vx_1$ is strongly left $T$-bordered.
\end{enumerate}
\end{lemma}
\begin{proof}
  First, let $x = y_1x'$ where $y_1$ is the maximal $T$-disjoint prefix of $x$.
  Then $x'$ is $T$-bordered.
  Next, let $u$ be the subtrace of $x'$ defined by $\pos(u) = \set{ q \in \min(x')}{ \rho(q) \ssne T }$.
  Clearly $u$ is a step as $\pos(u) \sse \min(x')$.
  We write $x' = u x''$.
  If $x'' = 1$, then we set $v = 1$ and $x_1 = 1$ and are done.

  Otherwise, we $x''$ is strongly left $T$-bordered.
  To see this, consider $q' \in \min(x'')$.
  If $q' \in \min(x')$, then $\rho(q') \cap T \neq \es$ and $q' \not\in \pos(u)$ and so $\rho(q') \sm T \neq \es$.
  If $q' \not\in \min(x')$, then there exists a position $q \in \pos(u) \sse \min(x')$ such that $q \to q'$ is a \HA in $x$.
  Then $\rho(q') \cap T \ssupe \rho(q') \cap \rho(q) \neq \es$ since $\rho(q) \ssne T$.
  Moreover, we have $\rho(q) < T$ and, hence, $\rho(q') > T$ as $x$ contains no $T$-position and no \SSparc with $S,S' < T$.
  It follows that $\rho(q') \sm T \neq \es$.
  We let $v$ and $x_1$ be defined by $\pos(v) = \min(x'')$ and $\pos(x_1) = \pos(x'') \sm \min(x'')$.
\end{proof}
The input for the following procedure is a standard state $E=(W,B,\cX,\rho,\mu,T)$ with an \esolu $(\alp, \sig)$.
The procedure \textsc{BabySplit} is potentially executed multiple times for each $T \sse \Rs$.
It turns every variable without a $T$-position into a $T$-immune variable.\\

\noindent\textsc{\textbf{BeginProcedure}} \textsc{BabySplit}\\
\noindent\textsc{\textbf{ForAll}} $\os{X, \bar X} \sse \cX$ where $\sig(X)$ has no $T$-position, but $X$ is not $T$-immune \textsc{\textbf{Do}}
\begin{enumerate}
  \item Let $b(X) = (y,u,v,x')$ be a baby splitting of $\sig(X)$ according to the notation in \prref{lem:bsp}, and let $b(\bar x') = (\bar y', \bar u', \bar v', \bar x'')$ be a baby splitting of $\bar x'$.
    Follow an \epstra defined by the substitution $X \mapsto YuvXv'u'Y'$ for fresh variables $Y$ and $Y'$.
    This leads to a new entire state $(E', \alp', \sig')$ such that $\sig'(Y) = y$, $\sig'(Y') = y'$, and $\sig'(X) = x''$.
    Note that $Y$ and $Y'$ are $T$-disjoint, and that $X$ is $T$-immune \wrt $\sig'(W')$ where $W'$ is the equation at $E'$.
  \item Rename $(E', \alp', \sig')$ as $(E, \alp, \sig)$.
\end{enumerate}
\noindent\textsc{\textbf{EndForAll}}\\
\noindent\textsc{\textbf{EndProcedure}}\\

We execute the procedure \textsc{BabySplit} for the first time right after executing the procedure \textsc{GiantSplit}.
At this point every variable $X \in \cX$ is either $T$-protected (in case $\sig(X)$ has $T$-position) or $T$-immune (otherwise), and it is now safe to perform basic $T$-reductions.
According to \prref{lem:splinv}, all $T$-immune variables remain $T$-immune (unless they are removed).
Similarly, the status of being $T$-protected can only fail in a controllable way, namely it might happen that we can split the variable into a left or right $T$-conic variable and a fresh variable $Y$ with $T \nsse \rho(Y)$.
We perform such splittings as necessary and, afterwards, turn the new fresh variables into $T$-immune variables using the procedure \textsc{BabySplit}.
Finally, we also decompose all decomposable $T$-immune variables.

The order in which we perform these operations is irrelevant, but performing any one such operation can cause others to be necessary again.
Whenever possible, we follow the transitions of \prref{ruli:tconic}.
Its first item uses \prref{lem:esolu} without an explicit reference.

\begin{ruli}\label{ruli:tconic}
As long as one of the following items leads to a new state, execute that item and rename the resulting state $E'$ and its \esolu $(\alp',\sig')$ as $E=(W,B,\cX,\rho,\mu,T)$ and $(\alp, \sig)$.
\begin{enumerate}
\item If $\sig$ is not reduced, then perform all basic $T$-reductions.
\item If $X\in \cX$ such that $\sig(X)$ contains a $T$-position but is not left $T$-conic, but there is a factorization $\sig(X)=yx'$ such that $T \nsse \rho(y)$ and $x$ is left $T$-conic, then follow the \epstra defined by $X\mapsto YX$ where $Y$ is a fresh variable (with $T \nsse \rho'(Y)$ and $\rho'(Y) \ssneq \rho(X)$).
This leads to a new entire state $(E', \alp', \sig')$ with 
 $\sig'(Y)=y$ and $\sig'(X)=x'$. 
 Rename $(E', \alp', \sig')$ as $(E, \alp, \sig)$ and repeat such substitutions until no longer possible.
 Then call the procedure \textsc{BabySplit}.
\item If there is some $X \in \cX$, which is $T$-immune but $x = \sig(X)$ is decomposable as $x = yz$ with $\rho(y), \rho(z) \ssne \rho(x)$, then follow an \epstra defined by $X \mapsto Y Z$ where $Y$ and $Z$ are fresh variables.
  This leads to a new state $E'=(W',B,\cX',\rho',\mu',T)$ and a new \esolu $(\alp', \sig')$ with $\sig'(Y)=y$ and $\sig'(Z) = z$. 
  Note that the variables $Y$ and $Z$ are $T$-immune \wrt $\sig'(W')$.
\end{enumerate}
\end{ruli}
\begin{remark}\label{rem:Tconic}
Applying \prref{ruli:tconic} to an entire state $(E,\alp,\sig)$ 
leads to an \esolu which is reduced. 
If a variable  $X$ is left $T$-conic, then basic $T$-reductions do not destroy this property, unless $X$ vanishes because of an \epstra which is defined 
$X\mapsto \sig(X)$. If a variable $X$, which is not left $T$-conic, becomes
left $T$-conic at some point, then it produces exactly two fresh variable $Y$ and $\bar Y$ such that $T \nsse \rho(Y)$. \Ip it does not contain any $T$-position.
This can happen only finitely many times.
Finally, if a $T$-immune variable $X$ is decomposable, then we split it into two fresh variables $Y$ and $Z$ which are still $T$-immune and use fewer resources.
\hspace*{\fill}$\diamond$\end{remark}
\begin{lemma}\label{lem:finsplit}
  Let $(E, \alp, \sig)$ be the entire state directly after executing the procedures $\textsc{GiantSplit}$ and $\textsc{BabySplit}$.
  Then applying \prref{ruli:tconic} exhaustively eventually terminates.
  Moreover, the resulting entire state $(E', \alp', \sig')$ where $E' = (W', B', \cX', \rho', \mu', T)$ has the following properties.
  \begin{enumerate}
  \item The \esolu $(\alp', \sig')$ is reduced.
  \item Every variable $X \in \cX'$ is either $T$-protected, or $T$-immune and indecomposable. 
  \end{enumerate}
  \Ip the \esolu $(\alp', \sig')$ is $T$-perfect according to \prref{def:Tper}.
\end{lemma}
\begin{proof}
  Let $E = (W, B, \cX, \rho, \mu, T)$.
  By construction every variable $X \in \cX$ is $T$-immune or $T$-protected.
  According to \prref{lem:splinv}, all $T$-immune variables remain $T$-immune during the entire process unless they are removed.
  A $T$-protected variable may fail to be left (or right) $T$-protected after performing basic $T$-reductions.
  This can happen only if it is not left (resp.\ right) $T$-conic.
  The second item in \prref{ruli:tconic} covers this case by making the variable left (resp.\ right) $T$-conic by splitting off a variable without $T$-positions.
  These variables are then made $T$-immune by a successive call to the procedure \textsc{BabySplit}.
  As no new variables with $T$-positions are created, the second item is only applicable a finite number of times.
  The third item in \prref{ruli:tconic} decomposes decomposable $T$-immune variables.
  Since the new variables created thereby use fewer resources and only finitely many new $T$-immune variables are created by the second item, the third item is only applicable a finite number of times as well.
  Therefore, the process eventually terminates in an entire state $(E', \alp', \sig')$ with a reduced entire solution $(\alp', \sig')$ such that every variable $X \in \cX'$  is either $T$-protected, or $T$-immune and indecomposable.
  As such, $(\alp', \sig')$ is $T$-perfect by \prref{lem:good}.
 \end{proof}

\subsubsection{The number of variables and constants due to splittings}\label{sec:maxvar} 

To bound the number of variables and constants due to splitting, let us first consider the fate of a variable $X \in \cX$ during the procedure.
There are three distinct cases.
First, if $X$ it $T$-disjoint, then $X$ remains $T$-disjoint throughout and is never split (but is potentially decomposed or removed during basic $T$-reductions).
Next, suppose that $X$ is not $T$-disjoint, but $\sig(X)$ is without $T$-positions.
Such a variable will become $T$-immune, and remains so until the end of the procedure.
The variable $X$ becomes $T$-immune during a call to \textsc{BabySplit}.
Here, the general splitting is according to the rule
\begin{equation}\label{eq:imfate}
  X \mapsto YuvXv'u'Y'
\end{equation}
wherein $Y$ and $Y'$ are fresh $T$-disjoint variables (or possibly dummies) using fewer resources, and the constants $u$, $v$, $v'$, and $u'$ are (possibly empty) steps with $\abs{uvv'u'} \leq 4 \abs{T \cap \rho(X)}$.

Finally, if $\sig(X)$ contains a $T$-position, then $X$ becomes $T$-protected during \textsc{GiantSplit}.
It might happen that, at a later point, $X$ needs to be split again to become left and/or right $T$-conic as part of the second item of \prref{ruli:tconic}.
Accounting for that possibility, the general combined splitting is
\begin{equation}\label{eq:prfate}
  X \mapsto Y_1uvY_2XY_2'v'u'Y_1'
\end{equation}
where $Y_1$, $Y_2$, $Y_2'$, and $Y'_1$ are fresh variables (or possibly dummies) using fewer resources, and the constants $u$, $v$, $v'$, and $u'$ are (possibly empty) steps with $\abs{uvv'u'} \leq 4 \abs{T \cap \rho(X)}$.

Let us now derive an upper bound on the number $\Delta V(r)$ of fresh variables that are created from some variable $X$ with $\abs{\rho(X)} \leq r$ during all splittings (recursively,
	i.e.\ if~$Y$ is split from $X$ and $Z$ from $Y$ then we count $Z$ towards $X$ and towards $Y$).
Such a variable may partake in the $T$-splitting process for all $\es\neq T \ssneq \Rs$, \ie less than $2^{\abs{\Rs}}$ such processes.
According to the above analysis, for fixed $T$, at most $4$ fresh variables are created directly from $X$ in order to ensure that $X$ is (and remains) $T$-immune or $T$-protected.
Finally, the variable $X$ may at some point be decomposable and, as such, be replaced by $2$ fresh variables.
Crucially, each fresh variable uses fewer resources than $X$.

The above discussion directly translates into the bound
\begin{equation}
  \Delta V(r) < 2^{\abs{\Rs}} \cdot 4 \cdot (1 + \Delta V(r-1)) + 2 \cdot (1 + \Delta V(r-1)).
\end{equation}
We obtain $1 + \Delta V(r) \leq 2^{\abs{\Rs} + 3} \cdot (1 + \Delta V(r-1))$ where 
$\abs{\Rs}\geq 2$.
Since $\Delta V(1) = 0$ and $r<|\Rs|$, this yields
\begin{equation}
  1 + \Delta V(r) \leq 2^{(\abs{\Rs} + 3)(r-1)} \leq 2^{(\abs{\Rs} + 3)(\abs{\Rs}-2)} \leq 2^{\abs{\Rs}^2 + 2\abs{\Rs}} 
\end{equation}
Additional variables are created during elimination of all $T$-$T$-arcs for fixed $T \sse \Rs$ as detailed in \prref{sec:eliTT}.
However as we will see later, all such variables will have been removed as dummies once all $T$-$T$-arcs are eliminated.
As such, these additional variables do not partake in the $T'$-splitting process for any $T'$.
This allows us to bound the total number $V_{\mathrm{split}}$ of variables present during splittings in terms of $\abs{\cX_{\mathrm{init}}}$:
\begin{equation}\label{eq:spVmax}
  V_{\mathrm{split}} 
  \leq (1 + \Delta V(\abs{\Rs})) \cdot \abs{\cX_{\mathrm{init}}}
  \leq 2^{\abs{\Rs}^2 + 2\abs{\Rs}} \cdot \abs{\cX_{\mathrm{init}}}
\end{equation}
The factor $2^{\abs{\Rs}^2 + 2\abs{\Rs}}$  
grows pretty
fast with growing $\Rs$. For $\abs{\Rs}=5$ it is $2^{35}$.
\begin{remark}\label{rem:fixwhcX}
As we will show later, the estimation (\ref{eq:spVmax}) will make it possible to bound the size $\wh\cX$ of the ambient set of variables by 
\begin{align}\label{eq:remfixwhcX}
|\wh\cX|\leq 7\cdot 2^{\abs{\Rs}^2 + 2\abs{\Rs}} \cdot\abs{\cX_{\mathrm{init}}}
\end{align}
Since we require that all sets of variables $\cX$ at states in $\cU$ 
are subsets of $\wh\cX$, fixing $\wh\cX$ by (\ref{eq:remfixwhcX}) has direct influence on $\cU$ and $\wh \cU$. Below we put more restrictions on their states. 
\hspace*{\fill}$\diamond$\end{remark}
The splitting procedure also produces a number of words over constants which are popped out by variables.
Next we give an upper bound for the total length for all of these words.
\begin{lemma}\label{lem:Totnb}
  Let $(E, \alp, \sig)$ be an entire state with equation $W$ when we enter the splitting procedure.
  Then there is path in~$\wh \cU$ {}from $(E,\alp,\sig)$ to $(E', \alp', \sig')$ with a reduced $T$-perfect \esolu $(\alp',\sig')$ such that for each equation $\wt W$ of an entire state along the path it holds that
\begin{equation}\label{eq:locnb}
  \abs{\wt W} - \abs{W} \leq 4 \cdot 2^{\abs{\Rs}^2 + 3\abs{\Rs}}\cdot\abs{\Winit}
\end{equation}
\end{lemma}
\begin{proof}
The factor $2^{\abs{\Rs}^2 + 4\abs{\Rs}} \cdot \abs{\Winit}$ appears because every position which is 
labeled by an initial variable leads to at most $2^{\abs{\Rs}^2 + 2\abs{\Rs}}$ positions in $W$ which are labeled by a variable in $\cX$ thanks to the upper bound \eqref{eq:spVmax}. 
During the splitting, the increase in equation length caused by a single variable can be bounded above by $4 + 4 \abs{\Rs} \leq 4 \cdot 2^{\abs{\Rs}} - 1$ as $\abs{\Rs} \geq 2$, with the longest increase happening in \prref{eq:prfate}.
Combining the two bounds yields the estimation \eqref{eq:locnb}.
\end{proof}

\subsection{Lifting $T$-perfect solutions}

We are now ready to discuss our sufficient condition for $T$-lifting anounced in \prref{sec:prospli}.

\begin{proposition}\label{prop:Tper}
  Let $E = (W, B, \cX, \rho, \mu, T)$ be a standard state.
  If $(\sig, \alp)$ is a $T$-perfect \esolu at $E$ such that $\sig(W)$ contains an \STarc with $T \nsse S$, then $(\sig, \alp)$ is not reduced.
\end{proposition}
The proof of \prref{prop:Tper} requires further preparation and will be carried out in \prref{sec:diam}.
For now, let us discuss some basic properties ensured by $T$-perfectness (see \prref{def:Tper}).

\medskip

The first part of \prref{def:Tper} is crucial to lifting.
By \prref{cor:TimX}, it implies that, for every \Tarc $p \to q$ in $\sig(W)$ with both endpoints visible, the corresponding positions in the equation $W$ are also connected by a \HA $\sus{p} \to \sus{q}$.
We will use this observation repeatedly.

The second part of \prref{def:Tper} says that every \STarc $p \to q$ in $\sig(W)$ with $T \nsse S$ is \emph{almost} noncrossing.
For technical reasons we have to allow for the possibility that the $S$-position $q$, say, is visible whereas the $T$-position $p$ is only semi-visible.
During $T$-lifting we only intend to modify the $S$-position anyway and, hence, this requirement is sufficient for our cause.
To avoid unnecessarily lengthening arguments with case distinctions, we will assume at the appropriate points that all \STarcs with $T \nsse S$ are actually noncrossing in the sense of \prref{def:fiv}.
Moreover, we assume that no $T$-position has both an incoming and an outgoing \HA to $S$-positions with $T \nsse S$: our justification for this assumption is just a `shadow away'.\footnote{Maybe it depends on the mood of a potential reader to see a shadow or to ignore it. We leave it to the interested reader to transform our proofs using this assumption into a shadowless landscape. To aid in such an endeavor we have also stated the essential ingredient, \prref{cor:TimX}, at the appropriate level of generality.}

\begin{remark}\label{rem:shadowWay}
  Consider a $T$-perfect \esolu $(\alp, \sig)$ at a state $E = (W, B, \cX, \rho, \mu, T)$.
  Let us now imagine a world where every $T$-position $q$ in $\sig(W)$ casts a shadow to its left and right, and where every incident \HA in $\sig(W)$ is attached to the corresponding shadow (instead of $q$ itself).
  Formally, let $\tilde B = B \cup \os{\tilde s}$ for some selfie $\tilde s \not\in \wh C$ with $\tilde\alp(\tilde s) = 1$, $\tilde\rho(\tilde s) = T$, and $\tilde\mu(\tilde s) = 1$.
  We refer to $\tilde s$ as a \emph{shadow letter}.
  For every $T$-letter $a \in B$, the trace $\tilde s a \tilde s$ is self-involuting if and only if $a$ is so, and we have $\tilde\alp(\tilde s a \tilde s) = \alp(a)$, $\tilde\rho(\tilde s a \tilde s) = T = \rho(a)$, and $\tilde\mu(\tilde s a \tilde s) = \mu(a)$.
  Therefore we can simply replace every occurrence of every $T$-letter $a \in B$ in $W$ by the trace $\tilde s a \tilde s$.
  This transformation yields a new state $\tilde E = (\tilde W, \tilde B, \cX, \tilde \rho, \tilde \mu, T)$ which is \emph{not} part of the automaton $\cU$ (as, e.g., $\tilde s \not\in \wh C$), but that should not hinder our imagination.
  The same transformation applied to $\sig(W)$ and to $\sig(X)$ for every $X \in \cX$ yields an \esolu $(\tilde\alp, \tilde\sig)$ at $\tilde E$.
  It is easy to see, that $(\tilde\alp, \tilde\sig)$ is $T$-perfect at $\tilde E$ and no $T$-position in $\tilde\sig(\tilde W)$ has incoming as well as outgoing \HA{s} to $S$-positions with $T \nsse S$.

  If the solution $\sig(X)$ of a variable $X \in \cX$ has a minimal $T$-position, then the transformed solution $\tilde\sig(X)$ has a minimal $T$-position labeled by the shadow letter $\tilde s$.
  For each variable $X \in \cX$ as above we pop out this shadow letter, making it visible on the outside.
  This leads to a new state $\tilde E' = (\tilde W', \tilde B, \cX, \tilde \rho, \tilde \mu, T)$ and a new \esolu $(\tilde\alp, \tilde\sig')$.
  Once again $(\tilde\alp, \tilde\sig')$ is $T$-perfect at $\tilde E'$ and no $T$-position in $\tilde\sig(\tilde W)$ has incoming as well as outgoing \HA{s} to $S$-positions with $S < T$.
  Moreover, this time every \STarc in $\tilde\sig'(\tilde W')$ with $T \nsse S$ is noncrossing.
  To see this, note that the $T$-position of such an arc is labeled by $\tilde s$, but a semi-visible position labeled by $\tilde s$ in $\tilde\sig'(\tilde W')$ is visible.

  Crucially, if we can perform a $T$-lifting at $(\tilde E', \tilde\alp', \tilde\sig')$, then we can perform a $T$-lifing at $(E, \alp, \sig)$ since we can always remove all positions labeled by the shadow letter $\tilde s$ afterwards.
  \hspace*{\fill}$\diamond$
\end{remark}
An $S$-position with $T \nsse S$ can have at most one incoming and at most one outgoing \Tarc.
\begin{lemma}\label{lem:STcnt}
  Let $E = (W, B, \cX, \rho, \mu, T)$ be a standard state with an \esolu $(\alp, \sig)$, and let $q \in \pos(\sig(W))$ be an $S$-position with $T \nsse S$.
  If $(\alp, \sig)$ is $T$-perfect, then all positions $q' \equiv q$ are incident with the same number of \Tarc{s} in $\sig(W)$.
\end{lemma}
\begin{proof}
  This is obvious in case $q \darc q'$.
  If $q \sim q'$, then both $q$ and $q'$ are invisible.
  Given that the \esolu $(\alp, \sig)$ is $T$-perfect, this implies that $\sus{p} = \sus{q}$ for every \Tarc $p \to q$ (or $q \to p$) in $\sig(W)$.
  Hence there exists a position $p' \sim p$ with $\sus{p'} = \sus{q'}$ such that $p' \to q'$ (or $q' \to p'$) is also a \HA in $\sig(W)$.
  It follows that the claim also holds in case $q \sim q'$.
  But this already shows the general case as ${\equiv}$ is the transitive closure of ${\darc} \cup {\sim}$ and the relation of `having the same number of incident \Tarc{s} in $\sig(W)$' is clearly transitive.
\end{proof}

\subsubsection{Lifting unbalanced \STarc{s}}\label{sec:plif}

\begin{definition}\label{def:plif}
Let $E=(W,B,\cX,\rho,\mu,T)$ be a standard state with an \esolu $(\alp,\sig)$.
Further, let $p \to q$ (resp.~$q \to p$) be a \Tarc in $\sig(W)$ with $\rho(p) = T$.
We say that $p\to q$ (resp.~$q\to p$) is \emph{$T$-liftable}\index{T-liftable@$T$-liftable}
in $\sig(W)$ if $T \nsse \rho(q)$ and there is no outgoing (resp.~incoming) \Tarc at $q$.
\end{definition}
The definition requires $T \nsse \rho(q)$ because we intend to lift the position $q$, by assigning more resources to $q$, thus we need $\rho'(q) = T \cup \rho(q) \neq \rho(q)$.
Furthermore, a $T$-liftable \HA is by definition unbalanced.
Such an arc $p \to q$ or $q \to p$ is uniquely determined by the set $\os{p,q}$.

\begin{lemma}\label{lem:plif}
  Let $E=(W,B,\cX,\rho,\mu,T)$ be a standard state, and let $(\alp, \sig)$ be a $T$-perfect \esolu at $E$.
  If $\sig(W)$ contains a $T$-liftable \HA, then $(\alp, \sig)$ is not reduced.
\end{lemma}
\begin{proof}
  Suppose that $p \to q$ is a $T$-liftable \HA in $\sig(W)$ with $\rho(p) = T$ and $T \nsse \rho(q) = S$.
  We claim that we can lift all positions $q'$ in $\sig(W)$ with $q' \equiv q$ so as to then have $S \cup T$ as resource set.
  Since such a transformation can then be realized by basic $T$-reductions as explained in \prref{sec:steplift}, this claim implies that $(\alp, \sig)$ is not reduced.
  To see why we can perform such a lifting note that, by \prref{lem:STcnt}, every position $q'$ with $q' \equiv q$ is part of a $T$-liftable \HA $\os{p',q'}$.
  Following \prref{rem:shadowWay} we may just as well assume that all such arcs $\os{p',q'}$ are noncrossing and pairwise disjoint. 
  Moreover, whenever $\os{p', q'}$ is (fully) visible, then $\os{\sus{p'}, \sus{q'}}$ is a \HA in $W$ by \prref{cor:TimX}.
  Hence, the subtraces $\os{p', q'}$ act as if they were a single position, and we may therefore redistribute their resources accordingly, i.e., replace $\rho(q') = S$ with $\rho(q') \cup \rho(p') = S \cup T$.
\end{proof}

\subsubsection{Lifting $T$-diamonds}\label{sec:diam}

Upon repeatedly lifting $T$-liftable \HA{s} we might eventually create a $T$-diamond as defined below and depicted in \prref{fig:diam}.
For an example of how this situation can occur see \prref{fig:ldiam}.

\begin{definition}\label{def:diam}
  Let $T \sse \Rs$. 
  A trace $v \in M(B, \rho)$ is called a \emph{$T$-diamond}\index{T-diamond@$T$-diamond}
  if the following hold.
  \begin{itemize}
  \item We can write $v=ab_1\cdots b_ra'$ for letters $a, b_1, \ldots, b_r, a' \in B$ and $r \geq 1$. 
  \item We have $\rho(a) = T = \rho(a')$ and $\es \neq \rho(b_i) \cap T \neq T$ for all $1 \leq i \leq r$.
  \item We have $\rho(b_i)\cap \rho(b_j) =\es$ for all $1\leq i<j\leq r$.
  \end{itemize}
\end{definition}
\begin{figure}[h] 
\begin{center} 
		\begin{tikzpicture}[node distance=8mm]

		\node[circle] (a) at (0,0) {$a$}; 
		\node[circle] (v) at (3,0) {$\vdots$};
		\node[circle] (b1) [above of=v] {$b_1$};
		\node[circle] (bs) [below of=v] {$b_r$};
		\node[circle] (c) at (6,0) {$a'$};
		\draw (a) edge[->, >=latex,thick] (v); 
		\draw (v) edge[->, >=latex,thick] (c); 
		\draw (a) edge[->, >=latex,thick] (b1); 
		\draw [thick,->, >=latex] (a) to (bs); 
		\draw (b1) edge[->, >=latex,thick] (c); 
		\draw (bs) edge[->, >=latex,thick] (c); 
		
		\end{tikzpicture}\end{center}
\vspace{-0.5cm}
  \caption{A $T$-diamond with an interior step $b_1\cdots b_r$ between~$a$ and~$a'$.}
\label{fig:diam}
\end{figure}

Given a trace $w \in M(B, \rho)$ as well as a subset $P=\os{p_1\lds p_m}\sse \pos(w)$, we write
$\gen{p_1\lds p_m}$ to denote the \emph{generated subtrace} of $P$, \ie $\gen{p_1\lds p_m}$ is the smallest subtrace in~$w$ containing all positions of~$P$.
Every $T$-diamond in~$w$, \ie every subtrace $v$ of~$w$ that is a $T$-diamond, is generated by its minimal and its maximal position.

\begin{lemma}\label{lem:diam}
  Let $E=(W,B,\cX,\rho,\mu,T)$ be a standard state, and let $(\alp, \sig)$ be a $T$-perfect \esolu at $E$.
  If $\sig(W)$ contains a $T$-diamond, then $(\alp, \sig)$ is not reduced.
\end{lemma}
\begin{proof}
  Following \prref{rem:shadowWay}, we assume that all \STarcs in $\sig(W)$ with $T \nsse S$ are noncrossing, and that no $T$-position in $\sig(W)$ has both an incoming and an outgoing \HA in $\sig(W)$ leading to $S$-positions with $T \nsse S$.
  This comes at no loss of generality since introducing shadow letters neither creates nor destroys $T$-diamonds.
  Suppose that $\sig(W)$ contains the $T$-diamond $v$, and let $p,p' \in \pos(v)$ be its minimal and maximal position, respectively.
  Furthermore, let $s$ be the interior step of $v$ which is defined by $\pos(s) = \pos(v) \sm \{p, p'\}$.

  If any position $q \in \pos(s)$ is invisible, then $\sus{p} = \sus{q} = \sus{p'}$ since $(\alp, \sig)$ is $T$-perfect and the solution $\sig(W)$ contains the \STarcs $p \to q$ and $q \to p'$ with $T \nsse S = \rho(q)$.
  This implies that $\sus{v} = \sus{q}$ as $v$ is generated by $\{p, p'\}$.
  If, on the other hand, some $q \in \pos(s)$ is visible, then so are $p$ and $p'$ since the \STarcs $p \to q$ and $q \to p'$ with $T \nsse S = \rho(q)$ are both noncrossing by our assumption on $\sig(W)$.
  Hence, by the above, every $q \in \pos(v)$ is visible.
  This argument shows that $v$ is noncrossing and, since $v$ is arbitrary, this in fact holds for every $T$-diamond.
  Moreover, every other $T$-diamond $v'$ is disjoint from $v$.
  Indeed, if $\min(v) = \min(v')$ or $\max(v) = \max(v')$, then $v = v'$ since, firstly, the (unique) minimal and maximal positions of a $T$-diamond are successive $T$-positions in $\sig(W)$ so that either one determines the other and, secondly, every $T$-diamond is generated by its minimal and its maximal position.
  On the other hand, $\max(v) = \min(v')$ (and, similarly, $\max(v') = \min(v)$) is impossible since we assumed that no $T$-position in $\sig(W)$ has both an incoming and an outgoing \HA to $S$-positions with $T \nsse S$.

  We claim that we can lift the interior step $s$ of $v$ together with its equivalent steps $s' \equiv s$ in $\sig(W)$.
  Since such a transformation can then be realized by basic $T$-reductions as explained in \prref{sec:steplift}, this claim implies that $(\alp, \sig)$ is not reduced.
  To see why we can perform such a lifting note that, by the above, for every $q \in \pos(s)$ and every $q' \equiv q$, there is a $T$-diamond $v' \equiv v$ with $q' \in \pos(v')$.
  It then follows that $q' \in \pos(s')$ and $s' \equiv s$ where $s'$ is the interior step of $v'$.
  In addition, $T$-diamonds are noncrossing and pairwise disjoint.
  Moreover, if $v'$ is a (fully) visible $T$-diamond, then $\sus{v'}$ is also a $T$-diamond in $W$ by \prref{cor:TimX}.
  Hence, the subtraces~$v' \equiv v$ behave as if they were single positions, and we may therefore redistribute their resources accordingly.
\end{proof}

\begin{figure}[ht] 
	\begin{center} 
		\begin{tikzpicture}[xscale=1, yscale=2]
		\node[circle] (a) at (-3,0)   {$a$}; 
		\node[circle] (b1) at (0,0.5) {$b_1$};
		\node[circle] (b2) at (-1.3,0) {$b_2$};
		\node[circle] (b3) at (0,-0.5){$b_3$};
		\node[circle] (b4) at (1.3,0) {$\bar b_2$};
		\node[circle] (c) at (3,0) {$\bar a$};
		\node[circle] (mc) at (0,0) {$c$};
\node[circle] (d) at (-3,0.5)   {$d$}; 
\node[circle] (d1) at (3,0.5)   {$d'$};
\node[circle] (d2) at (3,-0.5)   {$d''$};
\draw (a) edge[->, >=latex,thick] (b1);
\draw (a) edge[->, >=latex,thick] (b2);
\draw (a) edge[->, >=latex,thick] (b3);
\draw (b3) edge[->, >=latex,thick] (c);
\draw (b1) edge[->, >=latex,thick] (c);
\draw (b2) edge[->, >=latex,thick] (mc);
\draw (mc) edge[->, >=latex,thick] (b4);
\draw (b4) edge[->, >=latex,thick] (c); 

\draw (d) edge[->, >=latex,dashed] (b2);
\draw (b4) edge[->, >=latex,dashed] (d1);
\draw (b4) edge[->, >=latex,dashed] (d2);

\begin{scope}[shift={(-1,0)}]
\node[circle] (ra) at (5,0)   {$a$}; 
		\node[circle] (rb1) at (8,0.5) {$b_1$};
		\node[circle] (rb2) at (6.7,0) {$c_2$};
		\node[circle] (rb3) at (8,-0.5){$b_3$};
		\node[circle] (rb4) at (9.3,0) {$\bar c_2$};
		\node[circle] (rc) at (11,0) {$\bar a$};
		\node[circle] (rmc) at (8,0) {$c$};
\node[circle] (rd) at (5,0.5)   {$d$}; 
\node[circle] (rd1) at (11,0.5)   {$d'$};
\node[circle] (rd2) at (11,-0.5)   {$d''$};
\draw (rb2) edge[->, >=latex,thick] (rb1);
\draw (rb2) edge[->, >=latex,thick] (rb3);
\draw (ra) edge[->, >=latex,thick] (rb2);
\draw (rb1) edge[->, >=latex,thick] (rb4);
\draw (rb2) edge[->, >=latex,thick] (rmc);
\draw (rmc) edge[->, >=latex,thick] (rb4);
\draw (rb3) edge[->, >=latex,thick] (rb4);
\draw (rb2) edge[->, >=latex,thick] (rmc); 
\draw (rb4) edge[->, >=latex,thick] (rc);
\draw (rd) edge[->, >=latex,dashed] (rb2);
\draw (rb4) edge[->, >=latex,dashed] (rd1);
\draw (rb4) edge[->, >=latex,dashed] (rd2);
\end{scope}
		\end{tikzpicture}\end{center}
	\vspace{-0.5cm}
	\caption{Creating a $T$-diamond upon lifting $b_2$ to its $T$-clone $c_2$.}
\label{fig:ldiam}
\end{figure}

\begin{proof}[Proof of \prref{prop:Tper}]
  Suppose that the solution $\sig(W)$ contains a \HA $p \to q$ such that $\rho(p) = T$ and $T \nsse \rho(q) = S$.
  If there is no \Tarc $q \to p'$ in $\sig(W)$, then $p \to q$ is a $T$-liftable \HA and the statement follows from \prref{lem:plif}.
  We therefore assume that such an arc exists and consider the subtrace $v$ generated by $\os{p,p'}$.
  This subtrace contains no $T$-position except for its (unique) minimal position $p$ and its (unique) maximal position $p'$ for otherwise $p \to q$ and $q \to p'$ could not both be \HA{s} in $\sig(W)$.
  Let $u$ be the interior of $v$, \ie $\pos(u) = \pos(v) \sm \os{p,p'}$.
  Consider any position $q' \in \pos(\min(u))$.
  By definition of $u$ there is a \HA $p \to q'$ in $\sig(W)$.
  It follows that $T \nsse \rho(q')$ for otherwise $p \to q$ would not be a \HA in $\sig(W)$.
  If $p \to q'$ is not $T$-liftable, then $q'$ has an outgoing \Tarc which must be the \HA $q' \to p'$ since $p$ and $p'$ are successive $T$-positions in $\sig(W)$.
  Repeating this argument for all positions in $\min(u)$ shows that either $v$ contains a $T$-liftable \HA or $v$ is a $T$-diamond.
  In the first case $(\alp, \sig)$ is not reduced by \prref{lem:plif}, and in the second case it is not reduced by \prref{lem:diam}.
\end{proof}

\section{Elimination of all $T$-$T$-arcs}\label{sec:eliTT}

\subsection{The age of \Tflat variables}
\label{sec:tflatness}
The elimination of all \TTarcs uses the existence
of \Tflat variables according to the next definition.\footnote{
Some hairy musical text says: `The dawning of the Age of
Harmony and Understanding'.}
\begin{definition}\label{def:flatvar}
  Let $T \sse \Rs$.
  A trace $x \in M(B, \rho)$ is called \emph{$T$-flat}\index{T-flat@$T$-flat} if all positions of $x$ are $T$-positions.
  Moreover, given a state $E=(W,B,\cX,\rho,\mu,\theta,T)$ with an \esolu $(\alp,\sig)$, a variable $X \in \cX$ is \emph{$T$-flat} (\wrt $\sig$) if $\sig(X)$ is $T$-flat.
  We denote the set of $T$-flat variables by \[
    \cXTfl =\set{X\in \cX}{\sig(X)\in B_T^*}
  \]
\end{definition}
Every sequence of \TTarcs in  $M(B\cup \cX,\rho)$ defines a 
\Tflat subtrace; and if $X$ is a \Tflat variable, then  
$\sig(X)$ is a word in $T$-letters in $B$. The position $p$ of a \Tflat variable $X$ in $W$ satisfies  $\rho(p)=T$. 
For the rest of section we assume that 
every set of variables comes with a subset $\cXTfl$ of \Tflat variables. 
Throughout \prref{sec:eliTT} we assume the semantic condition that $\sig(W)$ is without any \Tarc
where one endpoint $q$ satisfies $T \nsse \rho(q)$.
By \prref{prop:Tper}, this condition is satisfied when we enter this section with at an entire state $(E, \alp, \sig)$ where $(\alp, \sig)$ is reduced and $T$-perfect.

\subsubsection{Small, medium, and large equations \wrt~$k$}\label{sec:le}

\begin{definition}\label{def:stawrtsig}
Let $k\in \N$ be a number and 
$E=(W,B,\cX,\rho,\mu,\theta,T)$ be a state.
The state is called \emph{$k$-small}\index{k-small@$k$-small} 
if $|\cX|\leq k$. 
It is called  \emph{$k$-medium}\index{k-medium@$k$-medium} if $|\cX|\leq 3k$ and 
$|\cXTfl|\leq 2k$. It is called 
 \emph{$k$-large}\index{k-large@$k$-large} if $|\cX|\leq 7k$ and $|\cXTfl|\leq 6k$.
\end{definition}
The idea behind \prref{def:stawrtsig} is to compute explicitly some
$k\in |\cXinit|\cdot 2^{\Oh(|\Rs|^2)}$ such that we can show completeness 
within a sub automaton of~$\cU$ such that all states on some path from the initial to a finial state remain $k$-large. Thus, we can avoid oversized states satisfying $|\cX| > 7k$.

We begin this section with standard state~$E$ and its \esolu $(\alp,\sig)$ such that there are at most~$k$ variables. That is,~$E$ is $k$-small. At some point, more
\Tflat variables appear by splitting $T$-conic variables. This forces us to deal with medium standard states because there are up to $2k$ fresh variables. During block compression, we have to add up to $4k$ more fresh (and typed) \Tflat variables. 
This makes the state $k$-large. 
{}From this point on we oscillate between $k$-large and $k$-medium states.
The oscillation eventually ends in a $k$-small state, and after that there are no more \SSparc{s} in the \solu where $S,S'\leq T$. 
This process will show the upper bound (\ref{eq:remfixwhcX}) in \prref{rem:fixwhcX}.

\subsubsection{The expected length of longest increasing prefixes}\label{sec:explip}
It is not enough to bound the maximal number of variables we also need an upper bound on the sum $\sum_{a\in B}|W|_a$. For 
this we deal linear orders chosen uniformly at random on various sets of size $m$, say $S=\os{1\lds m}$. Consider the set $S^n$ of all sequences of length $n \geq 1$. Of interest are only those sequences where 
all symbols all distinct. In linguistics these sequences  are \emph{heterograms} but we found various other names\footnote{For example, heterograms are called \emph{injective words} in \cite{ChacholskiLM20}.} as well.
So, we feel free to add another one inspired by some popular puzzle. 
Our notation is \emph{sudoku}-word\footnote{The term
Sudoko is an abbreviation of the Japanese expression that every digit must be single, and  
the first two Kanji symbols of this expression are romanized 
as `Su-Doku'.
More is on Wikipedia or in some Sudoku- and Suneko-books, \eg~in \cite{KobayashiT24}. }. Given a word $\oo\in B^+$ we let $\mathrm{ld}(\oo)$ to be  the length of the longest sudoku-prefix of $\oo$.  Actually, we are interested in the length $\li(\oo)$ of the \emphind{longest increasing prefix} of $\oo$, i.e., $\li(\oo)$ is the maximum number $k \leq n$ such that $\oo(i) < \oo(j)$ for all
$1 \leq i < j \leq k$. Clearly, $1 \leq \li(\oo) \leq \mathrm{ld}(\oo) \leq m$  and
$\li(\oo)$ and $\mathrm{ld}(\oo)$ are random variable for each $\oo\in B^+$.
We write a word $\oo\in B^n$ as $\oo=\oo(1)\cdots \oo(n)$ with $\oo(i)\in B$ or as 
a sequence $\oo=(\oo(1)\lds \oo(n))$.

The following lemma can be deduced from \cite{Jez16jacm_stacs}, and it is used in \cite{DiekertJK2016short}, too. Our proof is, to the best of our knowledge, more concise than earlier published proofs. We suppose that the result is folklore and/or shown elsewhere, but we lack any explicit reference.
\begin{lemma}
\label{lem:explip}
Let $e=2.718\ldots$ denote the Euler number.
For all $\oo \in B^n$ we have \[
  1 \leq \E{\li(\oo)} = \sum_{k = 1}^{\mathrm{ld}(\oo)}  1 / k! < e - 1.
\]
\end{lemma}
\begin{proof}
  Let $1 \leq k \leq \mathrm{ld}(\oo)$.
  Then $\li(\oo) \geq k$ holds if and only if $\oo(1) < \oo(2) < \dotsc < \oo(k)$.
  The probability of this event is $\Prob{\li(\oo) \geq k} = 1/ k!$, since we can restrict our randomly chosen order to the $k$-element set $\{\oo(1), \dotsc, \oo(k)\} \subseteq B$ and this restriction still follows a uniform distribution.
Hence, the expected length $\E{\li(\oo)}$ of the longest increasing prefix of $\oo$ is given by
  \[
    \E{\li(\oo)} = \sum_{k=1}^{\mathrm{ld}(\oo)} k \cdot \Prob{\li(\oo) = k} 
    = \sum_{k=1}^{\mathrm{ld}(\oo)} \Prob{\li(\oo) \geq k} 
    = \sum_{k = 1}^{\mathrm{ld}(\oo)}  1 / k! < e - 1 <2. \qedhere
  \]
\end{proof}

\subsection{Maximal \Tarc sequences}\label{sec:mts}
The following notion is crucial for getting rid of \TTarcs. 
\begin{definition}\label{def:maxrun}
Let $w\in M(B,\rho)$ be a trace with $\rho(\min(w))=\rho(\max(w))=\Rs$ and 
$p_1\to \cdots \to p_m$ be a sequence of \TTarc{s} with $m\geq 2$. 
It is called a \emph{maximal \Tarc sequence} in~$w$ if there is neither a \TTarc $p\to p_1$ nor a \TTarc $p_m\to p'$.
\end{definition}
Since we assume $T\neq \Rs$ every maximal \Tarc sequence $p_1\to \cdots \to p_m$ is surrounded by two different positions $q$ and $q'$ in~$\sig(W)$
such that there are \HAs $q\to p_1\to p_2$ and $p_{m-1}\to p_m\to q'$.
Moreover, both sets 
$\os{q,p_1,p_2}$, $\os{p_{m-1},p_m,q'}$ define subtraces in~$\sig(W)$; and we have $T<\rho(q)$, $T<\rho(q')$ because $\sig(W)$ (being a \solu at $E$) is without any \STarc where $S<T$.
\begin{lemma}\label{lem:TTR}
Let $E=(W,B,\cX,\rho,\mu,T)$ be a standard state with an \esolu $(\alp,\sig)$
such that $\sig(W)$ without any \STarc where $S<T$ and $(\alp,\sig)$ is a reduced and $T$-perfect \esolu. 
Then 
$\wh \cU$ contains an entire path from $(E,\alp,\sig)$
to some $(E,\alp,\sig')$ such that if $p_1\to p_2 \to \cdots \to p_m$ is a maximal $T$-arc sequence in $\sig'(W)$, then $p_1$ and $p_{m}$ each have equivalent semi-visible positions.
\end{lemma}
\begin{proof}
By contradiction assume that $p_1$ has no equivalent position which is semi-visible.
This implies that there is a subtrace $q\to p_1\to p_2$ in $\sig(W)$ where
$\rho(q)\neq T$ and $p_1$ is fully invisible. 
Since $(\alp,\sig)$ is $T$-perfect we know by \prref{prop:Tper}
that $T\ssneq \rho(q)$. As a consequence, no other \HA ends in $p_1$.  
Hence, if $p_1'\equiv p_1$, then 
$(q\to p_1\to p_2)\equiv (q'\to p'_1\to p'_2)$.
Since $(\alp,\sig)$ is reduced, the label $a'$ of $p'_1$ is a letter in $\wh A$.
Moreover we can relabel every $p'_1$ by some $(a,\rho(q)) \in \wh A$ such that $\wh \pi (a)=\wh \pi (a')$.
Thereby we obtain a new \solu $\sig'$ at $E$; see also \prref{fig:shapy}.\footnote{The lifting of the $T$-letter at $p_1$ also is a $T'$-lifting with $T\ssneq T'=\rho(q)$ according to \prref{sec:steplift}.}
Since the number of \TTarcs in $\sig'(W)$ is less than in $\sig(W)$, we are done by induction.
\end{proof}
\begin{figure}[t]
\begin{center} 
		\begin{tikzpicture}[xscale=1.45, yscale=0.5]
		\node[circle] (q'') at (0,1) {$q$};
		\node[circle] (pp) at (-0.1,1) {\phantom{q}};
	
		\node[circle] (p) at (1.3,0) {$p_{1}$};
		\node[circle] (u) at (2,2.0) {$u$};
	\node[circle] (+) at (2,0.48) {$+$};
		\node[circle] (c0) at (2,0) {};
		\node[circle] (c1) at (2,-0.4) {};
		\node[circle] (a1) at (2.7,0) {$p_m$};
		\node[circle] (a2) at (3.9,0.5) {$q'$};
		\node[circle] (r) at (3.95,0.5) {\phantom{q'}};
		
		\node[circle] (sq') at (5,1) {$q$};
		\node[circle] (spp) at (4.9,1) {\phantom{q}};
		\node[circle] (sp1) at (6.3,1.45) {$p_{1}$};
		\node[circle] (su) at (7.4,2) {$u$};
		\node[circle] (sc1) at (7.07,1.07) {+};
		\node[circle] (sa1) at (7.7,0) {$p_m$};
		\node[circle] (sa2) at (8.9,0.6) {$q'$};
		\node[circle] (sr) at (8.9,0.6) {\phantom{q'}};
		\draw (q'') edge[->, >=latex,dashed] (u);
		\draw (p) edge[->, >=latex,thick] (a1);
		\draw (u) edge[->, >=latex,dashed] (a2); 
		\draw (c0) edge[->, >=latex,thick] (a1);
		\draw (a1) edge[->, >=latex,thick] (a2);
		\draw (sp1) edge[->, >=latex,thick] (sa1);

		\draw (sq') edge[->, >=latex,thick] (sp1);
		\draw (su) edge[->, >=latex,dashed] (sa2);
		\draw (sp1) edge[->, >=latex,dashed] (su);
		\draw (sa1) edge[->, >=latex,thick] (sa2);
		\draw (q'') edge[->, >=latex,thick] (p);
		
\draw[rounded corners,gray] (pp.south west) -- (pp.north west) --(u.north west) -- (u.north east) -- (r.north east) -- (r.south east) -- (a1.south east) -- (a1.south west)
 -- (p.south west) -- cycle;
\draw[rounded corners,gray] (spp.south west) -- (spp.north west) --(su.north west) -- (su.north east) -- (sr.north east) -- (sr.south east) 
-- (sa1.south east) -- (sa1.south west) --(sp1.south west) -- cycle;
\end{tikzpicture}
\end{center}
	\vspace{-0.5cm}
	\caption{The effect of lifting the $T$-position $p_1$
	inside the subtrace $\gen{q,q'}$ where  $p_1\arc +p_m$ is a maximal 
	$T$-sequence inside $\sig(W)$. Lifting~$p_1$ breaks a symmetry.
	}\label{fig:shapy}
\end{figure}

\subsection{Creating $T$-flat variables by splitting}\label{sec:createTflat}
Let $E=(W,B,\cX,\rho,\mu,T)$ be a standard state with a reduced \esolu $(\alp,\sig)$ such that
there are no \STarc{s} where $T \nsse S$ in $\sig(W)$ and where the endpoints in every maximal \Tarc-sequence have equivalent semi-visible positions, see \prref{lem:TTR}. 
Next, we wish to split \Tflat variables off of left- and right-$T$-conic variables. 
The possibility to do so is part of the following lemma.
\begin{lemma}\label{lem:splitflat}
Let $E=(W,B,\cX,\rho,\mu,T)$ be a standard state with an 
\esolu $(\alp,\sig)$ such that there are no \STarcs with $T \nsse S$ in $\sig(W)$.
In addition suppose that both endpoints in every maximal \Tarc-sequence have equivalent semi-visible positions.
Then there is an entire path of \epstras in~$\wh \cU$ from $(E, \alp, \sig)$ to $(E', \alp', \sig')$ for a state $E'=(W',B,\cX',\rho',\mu',T)$ with a reduced \esolu $(\alp',\sig')$ such that the following conditions are satisfied. 
\begin{enumerate}
\item If $\sig'(X)$ contains a minimal (or maximal) $T$-position for some $X \in \cX'$, then $X$ is \Tflat.
\item Both endpoints in every maximal \Tarc-sequence in $\sig'(W')$ have equivalent semi-visible positions.
\item We have $|\cX'\sm \cX|\leq 2|\cX|$ and $|W'|-|W|\leq 2\cdot\sum_{X\in \cX}|W|_X$.
\item  If~$E$ is $k$-small, then every state on the path from~$E$ to~$E'$ is $k$-medium.
\end{enumerate}
\end{lemma}
\begin{proof}
  Suppose that there is a variable $X \in \cX$ such that $x = \sig(X)$ contains a minimal $T$-position but $X$ is not $T$-flat.
  We then factorize $x = zx'$ such that $z$ is the maximal prefix of $x$ consisting solely of $T$-positions.
  Then we follow an \epstra defined by a substitution $X \mapsto ZX$ with a fresh variable $Z$.
  This transition leads to a state $E' = (W', B, \cX', \rho', \mu', T)$ with an \esolu $(\alp', \sig')$ such that $\sig'(X) = x'$ and $\sig'(Z) = z$.
  Observe that $\sig'(X)$ does not contain a minimal $T$-position, and that the fresh variable $Z$ is $T$-flat.
  If necessary we rename the entire state $(E', \alp', \sig')$ as $(E, \alp, \sig)$ and repeat the above process until the first item of the assertions is satisfied.
  This requires at most $\abs{\cX}$ iterations and, therefore, the third item is also satisfied at this point.
  As all fresh variables thereby created are $T$-flat, the fourth item also holds.
  As these substitutions don't interfere with the equivalence of $T$-positions, the second item holds, since it holds at $(E, \alp, \sig)$ by assumption.

  Finally, performing basic $T$-reductions if necessary, we may also assume that $(\alp',\sig')$ is reduced.
  This does not destroy the any of the previously established properties.
\end{proof}
Eventually, we will remove all \Tflat variables using compressions.
For that we need some `progress' 
to guarantee termination. 
This is achieved by popping out a letter out of \Tflat variables. \\
 
\noindent\textsc{\textbf{BeginProcedure}} \textsc{InitSplit}
\begin{enumerate}
\item \textsc{\textbf{ForAll}} \Tflat variables $X\in \cX$ \textsc{\textbf{Do}}

Follow the \epstra defined by $X\mapsto Xa$ where~$a$ is the last letter in $\sig(X)$. 

\noindent\textsc{\textbf{EndForAll}}

\item Perform all basic $T$-reductions and rename the resulting entire state $(E', \alp', \sig')$ as $(E, \alp, \sig)$.
\end{enumerate}
\noindent\textsc{\textbf{EndProcedure}}\\

Observe that popping out a minimal position does not maintain the equivalence relation between $T$-arc sequences, in general.
Nonetheless, the next lemma shows that the effect of the procedure \textsc{InitSplit} is that the endpoints of all maximal \Tarc-sequences have equivalent visible positions.
\begin{lemma}\label{lem:maxvis}
Let $E=(W,B,\cX,\rho,\mu,T)$ be a standard state with an \esolu $(\alp,\sig)$ such that every maximal $T$-arc sequence
$p_1\to p_2 \to \cdots \to p_m$ in $\sig(W)$ the position $p_1$ has an equivalent position which is semi-visible.
Then executing the procedure \textsc{InitSplit}
leads to a state~$E'$ with a reduced entire \esolu $(\alp',\sig')$ such that in every maximal $T$-arc sequence both endpoint have equivalent visible positions.
\end{lemma}
\begin{proof}
We may assume that $p_1\to p_2 \to \cdots \to p_m$ is a maximal $T$-arc sequence in $\sig(W)$ where
 $p_1$ is not semi-visible \wrt~$\sig$ because 
otherwise $p_1$ becomes visible \wrt~$\sig'$.
Let $2\leq k\in \N$ be minimal that \wrt~$\sig$ we have 
a chain $p_1=p'_1\approx p'_{2}\approx\cdots\approx p'_k$ and $p'_k$ semi-visible \wrt~$\sig$. For each $1\leq i <k$ the position 
$p'_i$ is not semi-visible and hence $\sus{p_i}$ is a variable 
$X_i$ where $p'_i$ is not semi-visible in $\sig(X_i)$. 
Now, consider the situation \wrt~$\sig'$. 
First of all $p'_k$ is visible in $\sig'(W)$. 
However, since none of the $p'_i$ was semi-visible \wrt~$\sig$, 
the chain  $p_1=p'_1\approx p'_{2}\approx\cdots\approx p'_k$ is not broken \wrt~$\sig'$. Thus, $p_1$ is equivalent to the visible position 
$p'_k$ in $\sig'(W)$.
\end{proof}

From now on, we maintain the following invariant which is preserved by basic $T$-reductions, popping out $T$-letters from $T$-flat variables, or compression of $T$-arc sequences into single positions.

\begin{invariant}\label{inva:nice}\mbox{}
\begin{enumerate}
\item If $\sig(X)$ has a minimal (or maximal) $T$-position, then $X \in \cXTfl$, i.e., $X$ is \Tflat.
\item Both endpoints of every maximal \Tarc-sequence in $\sig(W)$ have equivalent visible positions. 
  \end{enumerate}
\end{invariant}

\subsection{The alphabet $\Bold$ of `old letters' and uncrossing}\label{sec:unold}
We begin this section with an entire state $(E,\alp,\sig)$ where
$E=(W,B,\cX,\rho,\mu,T)$ is a standard state and $(\alp,\sig)$ is reduced such that there are no \STarcs with $T \nsse S$ in $\sig(W)$.
Moreover, we assume that \prref{inva:nice} holds.
The target\footnote{As the Tipperary Song as early as  1912 predicted: `It's a long way, it's a long way to go.'} at the horizon is to reach an entire state $(E_t, \alp_t, \sig_t)$ with a standard state $E_{t}=(W_{t},B_{t},\cX_{t},\rho_{t},\mu_{t},T)$ and a reduced \esolu $(\alp_{t},\sig_{t})$ satisfying the same properties as stated for $(E,\alp,\sig)$ such that we have
\begin{equation}\label{eq:ET}
\set{X\in \cX_{t}}{X \text{ is \Tflat}}= \es  \quad \text{ and }\quad
  |W_{t}|\leq |W|+ 2\cdot\sum_{X \text{ is \Tflat}} \abs{W}_X
\end{equation}

Moreover, by following an \entipa in $\wh \cU$ we shrink and we expand the lengths of maximal \Tarc-sequences.
We have to reenter the procedure described in this section over and over again and there is no a priori bound depending only on the state~$E$ (and therefore independent of $\sig$) on the number of iterations required.

The main idea is to compress all \Tarc sequences into a single letter.
This is easy once we reach a state~$E$ with reduced \esolu $(\alp,\sig)$
such that all maximal \Tarc sequences are either fully visible or invisible.
Then every maximal \Tarc sequence has an equivalent fully visible \Tarc sequence which can be reduced into a single letter without increasing the length of the equation.
After that we can increase the $T$-component in the state to some $T'$ with $T<T'$. 

At this specific point, after having popped out by \textsc{InitSplit} one $T$-letter on both sides of all \Tflat variables, we define a set of \emph{old} letters $\Bold$ by 
\begin{equation}\label{eq:bold}
  \Bold = B_T \sm \wh A
\end{equation}
We redefine $\Bold$ by \prref{eq:bold} again after each call to the procedure \textsc{InitSplit}. 
We will be able to maintain the invariant that $\Bold\sse B_T$ and 
$\Bold \cap \wh A= \es$. 
In the following subsections $\Bold$ might shrink, but when we redefine $\Bold$ there might be (much) more letters in $B_T$ than before.
However, we will able to guarantee that the corresponding equation $W$ will not get too long.
We will repeat this process over and over again, but eventually our goal is achieved, \ie all \Tflat variables are gone with an equation $W_{t}$ satisfying \prref{eq:ET}.

In each round we intend to compress equivalent subtraces in $\sig(W)$ labeled by a word in $\Bold^{\vphantom{+}}\Bold^+$ into a fresh $T$-letter.
For that we have to split variables and to pop out old letters; and after the splitting we obtain a new equation $W'$ which is typically longer than~$W$.
If the equation is too long, compression will dominate to make the equation shorter again.\footnote{That's the plan. But, unfortunately, making equations shorter did not make the paper shorter.}

Throughout this entire process, we maintain the following invariant for \Tflat variables.
\begin{invariant}\label{inva:Tos}\mbox{}
\begin{enumerate}
\item If $X \in \cXTfl$, then $|\sig(X)|\geq 2$ and every occurrence of $X$ in $W$ is inside 
a factor $B_TXB_T$.
\item If $Xd$, $Xe$ are factors of $W$ with $X \in \cXTfl$ and $d,e\in \Bold$, then $d=e$. 
\item If $XuY$ is a factor of $W$ with $X,Y \in \cXTfl$ and $u \in \Bold^\ast$, then $\abs{u} \geq 2$.
\end{enumerate}
\end{invariant}

The first and last item hold because we perform \textsc{InitSplit} with an 
output such that every \Tflat variable $X$ occurs in the equation inside a factor
$B_TXB_T$ and with $|\sig(X)|\geq 2$. After that we define $\Bold$ and in the beginning we have $B_T=\Bold$. 
Later on we pop out letters from $\Bold$ only, unless we execute \textsc{InitSplit} again, which redefines the set of `old letters' $\Bold$.
At some stage all $T$-letters become old in this sense. 

\subsection{Uncrossing and compression of short factors in $\Bold^+$}\label{sec:ucr}
The input for the following procedure \textsc{Uncross}($w$)
is a state $E=(W,B,\cX,\rho,\mu,\theta,T)$ where $\Bold \sse B_T$ is defined with a reduced \esolu $(\alp,\sig)$ and a word $w\in \Bold^+$. 
The intention is to enable the possibility to compress every (short) factor~$w$ in $\sig(W)$ (and simultaneously every factor $\ov{w}$) into a single letter. 
Whether or not such a compression is possible
depends on~$w$ and $\sig(W)$. 
In the applications we uncross a word only if a compression is possible.
The following notion is useful in ensuring this.
\begin{definition}\label{def:nsovl}
  Let $u, w \in M(B, \rho)$.
  We say that $u$ has \emph{no non-trivial \index*{overlap}s} in~$w$ if all subtraces $v_1,v_2$ of~$w$ labeled by $\lambda(v_1), \lambda(v_2) \in \{u, \bar u\}$ have equal or disjoint sets of positions, i.e., $\pos(v_1) = \pos(v_2)$ or $\pos(v_1) \cap \pos(v_2) = \emptyset$.
\end{definition}
We assume throughout that \prref{inva:Tos} holds. 
The following procedure is applied and defined to uncross words crossing \Tflat variables.
\\
\noindent\textsc{\textbf{BeginProcedure}} \textsc{Uncross}($w$)\\
\noindent\textsc{\textbf{ForAll}} \Tflat variables $X\in \cX$ \textsc{\textbf{Do}}
\begin{enumerate}
  \item If $\sig(X)\in B_T^* u$ and $W$ has a factor $Xu'$ such that $uu' \in \os{w, \bar w}$ and $|u|, |u'| \geq 1$, then follow the \epstra defined by $X\mapsto Xu$.
\item Perform all basic $T$-reductions and rename the resulting state $E''$ and its reduced \esolu $(\alp'',\sig'')$ as $E=(W,B,\cX,\rho,\mu,T)$ and $(\alp, \sig)$.
\end{enumerate}
\noindent\textsc{\textbf{EndForAll}}\\
\noindent\textsc{\textbf{EndProcedure}}
\begin{lemma}\label{lem:stfl}
Let $E=(W,B,\cX,\rho,\mu,T)$ with $(\alp,\sig)$ as its \esolu and suppose that~$E$ and $(\alp, \sig)$ satisfy \prref{inva:Tos}.
  Furthermore, let $w \in \Bold^+$ with $2 \leq \abs{w} \leq 3$   
such that~$w$ has no non-trivial overlaps in $\sigma(W)$.
  Then the state $E'=(W',B,\cX',\rho',\mu',T)$ and its \esolu $(\alp',\sig')$ after the procedure \textsc{Uncross}($w$) have the following properties.
  \begin{enumerate}
    \item \prref{inva:Tos} holds for the entire state $(E',\alp',\sig')$ and $(\alp', \sig')$ is reduced.
    \item All subtraces of $\sig'(W')$ labeled by~$w$ or $\bar w$ are either fully visible or fully invisible.
    \item We have $\cX' \sse \cX$, $\cXpTfl \sse \cXTfl$, and 
\begin{equation}\label{eq:lunc}
  \abs{W'} - \abs{W} \leq 2 \cdot (\abs{w} - 1) \cdot \sum_{X \in \cXTfl} \abs{W}_X \leq 4 \cdot \sum_{X \in \cXTfl} \abs{W}_X
\end{equation}
  \end{enumerate}
\end{lemma}

\begin{proof}
  \prref{inva:Tos} cannot be destroyed by popping out letters (whether old or new), which includes removing dummies.
  Hence, the invariant is maintained by \textsc{Uncross}($w$).

  For the second item, recall that each variable $X \in \cX$ satisfies $\abs{\sig(X)} \geq 2$ and that the length of relevant gaps between variables in $W$ is at least two by \prref{inva:Tos}.
  As such, any subtrace $v$ of $\sig(W)$ labeled by~$w$ or $\bar w$ crosses at most one variable and only in either an initial or terminal segment of $v$.
  As such, if $v$ crosses a variable $X$, then $v$ is uncrossed when either $X$ or $\bar X$ is considered.
  Hence, if $v$ has a visible position, then the corresponding subtrace $v'$ of $\sig'(W')$ will be fully visible.
  Since~$w$ has no non-trivial overlaps in $\sig(W)$, if $v$ is fully invisible in $\sig(W)$, then the same is true for $v'$ in $\sig'(W')$. We obtain Inequality~\eqref{eq:lunc} by noting that at most $2(\abs{w} - 1)$ letters are popped out of each variable $X \in \cXTfl$. 
  Finally, the inclusions $\cX' \sse \cX$ and $\cXpTfl \sse \cXTfl$ are trivial.
\end{proof}
The second item of \prref{lem:stfl} guarantees that every subtrace in $\sig(W)$ labeled by~$w$ or $\bar w$ is fully visible or fully invisible. 
Therefore a compression of all factors of $\sig(W)$ labeled by~$w$ or $\bar w$ as in the following procedure is indeed possible (this statement includes that possibly there are none of them).
Note that, for example, the words $ab$ and $ba$ might have no non-trivial overlap in 
$\sig(W)$ although they have an obvious non-trivial overlap in 
the word $aba$.
\\

\noindent\textsc{\textbf{BeginProcedure}} \textsc{Compress($w$)}\\
\indent\textsc{\textbf{Condition:}} $w\in \Bold^+$, $2 \leq |w| \leq 3$, and~$w$ has no non-trivial overlaps in $\sigma(W)$.
\begin{enumerate}      
  \item \textsc{\textbf{Call}} \textsc{Uncross}($w$). (Recall that \textsc{Uncross}($w$) uncrosses~$w$ and $\bar w$.)
  \item Define a fresh clone $c_w\in \wh C\sm B$ of~$w$ with $\ov{c_w} = c_w$ if $\bar w = w$ and replace every occurrence of~$w$ in $\sigma(W)$ by $c_w$ and every occurrence of $\bar w$ in $\sigma(W)$ by $\ov{c_w}$.
    Follow the \comp \tra defined by $h(c_w) = w$.
  \item Perform all basic $T$-reductions and rename the resulting state $E'$ and its reduced \esolu $(\alp',\sig')$ as $E=(W,B,\cX,\rho,\mu,T)$ and $(\alp, \sig)$.
\end{enumerate}
\noindent\textsc{\textbf{EndProcedure}}\\

Let us explain why we restrict the compression of 
words which use an alphabet without any letter in $\wh A$. Imagine we are at a state 
$E$ with a reduced \esolu $(\alp,\sig)$ and ran the compression  for 
$w=aa$ with  $a \in \wh A$. Then 
$aa$ is fully invisible in $\sig(W)$. 
Compression would first introduce a fresh letter $c_w$, but the letter is fully invisible, so an alphabet reduction leads us back
to the same state~$E$ without changing the \morph $\alp\sig$.

\begin{lemma}\label{lem:stfl2}
  Let $E=(W,B,\cX,\rho,\mu,T)$ with $(\alp,\sig)$ as its \esolu and suppose that~$E$ and $(\alp, \sig)$ satisfy \prref{inva:Tos}.
  Furthermore, let $w \in \Bold^+$ with $2 \leq \abs{w} \leq 3$ such that~$w$ has no non-trivial overlaps in $\sigma(W)$.
  Then after the procedure \textsc{Compress}($w$) the state $E'=(W',B',\cX',\rho',\mu',T)$ and its \esolu $(\alp',\sig')$  have the following properties.
\begin{enumerate}
\item \prref{inva:Tos} holds for $E'$ and $(\alp',\sig')$.
\item There are no subtraces of $\sig'(W')$ labeled by~$w$ or $\bar w$.
\item The third item in \prref{lem:stfl2} still holds:
we have $\cX' \sse \cX$, $\cXpTfl \sse \cXTfl$, and 
\begin{equation}\label{eq:lcom}
  \abs{W'} - \abs{W} \leq 2 \cdot (\abs{w} - 1) \cdot\hspace{-.3cm} \sum_{X \in \cXTfl} \abs{W}_X \leq 4 \cdot\hspace{-.3cm} \sum_{X \in \cXTfl} \abs{W}_X
\end{equation}
\end{enumerate}
\end{lemma}
\begin{proof}
  By the properties of \textsc{Uncross}($w$) given in \prref{lem:stfl}, we maintain \prref{inva:Tos} in the first step of \textsc{Compress}($w$).
  As the fresh letter $c_w$ introduced is not contained in $\Bold$, the second step also maintains \prref{inva:Tos}; as it does the third step of the procedure.
  The second and third assertion of the lemma are obvious.
\end{proof}

\subsection{Compression of $ab\bar a$ for $a\neq \bar a\in \Bold$ and $b=\bar b\in \os{1} \cup\Bold$}\label{sec:abovaf}
We consider a standard state $E=(W,B,\cX,\rho,\mu,T)$ with a reduced \esolu $(\alp,\sig)$; and we run the following procedure, directly after we finished
\textsc{InitSplit}. Hence, \prref{inva:Tos} holds when we call 
\textsc{$aS\bar a$-Compression}. 
\\

\noindent\textsc{\textbf{BeginProcedure}} \textsc{$aS\bar a$-Compression}\\
\mbox{}\quad\textsc{\textbf{ForAll}}  $a\in \Bold$ with $\bar a\neq a$ in any order
\textsc{\textbf{Do}}\\
\mbox{}\quad\quad\textsc{\textbf{ForAll}} $b = \bar b \in \os{1}\cup\Bold$  in any order 
\textsc{\textbf{Do}}\\
\mbox{}\quad\qquad \textsc{\textbf{Call}} \textsc{Compress($ab\bar a$)}\\
\mbox{}\quad\quad\textsc{\textbf{EndForAll}}\\
\mbox{}\quad\textsc{\textbf{EndForAll}}\\
\noindent\textsc{\textbf{EndProcedure}}\\

Let us argue that every variable is affected (by uncrossing) at most twice (per side). Consider the outcome of the procedure \textsc{Compress($ab\bar a$)}. 
If $\sig(X)$ ends in~$a$ (resp.~$ab$) for some 
variable $X$, then uncrossing is not necessary unless
$W$ has a factor $X\bar a$ (resp.~$Xb\bar a$). 
In both cases, if $Xc$ is a factor of $W$ after \textsc{Compress($ab\bar a$)}, then either $c$ is fresh and not in $\Bold$ or we have $c = a$.
During further calls to \textsc{Compress($a b' \bar a$)} in the inner loop, where~$a$ is fixed, there is no further uncrossing. 
The only time we can touch the variable~$X$ again is when, for some $b'=\ov{b'}\in \os{1} \cup\Bold$, we call \textsc{Compress($\bar a b' a$)} and $\sig(X)$ ends in $\bar ab'$. 
Then, an uncrossing of $\bar ab' a$ involving a variable~$X$ leads, after the compression of $\bar ab' a$ into a fresh letter, to some $W'$ such that every occurrence of~$X$ in $W'$ has either the letter~$\bar a$ as a right neighbor or some fresh letter. 
As the outer loop has already completed the iteration for~$a$ (since it is now at $\bar a$ and $\bar a \neq a$), no further uncrossing can affect $X$ on the right.
This fact is used in the proof of the next lemma.
\begin{lemma}\label{lem:abovacomp}
Let $E=(W,B,\cX,\rho,\mu,T)$ and $(\alp,\sig)$ its reduced \esolu 
before and 
$E'=(W',B',\cX',\rho',\mu',\theta')$ be the state and $(\alp',\sig')$ its \esolu directly after the run of \textsc{$aS\bar a$-Compression}.
Then the following properties holds.
\begin{enumerate}
\item For all $a\neq \bar a\in \Bold$ the \esolu is reduced and 
$\sig'(W')$ is without any factor $ab\bar a$ 
where $b=\bar b\in \os 1 \cup \Bold$.
\item If for $X\in \cXTfl'$ there are factors $Xd$ and $Xe$ in $W'$ with $d,e\in \Bold$, then $d=e$. 
\item We have $\cX' \sse \cX$, $\cXTfl' \sse \cXTfl$, and 
$ 
\abs{W'} - \abs{W} \leq 8 \cdot \sum_{X \in \cXTfl} \abs{W}_X
$.
\end{enumerate}
\end{lemma}
\begin{proof}
The first item follows because $ab\bar a= \ov{ab\bar a}$ has no non-trivial overlap with itself. 
The second item is also clear because it was correct before we called \textsc{$aS\bar a$-Compression}.
The third item follows because uncrossing pops out at most four letters (on each side) as we argued just above. 
The assertions $\cX'\sse \cX$ and $\cXTfl' \sse \cXTfl$ are trivial. 
\end{proof}
As a result of the procedure the following \prref{inva:hugo} holds; and we keep it for all states~$E$ and \esolu{s} $(\alp,\sig)$. 
until either there are no \TTarcs in $\sig(W)$ anymore or \prref{sec:unold} is visited again, where $\Bold$ is redefined as the current set~$B_T$.
\begin{invariant}\label{inva:hugo}{\phantom x}
Let $E=(W,B,\cX,\rho,\mu,\theta, T)$ be a state with an \solu $\sig$.
\begin{enumerate}
\item 
If $a,b\in \Bold$ with $a\neq \bar a$ and $b=\bar b$, 
then $\sig(W)$ is without factors $a\bar a$ or $ab\bar a$.
\item If for $X\in \cXTfl$ there are factors $Xd$ and $Xe$ in $W$ with $d,e\in \Bold$, then $d=e$. 
\item We have $|\sig(X)|\geq 2$ for all $X\in \cX$.
\item If $XuY$ is a factor of $W$ with $X,Y \in \cXTfl$ and $u \in \Bold^\ast$, then $\abs{u} \geq 2$.
\end{enumerate}
\end{invariant}
\subsection{Block compression}\label{sec:blocked}
Let $E=(W,B,\cX,\rho,\mu,T)$ be a standard state
with a reduced \esolu $(\alp,\sig)$ with a notion of `old letters': $\Bold \sse B_T =\set{a\in B}{\rho(a)=T}$ according to \prref{eq:bold}.

We assume that \prref{inva:nice} and \prref{inva:hugo} hold. \Ip $\sig(W)$ is without any factor 
$ab\bar a$ for all $a\neq \bar a\in \Bold$ and $b=\bar  b\in \os 1\cup \Bold$.

\subsubsection{(Maximal) $(a,b,\ell)$-blocks}\label{sec:maxblock}

The goal of 
\prref{sec:blocked} is to compress all factors $ab\bar a$ for all $a\neq \bar a\in \Bold$ and where $b=1$ or $b=\bar  b\in \Bold$. However, these may appear inside arbitrarily long factors of $\sig(W)$ labeled in $aaa^\ast$ or $aba(ba)^\ast$, respectively, which we compress first.\footnote{The compression yields an \edtol representation for the full \solu set for systems of linear Diophantine equations. No surprise:
these are systems of word equations over a unary alphabet.} 

This motivates the following definition.

\begin{definition}\label{def:Tblock}
  Let $a \in B_T\sm \wh A$ and $b=\bar b \in \os 1\cup B_T\sm \wh A$ with $a \neq b$.
  A word $w \in (B_T\sm \wh A)^+$ of the form $w = aba\, (ba)^\ell$ with $\ell \in \N$ is called an \emph{$(a,b,\ell)$-block}\index{a,b,l-block@$(a,b,\ell)$-block}.\footnote{The notation $(a,b,\ell)$ can be be viewed a  \emphind{run length encoding} of the word $aba\, (ba)^\ell$.} The term $(a,b,\ell)$-block refers to a subtrace in $\sig(W)$ given by 
its set of positions; and it is also used to denote its label in $(B_T\sm \wh A)^+$.

An $(a,b,\ell)$-block $v$ in $\sig(W)$ is called \emph{maximal\index{maximal block}}
if there is no $\ell'>\ell$ such that $v$ (which is defined by its set of positions) is a subtrace of
an $(a,b,\ell')$-block in $\sig(W)$. 
\end{definition}
\begin{remark}\label{rem:nover}
Typically $\sig(W)$ contains many maximal $(a,b,\ell)$-blocks 
for various~$a$'s, $b$'s, and~$\ell$'s.
By definition an $(a,b,\ell)$-block is self-involuting if and only if $a = \bar a$.
The reason for considering maximal $(a,b,\ell)$-blocks is that if $v_i$ is a maximal $(a,b,\ell_i)$-block in $\sig(W)$ for $i = 1,2$, then either $\pos(v_1) = \pos(v_2)$ (in which case $\ell_1 = \ell_2$) or $\pos(v_1) \cap \pos(v_2) = \emptyset$.
Moreover, we restrict the notion of $(a,b,\ell)$-block to words without any letter in $\wh A$, and we work with reduced \solu{s}.
Thus, every maximal $(a,b,\ell)$-block has an equivalent (but not necessarily maximal) block with a visible position.
Because maximal $(a,b,\ell)$-blocks are without any non-trivial overlap, this implies that the total number of equivalence classes (generated by) maximal $(a,b,\ell)$-blocks can be bounded above in terms of the equation $W$.
\hspace*{\fill}$\diamond$\end{remark}

\subsubsection{The $T$-Block-Compression procedure}\label{sec:blopro}

We call \textsc{$T$-BlockCompression} below on input $E=(W,B,\cX,\rho,\mu,T)$ and its \esolu $(\alp,\sig)$. 
A block compression over alphabets without selfies has been has been considered in \cite{CiobanuDiekertElder2016ijac,Jez16jacm_stacs}.
Without selfies it is less technical because $(a,b,\ell)$-blocks and $(\bar a,b,\ell')$-blocks never overlap. 
The construction here is a simplification of the  method in \cite{DiekertJP16}. 
Still, the \textsc{$T$-BlockCompression} procedure is the most technical
part of our paper.
It involves several steps including typing and cloning of letters which have to be preformed in a carefully chosen order.
Even in the simplest case where typing is not necessary, we might need a huge number of rounds  before a single maximal $(a,b,\ell)$-block is compressed into a fresh letter.
Such a case is worked out in \prref{ex:cloony} which is also quite involved.
So, the reader might skip it in a first reading.

\begin{example}\label{ex:cloony}
Consider a situation where we have a single 
\inner equation as in (\ref{eq:cdcbc}) where $X$ is the only variable and $b,d$ are selfies with $a\neq b<d\neq a$ and where $\leq$ is a liner order on the set of selfies and where $2\leq  \ell$ is even. 
\begin{align}\label{eq:cdcbc}
Xd\,Xd\, (aba)\,(ba)^\ell&=a\ d\,a\, d\, (aba)\,(ba)^\ell
\\
\label{eq:cdc_b}
X\,d\,Xd\, (c_{\kap})\,(ba)^\ell&=ada\, d\, (c_{\kap})\,(ba)^\ell
&h(c_\kap)= aba
\\
\label{eq:c2a}
X d\, Xd\, (c_{\kap})\,(bc)^{\ell}&=ada\,d\, (c_{\kap})\,(bc)^{\ell} 
&c\text{-cloning: } h(c)=a
\\
\label{eq:cal}
Xd\,Xd\,  (c_{\kap})\,(bc)^{\frac{\ell}{2}}&=ada\, d\, (c_{\kap})\,(bc)^{\frac{\ell}{2}}
&c\text{-halfing: } h(c)= cbc
\\
\label{eq:leven}
Xd\,Xd\,  (c_{\kap})\,(bc)^{\frac{\ell}{2}-1}&=ada\, d\, (c_{\kap})\,(bc)^{\frac{\ell}{2}-1}
&\text{Eventually $\ell/2$ becomes odd}
\\
\label{eq:lend}
Xd\,Xd\,  (c_{\kap})&=ada\, d\, (c_{\kap})
&\big((\ref{eq:cal})^*\cdot(\ref{eq:leven})\big)^* \text{ removes $c$}
\\
\label{eq:c_d}
ada \,d\, (c_{\kap})\,(ba)^\ell&=ada\, d\, (c_{\kap})\,(ba)^\ell
&\sig(X)=a
\\
\label{eq:dlam}
(d_{\lam})\, d\, (c_{\kap})&=(d_{\lam})\, d\, (c_{\kap})
&h(d_\lam)= ada
\end{align}
There is no other \solu than $\sig(X)=a$. Every position in 
(\ref{eq:cdcbc}) is visible, and on each side of the equation 
we intend to compress first one factor $aba$ into fresh constants~$c_{\kap}$. Since $b<d$, there is no uncrossing to obtain the shorter \inner 
equation as in (\ref{eq:cdc_b}). 
In order to compress the remaining maximal $(a,b,\ell)$-block $a(ba)^\ell$,
we relabel of all $a$-positions in every $(a,b,\ell)$-block by fresh clone~$c$ of~$a$. 
The reason for cloning is to 
use a \morph $h$ defined by $h(c)=cbc$ as the label a \comp \tra.
If $U=V$ is the \inner equation in (\ref{eq:c2a}), then we have $U=h(U')$ and $V=h(V')$
for the \inner equation $U'=V'$ in (\ref{eq:cal}). For that 
it is crucial that $c$ in (\ref{eq:c2a}) is different from the letter~$a$. Otherwise we could not use this method: uncrossing before cloning is crucial! Note also that the length of $UV$  
is almost twice as long as the length of $U'V'$ which is giant step
for compressing.
Finally suppose that $\ell/2$ is odd, then we can use 
a \morph $h$ defined by $h(c_{\kap})=c_{\kap}ba$
to obtain the \inner equation in (\ref{eq:leven}). Alternating between 
steps leading to equations in  (\ref{eq:cal}) and (\ref{eq:leven}), we eventually achieved the goal which is 
the \inner equation in (\ref{eq:lend}). The letter~$c$ is gone: its lifetime is temporary during the process of compressing maximal $(c,b,\ell)$-blocks. 
After that $ada$ is uncrossed and compressed 
into a fresh constant~$d_{\lam}$. This leads to 
the equation without variables as in (\ref{eq:c_d}). 
\end{example}

\subsubsection{Preparing the block compression}\label{sec:bloprp}

To prepare for block-compression, we apply the following procedure to a standard state 
with a reduced \esolu. 
Its execution leads to a new state, and this state will then be a typed state, only if $\sig(W)$ contains an $(a,b,\ell)$-block.\\

\noindent\textsc{\textbf{BeginProcedure}} 
\textsc{PrepareBlocks($aba$)}\\
\mbox{}\quad\textbf{If} $\sig(W)$ contains an $(a,b, \ell)$-block \textbf{then}\\
\mbox{}\quad\quad\textsc{\textbf{Call}} \textsc{UncrossBlocks($aba$)}\\
\mbox{}\quad\quad\textsc{\textbf{Call}} \textsc{TypeBlocks($aba$)}\\
\mbox{}\quad\textbf{EndIf}\\
\textsc{\textbf{EndProcedure}}\\

The following procedure uncrosses all maximal $(a,b,\ell)$-blocks in the sense that, after the call, every visible position of such a block resides inside a visible $aba$ factor.\\

\noindent\textsc{\textbf{BeginProcedure}} 
\textsc{UncrossBlocks}($aba$)\\
\indent\textsc{\textbf{Condition:}} 
We have $a \in \Bold$ and $b = \bar b \in \os{1} \cup \Bold$ with $a \neq b$ and the \esolu is reduced according to \prref{def:stawrtsig}. 
Moreover,~$E$ is $k$-medium and ignoring all \Tflat variables would make the state $k$-small. 
\begin{enumerate}
\item 
For each \Tflat variable $X$ and $\wt a \in \os{a, \bar a}$ such that $\sig(X) \in B_T^\ast \wt a$ and $Xb\wt a$ is a factor of $W$ (or  $\sig(X) \in B_T^\ast \wt a b$ and $X\wt a$ is a factor of $W$) follow an \epstra leading to a \solu $\sig'$ which is defined by 
$X \mapsto Xu$ where $u$ is the longest common suffix of $\sig(X)$ and of $b \wt a b \wt a$ (or $b \wt a b \wt a b$, resp.).
If $\abs{\sig'(X)} \leq 1$, then follow an \epstra defined by $X \mapsto \sig'(X)$.\\[1ex]
    (At this point, every maximal $(\wt a,b, \ell)$-block of $\sig(W)$ with $\wt a \in \os{a, \bar a}$ is either fully invisible or has a visible subtrace labeled by $\wt ab \wt a$.)
\item Perform all basic $T$-reductions. 
Rename the result as a state 
$E=(W,B,\cX,\rho,\mu,T)$ with an \esolu $(\alp, \sig)$.
 
\end{enumerate}
\textsc{\textbf{EndProcedure}}

\begin{lemma}\label{lem:preaba}
Let $E = (W, B, \cX, \rho, \mu, T)$ be the state and $(\alp, \sig)$ its \esolu when we call the procedure \textsc{UncrossBlocks}($aba$),  and 
$(E',\alp',\sig')$ be the entire state directly after the run where
$E'=(W',B',\cX',\rho',\mu',T)$. Suppose that \prref{inva:hugo} holds for the entire state $(E,\alp, \sig)$.
Then the following holds. 
\begin{enumerate}
  \item \prref{inva:hugo} holds for $E'$ and $(\alp', \sig')$.
  \item Every maximal $(\wt a, b, \ell)$-block in $\sig'(W')$ with $\wt a \in \os{a, \bar a}$ is either invisible or it has a visible factor $\wt a b \wt a$.
  In either case there exists an equivalent maximal $(a,b,\ell)$-block or $(\bar a, b, \ell)$-block where a factor $a b a$ or $\bar a b \bar a$ is visible, respectively.
  \item We have $\cX' \sse \cX$, $\cXTfl' \sse \cXTfl$, and
  \begin{align}\label{eq:precbc}
    \abs{W'}-\abs{W}\leq 2|babab| \,\cdot\hspace{-0.3cm} \sum_{X \in \cXTfl} \abs{W}_X
    \leq 10 \,\cdot \hspace{-0.3cm} \sum_{X \in \cXTfl} \abs{W}_X
  \end{align}
\end{enumerate}
\end{lemma}
\begin{proof}
  \prref{inva:hugo} cannot be destroyed by popping out letters (whether old or new) or by basic $T$-reductions. Hence, the invariant is maintained.
The second item is a consequence of the splitting $X\mapsto Xu$ and the 
definition of $u$ which uncrosses some factor $\wt ab\wt a$ of every maximal $(\wt a,b,\ell)$-block. 
More precisely, let us consider a maximal $(\wt a, b, \ell)$-block $v$ in $\sig(W)$.
  If $v$ has a visible factor $b$, then at least one of the adjacent positions labeled $\wt a$ is visible as well due to the fourth item in \prref{inva:hugo}.
  The other position labeled $\wt a$ is either visible (in which case we are done), or becomes visible in $\sig'(W')$ due to uncrossing.
  Similarly, if $v$ contains a visible factor $\wt a$, then $v$ contains an adjacent factor $\wt a b$ to the left of $\wt a$ or a factor $b \wt a$ to its right.
  We can assume that the corresponding position labeled $b$ is invisible for otherwise we are in the situation we just dealt with. 
  By the third item of \prref{inva:hugo}, the entire adjacent factor $\wt a b$ or $b \wt a$ is invisible.
  However, it becomes visible in $\sig'(W')$ due to uncrossing.
  Next, suppose that every position in $v$ is invisible but one of the endpoints of $v$ is semi-visible which, by symmetry, we assume to be the rightmost position.
  Let $X \in \cXTfl$ be the variable labeling $\sus{v}$.
  Even though $v$ is not part of the solution a factor $Xb\wt a$ of $W$ since $v$ is a maximal $(\wt a, b, \ell)$-block, the equation $W$ might still contain such a factor.
  In that case our uncrossing is designed in such way that the suffix of $v$ labeled $\wt a b \wt a$ becomes visible in $\sig'(W')$ as well. (This is why we pop out the largest common suffix of $\sig(X)$ and $b\wt a b \wt a$ instead of $b \wt a$.)
  A similar situation arises when no position of $v$ visible but when $v$ is contained in a factor $bv$ or $vb$ of $\sig(W)$ with the position labeled $b$ being semi-visible. 
  Once again our uncrossing either keeps $v$ invisible or makes a prefix or suffix of $v$ labeled $\wt a b \wt a$ visible.
  This shows that every maximal $(\wt a, b, \ell)$-block in $\sig'(W')$ is either invisible or has visible factor $\wt a b \wt a$.
  It follows readily that every maximal $(\wt a, b, \ell)$-block is equivalent in $\sig'(W')$ to an $(a,b,\ell)$-block or $(\bar a, b, \ell)$-block with a visible $aba$ or $\bar a b \bar a$ factor, since every position of such a block is labeled by a letter in $\Bold$ and, as $\Bold\sm\wh A = \es$, is therefore equivalent in $\sig(W)$ to a visible position.
  The fact there is such an equivalent block in $\sig'(W')$ which is also a maximal follows from a case distinction similar to the above.
  We leave the details to the reader.

To see the third item, note that the maximal length of $u$ is given by $\abs{babab}$.
Since we also split $\bar X$ we have to count this length twice.
The assertions $\cX'\sse \cX$ and $\cXTfl' \sse \cXTfl$ are trivial.
\end{proof}
The following procedure is called directly after \textsc{UncrossBlocks($aba$)}.
It leads to a typed state in which block compression can be performed.\\

\noindent\textsc{\textbf{BeginProcedure}} 
\textsc{TypeBlocks}($aba$)
\begin{enumerate}
\item Create a fresh clone $c$ for the old letter~$a$; and replace~$a$ (and $\bar {a}$) by $c$ (and $\bar c$, resp.) in every maximal $(\wt a, b, \ell)$-block of $\sig(W)$.
Realize this modification by following a \comp \tra $E\arc h E'$ where $h$ is defined by $h(c)=a$. 
It leads to a state $E'=(W',B',\cX',\rho',\mu',T)$ with an \esolu $(\alp,\sig')$ such that $W=h(W')$.\\[1ex] 
(Note that the fresh letter~$c$ is visible in the equation $W'$.)
\end{enumerate}
\quad\textsc{\textbf{ForAll}} $\wt c\in \os{c,\bar c}$ and $X\in\cXTfl$ where $\sig'(X) \in B^{\prime\ast}_T\wt c$ and $W'$ contains $Xb\wt c$ \textsc{\textbf{Do}}
\begin{enumerate}\setcounter{enumi}{1}
\item Create fresh variables $X_{cbc}$ and $\ov{X_{cbc}}$.
Define $u$ to be the maximal suffix of $\sig'(X)$ with $u \in \wt c (b \wt c)^\ast$.
If $\wt c=c$, then follow an \epstra $X\mapsto XX_{cbc}$ with $\sig'(X_{cbc}) = u$.
Otherwise follow an \epstra $X\mapsto X\ov{X_{cbc}}$ with $\sig'(X_{cbc}) =\bar u$.
Then rename the resulting state as $E' = (W', B', \cX', \rho', \mu', T)$ and its \esolu as $(\alp, \sig')$.
\\[1ex]
(When considering $\bar X$ we may create additional fresh variables. 
As such, we may then have up to four \emph{different} fresh variables: $X_{c b c}$, $\ov{X_{c b c}}$, $\ov{X}_{c b c}$, and $\ov{\ov{X}_{c b c}}$.)
\end{enumerate}
\quad\textsc{\textbf{EndForAll}}
\begin{enumerate}\setcounter{enumi}{2}

\item Define a set of positions $K\sse \set{\kap\in \pos(\sig'(W'))}{\kap \text{ is visible}}$ such that every 
  maximal $(c,b,\ell)$-block with a visible position has at most one position in $K$, but for every such block there is an equivalent block with a visible position in $K$.\\[1ex]
(Thus, there is a natural bijection between $K$ and the set of 
equivalence classes of maximal $(c,b,\ell)$-blocks with a visible position. Note that for every $(c,b,\ell)$-block with a visible position there is dual $(\bar c,b,\ell)$-block where the 
dual position is visible. Thus, it is enough to choose the 
visible positions in $K$ inside $(c,b,\ell)$-blocks.
) 
\item 
Define fresh letters $c_{\kap}$ and $\ov{c_{\kap}}$ for all $\kap \in K$ with 
$\rho''(c_{\kap})=T$ and $\mu''(c_{\kap})=\mu(cbc)$.
Let $B'' = B' \cup \set{c_{\kap},\,\ov{c_{\kap}}}{\kap \in K}$.
Call the variables $\wt{X_{cbc}} \in \cX' \sm \cX$ as well as the letters $\wt{c_\kap} \in B'' \sm B'$ \emph{typed} and introduce the following set $\theta$ of defining equations.
\begin{align}
\label{eq:akap}
\begin{split}
\theta=&\,\set{c\,b\, c_{\kap} = c_{\kap}\,b\,c}{c_{\kap} \text{ is typed}} \\
\cup &\, \set{c_{\kap}\, b \, X_{cbc} =X_{cbc}\,  b\, c_{\kap}}{X_{cbc}\text{ and $c_{\kap}$ are typed}}\\
\cup &\, \set{c\,b \, X_{cbc} =X_{cbc}\,b\,c}{X_{cbc}\text{ is typed}}
\end{split} 
\end{align}
\item
For each maximal $(c,b,\ell)$-block in $\sig'(W')$ (whether or not it has a visible position) choose exactly one subtrace labeled by the word $cbc$.
If a position of the block is visible, then we require to choose the subtrace such that all positions of $cbc$ are visible. 
After the choice of all these factors $cbc$ we compress the chosen subtrace $cbc$ into a single position labeled by the clone $c_{\kap}$ of $cbc$.
Define the typed state $E'' = (W'', B'', \cX', \rho'', \mu'', \theta, T)$ where $W''$ is the resulting equation and let $\sig''$ be its resulting solution.\\[1ex]
(Recall that $c_{\kap}$ floats freely inside its maximal
$(c, b,\ell)$-blocks thanks to (\ref{eq:akap}). 
This is why $\sig''$ is a \solu at the state \emph{with type $\theta$}.
Without restriction, each $\kap$ survives the compression as a visible position in
$\sig''(W'')$ such that $\kap$ has the label $c_{\kap}$.) 
\item
Realize the above modifications by following a sequence of \comp \tras $E'\arc h E''$ with the $(B''\cup \cX')$-\morph $h:M(B''\cup \cX',\rho'',\mu'',\theta,T) \to M(B''\cup \cX',\rho'',T)$ defined by $h(c_\kap)= cbc$.
Moreover, as $W' = h(W'')$ and $\sig' = h \sig''$, we obtain an \esolu $(\alp,\sig'')$ at the typed state $E''$.
\end{enumerate}
\noindent\textsc{\textbf{EndProcedure}}

\subsubsection{Executing the block compression}\label{sec:bloprf}

\noindent\textsc{\textbf{BeginProcedure}} \textsc{CompressBlocks($aba$)}\\
\noindent\textsc{\textbf{Call}} \textsc{PrepareBlocks($aba$)} \\
(The resulting state $E=(W,B',\cX',\rho,\mu,\theta,T)$ is either typed, with $\theta$ being a $(c,b,\Delta)$-type for some fresh clone $c$ of the old letter~$a$, or a standard state without $(a,b, \ell)$-blocks.)\\
\noindent\textsc{\textbf{While}} the state~$E$ is typed 
\textsc{\textbf{Do}}
\begin{enumerate}
\item 
  \noindent\textsc{\textbf{ForAll}} typed variables $X_{c b c}$ such that $|\sig(X_{cbc})|_{c}$ is odd 
\textsc{\textbf{Do}}
\begin{itemize}
\item 
Follow an \epstra 
defined by
$X_{c b c}\mapsto X_{c b c}\, b c$  if  $|\sig(X)|_{c}\geq 3$ and
$X_{c b c}\mapsto c$  if  $|\sig(X)|_{c} = 1$ such that we obtain a \solu $\sig'$ where $|\sig'(X_{c b c})|_{c}$ is even. 
\item Rename the new state and \solu as $E=(W,B,\cX,\rho,\mu,\theta,T)$ and $(\alp,\sig)$.
\end{itemize}
\noindent\textsc{\textbf{EndForAll}} 
(The $c$-length $|\sig(X_{cbc})|_{c}$ is even for all typed variables $X_{cbc}$.)
\item \textsc{\textbf{ForAll}} $\kap\in K$ where $\abs{u}_c$ is odd for a maximal factor $u\in c_{\kap}(bc)^\ast$
  in $\sig(W)$  
\textsc{\textbf{Do}}\\
(This implies that $\sig(W)$ has a factor $c_{\kap} b u'$ where $c_{\kap}$ is the unique letter with this name in this factor. 
Moreover, since $\abs{\sig(X_{cbc})}_c$ is even for all typed variables $X_{cbc}$, the factor $c_{\kap} bc$ is visible in $W$, thanks to the floating of $c_{\kap}$.)
\begin{itemize}
\item Follow a \comp \tra labeled by the \morph $g$ which is defined by
$g(c_{\kap})=c_{\kap} bc$
leading to a state 
$E'=(W',B,\cX,\rho',\mu',\theta',T)$ and \esolu $(\alp,\sig')$ such that
$g(W')=W$ and $\sig=g\sig'$. 
\item Rename the new state and \solu as $E=(W,B,\cX,\rho,\mu,\theta,T)$ and $(\alp,\sig)$.
\end{itemize}
\textsc{\textbf{EndForAll}}
(The $c$-length $|u|_c$ us even for all maximal $(c,b,\ell)$-blocks $c$.)
\item Follow a \comp \tra labeled by a \morph $h$ which is defined by $h(c)=cbc$ leading to a state 
$E'=(W',B,\cX,\rho',\mu',\theta,T)$ and \esolu $(\alp,\sig')$ 
such that $W=gh(W')$ and $\sig = gh\sig'$. 
Rename $E'=(W',B,\cX,\rho',\mu',\theta,T)$ and $(\alp,\sig')$
as $E=(W,B,\cX,\rho,\mu,\theta,T)$ and~$(\alp,\sig)$.
\item If the fresh clone $c$ does not appear in $\sig(W)$, which implies that there are no typed variables anymore, then replace 
$B$ by $B'=B\sm \os{c, \bar c}$.\\ (Note that in contrast all fresh letters $c_\kap$ 
remain visible.)
\item 
Follow 
an \epstra from~$E$ to the standard state $E'=(W,B',\cX,\rho,\mu,T)$.
The state $E'$ has the \esolu~$(\alp,\sig')$ where $\sig'(x)=\sig(x)$ for all 
$x\in B\cup \cX$.
\\
(Note that the free resource monoid $M(B'\cup \cX,\rho)$ embeds into the 
typed monoid $M(B\cup \cX,\rho,\theta)$ in this case; and we have
$W\in M(B'\cup \cX,\rho)$ and $\sig(W)\in M(B',\rho)$.)
\item Perform all basic $T$-reductions; and rename the standard state $E'$ and $(\alp',\sig')$ as $E=(W,B,\cX,\rho,\mu,T)$ and 
$(\alp,\sig)$.
\end{enumerate}
\noindent\textsc{\textbf{EndWhile}} (The state~$E$ is $k$-medium since all typed variables are gone.)\\
\noindent\textsc{\textbf{EndProcedure}}

Note that the procedure terminates since in the first inner loop the
length of every typed variable gets shorter, so eventually all typed variables are removed. If there are no typed variables anymore, then each outer loop makes the equation shorter.)
\begin{lemma}\label{lem:Comaba}
Let $a\in \Bold$ and $b\in B_S = \set{b\in \Bold}{b=\bar b}$ with $a \neq b$.
Suppose that $E=(W,B,\cX,\rho,\mu,T)$ is the $k$-medium state and $(\alp,\sig)$ its \esolu satisfying \prref{inva:Lugo} when \textsc{CompressBlocks($aba$)} is called.
Then~$\cU$ contains a path $\pi$ following $(\alp,\sig)$ {}from~$E$ to 
a state $E'=(W', B', \cX', \rho', \mu', T)$ and its \esolu $(\alp',\sig')$ such that the following holds.
\begin{enumerate}
\item The state $E'$ is $k$-medium, it satisfies \prref{inva:hugo}, and in addition, 
$\sig'(W')$ is without any factor $aba$. 
\item Every state $\wt E$ on the path $\pi$ from~$E$ to $E'$ is $k$-large and its equation $\wt W$ satisfies 
  \[
    \abs{\wt W} - \abs{W} \leq 20\cdot\hspace{-0.3cm} \sum_{X \in \cXTfl} \abs{W}_X
  \]
\item We have $\cX' \subseteq \cX$, $\cXTfl' \subseteq \cXTfl$, and 
  \[
    \abs{W'} - \abs{W} \leq 4\cdot\hspace{-0.3cm} \sum_{X \in \cXTfl} \abs{W}_X
  \]
\end{enumerate}
\end{lemma}
\begin{proof}
The absence of $aba$ factors and $k$-largeness holds 
after the call to \textsc{PrepareBlocks($aba$)} and is maintained throughout.
The state $E'$ is $k$-medium again as all typed variables, introduced during \textsc{PrepareBlocks($aba$)}, have been removed.

The estimation $\abs{\wt W} \leq \abs{W} + 20\cdot \sum_{X \in \cXTfl} \abs{W}_X$ results from two steps. 
First, the estimation (\ref{eq:precbc}) in \prref{lem:preaba} for the uncrossing of maximal $(a,b,\ell)$-blocks and the temporary existence of typed variables leads to an increase of at most $12\cdot \sum_{X \in \cXTfl} \abs{W}_X$.
Then, in each iteration of the while-loop of \textsc{CompressBlocks($aba$)}, we pop out at most $4 \cdot \sum_{X \in \cXTfl} \abs{W}_X$ letters in the first step and compress at least half of the letters popped out in this way (including ones from previous iterations) during third step.
As such, we can bound the increase by $8 \cdot \sum_{X \in \cXTfl} \abs{W}_X$ throughout the entire while-loop.

For the final item, note that at most two of the letters popped out of each variable survive compression per side.
We leave the details to the reader.
\end{proof}

\subsubsection{Putting it all together}

The input to the following procedures is
a standard $E=(W,B,\cX,\rho,\mu,T)$ with a reduced \esolu $(\alp,\sig)$
such that~$E$ is firstly $k$-medium for some $k\in \N$ and secondly, ignoring all \Tflat
variables $X$ would make the corresponding state $k$-small.
We consider several cases separately, although they differ only in details. 
We also crucially have to consider several times random linear orders over different sets. 
The effect of choosing linear orders uniformly add random will give us (later) an expected upper bound in $\Oh(1)$ on the length increase caused by uncrossing. 
So we can guess the existing good order which realizes that upper bound.
Phrased differently, we can follow nondeterministically the correct path in~$\cU$ which does not increase the length of an equation more than expected.\\

\noindent\textsc{\textbf{BeginProcedure}}
\textsc{$T$-BlockCompression}\\
\mbox{}\quad\textsc{\textbf{Condition:}} \prref{inva:hugo} holds. 
The \esolu $(\alp,\sig)$ is reduced.\\
\mbox{}\quad\textsc{\textbf{Call}} \textsc{L-BlockCompression}\\
\mbox{}\quad\textsc{\textbf{Call}} \textsc{NSN-BlockCompression}\\
\mbox{}\quad\textsc{\textbf{Call}} \textsc{2S-BlockCompression}\\
\mbox{}\quad Perform all basic $T$-reductions and rename the result as $E=(W,B,\cX,\rho,\mu,T)$ and $(\alp,\sig)$.\\
\noindent\textsc{\textbf{EndProcedure}}\\

The procedure above defines the order of compression using three subroutines
which are disjoint in the sense that they address pairwise disjoint factors $ba$ of maximal $(a,b,\ell)$-blocks. 
Each subroutine may introduce temporally fresh typed variables which vanish inside the same subroutine again. 
Whether or not typed variables appear at all depends on the current \solu{s}. 
In general it does not depend on the state alone.

In the following procedure the acronym L refers to `Letter' since we compress all $(\wt a, 1,\ell)$-blocks where~$a$ is a letter in $\Bold$.\\

\noindent\textsc{\textbf{BeginProcedure}}
\textsc{L-BlockCompression}\\
\mbox{}\quad\textsc{\textbf{Condition:}} \prref{inva:hugo} holds.\\
\mbox{}\quad\textsc{\textbf{ForAll}} $a\in \Bold$ in any order  \textsc{\textbf{Do}}\\
\mbox{}\quad\quad\textsc{\textbf{Call}} \textsc{CompressBlocks($aa$)}\\
\mbox{}\quad\textsc{\textbf{EndForAll}}\\
\noindent\textsc{\textbf{EndProcedure}}\\

The purpose of \textsc{L-BlockCompression} is to replace \prref{inva:hugo}
by the stronger \prref{inva:Lugo}. As above, $B_S=\set{b\in \Bold}{b=\bar b}$.
\begin{invariant}\label{inva:Lugo}{\phantom x}
\begin{enumerate}
\item If $a\in \Bold$, then 
$\sig(W)$ is without factors $aa$ or $ab\bar a$ with $b\in \os{1}\cup B_S$.
\item If for $X\in \cXTfl$ there are factors $Xd$ and $Xe$ in $W$ with $d,e\in \Bold$, then $d=e$. 
\item We have $|\sig(X)|\geq 2$ for all $X\in \cX$.
\item If $XuY$ is a factor of $W$ with $X,Y \in \cXTfl$ and $u \in \Bold^\ast$, then $\abs{u} \geq 2$.
\end{enumerate}
\end{invariant}
 
In the following we say that a procedure \emphind{affects a variable on the right}
(\wrt a state $E=(W,B,\cX,\rho,\theta,\mu,T)$ and its \esolu $(\alp,\sig)$)
if the procedure pops out at least one letter on the right of that variable. Clearly, a procedure 
affects a variable~$X$ on the right \IFF it affects~$\bar X$ on the left.
Affecting $X$ leads to a state with an \esolu $(\alp',\sig')$ such that 
$|\alp'\sig'(X)|<|\alp\sig(X)|$. 
\begin{lemma}\label{lem:Lugo}
Let $E=(W,B,\cX,\rho,\mu,T)$ be the standard state with its \esolu $(\alp,\sig)$
satisfying \prref{inva:hugo} before we call \textsc{L-BlockCompression} such that~$E$ is $k$-medium.
Then~$\cU$ contains a path $\pi$ following $(\alp,\sig)$ {}from~$E$ to 
a state $E'=(W', B', \cX', \rho', \mu', T)$ and its \esolu $(\alp',\sig')$ such that the following holds.
\begin{enumerate}
\item The state $E'$ is $k$-medium, it satisfies \prref{inva:Lugo}.
\item Every state $\wt E$ on the path $\pi$ from~$E$ to $E'$ is $k$-large and its equation $\wt W$ satisfies 
  \[
    \abs{\wt W} - \abs{W} \leq 20\cdot\hspace{-0.3cm} \sum_{X \in \cXTfl} \abs{W}_X
  \]
\item We have $\cX' \subseteq \cX$, $\cXTfl' \subseteq \cXTfl$, and 
  \[
    \abs{W'} - \abs{W} \leq 2\cdot\hspace{-0.3cm} \sum_{X \in \cXTfl} \abs{W}_X
  \]
\end{enumerate}
\end{lemma}
\begin{proof}
  The assertions follow from \prref{lem:Comaba} and two additional observations.
  Firstly, a variable $X$ can only be affected on the right during a call to \textsc{CompressBlocks($aa$)} once throughout the entire procedure, \ie for at most one $a \in \Bold$.
  Indeed, in order for $X$ to be affected on the right, $W$ must contain a factor $Xa$.
  The equation $\wt W$ immediately after the call to \textsc{CompressBlocks($aa$)} then either contains the factor $Xa$ as well or every factor $Xd$ in $\wt W$ with $d \in B_T$ satisfies $d \not\in \Bold$.
  In either case, $Xa'$ with $a' \in \Bold$ and $a' \neq a$ can never be a factor of $\wt W$ thanks to \prref{inva:hugo}.
  Secondly, the improved bound on the increase in length of the third item is due to the observation that at most one letter popped out of a variable per side survives compression.
\end{proof}
 
The following acronyms NSN and 2S the symbol $N$ refers to a non-selfie in $\Bold$ and $S$ refers to a selfie in $\Bold$.\\

\noindent\textsc{\textbf{BeginProcedure}}
\textsc{NSN-BlockCompression}\\
\mbox{}\quad\textsc{\textbf{Condition:}} \prref{inva:Lugo} holds.\\
\mbox{}\quad\textsc{\textbf{ForAll}} $\{a, \bar a\} \subseteq \Bold$ with $a\neq \bar a$ in any order \textsc{\textbf{Do}}\\
\mbox{}\quad\quad Choose uniformly a random linear order $\leq$ on $B_S=\set{b\in \Bold}{\bar b = b}$.\\
\mbox{}\quad\quad\textsc{\textbf{ForAll}} $b \in B_S$ in the chosen order $\leq$ \textsc{\textbf{Do}}\\
\mbox{}\quad\quad\quad\textsc{\textbf{Call}} \textsc{CompressBlocks($aba$)}\\
\mbox{}\quad\quad\textsc{\textbf{EndForAll}}\\
\mbox{}\quad\textsc{\textbf{EndForAll}}

\noindent\textsc{\textbf{EndProcedure}}\\

\begin{remark}\label{rem:NSN}
After executing \textsc{NSN-BlockCompression} $\sig(W)$ is without $aba$ factors where $a\neq \bar a\in \Bold$ and $b\in B_S$.
However there might be factors $ab$, $ba$, and $bab$. 
\hspace*{\fill}$\diamond$\end{remark}
\begin{lemma}\label{lem:incrNSN}
Let $E=(W,B,\cX,\rho,\mu,T)$ be the standard state with its \esolu $(\alp,\sig)$ which satisfies \prref{inva:Lugo}
before we call \textsc{NSN-BlockCompression}. 
Suppose that~$E$ is $k$-medium.
Then~$\cU$ contains a path $\pi$ following $(\alp,\sig)$ {}from~$E$ to 
a state $E'=(W', B', \cX', \rho', \mu', T)$ and its \esolu $(\alp',\sig')$ such that the following holds.
\begin{enumerate}
\item The state $E'$ is $k$-medium, it satisfies \prref{inva:Lugo}, and in addition, $\sig'(W')$ is without any factor $aba$ for all $a\neq a\in \Bold$ and $b=b\in \Bold$.  

\item Every state $\wt E$ on the path $\pi$ from~$E$ to $E'$ is $k$-large and its equation $\wt W$ satisfies 
  \[
    \abs{\wt W} - \abs{W} \leq 28\cdot\hspace{-0.3cm} \sum_{X \in \cXTfl} \abs{W}_X
  \]
\item We have $\cX' \subseteq \cX$, $\cXTfl' \subseteq \cXTfl$, and 
  \[
    \abs{W'} - \abs{W} \leq 8\cdot\hspace{-0.3cm} \sum_{X \in \cXTfl} \abs{W}_X
  \]
\end{enumerate}
\end{lemma}
\begin{proof}
The assertions follow from \prref{lem:Comaba} and the following observations on the fate of a single variable $X \in \cX$.
Suppose that $X$ is first affected on the right during a call to the procedure \textsc{CompressBlocks($aba$)}.
Then $X$ is \Tflat and moreover there are two disjoint cases.
Either $\sig(X)\in B^\ast \wt a$ and $W$ has a factor $Xb\wt a$ or $\sig(X)\in B^*\wt ab$ and $W$ has a factor $X\wt a$ where $\wt a \in \{a, \bar a\}$.
Thus, for each $X$ there the pair $(\{a, \bar a\},b)$ with $a\neq \bar a\in \Bold$ and $b= \bar b\in \Bold$, if it exists, is uniquely determined.
Moreover, when $X$ is affected on the right, then \textsc{UncrossBlocks($aba$)} pops out of $X$ at most two letters from $\Bold$.
An easy reflection show that after that point $X$ always has either~$\wt a$ as a direct right neighbor, or a fresh letter $c$, or $bc$ where $c$ is a fresh letter.
Only for as long as some occurrence of $X$ in the equation has~$\wt a$ as a direct right neighbor, it is possible that $X$ is affected on the right again.
If so, it must happen in the inner for-all loop where we consider all $b\in B_S=\set{b\in \Bold}{b=\bar b}$ in the chosen random order. 

Let $b_1, \dotsc, b_m \in B_S$ be such that $X$ is affected on the right by 
\begin{align}\label{eq:califa}
\textsc{CompressBlocks}(ab_1a)\lds \textsc{CompressBlocks}(ab_ma)
\end{align}
in that order, \ie with $b_1 < \cdots < b_m$ in the chosen random order on $B_S$.
Then, due to the above reasoning, $\sig(X) \in B_T^\ast (\wt a b_m)^+ \dotsc (\wt a b_2)^+ u$ where $u$ is a prefix of a word in $(\wt ab_1)^+$ and $b_1 = b$.
As such, if we consider the word $\oo$ obtained from $\sig(X)b$ by removing all letters not in $B_S$ as well as all repetitions, then $\oo$ has the decreasing suffix $b_m \ldots b_1$ of length $m$.
By \prref{lem:explip}, the expected value $\E {m_X}$ of $m_X = m$ is less than~$2$.
As such, we can now bound the expected the length increase $I'_W = \abs{W'} - \abs{W}$ for the final state and $\wt I_W = \abs{\wt W} - \abs{W}$ for any intermediate state by
\begin{align*}
  \E {I'_W} &\leq 2 \cdot \hspace{-0.3cm}\sum_{X \in \cXTfl} \E {m_X + m_{\bar X}} \cdot \abs{W}_X \leq 8 \cdot \hspace{-0.3cm}\sum_{X \in \cXTfl} \abs{W}_X\\
  \E {\wt I_W} &\leq \E {I'_W} + 20 \cdot \hspace{-0.3cm}\sum_{X \in \cXTfl} \abs{W}_X \leq 28 \cdot \hspace{-0.3cm}\sum_{X \in \cXTfl} \abs{W}_X \qedhere
\end{align*}
\end{proof}
\noindent\textsc{\textbf{BeginProcedure}}
\textsc{2S-BlockCompression}\\
\mbox{}\quad\textsc{\textbf{Condition:}} \prref{inva:hugo} holds.\\
\mbox{}\quad Choose uniformly at random a linear order $\leq$ on $\binom{B_S}{2}$ where $B_S = \set{a \in \Bold}{a = \bar a}$.\\
\mbox{}\quad\textsc{\textbf{ForAll}} $\{a, b\} \in \binom{B_S}{2}$ in the chosen order $\leq$ \textsc{\textbf{Do}}\\
\mbox{}\quad\quad \textsc{\textbf{Call}} \textsc{CompressBlocks($aba$)}\\
\mbox{}\quad\quad \textsc{\textbf{Call}} \textsc{Compress($bab$)}\\
\mbox{}\quad\quad \textsc{\textbf{Call}} \textsc{Compress($ab$)}\\
\mbox{}\quad \textsc{\textbf{EndForAll}}\\
\noindent\textsc{\textbf{EndProcedure}}\\

The procedure  \textsc{2S-BlockCompression} and
\textsc{NSN-BlockCompression} are very similar. They differ whether $a=\bar a$ or not and in the choice of the  to compress 
all factors $aba$ with $a,b\in \Bold$ with $b=\bar b$. But they share important features.
\begin{enumerate}
\item Maximal $(a,b,\ell)$-blocks do not overlap.
\item A variable $X$ is affected on the right \IFF 
there are $a,b\in \Bold$ (with $a\neq a$ for NSN and $a= a$ for 2S) and $b=\bar b$  such that either 
$\sig(X)\in B_T^*ab$ and 
$W$ has a factor $Xa$ or $\sig(X)\in B_T^*a$ and 
$W$ has a factor $Xba$. In either case, if $X$ is affected on the right, then $(a,b)$ is determined by $\sig(X)$ and $W$, but not by the position of $X$ in $\pos(W)$. 
\item Executing the inner body \textsc{CompressBlocks}($aba$) can be realized by a path in~$\cU$ to a 
state $E'=(W',B',\cX',\rho',\mu',T)$ and \esolu
$(\alp',\sig')$ such that firstly, $B'\sse B$ and secondly, $\sig'(W')$ does not have any factor 
$aba$. However, executing \textsc{2S-BlockCompression} we have more, we can exclude factors $ab$ where~$a$ and $b$ are selfies in $\Bold$. 
\end{enumerate}
\begin{lemma}\label{lem:2Scomp}
Let $E=(W,B,\cX,\rho,\mu,T)$ be the standard state with its \esolu $(\alp,\sig)$ which satisfies \prref{inva:Lugo}
before we call \textsc{2S-BlockCompression}.
Suppose that~$E$ is $k$-medium.
Then~$\cU$ contains a path $\pi$ following $(\alp,\sig)$ {}from~$E$ to 
a state $E'=(W', B', \cX', \rho', \mu', T)$ and its \esolu $(\alp',\sig')$ such that the following holds.
\begin{enumerate}
\item The state $E'$ is $k$-medium, it satisfies \prref{inva:Lugo}, and in addition, $\sig'(W')$ is without any factor $ab$ for all $a,b\in B_s$.  

\item Every state $\wt E$ on the path $\pi$ from~$E$ to $E'$ is $k$-large and its equation $\wt W$ satisfies 
  \[
    \abs{\wt W} - \abs{W} \leq 28\cdot\hspace{-0.3cm} \sum_{X \in \cXTfl} \abs{W}_X
  \]
\item We have $\cX' \subseteq \cX$, $\cXTfl' \subseteq \cXTfl$, and 
  \[
    \abs{W'} - \abs{W} \leq 8\cdot\hspace{-0.3cm} \sum_{X \in \cXTfl} \abs{W}_X
  \]
\end{enumerate}
\end{lemma}
\begin{proof}
Most of the assertions are immediate from the corresponding properties of the subprocedures as detailed in \prref{lem:Comaba} and \prref{lem:stfl2}.
To see the estimates on the length increase $I'_W = \abs{W'} - \abs{W}$ and $\wt I_W = \abs{\wt W} - \abs{W}$, let us consider (as in the previous proofs) the fate of a single variable $X \in \cX$ which is affected by \textsc{2S-BlockCompression} on the right.
During the procedure at least one of the 
factors $aba$, $bab$, $ab$ or $ba$ with $\os{a,b}\in \binom{B_S}{2}$ is uncrossed on the right for at least one occurrence of~$X$. 
Then $X$ is \Tflat and, by possibly interchanging the role of~$a$ and~$b$, we may assume that $\sig(W)$ ends in the selfie~$b$ 
and there is a factor $Xa$ in $W$, too. 
This implies that $X$ is uncrossed for the first time when the for-all loop considers the unordered pair~$\os{a,b}$.
By \prref{inva:Lugo} we know that there is no factor $Xd$ in $W$ with $d\in \Bold$ unless $a=d$. However, the uncrossing at $X$ changes the situation.
We see an equation $W_1$ and a corresponding \solu $\sig_1(W)$ such that 
the right neighbor of $X$ is either a fresh letter or it is~$b$.
By the description of \textsc{2S-BlockCompression} $\sig_1(X)$ cannot end in the letter~$a$ since  \textsc{Compress}($ab$) compresses $ab$ and $ba$. Say, $\sig_1(X)$  ends in a selfie $d\in B_S$. Then 
$X$ is affected on the right again for the unordered pair~$\os{b,d}$.
A third time is possible for an unordered pair~$\os{d,e}\in \binom{B_S}{2}$, etc.
Thus, for each $X\in \cX$ there is maximal number $m$ such that $X$ is affected on the right by executing \textsc{2S-BlockCompression} when the body of the for-all loop 
defines an unordered pair $\os{a_i,b_i}$. 
Similar as in (\ref{eq:califa}) 
we obtain a sequence 
\begin{align}\label{eq:2Scali}
  \os{a_1,b_1} <\cdots < \os{a_m,b_m}
\end{align}
The interpretation is that $X$ is affected on the right exactly during the loop iteration for these unordered pairs in that order.
As in the proof of \prref{lem:incrNSN}, we the word $\os{a_m,b_m} \ldots \os{a_1,b_1}$ over the alphabet $\binom{B_s}{2}$ as a suffix of a word $\oo$ determined entirely by $W$ and $\sig$.
Indeed, we can first remove from $\sig(X)a$ all letters not in $B_S$ and then all repetitions.
Forming unordered pairs of consecutive letters and removing repetitions again yields such a word $\oo$.
By \prref{lem:explip}, the expected value of $m = m_X$ is less than~$2$.

For the estimating $I'_W = \abs{W'} - \abs{W}$ note that, in the worst case, $X$ is affected on the right by \textsc{CommpressBlocks($a_ib_ia_i$)}, by \textsc{Compress($b_ia_ib_i$)}, and by \textsc{Compress($a_ib_i$)} for each $1 \leq i \leq m$.
Note, however, that of the letters popped out during these calls for fixed $i$, at most two letters survives compression for each occurrence of $X$.
Hence, 
\begin{align*}
  \E {I'_W} &\leq 2 \cdot\hspace{-0.3cm}\sum_{X \in \cXTfl} \E {m_X + m_{\bar X}} \cdot \abs{W}_X \leq 8 \cdot\hspace{-0.3cm} \sum_{X \in \cXTfl} \abs{W}_X \\
  \E {\wt I_W} &\leq \E {I'_W} + 20 \cdot\hspace{-0.3cm}\sum_{X \in \cXTfl} \abs{W}_X \leq 28 \cdot\hspace{-0.3cm}\sum_{X \in \cXTfl} \abs{W}_X \qedhere
\end{align*}
\end{proof}
The following proposition summarizes the efforts undertaken so far.
\begin{proposition}\label{prop:TBC}
Let $E=(W,B,\cX,\rho,\mu,T)$ and  $(\alp,\sig)$ its \esolu  
when we enter \prref{sec:unold} which defines $\Bold$. 
Let~$E$ be $k$-medium.
Then there is path $\pi$ in~$\cU$ {}from 
$E$ to a $k$-medium standard state $E'=(W',B',\cX',\rho',\mu',T)$ and a reduced \esolu $(\alp',\sig')$ 
which follows $(\alp,\sig)$ such that $\cX' \sse \cX$, $\cXTfl' \sse \cXTfl$, and
\begin{align}\label{eq:TBC}
  \abs{W'} - \abs{W} &\leq 28\cdot\hspace{-.3cm}\sum_{X \in \cXTfl} \abs{W}_X
\end{align}
Moreover, the factors of the form $aa$, $a\bar a$, $aba$, $ab\bar a$, and $bb'$ 
with $a\in \Bold$ and selfies  $b,b'$ in $\Bold$ do not appear in $\sig'(W')$ anymore.
If $\wt E$ is any state on $\pi$ then the state $\wt E$ is $k$-large and satisfies 
\begin{align}\label{eq:propTBC}
  \abs{\wt W} - \abs{W} \leq 48 \cdot\hspace{-.3cm} \sum_{X \in \cXTfl} \abs{W}_X.
\end{align}
\end{proposition}
\begin{proof}
  Consider the following realization of the path $\pi$ beginning with 
$E$ and $(\alp,\sig)$. 
The first modifications of~$E$ and $(\alp,\sig)$ are due to \textsc{InitSplit} which pops out for each \Tflat variable at most two letters in $\Bold$ and establishes \prref{inva:Tos}.
Then we call the procedures $aS\bar a$-\textsc{Compression} which establishes \prref{inva:hugo}; see \prref{lem:abovacomp}.
Finally, the procedure $T$-\textsc{BlockCompression} comprising L-, NSN-, and 2S-\textsc{BlockCompression} removes all factors of the form $aa$, $aba$, and $bb'$ with $a\in \Bold$ and selfies  $b,b'$ in $\Bold$; see Lemma~\ref{lem:Lugo}, \ref{lem:incrNSN}, 
and \ref{lem:2Scomp}.
Since $T$-\textsc{BlockCompression} finishes by performing all basic $T$-reductions, the resulting entire solution is reduced.
\end{proof}
Thanks to \prref{prop:TBC} we are able to sharpen 
\prref{inva:Lugo}. The following invariant holds 
after each execution of $T$-\textsc{BlockCompression}
until either we visit 
\prref{sec:eliTT} with the same subset $T\ssneq \Rs$ 
again or all \TTarcs are eliminated or we can change $T$ to a set $T'\sse \Rs$ with $T<T'$. 
\begin{invariant}\label{inva:afterT}{\phantom x}
Let $E=(W,B,\cX,\rho,\mu,\theta, T)$ be a state  and $(\alp,\sig)$ be its \esolu. 
Then:
\begin{enumerate}
\item 
The trace $\sig(W)$ is without any factor $ab\wt a$ for all 
$a\in \Bold$, $\wt a\in \os{a,\bar a}$, and $\bar b=b\in \os{1} \cup \Bold$. It is also without any factor $ab$  where
$a$ and $b$ are selfies in $\Bold$. 
\item If $Xd$ and $Xe$ are factors in $W$ with $d,e\in \Bold$, then $d=e$. 
\item The state~$E$ is a standard state which is $k$-medium. If~$E$ is a state without \Tflat variables, then it is $k$-small. 
  The \esolu $(\alp,\sig)$ is reduced.
\end{enumerate}
\end{invariant}
\Ip \prref{inva:afterT} holds when we consider \textsc{PairCompression} next. 

\subsection{Pair compression}\label{sec:pair}
We continue with a standard state  $E=(W,B,\cX,\rho,\mu,T)$ and its \esolu $(\alp,\sig)$ which is the output of the procedure \textsc{$T$-BlockCompression} 
We are interested to compress many factors $ab$ with $a,b\in \Bold$ into a single position where one at least one endpoint is visible. 
If there is still such a factor, then we have $a\notin\os{b,\bar b}$
because there are no factors violating this condition anymore.

To this end we begin by choosing for each $a\neq \bar a \in \Bold$ 
uniformly at random a linear order $\leq$ 
for $\os{a,\bar a}$. 
This defines a set of \emph{positive\index{positive letters}} and \emph{negative\index{negative letters}} letters $\Bpos$ and $\Bneg$ by 
\begin{equation}\label{eq:Bpos}
\Bpos=\set{a\in \Bold}{a< \bar a} \text{ and } \Bneg=\set{a\in \Bold}{a> \bar a}
\end{equation}
As usual, we let $B_S=\set{b\in \Bold}{b=\bar b}$ be the set of `old selfies'.
Next we choose the linear order on $\Bpos$ uniformly at random and extend it arbitrarily to 
$\Bpos\cup B_S$ such that $a<b$ if $a\neq \bar a$ and 
$b= \bar b$. 
Finally, we extend the linear order to a linear 
order $\leq$ of $\Bold$
by forcing $a\leq b\iff \bar b \leq \bar a$ for all $a\in \Bpos$. 
This is easy to achieve; 
and for some bijection $\zeta$ {}from the set $\os{1\lds \ell}$ to  $\Bpos\cup B_S$ the final order on $\Bold$ can be written for some  $0\leq k\leq \ell$ as
\begin{equation}\label{eq:finord}
\underbrace{{\zeta(1)}< \cdots < {\zeta(k)}}_{\Bpos} < \underbrace{{\zeta(k+1)} <\cdots < {\zeta(\ell)}}_{\text{The set of selfies}} < \underbrace{\ov{{\zeta(k)}} < \cdots < \ov{{\zeta(1)}}}_{\Bneg}
\end{equation}
 \noindent\textsc{\textbf{BeginProcedure}} \textsc{PairCompression}
 
 \textbf{Condition:}~$E$ and $(\alp,\sig)$ satisfy \prref{inva:afterT}, and  
$(\Bold,<)$ is listed in (\ref{eq:finord})

\textsc{\textbf{ForAll}} $a\in \Bpos$ in the linear order $\leq$ \textsc{\textbf{Do}}

\quad\textsc{\textbf{ForAll}} $b\in \set{b\in \Bpos\cup B_S}{a<b}$ 
in the linear order $\leq$ 
 \textsc{\textbf{Do}}

\qquad\textsc{\textbf{Call}} \textsc{Compress($ab$)}

\quad\textsc{\textbf{EndForAll}}

\textsc{\textbf{EndForAll}}

Perform all basic $T$-reductions according to \prref{sec:stproc}.

\noindent\textsc{\textbf{EndProcedure}}\\

Let us list some properties of the procedure. 
\begin{enumerate}
\item If $ab$ is compressed, then the subtraces 
$ab$ and $\bar b \bar a$ do not have any non-trivial overlap in $\sig(W)$ thanks to 
\prref{inva:afterT}. 
\item Clearly, a factor $ab$ is compressed \IFF $\bar b \bar a$ is compressed.
Moreover, if $ab$ is compressed, then $a<b$ and $\bar b <\bar a$.
\item For $a\in \Bpos$ and $a<b$ the factor $ba$ is not compressed, \eg~if 
$a_1$ is the first letter in $\Bpos$ and $b$ is any other letter in $B$, then $ba$ is not compressed. 
\item For every $(a,b)$ with $a\neq \bar a\in \Bold$ and 
$a\neq b$ the probability that $ab$ is compressed is $1/2$ whether or not $b\in B_S$. Recall that $\sig(W)$ is without any factor $ab\in B_S^{\;2}$.
\end{enumerate}
\begin{lemma}\label{lem:paircomp}
Let $E=(W,B,\cX,\rho,\mu,T)$ be the standard state with its \esolu $(\alp,\sig)$ which satisfies \prref{inva:afterT}
before we call \textsc{PairCompression}.
Suppose that~$E$ is $k$-medium and let $H$ denote the set of \TTarc{s} in $W$ with both endpoints a constant labeled in $\Bold$.
Then~$\cU$ contains a path $\pi$ following $(\alp,\sig)$ {}from~$E$ to 
a state $E'=(W', B', \cX', \rho', \mu', T)$ and its \esolu $(\alp',\sig')$ such that the following holds.
\begin{enumerate}
\item The state $E'$ and its \esolu $(\alp',\sig')$ satisfy \prref{inva:afterT}.
  In addition, $\sig'(W')$ has no factors $ab$ where $a\in \Bpos$ with $a<b\in \Bold$. 
  Moreover,
  \[
    \abs{W'} - \abs{W} \leq {- \tfrac 16\abs{H} + {}}5 \cdot\hspace{-.3cm}\sum_{X \in \cXTfl} \abs{W}_X 
  \]
\item Every state $\wt E$ on the path $\pi$ satisfies $\wt\cX \sse \cX$, $\wt\cX_{\textup{Tflat}} \sse \cXTfl$, and
  \[
    \abs{\wt W} - \abs{W} \leq + \tfrac16\abs{H} + {}4 \cdot\hspace{-.3cm}\sum_{X \in \cXTfl} \abs{W}_X  \qedhere
  \]
\end{enumerate}
\end{lemma}
\begin{proof}
Most assertions are immediate from the description of the procedure \textsc{PairCompression} and the corresponding properties of the subprocedure \textsc{Compress}; see \prref{lem:stfl2}.
To see the estimate on the length change $\abs{W'} - \abs{W}$, let us consider 
(as in the previous proofs) 
the fate of a single variable $X \in \cX$ which is affected by 
\textsc{PairCompression} on the right.
The increase in length of the equation caused by uncrossing 
\TTarcs $a \to b$ where $a < b \in \Bold$ and not both letters in $\os{a,b}$ are selfies.
Thus, for every variable $X$ there is some $m\in \N$ such that the sequence of 
calls 
\begin{align}\label{eq:compabi}
\textsc{Compress($a_{1} b_{1}$)}, \textsc{Compress($a_{2} b_{2}$)}\lds \textsc{Compress($a_{m} b_{m}$)}
\end{align}
affect $X$ on the right in that order and $m$ is maximal. 
Due to the order, we have $a_1 \leq a_2 \leq \dotsc \leq a_m$.
We claim that $a_1 < a_2 < \dotsc < a_m$.
To see this, let us show that $a_i < a_{i+1}$ for all $1 \leq i < m$.
There are two cases. 

In the first case we assume that $\sig(X)$ ends in $a_{i}$ and $W$ contains a factor $Xb_{i}$. 
This means that \textsc{Compress($a_{i} b_{i}$)} pops out the letter $a_{i}$ and so a factor $Xa_{i}$ appears in the new equation $\wt W$.
Since $a_{i} \in \Bpos$ and the variable $X$ is next affected on the right by \textsc{Compress($a_{i+1} b_{i+1}$)}, we must have $b_{i+1} = a_{i}$.
But this implies $a_{i+1} < b_{i+1} = a_{i}$ contradicting $a_i \leq a_{i+1}$.
So this case cannot occur.

In the second case $\sig(X)$ ends in $\ov{b_i}$ and $W$ contains a factor $X\ov{a_{i}}$.
The call to \textsc{Compress($a_{i} b_{i}$)} pops out the letter 
$\ov{b_{i}}$.
For the new equation $\wt W$ and \solu $\wt \sig$ we see a factor $X\ov{b_{i}}$ in $\wt W$ and $\wt \sig(X)$ ends in some letter $c_{i+1}$.
The variable $X$ is next affected on the right by \textsc{Compress($a_{i+1} b_{i+1}$)}.
If $b_{i+1} = \ov{b_{i}}$ and $a_{i+1} = c_{i+1}$, then $b_{i+1} = b_{i} \in B_S$.
Hence, $a_{i} < a_{i+1}$.
Otherwise, $\ov{a_{i+1}} = \ov{b_{i}}$ and, thus, $a_i < b_i = a_{i+1}$.

  Suppose now, that $m \geq 1$ and thus, that $W$ contains a factor $Xa$ with $a \in \Bold$.
Let us consider the word $\oo$ obtained from $\overline{a\sig(X)}$ by removing all occurrences of selfies.
Then the above argument shows that $\omega$ starts with $\wt a_1 \wt a_2 \dotsc \wt a_m$ where $\wt a_i \in \{a_i \bar a_i\}$ for each $1 \leq i \leq m$.
Applying \prref{lem:explip} shows that the expected value of the random variable $m = m_X$ is less than~$2$. 

Let us now turn our attention to the \TTarc{s} contained in $H$.
As~$(E, \alp, \sig)$ satisfies \prref{inva:afterT}, each such arc $a \to b$ has at most one endpoint in $B_S$ and $b \not\in \{a,\bar a\}$.
Such an arc awaits one of three possible fates during the procedure.
\begin{enumerate}
  \item It is compressed into a single position labeled by a fresh letter.
  \item At least one endpoint becomes a fresh letters due to an adjacent compression and, as such, the arc will not become compressed into a single position.
  \item It is not affected by any form of compression whether directly or adjacent.
\end{enumerate}
To account for these three fates we use the following amortization strategy for the savings in length due to compression of arcs from $H$.
Firstly, we count an arc with the first fate as $\tfrac23$ arcs even though it is entirely removed.
This allows us to distribute $\tfrac13$ arcs worth of savings to both adjacent arcs (if there is only one adjacent arc from $H$, or none at all, we simply ignore the contribution).
Similarly, each occurrence of a \Tflat variable in $W$ distributes $\tfrac13$ arcs worth of savings to adjacent arcs from $H$ from a separate pool of $\sum_{X \in \cXTfl} \abs{W}_X$ arcs.
Following this amortization strategy permits us to count each arc awaiting the second fate as only $\tfrac23$ arcs even though it is not compressed, since such arc has received at least $\tfrac13$ arcs worth of savings from an adjacent arc or \Tflat variable.

Denoting by $h_1$, $h_2$, and $h_3$ the number of arcs from $H$ that await the first, second, and third fate, respectively, the above argument shows that
\[
  \abs{W'} - \abs{W} \leq - \tfrac13 (h_1 + h_2) + (1 + m_X + m_{\bar X})\cdot \hspace{-.3cm}\sum_{X \in \cXTfl} \abs{W}_X 
\]

We already know that the expected value of $m_X$ satisfies $\E{m_X} \leq 2$ for all $X \in \cXTfl$.
Moreover, we have $\abs{H} = h_1 + h_2 + h_3$ and, since the probability of an arc from $H$ to await the third fate is clearly at most $\tfrac12$, we deduce that $\E{h_1 + h_2} = \abs{H} - \E{h_3} \geq \tfrac12 \abs{H}$.
We can thus choose the path $\pi$ in~$\cU$ with the expectation that
\[
  \abs{W'} - \abs{W} \leq - \tfrac16\abs{H} + 5\cdot \hspace{-.3cm}\sum_{X \in \cXTfl} \abs{W}_X 
\]
To see the bound for the length of any intermediate equation $\wt W$, simply note that
\[
  \abs{\wt W} - \abs{W} \leq (m_X + m_{\bar X})\cdot \hspace{-.3cm}\sum_{X \in \cXTfl} \abs{W}_X
  \qedhere
\]
\end{proof}
\begin{proposition}\label{prop:eliTT}
Let $E=(W,B,\cX,\rho,\mu,T)$ and $(\alp,\sig)$ be a reduced $T$-perfect \esolu when we enter \prref{sec:eliTT} where we started to eliminate all \TTarcs. 
If~$E$ is $k$-small, then there is path $\pi$ in~$\wh \cU$ {}from $(E,\alp,\sig)$ to $(E',\alp',\sig')$ for  a $k$-small standard state $E'=(W',B',\cX',\rho',\mu',T')$ with $T' > T$ and a reduced \esolu $(\alp',\sig')$.
Moreover, we have
\begin{equation}\label{eq:elimTT}
  \abs{W'} \leq \abs{W} + 2\sum_{X \in \cX} \abs{W}_X
\end{equation}
and 
\begin{equation}\label{eq:deceliTT}
  \abs{W'} < \abs{W} \quad\text{or}\quad 
  \sum_{X\in \cX'}|\alp'\sig'(X)|\cdot |W'|_X < \sum_{X\in \cX}|\alp\sig(X)|\cdot|W|_X
\end{equation}
If $(\wt E, \wt \sig, \wt \alp)$ is any entire state on $\pi$ with an equation~$\wt W$, then $\wt E$ is $k$-large and $\wt W$ satisfies
\begin{align}\label{eq:elias}
  \abs{\wt W} \leq \tfrac76 \abs{W} + 579\cdot\sum_{X \in \cX} \abs{W}_X
\end{align}
\end{proposition}
\begin{proof}
Since $(\alp,\sig)$ is reduced, $\sig(W)$ contains a \TTarc.
If every position in every maximal \Tarc sequence is visible, we can choose any (fully visible) \TTarc and compress it. 
This shows \eqref{eq:elimTT} and \eqref{eq:deceliTT}; \eqref{eq:elias} holds trivially in this case.
By repeating this process, we end up in a state $E'$ with an \esolu $(\alp', \sig')$ that has the desired properties.

Otherwise, we create \Tflat variables by splitting as in \prref{lem:splitflat}.
This results in a $k$-medium state $E'_0$ with an equation $W'_0$ such that $\abs{W'_0} \leq \abs{W} + 2\sum_{X \in \cX} \abs{W}_X$.
From here on, we apply \prref{prop:TBC} to obtain a path resulting in a $k$-medium state $E''_0$ with an equation $W''_0$ and then apply \prref{lem:paircomp} to obtain a path to a $k$-medium state $E'_1$ with an equation $W'_1$.
All states on this combined path are $k$-large.
We repeat this step to obtain states $E''_1, E'_2, E''_2, \dotsc$ with respective equations $W''_1, W'_2, W''_2, \dotsc$ until we eventually reach a state $E'_m$ with an equation $W'_m$ such that $E'_m$ does not contain any \Tflat variables.
The see that we eventually reach such a state, note that at the beginning of each step we pop out at least one letter from every $T$-flat variable $X$ and, as such, $X$ is eventually removed once $\abs{\wt\sig(X)} \leq 1$. 
This also shows \eqref{eq:deceliTT}.
Once we reach the state $E'_m$, every maximal \Tarc sequence is visible and we can simply proceed as above to obtain a path to a state $E'$ with an \esolu $(\alp', \sig')$ with the desired properties.
 
It remains to bound the increase in length $\abs{\wt W} - \abs{W}$ for all intermediate states.
To this end, let us first bound the size of the set $H'_i$ of \TTarc{s} in $W'_i$ with both endpoints being constants.
We claim that the following holds for all $0 \leq i < m$.
\begin{align}\label{eq:longTTrec}
  \abs{H'_{i+1}} &\leq \tfrac56 \abs{H'_i} + 72 \cdot \sum_{X \in \cX} \abs{W}_X
\end{align}
Before we establish \eqref{eq:longTTrec}, let us first note that it implies the bound
\begin{align}\label{eq:longTTabs}
  \abs{H'_{i}} &\leq (\tfrac56)^i\abs{H'_0} + 432 \cdot \sum_{X \in \cX} \abs{W}_X
  \leq \abs{H'_0} + 420 \cdot \sum_{X \in \cX} \abs{W}_X 
\end{align}
for all $0 \leq i \leq m$.
Moreover, we also have $\abs{W'_i} - \abs{W'_0} \leq \abs{H'_i} - \abs{H'_0}$.
Indeed, popping out $T$-letters and compressing \TTarc{s} affects both the length of the equation and the number of \TTarc{s} with both endpoints being constants in the same way.
On the other hand, replacing \Tflat variables $X, \bar X$ by their solutions $\sig(X)$, $\sig(\bar X)$ when $\abs{\sig(X)} = \abs{\sig(\bar X)} \leq 1$ does not increase the length of the equation and does not decrease the number of \TTarc{s} with both endpoints being constants.
As such, 
\begin{align}\label{eq:longWabs}
\abs{W'_i} &\leq \abs{W'_0} + 432 \cdot \sum_{X \in \cX} \abs{W}_X \leq \abs{W} + 434 \cdot \sum_{X \in \cX} \abs{W}_X 
\end{align}
and, thanks to the estimate in \prref{prop:TBC}, we also obtain
\begin{equation}
  \abs{W''_i} - \abs{W} \leq 490 \cdot \sum_{X \in \cX} \abs{W}_X  
\end{equation}

We now turn to proving \prref{eq:longTTrec}.
To this end, let us consider the fate of an arc in $H'_i$ along the path from $E'_i$ to $E''_i$.
As in the proof of \prref{lem:paircomp}, such an arc is either compressed into a single position, at least one of its endpoints is replaced by a fresh letter due to an adjacent compression, or is not affected by compression at all.
Let us denote by $K'_i$ the set of such arcs that are unaffected.
Using the same amortization strategy as in the proof of \prref{lem:paircomp}, we improve the bound from \prref{prop:TBC} to 
\begin{equation*}
  \abs{W''_i} - \abs{W'_i} \leq -\tfrac13\abs{H'_i \setminus K'_i} + 58\cdot\sum_{X \in \cX}\abs{W}_X
\end{equation*}
Since the number of \TTarc{s} in $W''_i$ with both endpoints being constants labeled in $\Bold$ is at least $\abs{K'_i}$, \prref{lem:paircomp} gives us the estimate
\begin{equation*}
  \abs{W'_{i+1}} - \abs{W''_i} \leq -\tfrac16\abs{K'_i} + 10\cdot\sum_{X \in \cX}\abs{W}_X
\end{equation*}
Combining the two bounds, we obtain 
\begin{equation}\label{eq:longWW}
  \abs{W'_{i+1}} - \abs{W'_i} \leq -\tfrac16\abs{H'_i} + 68\cdot\sum_{X \in \cX}\abs{W}_X
\end{equation}
Arguing similarly to the proof of \prref{eq:longWabs} above, but using that replacing \Tflat variables $X, \bar X$ by their solutions when $\abs{\sig(X)} = \abs{\sig(\bar X)} \leq 1$ increases the number of \TTarc{s} with both endpoints being constants by at most $2(\abs{W'_i}_X + \abs{W'_i}_{\bar X})$, we obtain 
\begin{equation*}
  \abs{H'_{i+1}} - \abs{H'_i} \leq \abs{W'_{i+1}} - \abs{W'_i} + 4\cdot\sum_{X \in \cX}\abs{W}_X
\end{equation*}
Together with \prref{eq:longWW}, this proves \prref{eq:longTTrec}.

It only remains to bound the lengths of all intermediate equations.
Thanks to \prref{eq:longTTabs} and \prref{prop:TBC}, the equation $W''_i$ and any equation $\wt W$ corresponding to a state $\wt E$ on the path from $E'_i$ to $E''_i$ satisfy 
\begin{equation}\label{eq:longWT}
  \abs{W''_i} - \abs{W} \leq 490\cdot \sum_{X \in \cX} \abs{W}_X
  \quad\text{and}\quad
  \abs{\wt W} - \abs{W} \leq 530\cdot \sum_{X \in \cX} \abs{W}_X
\end{equation}
Bounding the number of \TTarc{s} in $W''_i$ with both endpoints being constants labeled in $\Bold$ from above by $\abs{W''_i}$, we obtain
\begin{equation}
  \abs{\wt W} \leq \tfrac76 \abs{W} + 579\cdot\sum_{X \in \cX} \abs{W}_X
\end{equation}
for the equation $\wt W$ of any state $\wt E$ along the path from $E''_i$ to $E'_{i+1}$.
\end{proof}

\section{The proof of \prref{thm:uro}}\label{sec:compT}

The following statement summarizes the contents of \prref{sec:prospli} and \prref{sec:eliTT}.
\begin{proposition}\label{prop:doors}
Let $(E,\alp,\sig)$ be the entire state with $E=(W,B,\cX,\rho,\mu,T)$ 
when we enter \prref{sec:elist} where we start to remove all \STarcs with $T \nsse S$.
There is a path $\pi$ in~$\wh\cU$ from $(E,\alp,\sig)$ to $(E',\alp',\sig')$ for a standard $T'$-state $E'=(W',B',\cX',\rho',\mu',T')$ with $T<T'\sse \Rs$ and $(\alp',\sig')$ reduced.
We can choose the path $\pi$ such that the length of the equation $W'$ satisfies
\begin{equation}
  \abs{W'} - \abs{W} \leq 5 \cdot 2^{\abs{\Rs}^2 + 3 \abs{\Rs}} \cdot \abs{\Winit}
\end{equation}
For every state $\wt E$ with an equation $\wt W$ on $\pi$, we have $\abs{\wt \cX} \leq 7 \cdot 2^{\abs{\Rs}^2 + 3 \abs{\Rs}} \cdot \abs{\cXinit}$ and
\begin{equation}
  \abs{\wt W} \leq \tfrac76 \abs{W} + 150 \cdot 2^{\abs{\Rs}^2 + 3 \abs{\Rs}} \cdot \abs{\Winit}
\end{equation}
\end{proposition}
\begin{proof}
  The number of variables in the state $E$, as well as the state $E''$ obtained from it using \prref{lem:Totnb}, is bounded by $2^{\abs{\Rs}^2 + 2 \abs{\Rs}} \cdot \abs{\cXinit}$.
  As in the proof of \prref{lem:Totnb}, we observe that the number of occurrences of variables in $W$, as well as in $W''$, is bounded above by $2^{\abs{\Rs}^2 + 2 \abs{\Rs}} \cdot \abs{\Winit}$.
  With this in mind and using the fact that $\abs{\Rs} \geq 2$, the claimed estimates follow from \prref{lem:Totnb} and \prref{prop:eliTT} via a straight-forward calculation.
\end{proof}
\begin{corollary}\label{cor:doors}
Let $\cSinit$ the initial state and $(\id_A,\siginit)$ some initial \esolu.
There is path $\pi$ in~$\wh\cU$ {}from $(\cSinit, \id_A, \siginit)$ to $(E', \alp', \sig')$ for a semi-final state $E'=(W',B',\es,\rho',\mu',\Rs)$.
Moreover, for every entire state $(\wt E, \wt \alp, \wt \sig)$ on $\pi$, we have $\abs{\wt \cX} \leq 7 \cdot 2^{\abs{\Rs}^2 + 2 \abs{\Rs}} \cdot \abs{\cXinit}$ and the equation~$\wt W$ of $\wt E$ satisfies
\begin{equation}\label{eq:TTelias2}
  \abs{\wt W} \leq 37\cdot 2^{\abs{\Rs}^2 + 4 \abs{\Rs}} \cdot \abs{\Winit} \leq 2^{\abs{\Rs}^2 + 7 \abs{\Rs}}\cdot \abs{\Winit}
\end{equation}
\end{corollary}
\begin{proof}
  We repeatedly use \prref{prop:doors} to eliminate all \SSparc with $S, S' < \Rs$.
  This requires at most $2^{\abs{\Rs}} - 1$ applications.
  Using $\abs{\Rs} \geq 2$, \prref{eq:TTelias2} follows from
  \begin{equation*}
    \abs{\wt W} \leq \tfrac76(1 + 5\cdot (2^{\abs{\Rs}} - 1)\cdot 2^{\abs{\Rs}^2 + 3 \abs{\Rs}})\cdot \abs{\Winit} + 150\cdot 2^{\abs{\Rs}^2 + 3 \abs{\Rs}}\cdot \abs{\Winit}
    \qedhere
  \end{equation*}
\end{proof}
We now prove \prref{thm:uro} and begin by fixing the ambient (resource) alphabets $(\wh C,\rho)$ and $\wh \cX$. 
The transducer mentioned \prref{thm:uro}~can output
the names for $(\wh A,\rho)$ in the space bound because $A$ is part of the input 
and $|\wh A| \leq |A|^{|\Rs|} \cdot 2^{|\Rs|}$. Thus, $\Oh(|\Rs|\cdot n)$
bits suffice to describe a letter in $\wh A$.
Since we can recycle names in $\wh C\sm \wh A$ we do not need
more than $2^{\abs{\Rs}}\cdot 2^{\abs{\Rs}^2 + 7\abs{\Rs}}\cdot\abs{\Winit}$ names for $\wh C\sm \wh A$ based on (\ref{eq:TTelias2}). 
Similarly, \eqref{eq:remfixwhcX} in 
\prref{rem:fixwhcX} tells us that we can define the ambient resource alphabet of variables $(\wh \cX,\rho)$ by fixing 
$\abs{\wh \cX}= 7\cdot 2^{\abs{\Rs}^2 + 2\abs{\Rs}}$.
The description of the type $\theta$ in \prref{def:type} consists 
of specifying $a,b\in \wh C \cup \os{1}$ and 
a subset of $\wh \Gam= \wh C\cup \wh \cX$. Each relation in $\theta$
uses exactly six symbols from $\wh \Gam$. So, we can ignore the 
necessary space to specify $\theta$.

Consider a state $E=(W,B,\cX,\rho,\mu,\theta,T)\in \cU$. For each 
$x\in B\cup\cX$ we have to specify $\mu(x)$ and following a \tra 
we need to do various computations in a finite monoid (with \invol)
which is admissible. The statement of \prref{thm:uro} accounts for the space 
we need to do this.   
We use the above assumptions on $\wh C$ and $\wh \cX$ as restrictions on the tuple $(B,\cX,\rho,\mu,\theta)$; and we add another restriction 
on $W$: we only allow states such that $|W|\leq 2^{\abs{\Rs}^2 + 7 \abs{\Rs}}\cdot \abs{\Winit}$ because this  bound is justified by \prref{cor:doors}. 
Using this corollary we define a finite sub automaton $\cB$ of $\cU$ as follows.
\begin{enumerate}
\item The state set $V(\cB)$ consists of all states  
$E=(W,B,\cX,\rho,\mu,\theta,T)\in \cU$ which satisfy 
$(B,\rho)\sse (\wh C,\rho)$, $\cX\sse \wh \cX$, and
$|W|\leq 2^{\abs{\Rs}^2 + 7\abs{\Rs}}\cdot\abs{\Winit}$.
\item The \tra set $E(\cB)$ consists of all \tras in $\cU$ which are incident with two different state in  $V(\cB)$.
  In particular, $\cB$ is without loops.
\end{enumerate}
Since $\cU$ is sound, the NFA $\cB$ is sound, too. \prref{cor:doors} shows that 
the NFA $\cB$ is complete. This shows \prref{eq:EQh}
in \prref{thm:uro}, which is a key part of statement. 
The theorem also says that there are infinitely many \solu{s} \IFF the NFA $\cA$ we aim for is without cycles. 
The NFA $\cB$ is not good enough to show this.
For example, if the initial system as a single \inner equation $Xab=aXc$ with $b\neq c$, then there are no \solu{s} but~$\cB$ has a cycle reachable from the initial state.
To solve this problem we use the subautomaton $\cA$ of $\cB$ obtained by trimming.

More precisely we define
$V(\cA)$ to be the set of states $E=(W,B,\cX,\rho,\mu,\theta,T)\in V(\cB)$ which are on some accepting path from $\cSinit$ to $\cent$ in $\cB$. 
The construction of $\cA$ can be carried out on the fly while producing its set of \tras as follows.
One after another we consider the \tras in $\cB$.
For each such \tra $E\arc h E'$ in $E(\cB)$ we check nondeterministically whether there is path in $\cB$ from $\cSinit$ to 
$E$ and a path from $E'$ to a semi-final state. If both answers are positive, then we output $E\arc h E'$ as a \tra in $\cA$, otherwise we discard the \tra. 

Since $\cB$ is sound and complete, the NFA $\cA$ is sound and complete.
Since nondeterministic space is closed under complementation by \cite{Immerman88siam,Szelepcsenyi88ai}, (c.f.~\cite{pap94}), the construction of $\cA$ is effectively possible in the required space complexity. 
It remains to show that $\cSinit$ has infinitely many 
\solu{s} \IFF $\cA$ contains a cycle. 
 
If there are infinitely many \solu{s}, then for every $m$ there are infinitely 
many initial  \solu{s} $\sig_{\init,m}$ with $|\sig_{\init,m}(\Winit)|>m$.
Since $\cA$ is finite there is a state~$E$ with an equation $W$ and path $\pi$ {}from~$E$ to~$E$ in $\cA$ with infinitely many \esolu{s} $(\alp,\sig)$.
This cycle cannot be induced by an $\eps$-loop in $\cU$ {}from~$E$ to~$E$ because if $(\alp,\sig)$ is an \esolu at $E$, then $|\alp\sig(W)|$ is not changed by this loop. So the cycle is a cycle in $\cA$.

For the other direction: let $\pi$ be a path in $\cA$ {}from a $T$-state $E=(W,B,\cX,\rho,\mu,\theta,T)$ to~$E$ in $\cA$. If the path was a nonempty path of \epstra{s}, 
then it would make $W$ longer by splitting of variables.
Hence, $\pi$ is a path which is labeled by some nontrivial 
\Endo  $h\in \End(B,\rho)$. We have 
$|h(c)|\geq 1$ for all $c\in B$. Moreover, by definition of 
a \comp \tra there is some 
$c\in B$ with $|W|_c\geq 1$ such that either $|h(c)|\geq 2$ or $|\rho(c)|> |\rho(h(c))|$
(or both). 
Since $\cA$ is trim, we can use \prref{thm:backstage}. The state is reachable from the initial state by some path labeled by $g\in \End(\wh C,\rho)$ and from~$E$ there is a path to the final state labeled by some $f\in \End(\wh C,\rho)$ such that \[
  \wh \pi g h^m f(\cent)= \wh \pi g h^m f(\Winit)= \wh \pi g h^m f(\ov{\Winit})
\]for all $m\in \N$.
Replacing $\pi$ by the path $\pi^{k}$ with $k=2^{\abs \Rs}$ we see that $|\rho(c)|> |\rho(h(c))|$ is impossible. Thus, we have 
$|h(c)|\geq 1$ for all $c\in B$ and $|h(c)|\geq 2$ for some 
$c\in B$ with $|W|_c\geq 1$. This implies 
$|\wh \pi g\alp h^{m+1}f (\Winit)|>|\wh \pi g\alp h^{m}f (\Winit)|$
for all $m\geq 1$.
Therefore there are infinitely many initial \solu{s} for $\cSinit$.
The proof of \prref{thm:uro} is complete.

\section{Conclusion and outlook}\label{sec:concl}

The main result in our paper is stated for solving systems of equations but using \prref{lem:inequ} it is straightforward to see by standard methods (and therefore left to the reader) that the existential theories of free partially commutative monoids and groups are decidable.
Our decidability for the existential theory generalizes a line of results in the following list: \cite{mak77,mak83a,schulz90,pla99focs,dmm99tcs,dgh05IC,dm06,Jez16jacm_stacs,DiekertJP16,CiobanuDiekertElder2016ijac}.
Unfortunately, although our decidability result is more general than in the papers occurring in this list we were not able to find way of showing a reduction of our main result to, say, Theorem~A in Paper~B. 
One reason is that most of the papers in the list did not aim to describe the full solution set.
The articles \cite{Jez16jacm_stacs,DiekertJP16,CiobanuDiekertElder2016ijac} do, but just for free monoids or free groups, and the term \edtol appears in this context for the first time in \cite{CiobanuDiekertElder2016ijac}.
The only paper dealing with partially commutative groups in the above list is \cite{dm06}, but it neither speaks about full solution sets nor does it allow for self-involuting letters. 
Including such letters led to many subtle difficulties in obtaining the results of this article, but in the end, it is a natural choice when dealing with monoids with involution.

In our opinion, it is regrettable that this paper reached a length of approximately~$80$ pages.
But there is also good news. Our proof is purely combinatorial and it does not rely on any deep theorem. Moreover, having read our 80~pages,\footnote{Remember the Makanin's seminal paper \cite{mak77} already has more than $80$~pages.} you get the important 
results of all the above papers for free!\footnote{One paper to rule them all, to bring them all, and in the lightness bind them?}

A natural next step is to generalize the results here to graph product over finitely many monoids $M_i$
where the full solution set for equations is \edtol over each $M_i$.
For the existential theory this was done in \cite{DiekertLohrey08}. 
Following the strategy in that paper, a reduction for graph products to the results of this paper seems within reach.
Another major open problem is the decidability of the elementary theory in free partially commutative groups (or more generally \wrt~taking graph products); for the state of the art in this line of research see \cite{CasalsK2011}. 
Regarding the elementary equivalence for graph products there has been significant recent progress in \cite{CassalsKN2024DissMath}.
So, this is not the end of this tale.\footnote{
`\emph{Non ignorabimus}' said Hilbert.} 

\printindex

\bibliographystyle{abbrv}
\bibliography{traces}
\end{document}